\documentclass{aa}  

\usepackage[utf8]{inputenc}
\usepackage{graphicx}
\usepackage{float}
\usepackage{multirow}
\usepackage{subfigure}
\usepackage[dvipsnames]{xcolor}
\usepackage{pdflscape}
\usepackage{afterpage}
\usepackage{upgreek}
\usepackage{amsmath}
\bibpunct{(}{)}{;}{a}{}{,}
\usepackage{aas_macros}
\usepackage{hyperref}

\newcommand\micron{\rm{\upmu m}}

\def\farcs{\hbox{$.\!\!^{\prime\prime}$}}

\newcommand{\neii}{\,\hbox{[\ion{Ne}{ii}]}}
\newcommand{\neiii}{\,\hbox{[\ion{Ne}{iii}]}}
\newcommand{\nev}{\,\hbox{[\ion{Ne}{v}]}}
\newcommand{\sitwo}{\,\hbox{[\ion{Si}{ii}]}}
\newcommand{\siv}{\,\hbox{[\ion{S}{iv}]}}
\newcommand{\siii}{\,\hbox{[\ion{S}{iii}]}}
\newcommand{\oiv}{\,\hbox{[\ion{O}{iv}]}}
\newcommand{\oi}{\,\hbox{[\ion{O}{i}]}}

\newcommand{\oiii}{\,\hbox{[\ion{O}{iii}]}}
\newcommand{\niii}{\,\hbox{[\ion{N}{iii}]}}
\newcommand{\nii}{\,\hbox{[\ion{N}{ii}]}}
\newcommand{\cii}{\,\hbox{[\ion{C}{ii}]}}

\newcommand{\feii}{\,\hbox{[\ion{Fe}{ii}]}}
\newcommand{\nevi}{\,\hbox{[\ion{Ne}{vi}]}}

\usepackage[flushleft]{threeparttable}
\usepackage{booktabs}

\usepackage[T1]{fontenc}
\usepackage{ae,aecompl}
\usepackage{newtxtext}
\usepackage[frenchmath,varg]{newtxmath}

\begin{document}

   \title{Calibration of mid- to far-infrared spectral lines in galaxies}

   \author{Sabrina Mordini\inst{1,2}\fnmsep\thanks{E-mail: \textsf{\href{mailto:sabrina.mordini@uniroma1.it}{sabrina.mordini@uniroma1.it}}}
          \and
          Luigi Spinoglio\inst{2}\fnmsep\thanks{E-mail: \textsf{\href{mailto:luigi.spinoglio@inaf.it}{luigi.spinoglio@inaf.it}}}
          \and
          Juan Antonio Fern\'andez-Ontiveros\inst{2}\fnmsep\thanks{E-mail: \textsf{\href{mailto:j.a.fernandez.ontiveros@gmail.com}{j.a.fernandez.ontiveros@gmail.com}}}
          }

   \institute{Dipartimento di Fisica, Universit\`a di Roma La Sapienza, P.le A. Moro 2, I--00185 Roma, Italy
         \and
             Istituto di Astrofisica e Planetologia Spaziali (INAF--IAPS), Via Fosso del Cavaliere 100, I--00133, Roma, Italy
             }

%   \date{Received yyy; accepted xxx}

% \abstract{}{}{}{}{} 
% 5 {} token are mandatory
 
  \abstract
  % context heading (optional)
  % {} leave it empty if necessary  
   {Mid- to far-infrared (IR) lines are suited to study dust obscured regions in galaxies, because dust extinction is strongly decreasing with wavelength and therefore IR spectroscopy allows us to explore the most hidden regions of galaxies, where heavily obscured star formation as well as accretion onto supermassive black-holes at the nuclei of galaxies occur. This is mostly important for the so-called ``Cosmic Noon'', i.e. at redshifts of 1$<$z$<$3, when most of the baryonic mass in galaxies has been assembled.}
  % aims heading (mandatory)
   {Our goal is to provide reliable calibrations of the mid- to far-IR ionic fine structure lines, the brightest H$_2$ pure rotational lines and the Polycyclic Aromatic Hydrocarbons (PAHs) features, that will be used to analyse current and future observations in the mm/submm range from the ground, as well as mid-IR spectroscopy from the upcoming \textit{James Webb Space Telescope}. }
  % methods heading (mandatory)
   {We use three samples of galaxies observed in the local Universe: star forming galaxies (196), active galactic nuclei (AGN; 90--150 for various observables) and low-metallicity dwarf galaxies (40). For each population we derive different calibrations of the observed line luminosities versus the total IR luminosities.}
  % results heading (mandatory)
   {Through the resulting calibrations, we derive spectroscopic measurements of the Star Formation Rate (SFR) and of the Black Hole Accretion Rate (BHAR) in galaxies using mid- and far-IR fine structure lines, H$_2$ pure rotational lines and PAH features. In particular, we derive robust star-formation tracers based on: the [CII]158$\, \rm{\micron}$ line; the sum of the two far-IR oxygen lines, the [OI]63$\mu$m line and the [OIII]88$\mu$m line;  a combination of the neon and sulfur mid-IR lines; the bright PAH features at 6.2 and 11.3$\, \rm{\micron}$, as well as -- for the first time -- the H$_2$ rotational lines at $9.7$, $12.3$ and $17\, \rm{\micron}$. 
   We propose the [CII]158$\mu$m line, the combination of the two neon lines ([NeII]12.8$\mu$m and [NeIII]15.5$\mu$m) and, for solar-like metallicity galaxies that may harbour an AGN, the PAH11.3$\mu$m feature as the best SFR tracers.
   On the other hand, a reliable measure of the BHAR can be obtained using the [OIV]25.9$\, \rm{\micron}$ and the [NeV]14.3 and 24.3$\, \rm{\micron}$ lines. For the most commonly observed fine-structure lines in the far-IR we compare our calibration with the existing ALMA observations of high redshift galaxies. We find an overall good agreement for the [CII]158$\, \rm{\micron}$ line for both AGN and star forming galaxies, while the [OIII]88$\, \rm{\micron}$ line in high-z galaxies is in better agreement with the low metallicity local galaxies (dwarf galaxies sample) than with the star forming galaxies, suggesting that high-z galaxies might have strong radiation fields due to low metal abundances, as expected.}
  % conclusions heading (optional), leave it empty if necessary 
   {}

   \maketitle
%
%-------------------------------------------------------------------

\section{Introduction}

   \label{sec:intro}

One of the major and still unsolved problems in astrophysics is related to the lack of a clear understanding of how galaxies evolve, from the time of structure formation till today, and what are the dominant processes that have influenced their evolution \citep{somerville2015, naab2017, bullock2017, Wechsler2018, foster-s2020}.
From the observational scenario that has been consolidated in the last tens of years, we know that the two main energy production mechanisms along galaxy evolution are, on one side, star formation and subsequent stellar evolution and, on the other side, accretion onto the super-massive black holes (SMBHs) that form at the centre of galaxies. 

The bulk of star formation and BH accretion took place at the so-called \textit{Cosmic Noon} ($1 < z < 3$), with a steep decline toward the present epoch, in heavily obscured environments, embedded in large amounts of gas and dust, where optical and UV detected radiation corresponds to only $\sim 10\%$ of the total emitted light \citep{madau2014} and with most of the radiated energy being absorbed by dust and re-emitted at longer wavelengths. 
Both star formation and black hole accretion contribute to the dust re-emission, and therefore measuring the integrated IR continuum, through photometric observations, does not discriminate easily between both components. The spatial resolution required to isolate the bulk of the nuclear IR emission from the host galaxy ($\lesssim 100\, \rm{pc}$) cannot be attained in high-$z$ galaxies, while spectral decomposition techniques \citep[e.g.,][]{berta2013} are ultimately dependent on templates of local active galactic nuclei (AGN), whose inner workings and dust distribution are still far from being understood (e.g. \citealt{lyu2020}).
Thus measuring star formation rate (SFR) and black hole accretion rate (BHAR) in galaxies across cosmic time, which is one of the major observational goals in galaxy evolution studies, has to be done through spectroscopic observations at wavelengths long enough to overcome dust absorption.

The mid to far-IR range is populated by a large number of atomic and molecular lines and features. In particular, the atomic and ionic fine-structure lines cover a wide range of physical parameters in terms of excitation, density and ionisation, as can be seen in fig.\,4 of \citet{spinoglio2012}, showing the critical density for collisional de-excitation versus the ionisation potential of the IR fine-structure lines. These lines can easily discriminate among the different gas excitation conditions, from the Narrow Line Regions (NLR) excited by the active nucleus, to the \textsc{Hii} and photo-dissociation regions (PDRs), whose origin is due to stellar excitation \citep[see, e.g.,][]{spinoglio1992, tommasin2010, spinoglio2015, fernandez2016}, offering an ideal tool to probe the highly opaque and dust obscured regions. Most of these lines can be assessed only from space IR telescopes, while only a few transitions happen to lie in the atmospheric windows, e.g. in the spectral region of $8$--$13\, \rm{\micron}$.

The intermediate ionisation fine-structure lines are good tracers of  the star formation and in particular the sum of the fluxes of [NeII] and [NeIII] has been shown to give a measure of the Star Formation Rate \citep{ho2007,zhuang2019} in galaxies. The low ionisation [CII] line at 158$\, \rm{\micron}$ is one of the brightest emission lines in star forming galaxies, and can thus trace star formation activity both in the local \citep{delooze2014, herrera2015} and high redshift \citep{leung2020} Universe.

The mid-IR spectra of star forming galaxies between $3$ and $19\, \rm{\micron}$ are often dominated by emission features attributed to Polycyclic Aromatic Hydrocarbons (PAHs), due to infrared (IR) fluorescence by these large molecules containing 50-100 C-atoms, pumped by single FUV photons \citep{allamandola1989,smith2007,tielens2008}. Because they trace the FUV stellar flux, they can be used to measure star formation. These features are strong when compared to fine structure lines in star forming galaxies, and are detected not only in local galaxies, but also at redshifts up to z$\sim$4 \citep{kirkpatrick2015,riechers2014,sajina2012}. Their emission accounts for $\sim 10-20\%$ of the total IR radiation by dust, and originate from PDRs near star forming regions. PAHs can be used to trace SFR not only in star forming galaxies \citep{shipley2016,xie2019}, but also in sources where an AGN contribution is present but not the dominant source of integrated light \citep{shipley2013}. In these AGN, the equivalent width of the PAH features can be used to estimate the star formation contribution to the total IR luminosity \citep{armus2007,tommasin2010}. However, in more extreme AGN, strong UV and X-ray radiation fields can also  suppress the PAH emission in the vicinity of the AGN by photo-dissociation, while increasing the mid-IR continuum emission \citep{lacy2013}. Analogously, also in Low Metallicity Galaxies (LMG, i.e. below 12 + log(O/H)$\sim$8.2), because of the reduced formation efficiency and the increased stellar radiation hardness, the strength of the PAH features is reduced and only detected at metallicities above 1/8-1/10 Z$_{\odot}$  \citep{engelbracht2008, cormier2015,galliano2021}.

Among the molecular lines in the mid-IR, the pure rotational transitions of H$_{2}$ are particularly important as the typical physical conditions of the gas associated with these lines can be found both in AGN and star forming galaxies. \citet{rigopoulou2002} found that, while in star forming environments the H$_{2}$ emission can originate from PDRs, with a small contribution from shocks, in AGN dominated galaxies the X-ray emission from the central AGN plays an important role in heating large amounts of gas boosting the H$_{2}$ emission. 

High ionisation lines trace AGN activity: all lines whose ionisation potential is higher than the one necessary to doubly ionise helium ($> 54.4\, \rm{eV}$) cannot be efficiently produced by stellar radiation in significant amounts. Therefore, the detection of the [NeV]14.3$\, \rm{\micron}$, [NeV]24.3 $\, \rm{\micron}$ or [NeVI]7.65$\, \rm{\micron}$ lines probes the presence of an AGN, while the [OIV]25.9$\, \rm{\micron}$ line, which can also be found in energetic starbursts and LMG, is however much stronger in AGN, having an equivalent width one order of magnitude larger compared to Star Forming Galaxies (SFG)  \citep{tommasin2010}. 
Typical IR line ratios used to measure the strength of the active nucleus with respect to the star formation component in a galaxy are: the [NeV]14.3$\, \rm{\micron}$ or 24.3$\, \rm{\micron}$ to [NeII]12.8$\, \rm{\micron}$ ratio, the [OIV]25.9$\, \rm{\micron}$/[NeII]12.8$\, \rm{\micron}$ line ratio and the [OIV]25.9$\, \rm{\micron}$/[OIII]52$\, \rm{\micron}$ or 88.35$\, \rm{\micron}$ ratio \citep{sturm2002,armus2007,tommasin2010,spinoglio2015,fernandez2016}.

The main goals of this work are: (i) to revise the calibration
of the mid- to far-IR lines including ionic fine-structure lines, the brightest H$_2$ pure rotational lines and the PAH features and (ii) to provide a local calibration of spectroscopic SFR and BHAR tracers, that can also be applied to  measurements at high-\textit{z}. We will follow the study presented in \citet{spinoglio2012}, update the IR spectroscopic observations presented there and include a sample of LMG, to extend the calibration to these objects. These latter are included to characterise the response of the lines to the conditions of low metallicity (average $1/5\, \rm{Z_\odot}$)  \citep{madden2013}. 
Low-metallicity AGN are not included because they are rare in the local Universe and only a few examples are found \citep[e.g. Circinus, ][]{oliva1999}, but the broad- and narrow-line regions in AGN show essentially little or no chemical evolution up to $z \sim 7$ \citep{nagao2006,juarez2009,onoue2020}, suggesting that the quasar phase appears mostly when galaxies are already chemically mature objects. Therefore including the metallicity dependence in the calibrations for AGN is not as relevant as in the case of star forming galaxies, where the chemical evolution with redshift is well known \citep{sanders2020}. This work extends and update a previous study \citep{spinoglio2021} aimed at the preparation of the SPICA mission \citep{roelfsema2018} where the authors give a calibration of the most important features that were used to plan spectroscopic observations with that mission.

A further motivation of this study resides in the need to exploit the great potential of extragalactic IR spectroscopy for high redshift galaxies, which is being already explored by the Atacama Large Millimeter/submillimeter Array \citep[ALMA,][]{wootten2009, carpenter2020}, and will have a dramatic boost with the next space IR telescopes, such as the \textit{James Webb Space Telescope} \citep[\textit{JWST},][]{gardner2006} and in a more far future, possibly by the
\href{https://asd.gsfc.nasa.gov/firs/docs/OriginsAPCwhitepaperassubmitted10July2019.pdf}{\it Origins Space Telescope}\footnote{\url{https://asd.gsfc.nasa.gov/firs}}.

The paper is organized as follows: section \ref{sec:data} describes the samples of galaxies, observed in the local Universe, that we have used to derive the correlations; section \ref{sec:results} reports our results, in particular: section \ref{sec:cor} presents the new correlations between the line luminosities and the total IR luminosities, while sections \ref{sec:SFR} and \ref{sec:BHAR} give simple recipes to measure the two main parameters of the SFR and BHAR. In section \ref{sec:discussion} we discuss our results: section \ref{sec:comp} compares our study to previous ones, in section \ref{sec:metallicity_discussion} we discuss the metallicity effect on the SFR tracers, section \ref{sec:high-z_data} presents how the observations at high redshift compare with the correlations we have derived and section \ref{sec:jwst_alma} shows how our results can be used to interpret present and future IR/(sub)-mm observations. Our conclusions are presented in section \ref{sec:conclusion}.
%--------------------------------------------------------------------
\section{The selected lines and features and the samples of galaxies}\label{sec:data}

We use the most representative samples of star forming galaxies (SFG), AGN, and low metallicity galaxies (LMG) in the local Universe for which IR spectroscopy is available, mainly from \textit{Spitzer}-IRS \citep{houck2004} and \textit{Herschel}-PACS \citep{pacs2010}, but also from the \textit{Infrared Space Observatory} (\textit{ISO}) SWS and LWS spectrometers \citep{degraauw1996,clegg1996}. For each of the three galaxy populations, we derive linear relations in logarithmic space between the line luminosity and the total IR luminosity. Then we derive the best tracers of the SFR and the BHAR, using the discussed IR lines and features, and compare them with what has been reported in the literature

  In this analysis we considered, in order of decreasing ionisation/excitation:
\begin{itemize}
    \item Four high-ionisation fine structure lines, typical of AGN: [NeVI]7.65$\, \rm{\micron}$, [NeV]14.32$\, \rm{\micron}$, [NeV]24.32$\, \rm{\micron}$ and [OIV]25.89$\, \rm{\micron}$;
    \item Ten intermediate ionisation fine structure lines, typical of stellar/\textsc{Hii} regions: [SIV]10.51$\, \rm{\micron}$, [NeII]12.81$\, \rm{\micron}$, [NeIII]15.55$\, \rm{\micron}$, [SIII]18.71$\, \rm{\micron}$, [SIII]33.48$\, \rm{\micron}$, [OIII]51.81$\, \rm{\micron}$, [NIII]57.32$\, \rm{\micron}$, [OIII]88.36$\, \rm{\micron}$, [NII]121.9$\, \rm{\micron}$ and [NII]205$\, \rm{\micron}$;
    \item Five low-ionisation/neutral fine structure lines, typical of PDR: [FeII]25.99$\, \rm{\micron}$, [SiII]34.81$\, \rm{\micron}$, [OI]63.18$\, \rm{\micron}$, [OI]145.5$\, \rm{\micron}$ and [CII]157.7$\, \rm{\micron}$;
    \item Four H$_{2}$ pure rotational lines at 9.67, 12.28, 17.03 and 28.22 $\, \rm{\micron}$;
    \item Five PAH features at 6.2, 7.7, 8.6, 11.3 and 17$\, \rm{\micron}$.
\end{itemize}

The fundamental parameters for the fine-structure lines considered in this analysis are reported in Table \ref{tbl_lines}.

In this analysis we have included those lines observed with the {\it Spitzer}-IRS high resolution (HR) channel in the 10-35$\mu$m range, for which good spectra are available in the literature. The only exception is the line of [NeVI]7.65$\mu$m, an exclusive AGN line,  for which we included ISO-SWS observations. Other well known fine-structure lines, such as [ArII]6.98$\mu$m, [ArIII]8.99$\mu$m and [NeIII]36.0$\mu$m, which could indeed have a relevant role for future observations, especially in view of the \textit{JWST} launch, at the present time do not have enough high quality spectra in the literature to be included in the analysis. We do not consider upper limits to derive our correlations. This is because, in general, our statistics are quantitatively appropriate, thus making the inclusion of upper limits not necessary. Moreover, where the statistics are less precise, upper limits are usually not available for the considered lines.

The definition of the various samples of galaxies chosen to compute the correlations are described in the following sections and summarised in Table\,\ref{tab:sample_references}, with the instruments used to observe the spectral lines and features, the total number of objects selected and the references for each sample.

\begin{table*}[hbtp]
\caption{Characteristics of the different samples used in this analysis. For each class of objects we report, for the different lines and features, the original sample (\textit{heterogeneous} indicates that the sources do not belong to a specific well-defined sample of galaxies), the facility and instrument that observed the sample of galaxies, the number of objects used in the analysis and the references where the original data can be found.}
\label{tab:sample_references}
\resizebox{\textwidth}{!}{
\begin{tabular}{clllcl}\hline 
\multicolumn{1}{l|}{    }&                &        &             & \multicolumn{1}{l}{}                &       \\
\multicolumn{1}{l|}{Type} & Lines/features & Sample & Observed by & \multicolumn{1}{l}{N. of objects} & References \\ \hline 
\multicolumn{1}{l|}{    }&                &        &             & \multicolumn{1}{l}{}                &       \\
\multicolumn{1}{c|}{\multirow{5}{*}{AGN}} & [NeVI]7.7$\mu$m & heterogeneous & ISO-SWS & 8 & \citet{sturm2002} \\
\multicolumn{1}{c|}{} & MIR lines in the 10-35$\mu$m range & \multirow{2}{*}{12MGS} & \multirow{2}{*}{{\it Spitzer}-IRS HR} & \multirow{2}{*}{88} & \multirow{2}{*}{\citet{tommasin2008,tommasin2010}} \\
\multicolumn{1}{c|}{} & including H$_{2}$ rotational lines &  &  &  &  \\
\multicolumn{1}{c|}{} & PAH features & 12MGS & {{\it Spitzer}-IRS LR} & 103 & \citet{wu2009} \\
\multicolumn{1}{c|}{} & FIR lines in the 50-205$\mu$m range & heterogeneous & {\it Herschel}-PACS and SPIRE & 149 & \citet{fernandez2016} \\ \hline 
\multicolumn{1}{l|}{    }&                &        &             & \multicolumn{1}{l}{}                &       \\
\multicolumn{1}{c|}{\multirow{4}{*}{SFG}} & MIR lines in the 10-35$\mu$m range & \begin{tabular}[c]{@{}l@{}}GOALS sample\\ + heterogeneous\\ + heterogeneous\end{tabular} & {{\it Spitzer}-IRS HR} & \begin{tabular}[c]{@{}c@{}}153\\ 10\\ 28\end{tabular} & \begin{tabular}[c]{@{}l@{}}\citet{inami2013}\\ \citet{bernardsalas2009}\\ \citet{goulding2009}\end{tabular} \\
\multicolumn{1}{c|}{} & H$_{2}$ rotational lines & \begin{tabular}[c]{@{}l@{}}GOALS sample\\ + heterogeneous\\ + heterogeneous\end{tabular} & {{\it Spitzer}-IRS HR} & \begin{tabular}[c]{@{}c@{}}153\\ 10\\ 28\end{tabular} & \begin{tabular}[c]{@{}l@{}}\citet{stierwalt2014}\\ \citet{bernardsalas2009}\\ \citet{goulding2009}\end{tabular} \\
\multicolumn{1}{c|}{} & PAH features & \begin{tabular}[c]{@{}l@{}}GOALS sample\\ + heterogeneous\end{tabular} & {{\it Spitzer}-IRS LR} & \begin{tabular}[c]{@{}c@{}}179\\ 12\end{tabular} & \begin{tabular}[c]{@{}l@{}}\citet{stierwalt2014}\\ \citet{brandl2006}\end{tabular} \\
\multicolumn{1}{c|}{} & FIR lines in the 50-205$\mu$m range & \begin{tabular}[c]{@{}l@{}}GOALS sample\\ + heterogeneous\\ + heterogeneous\end{tabular} & \begin{tabular}[c]{@{}l@{}}{\it Herschel}-PACS \\ {\it Herschel}-PACS and SPIRE\\ {ISO-LWS}\end{tabular} & \begin{tabular}[c]{@{}c@{}}153\\ 20\\ 23\end{tabular} & \begin{tabular}[c]{@{}l@{}}\citet{diazsantos2017}\\ \citet{fernandez2016}\\  \citet{negishi2001}\end{tabular} \\ \hline 
\multicolumn{1}{l|}{    }&                &        &             & \multicolumn{1}{l}{}                &       \\
\multicolumn{1}{c|}{\multirow{2}{*}{LMG}} & MIR lines in the 10-35$\mu$m range & DGS & {{\it Spitzer}-IRS HR and LR} & 40 & \citet{cormier2015} \\
\multicolumn{1}{c|}{} & FIR lines in the 50-158$\mu$m range & DGS & {\it Herschel}-PACS & 40 & \citet{cormier2015}
\\ \hline \\[0.02cm]
\end{tabular}
}
\end{table*}

\subsection{The AGN sample}
The AGN sample has been drawn from the $12\, \rm{\micron}$ selected active galaxies sample \citep[12MGS,][]{rms1993}, which is the brightest complete and unbissed sample of Seyfert galaxies in the local Universe. For the mid-IR fine structure lines and the H$_2$ rotational lines, we have used the sub-sample of the 12MGS observed by \textit{Spitzer} IRS at high spectral resolution (R$=600$) which contains 88 AGN \citep{tommasin2008,tommasin2010}. For the PAH features at 6.2 $\mu$m and 11.2 $\mu$m we have used the \textit{Spitzer}-IRS data at low spectral resolution \citep[R$\sim60$--$120$;][]{buchanan2006, wu2009} of the 12MGS (103 objects), because this setting matches better the intrinsic width of these features. We note here that \citeauthor{wu2009} measure the PAH features using a spline function to determine the continuum level. In order to make these measurements comparable to those obtained using automated fitting procedures \citep[e.g. PAHFIT,][]{smith2007}, we apply a correction factor of $1.7$ and $1.9$ to increase the fluxes of the $6.2\, \rm{\mu m}$ and $11.3\, \rm{\mu m}$ PAH in \citeauthor{wu2009}, respectively, following the differences found by \citet{smith2007}.

The adoption of the complete 12MGS as the main catalog, from which we cover about 75\% of the total sample, allows us to derive statistically robust calibrations.

For the [NeVI]7.65$\, \rm{\micron}$ line, we could not use the 12MGS, because this line was not detected by \textit{Spitzer} at low resolution and was outside its spectral range at high-resolution. Therefore we had to use the data from \citet{sturm2002},  which contain 8 detections of the [NeVI]7.65$\, \rm{\micron}$ line with \textit{ISO}-SWS at medium resolution (R $\sim$ 1500), and thus not contaminated by the PAH emission at 7.7 $\, \rm{\micron}$.

For the far-IR spectral range, namely the $50$--$205\, \rm{\micron}$, we used the catalogue of AGN assembled by \citet{fernandez2016}, which includes all the Seyfert galaxies and the quasars of the \citet{veron2010} catalogue which have far-IR spectra observed by \textit{Herschel}-PACS. The sample of the AGN observed in the far-IR lines counts 170 galaxies and contains about 50\% of objects from the 12MGS, while the others do not come from a complete sample.

\subsection{The SFG sample}

The SFG sample was constructed using the \textit{The Great Observatories All-Sky LIRG Survey} \citep[GOALS sample,][]{armus2009}, from which we extracted 158 galaxies, with data from \citet{inami2013}, who report the fine structure lines at high resolution in the $10$--$36\, \rm{\micron}$ interval, and \citet{stierwalt2014}, who include the detections of the H$_{2}$ molecular lines and the PAH features at low spectral resolution. For those galaxies in the GOALS sample that have a single IRAS counterpart, but more than one source detected in the emission lines, we have added together the line/feature fluxes of all components, to consistently associate the correct line/feature emission to the total IR luminosity computed from the IRAS fluxes. To cover also lower luminosity galaxies, as the GOALS sample only includes Luminous IR Galaxies (LIRGs) and Ultra-Luminous IR Galaxies (ULIRGs), we have included 38 galaxies from \citet{bernardsalas2009} and \citet{goulding2009}, to reach the total sample of 196 galaxies with IR line fluxes in the $5.5$--$35\, \rm{\micron}$ interval in which an AGN component is not detected. For the \citet{bernardsalas2009}, \citet{goulding2009} and the GOALS samples we excluded all the composite starburst-AGN objects identified as those with a detection of [NeV] either at $14.3$ or $24.3\, \rm{\micron}$. It is worth to note that the original samples from \citet{goulding2009} and \citet{bernardsalas2009} have spectra covering solely the central region of the galaxies. To estimate the global SFR, we corrected the published line fluxes of the  \textit{Spitzer} spectra 
by multiplying them by the ratio of the continuum reported in the IRAS Point Source Catalogue to the continuum measured on the \textit{Spitzer} spectra extracted from the CASSIS database \citep{lebouteiller2015}. We assume here that the line emission scales, at first order, with the IR brightness distribution.
In particular, we considered the continuum at $12\, \rm{\micron}$ for the [NeII]$12.8\, \rm{\micron}$ and [NeIII]$15.6\, \rm{\micron}$ lines, and the continuum at 25 $\, \rm{\micron}$ for the [OIV]$25.9\, \rm{\micron}$, [FeII]$26\, \rm{\micron}$, [SIII]$33.5\, \rm{\micron}$ and [SiII]$34.8\, \rm{\micron}$ lines. This correction was not needed for the AGN sample and the GOALS sample because of the greater average redshift of the galaxies in the 12MGS and GOALS samples. In particular, the 12MGS active galaxies sample has a mean redshift of 0.028 \citep{rms1993}, while the GOALS sample has a mean redshift of 0.026. The galaxies presented by \citet{bernardsalas2009} have instead an average redshift of 0.0074, while the sample by \citet{goulding2009} has an average redshift of 0.0044. For the other lines in the $10$--$36\, \rm{\micron}$ interval, \citet{goulding2009} do not report a detection, and we used the data presented in \citet{bernardsalas2009} for a total of 15 objects. Both \citet{bernardsalas2009} and \citet{goulding2009} report data from the high-resolution \textit{Spitzer}-IRS spectra. Data in the $50$--$205\, \rm{\micron}$ interval were taken from \citet{diazsantos2017} for the GOALS sample, 20 starburst galaxies from \citet{fernandez2016}, and 23 objects from the ISO-LWS observations of \citet{negishi2001}, resulting in a total sample of 193 objects. Lastly, the PAH features fluxes were measured from the low resolution \textit{Spitzer}-IRS spectra by \citet{brandl2006}, including 12 objects from the sample of \citet{bernardsalas2009} and 179 objects from \citet{stierwalt2014}.

\subsection{The LMG sample}
The LMG sample was selected from \citet{cormier2015}, where the $10$--$36\, \rm{\micron}$ interval was observed by \textit{Spitzer}-IRS in both high and low resolution, and the $50$--$158\, \rm{\micron}$ interval was observed by \textit{Herschel}-PACS. For the \textit{Spitzer}-IRS data, we only considered the high resolution results, for a total sample of 40 objects.

\section{Results}\label{sec:results}

%In the following Section 
We derive, for each line or feature in each galaxy sample, the correlation between the logarithms of the total IR luminosity in the $8$--$1000\, \rm{\micron}$ range --\,computed from the IRAS fluxes following \citet{sanders1996}\,-- and the line luminosity, according to the equation:
\begin{equation}
\log L_{\rm Line} = (a \pm \delta a) \log L_{IR} + (b \pm \delta b)
\end{equation}
with all luminosities expressed in units of $10^{41}\, \rm{erg\,s^{-1}}$. %This approach takes into account not only an overall factor between the IR and the line luminosities, but also a possible non-linear dependency between these two variables (see Section\,\ref{sec:cor}). 
We report in Table\,\ref{tab:calibrations} of the Appendix D the best-fit parameters obtained for each line/feature using the orthogonal distance regression fit \citep{boggs1990}, the number of objects $N$, and the Pearson correlation coefficient $r$. We use the orthogonal distance regression because the two variables are independent each other, instead of the ordinary least-square minimisation, where one variable is dependent from the other one. This is also particularly useful to derive the inverse relation between the two variables from the best-fit coefficients.

Using these correlations, we derive the tracers for the SFR (Section\,\ref{sec:SFR}) and the BHAR (Section\,\ref{sec:BHAR}). For the SFR, we used the [CII]158$\, \rm{\micron}$ luminosity, various combinations of the luminosities of the [NeII]12.8$\, \rm{\micron}$, [NeIII]15.6$\, \rm{\micron}$, [SIII]18.7$\, \rm{\micron}$ and [SIV]10.5$\, \rm{\micron}$ lines, the luminosity of the PAH features at 6.2$\, \rm{\micron}$ and 11.3$\, \rm{\micron}$ and the luminosity of the H$_2$ rotational lines at 9.7$\, \rm{\micron}$, 12.3$\, \rm{\micron}$ and 17.3$\, \rm{\micron}$. For the BHAR, we have used the luminosities of the [OIV]25.9$\, \rm{\micron}$, [NeV]14.3$\, \rm{\micron}$ and 24.3$\, \rm{\micron}$ lines.

\subsection{Spectral lines and features vs. total IR luminosity}\label{sec:cor}

\begin{figure*}
\centering
\includegraphics[width=0.66\columnwidth]{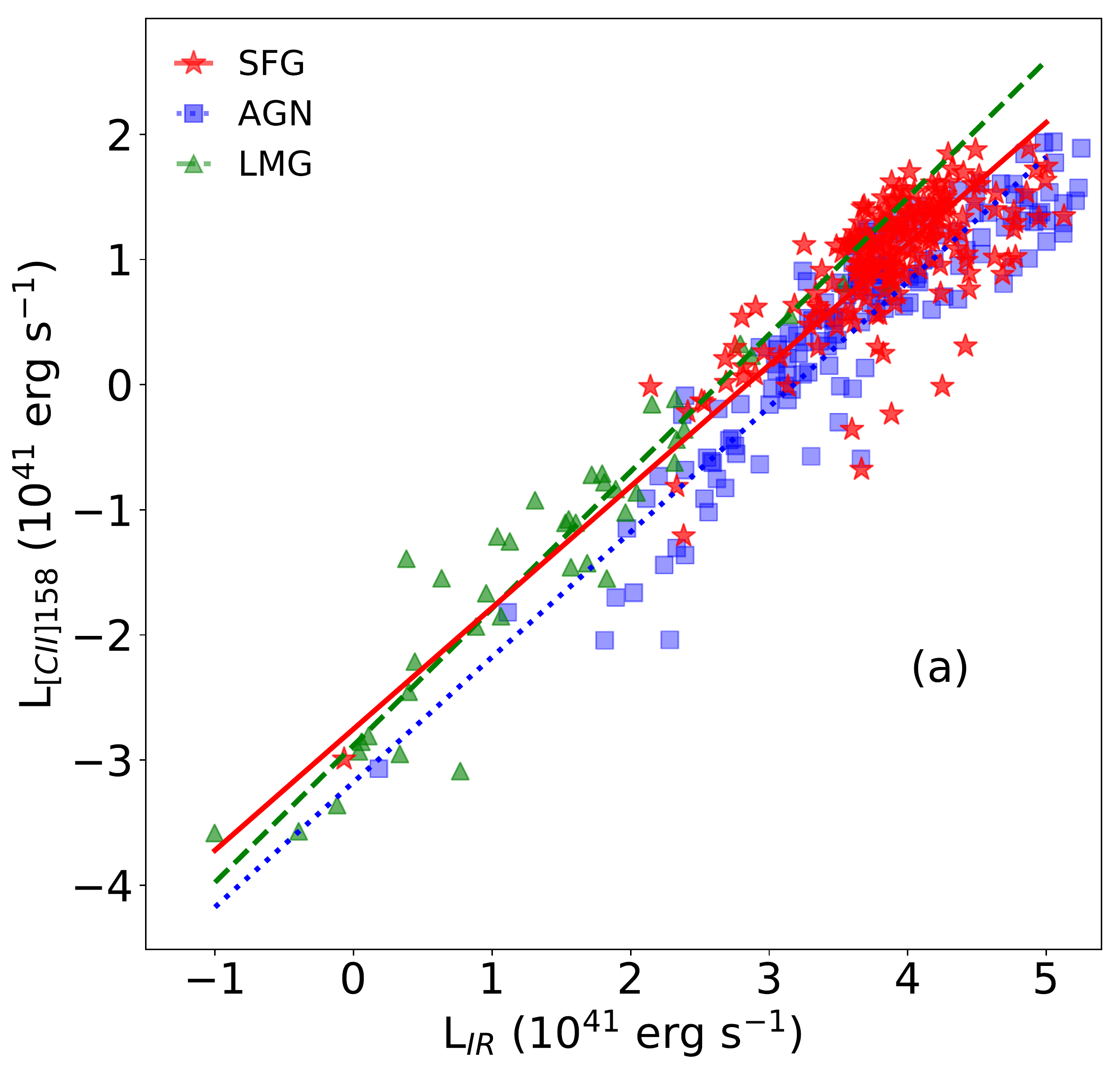}~
\includegraphics[width=0.66\columnwidth]{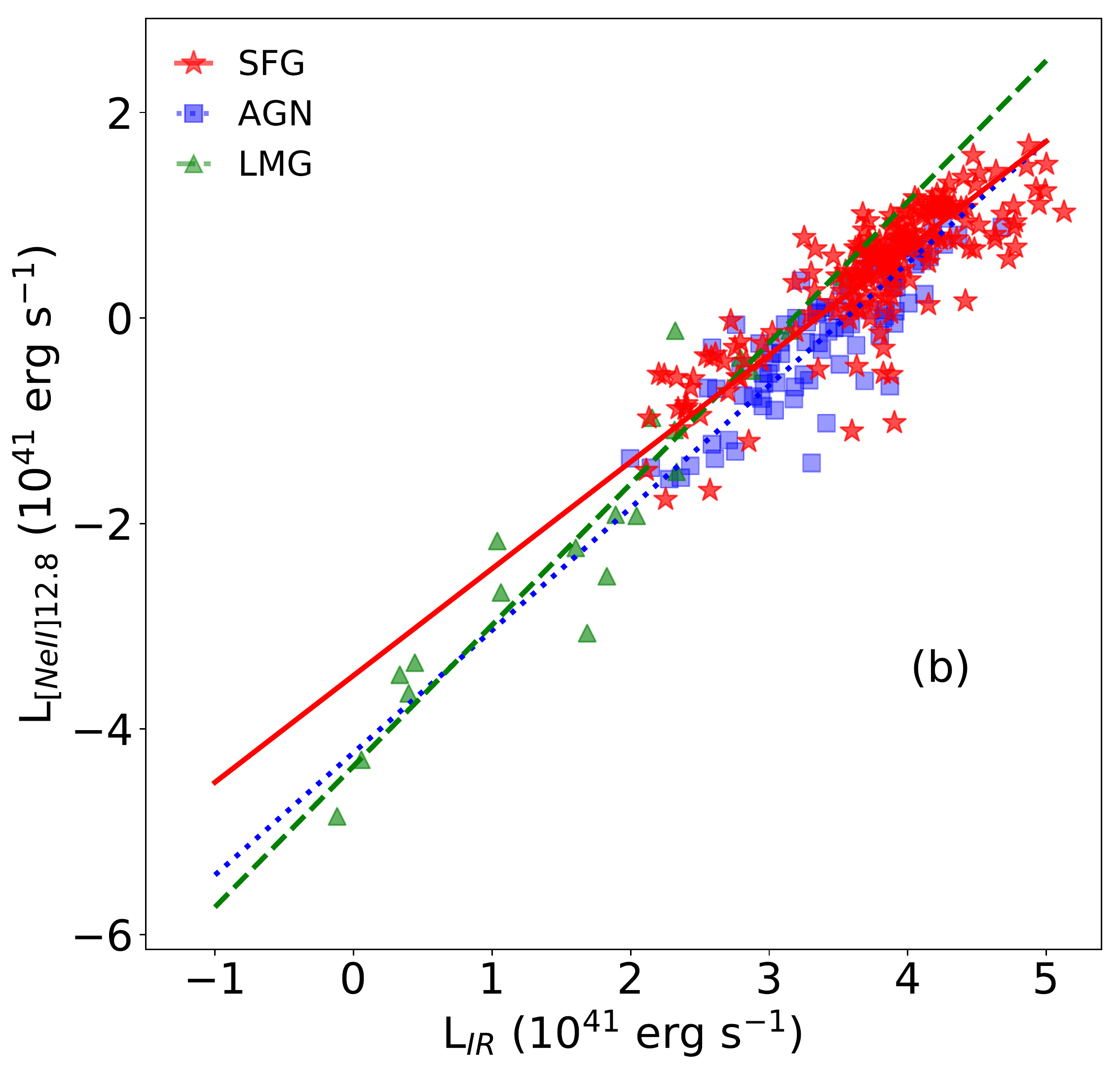}
\includegraphics[width=0.66\columnwidth]{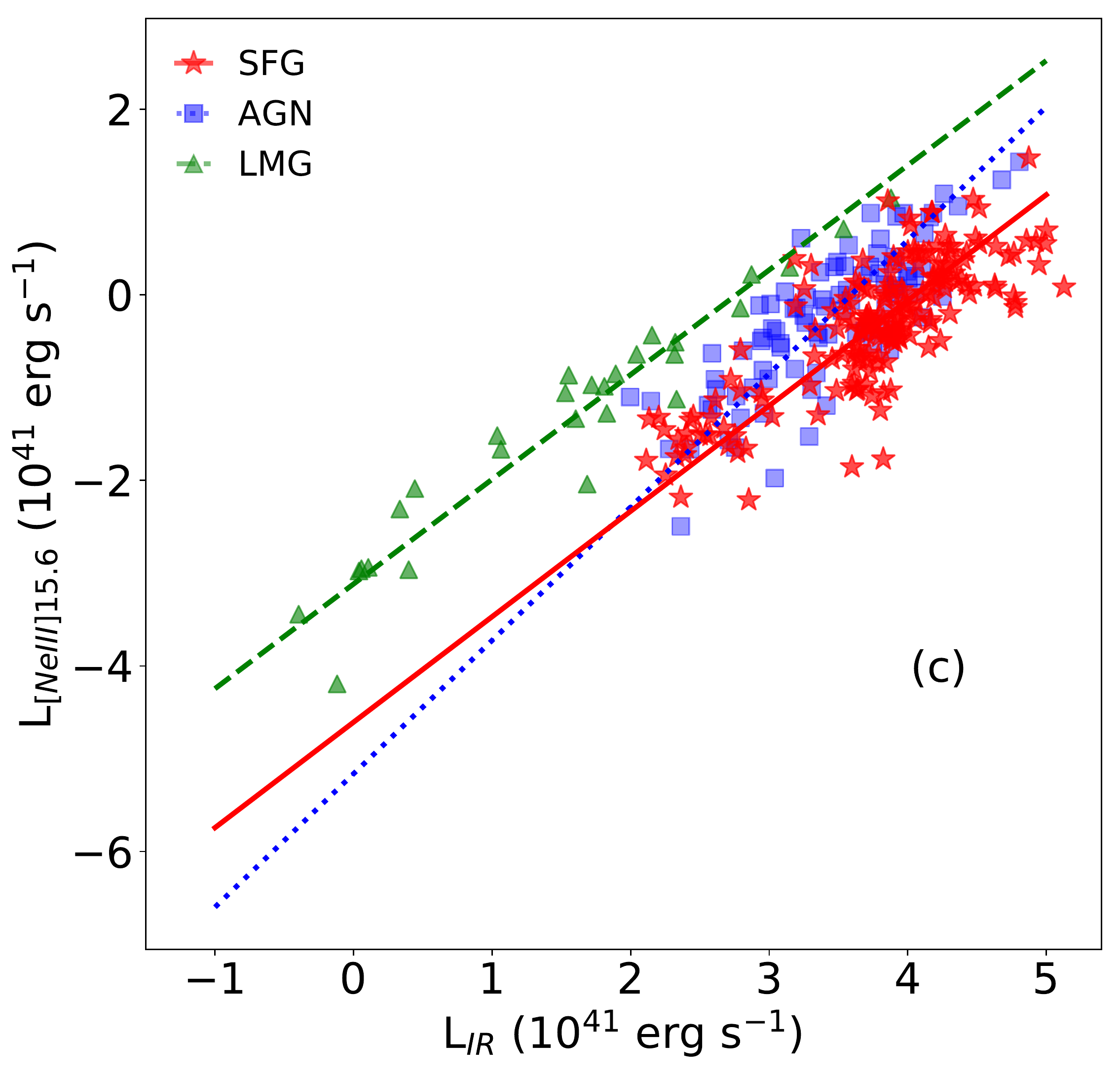}

\caption{{\bf (a: left)} The [CII]158$\, \rm{\micron}$ line luminosity as a function of the total IR luminosity. Blue squares represent detections in AGN, red stars indicate SFG and green triangles LMG. The solid red line represents the linear relation calculated for SFG, the blue dotted line shows the relation for AGN and the green dashed line the one for LMG. {\bf (b: centre)} The [NeII]12.8$\, \rm{\micron}$ line luminosity as a function of the total IR luminosity. {\bf (c: right)} The [NeIII]15.6$\, \rm{\micron}$ line luminosity as a function of the total IR luminosity. In {\bf (b)} and {\bf (c)} we use the same notations as in {\bf (a)}.}
\label{fig:corr_c2_ne2_ne3}
\end{figure*}

\begin{figure*}%[h!]
\centering
\includegraphics[width=0.66\columnwidth]{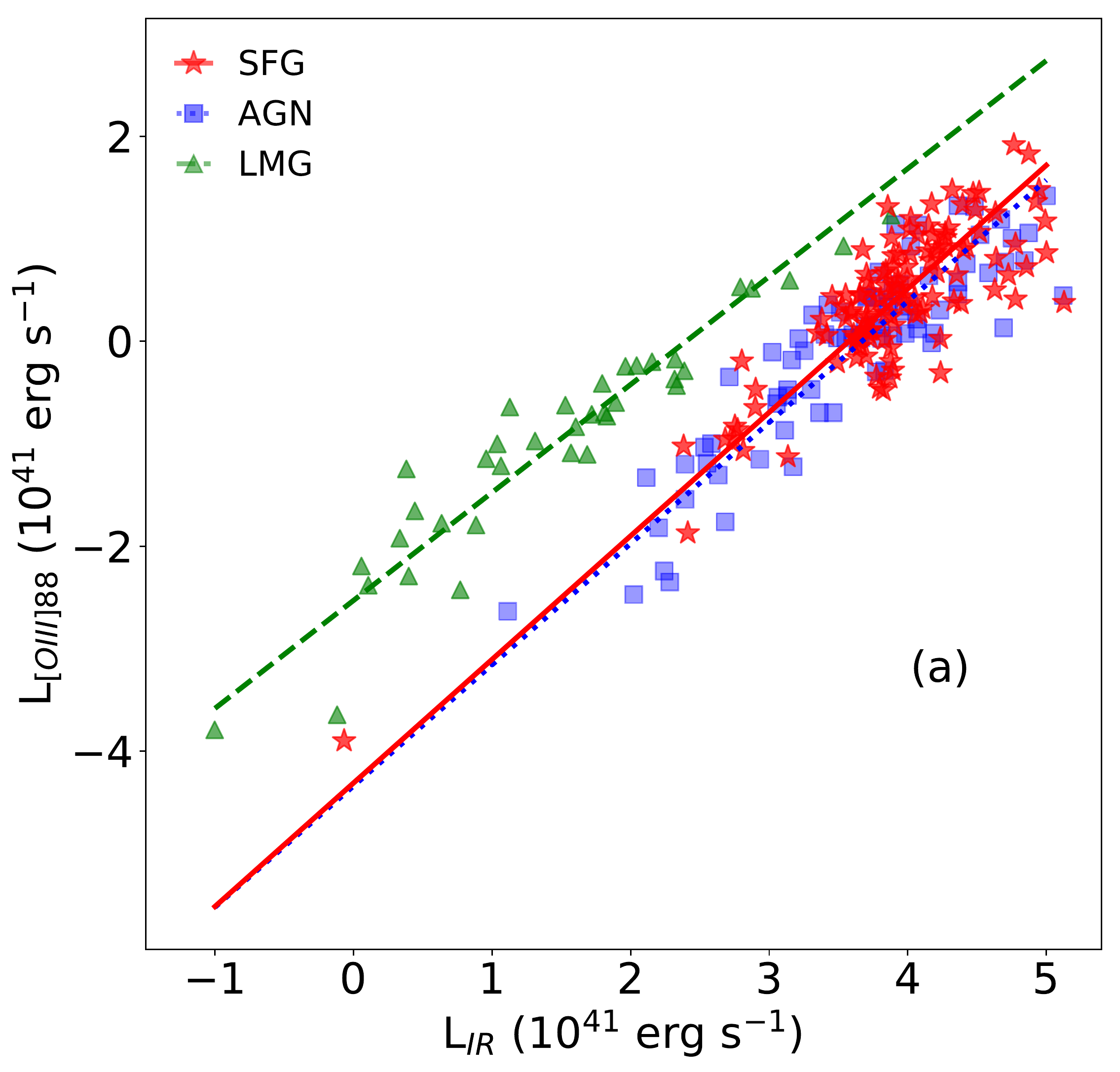}~
\includegraphics[width=0.66\columnwidth]{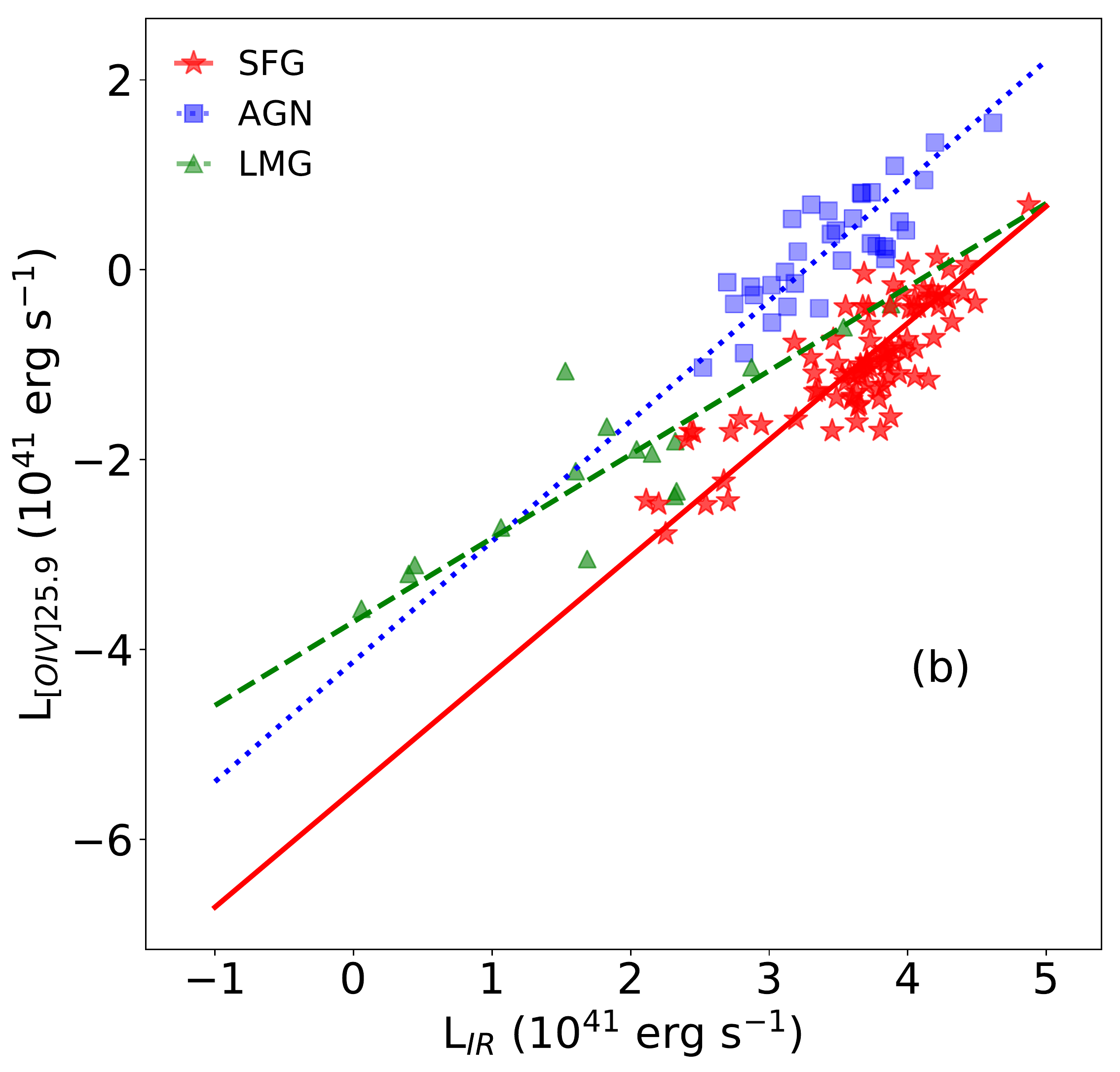}
\includegraphics[width=0.66\columnwidth]{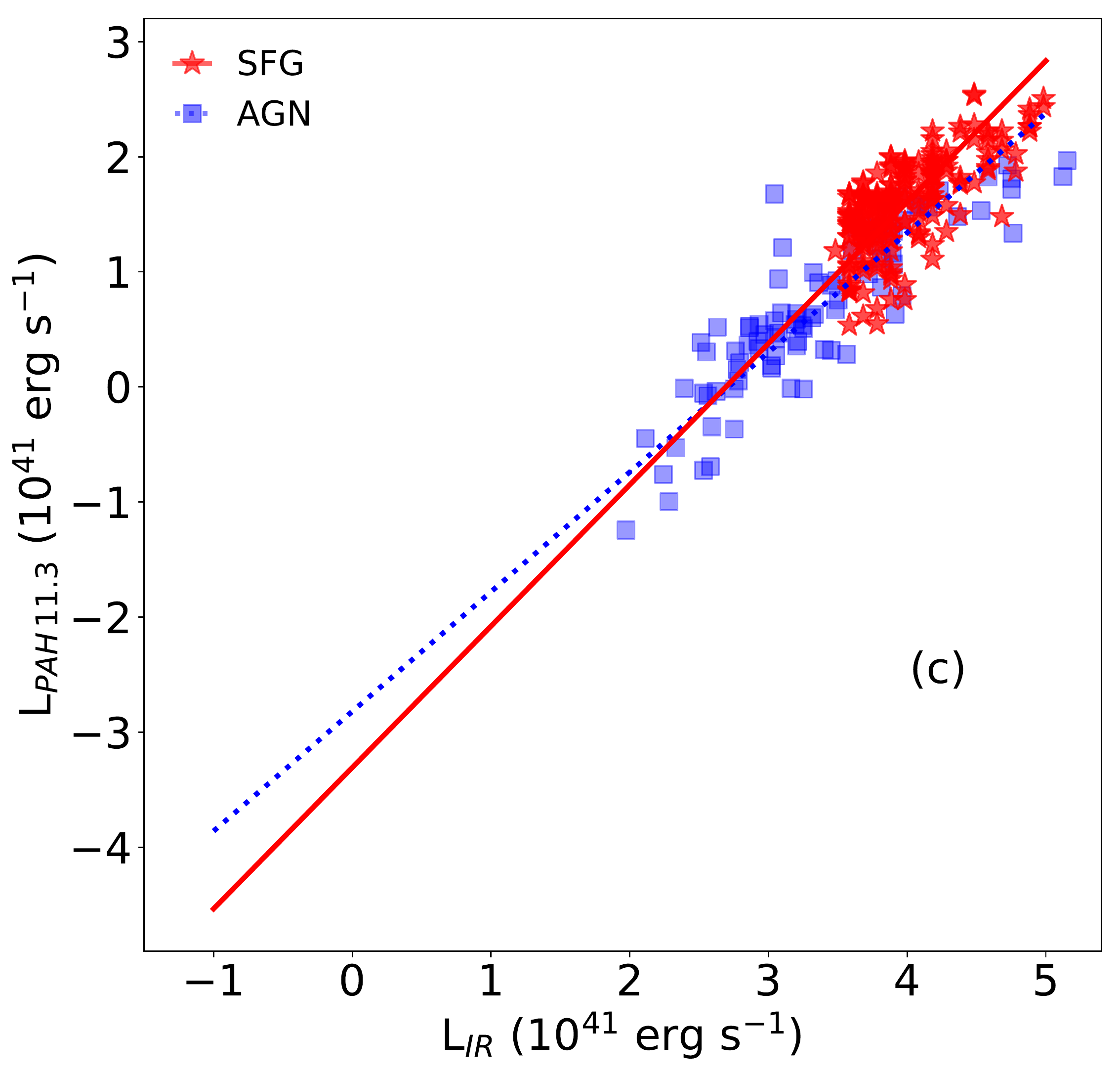}

\caption{{\bf (a: left)} The [OIII]88$\, \rm{\micron}$ line luminosity as a function of the total IR luminosity. {\bf (b: centre)} The [OIV]25.9$\, \rm{\micron}$ line luminosity as a function of the total IR luminosity. {\bf (c: right)} The luminosity of the PAH feature at 11.3 $\, \rm{\micron}$ as a function of the total IR luminosity. The same legend as in Fig.\,\ref{fig:corr_c2_ne2_ne3} was used.} 
\label{fig:corr_ne3_o3}
\end{figure*}

Among the different correlations derived in this work and presented in Table\,\ref{tab:calibrations} of the Appendix D, we highlight in this section the main results, while we refer the reader to Appendix\,\ref{sec:app} for the whole set of figures illustrating the correlations with the total IR luminosity. 

In Fig.\,\ref{fig:corr_c2_ne2_ne3}a, we present the correlation between the [CII]158$\, \rm{\micron}$ line luminosity and the total IR luminosity for the three galaxy populations considered. We find that the SFG and the LMG samples follow a tight relation over six orders of magnitude in L$_{IR}$ with a consistent regression slope and offset values within the uncertainties, from low-luminosity dwarf galaxies to LIRGs and ULIRGs, confirming the results of \citet{delooze2014}. This will be further discussed in Section\,\ref{sec:SFR-CII}. On the other hand, AGN have higher IR luminosities for a given [CII]158$\, \rm{\micron}$ line luminosity with respect to the SFG and LMG samples, likely due to the contribution from the AGN continuum emission to the total IR luminosity.This effect is also seen in low ionisation transitions where the AGN contribution to the line emission is expected to be small, such as the [NII]122$\mu$m line (see Fig.\,\ref{fig:corr_c2_ne2_ne3}c). However, the same effect is not observed for [OI]63$\mu$m (see Fig.\,\ref{fig:corr_app_7}b). This might be related to the higher critical density of the [OI]63$\mu$m line, which becomes an efficient coolant in X-ray dissociation regions in the presence of an AGN \citep{maloney1996,dale2004}. High-excitation lines where the AGN completely dominates the line emission show the opposite behaviour, that is, the sources are shifted towards higher line intensities at a given IR luminosity (e.g. [OIV]25.9$\mu$m in Fig.\,\ref{fig:corr_ne3_o3}b). This suggests that the lack of a noticeable shift between AGN and SFGs in intermediate excitation lines such as [NeII]12.8$\mu$m or [SIII]18.7,33.5$\mu$m (Figs.\,\ref{fig:corr_c2_ne2_ne3}b, \ref{fig:corr_app_4}b and \ref{fig:corr_app_6}a) might be caused by a comparable AGN contribution to both the IR luminosity and the line intensity.

We note that various authors \citep[e.g.][]{herrera2015,croxall2017} have reported an observed deficit in [CII] luminosity with the increase of total L$_{IR}$, in particular in ULIRGs. Different mechanisms have been proposed to explain the lower [CII] emission \citep[][and references therein]{sutter2019,sutter2021}. In particular, \citet{sutter2019} find that the [CII] deficit is particularly evident when the emission arises from ionised gas, while the effect is negligible when the emission comes predominantly from PDR regions.

%We obtain a similar difference based on galaxy type} 
For the [NeII]12.8 $\, \rm{\micron}$ line (Fig.\,\ref{fig:corr_c2_ne2_ne3}b) there is a slight difference between SFGs and the other samples, but the differences are not statistically relevant, being consistent within 3$\sigma$ of each other.

In Fig.\,\ref{fig:corr_c2_ne2_ne3}c, we report the correlation obtained for the [NeIII]15.6$\, \rm{\micron}$ line with the total IR luminosity, showing that the AGN  and the SFG have a comparable correlation, while LMG have the [NeIII]15.6$\, \rm{\micron}$ line more than one order of magnitude brighter, at a given IR luminosity \citep[see also ][]{cormier2012}.

In Fig. \ref{fig:corr_ne3_o3}a, b and c we report the correlations obtained for the  [OIII]88$\, \rm{\micron}$, the [OIV]25.9$\, \rm{\micron}$ line luminosities and the luminosity of the PAH feature at 11.3 $\, \rm{\micron}$ with the total IR luminosity, respectively.
For the [OIII]88$\, \rm{\micron}$ line, LMG are on average two orders of magnitude brighter, at a given IR luminosity \citep[see also ][]{cormier2012}, compared to the other two classes of galaxies, that are almost overlapping. This could be due to the differences in the ionising spectra and conditions in the ISM of LMG with respect to the SFG and the AGN. In LMG, an increased number of photons can ionise gas at greater distances from the star forming regions, thus facilitating cooling through ionised gas emission. Additionally, the lower dust-to-gas ratios in LMG is also expected to cause a decrease of the IR luminosities, at a fixed stellar mass, in these galaxies.

AGN have about one order of magnitude brighter [OIV]25.9$\, \rm{\micron}$ emission compared to SFG \citep[see also ][]{tommasin2010}, LMG have a shallower relation with IR luminosity, with a decreasing slope at higher luminosities (Fig.\,\ref{fig:corr_ne3_o3}b). 
%Even though the LMG sample presents a higher L$_{IR}$ to L$_{[OIV]}$ relation than the SFG sample, we note that for the LMG we have a significant lower number of objects, thus a less statistically significant result. 
The flatter slope could be related to the high ionisation potential of the [OIV]25.9$\mu$m line, that is, the few ionising photons beyond 54.4 eV produced by LMGs might not scale linearly with the overall luminosity. However, no firm conclusion can be drawn because of the relatively poor statistics of LMG, for which only 16 detections of the [OIV] line are available. 
We note here that in order to obtain a reliable correlation for the AGN sample, we considered only objects with an  AGN component at the 19$\, \rm{\micron}$ continuum, as defined in  \citet{tommasin2010}, greater than 85$\%$. We apply this limit only in this case, to minimise a possible contamination in this line from emission due to strong starburst activity \citep[see, e.g.][]{lutz1998}. This approach is motivated by the use of the [OIV]25.9$\mu$m line as a BHAR tracer in Sect.\,\ref{sec:BHAR}, where the same reduced sample is adopted.

%Fig.\,\ref{fig:corr_ne3_o3}c shows that the 11.3$\, \rm{\micron}$ PAH feature is present both in AGN and SFG and well correlates with the total IR luminosity, while it is absent in LMG. 
Fig.\,\ref{fig:corr_ne3_o3}c shows that the 11.3$\, \rm{\micron}$ PAH feature is present both in AGN and SFG and correlates well with the total IR luminosity, while in this analysis we do not consider the PAH detections in LMG, because detections are available only in a few cases, and thus are not enough to obtain a statistically significant result. 
As shown by different authors \citep{madden2000, engelbracht2005, wu2006, smith2007, calzetti2007}, in LMG there is evidence for a deficit of PAH emission and the available measurements present significantly weaker features than SFG. Moreover, the higher ionising continuum present in LMG contribute to destroy these features \citep{engelbracht2008,cormier2015}.

As will be shown in Section\,\ref{sec:pah_sfr}, while there is a difference of $\sim$0.3 dex between the emission of SFG and AGN, the slopes of the two correlations are comparable within the errors, with the difference linked to a higher L$_{IR}$ for equal PAH emission in AGN.

We note here that, while we report in Tab.\,\ref{tab:calibrations} the correlation derived for the [NeVI]7.6$\, \rm{\micron}$ line, this correlation was obtained for a sample of only 8 AGN, while for the majority of AGN this line is not detected. We therefore conclude that this calibration has to be taken with caution, because it may be bissed toward AGN with a high ionisation parameter (U$\sim$-1), while for a lower ionisation we expect the [NeVI] line to be  considerably less prominent, as discussed in  \citet[][in particular in their Fig. 5]{satyapal2021}.

\subsection{Star formation rate tracers}\label{sec:SFR}

 \subsubsection{Determination of SFR}

In this section, we propose different SFR tracers. In order calibrate the proposed  SFR tracers, we use two different methods. For SFG, we measure the SFR directly from the total IR luminosity L$_{IR}$, following \citet{kennicutt1998}:
\begin{equation}
    SFR= k_{IR}*L_{IR}
\end{equation}
where $k_{IR}=4.5 \times 10^{-44}$M$_{\odot}$yr$^{-1}$erg$^{-1}$s.  We apply the same conversion also to AGN sources (see Sect.\,\ref{sec:pah_sfr}), but limiting our sample to the sources with L$_{IR}\leq 10^{45}$erg\,s$^{-1}$. We apply this limit to minimise the AGN effect on the L$_{IR}$ and avoid overestimating the SFR because of AGN activity.

For the LMG, we adopt the SFR derived from H$_{\alpha}$ and corrected by the total IR luminosity, as reported in \citet{remyruyer2015}. The total IR luminosity alone does not accurately represent the total SFR in LMG due to the lower dust-to-gas ratio in their ISM. The inclusion of the optical component for computing the SFR is necessary to properly account for the emission not reprocessed by dust.

\subsubsection{L\texorpdfstring{$_{[CII]158 \rm{\micron}}$}{micron}--SFR relation}\label{sec:SFR-CII}

%da sistemare usanto nuove figure e formule

The [CII]158 $\, \rm{\micron}$ line can be used as a tracer of the SFR, as proposed by different authors \citep[see, e.g.,][]{delooze2014}. We compare the line intensity of SFG and LMG to the SFR. Applying the orthogonal distance regression fit, we find a strong correlation between the two quantities:
\begin{multline}\label{eq:c2_sfr}
\log\left ( \frac{SFR}{\rm M_{\odot}\,yr^{-1}}\right )=(0.62 \pm 0.02)\\ +(0.89 \pm 0.02) \left( \log\frac{L_{[CII]}}{\rm 10^{41}\,erg\,s^{-1}}\right )
\end{multline}

\begin{figure*}
 \centering
 \includegraphics[width=0.66\columnwidth]{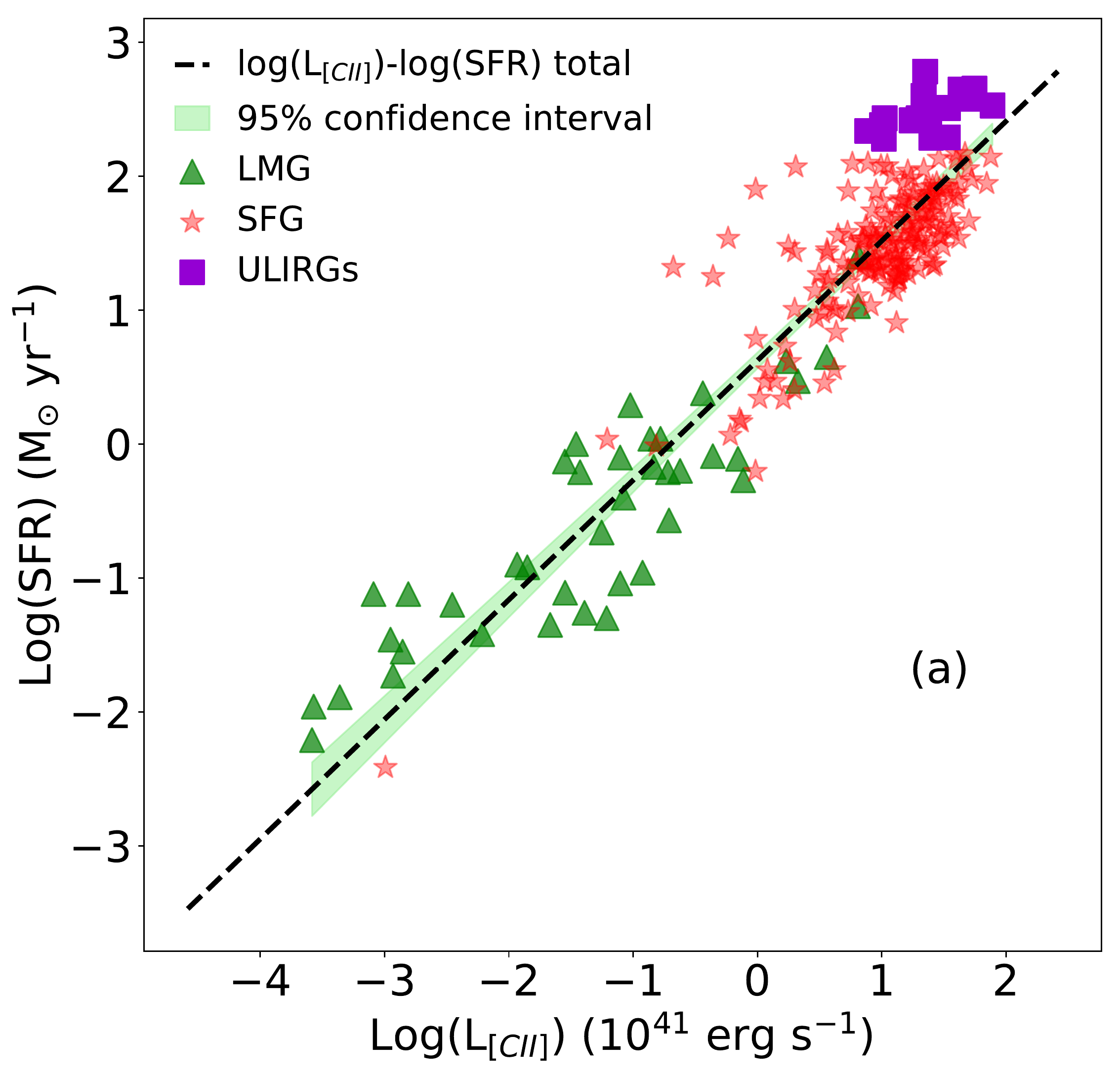}~
 \includegraphics[width=0.66\columnwidth]{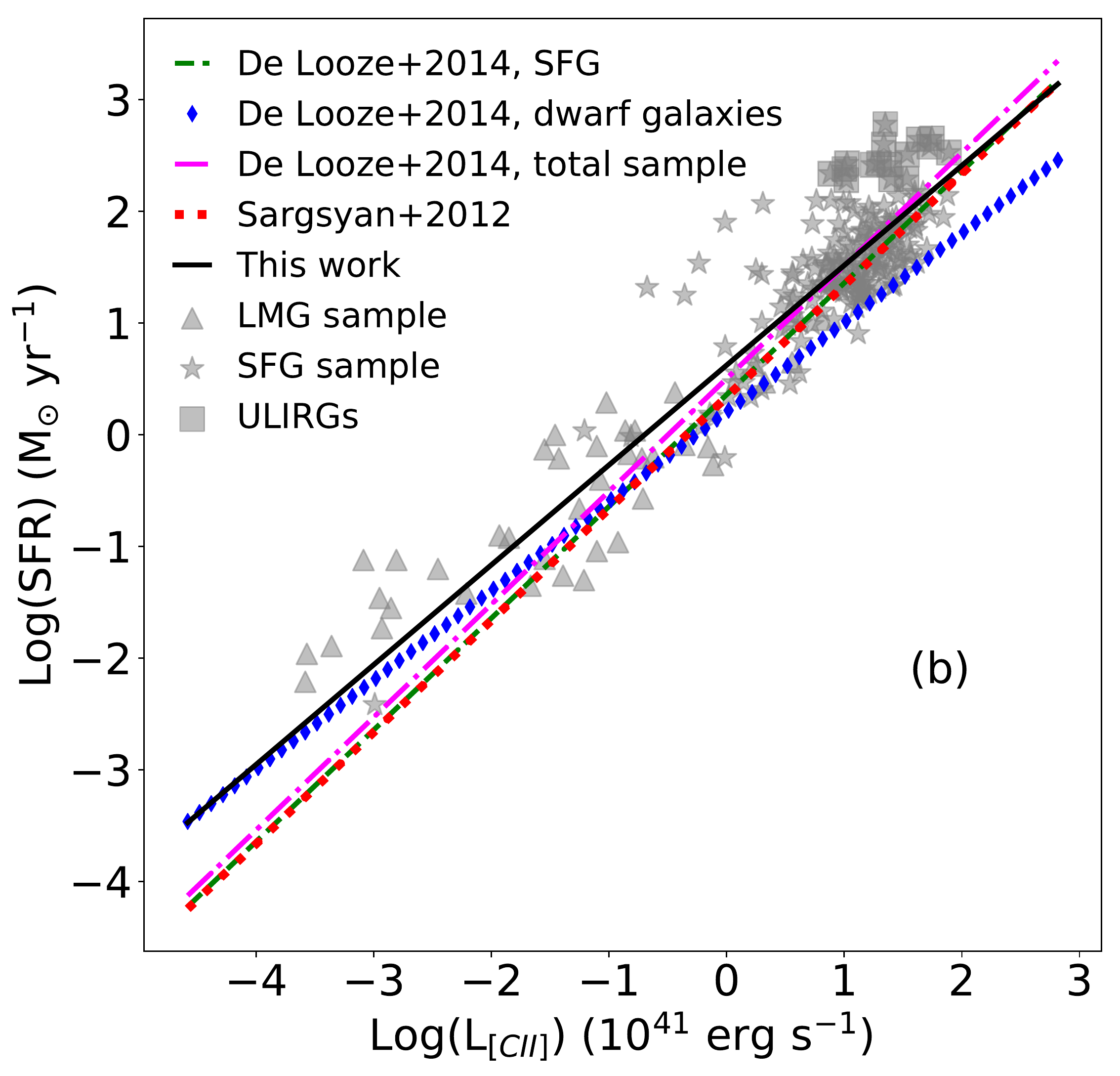}
 \includegraphics[width=0.66\columnwidth]{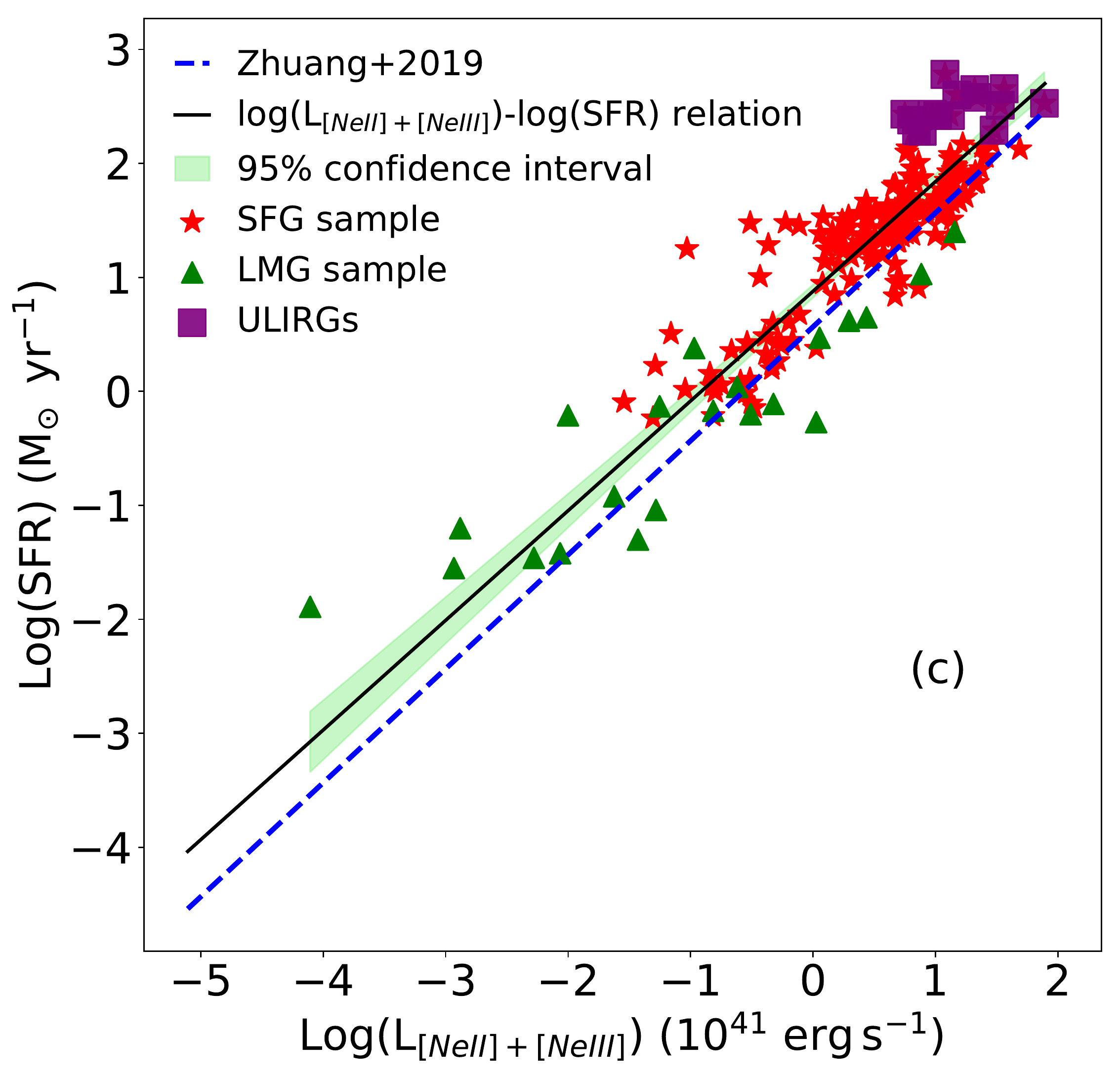}
 \caption{{\bf (a: left)} Correlation between the [CII]158 $\, \rm{\micron}$ line luminosity and the SFR derived from the total L$_{IR}$. Red stars represent SFG, green triangles the LMG. Purple squares show the ULIRG population of SFG, and the shaded green area indicates the 95$\%$ confidence interval.
 {\bf (b: centre)} Comparison of the logL$_{[CII]}$-log(SFR) relation obtained in this work (black solid line) with the results obtained by \citet{delooze2014}: the green dashed line represents a sample of HII/star forming galaxies, the blue diamond line shows the results for the low metallicity dwarf sample, and the pink dash-dotted line considers the whole sample. The red dotted line shows the results obtained by \citet{sargsyan2012}.
 {\bf (c: right)} Correlation between the [NeII]12.8$\, \rm{\micron}$ and [NeIII]15.6$\, \rm{\micron}$ summed emission lines luminosity, in units of $10^{41}\, \rm{erg\,s^{-1}}$ and the SFR derived from the total IR luminosity (black dashed line) for SFG (red star) and from the H$_{\alpha}$ luminosity (corrected for the IR luminosity) for LMG (green triangles). The shaded green area indicates the 95$\%$ confidence interval.  The purple squares highlight the ULIRG population in the SFG sample. The blue dashed line shows the results obtained by \citet{zhuang2019} for the same relation .
 }
 \label{fig:SFR1}

\end{figure*}

In Fig.\,\ref{fig:SFR1}a we show the correlation obtained for 227 galaxies, of which 37 are LMG and 190 are SFG. The relation is nearly linear, with a Pearson r-value of $r$ = 0.92, and covers six orders of magnitude in SFR, indicating that the [CII] line emission is an overall good tracer of SFR for local star forming galaxies independently of their metallicity. We excluded the AGN from this correlation, since we derive the SFR from the L$_{IR}$ and AGN have an excess IR continuum emission not due to star formation.

In Fig.\,\ref{fig:SFR1}b, we compare our results with those obtained in \citet{delooze2014} and \citet{sargsyan2012}. We note that, while in this work and in the work by \citeauthor{sargsyan2012} the SFR was determined following \citet{kennicutt1998}, \citeauthor{delooze2014} trace the SFR using the GALEX FUV emission \citep{cortese2012} and the {\it Spitzer}-MIPS 24$\, \rm{\micron}$ emission \citep{rieke2004}. When compared to the results obtained by \citet{delooze2014} for their total sample, we find good agreement at luminosities  of L$_{[CII]}>10^{41}$ erg\,s$^{-1}$, while there is a difference of $\sim$0.3 dex for lower luminosities.
In \citeauthor{delooze2014} the dwarf sample, when considered alone, shows a flatter slope than the total sample, equal to $0.80\pm 0.5$ (see Fig.\,\ref{fig:SFR1}b, the blue diamond line). The authors link the flatter slope to an underestimation of the SFR based on the far-UV emission. We observe a similar flattening of the slope when considering the LMG sample alone, equal to $0.69 \pm 0.05$, consistent with the result by \citeauthor{delooze2014} within 3$\sigma$ of each other.
% The metal-poor dwarf galaxies sample shows a flatter slope, due, according to these authors, to an underestimation of the SFR based on the far-UV emission

\citet{delooze2014} find that the ULIRG population presents a scatter of almost one order of magnitude in the [CII]-SFR relation, when compared to the total sample. We find the same result in our correlation, with the ULIRG population (composed of 16 objects) lying between 0.2 and 1.0 dex above the correlation derived for the total sample. Given the small number of sources, we do not derive a specific [CII]-SFR correlation for the ULIRG sample.

\subsubsection{Oxygen based SFR tracer}

\begin{figure}
    \centering
    \includegraphics[width=0.85\columnwidth]{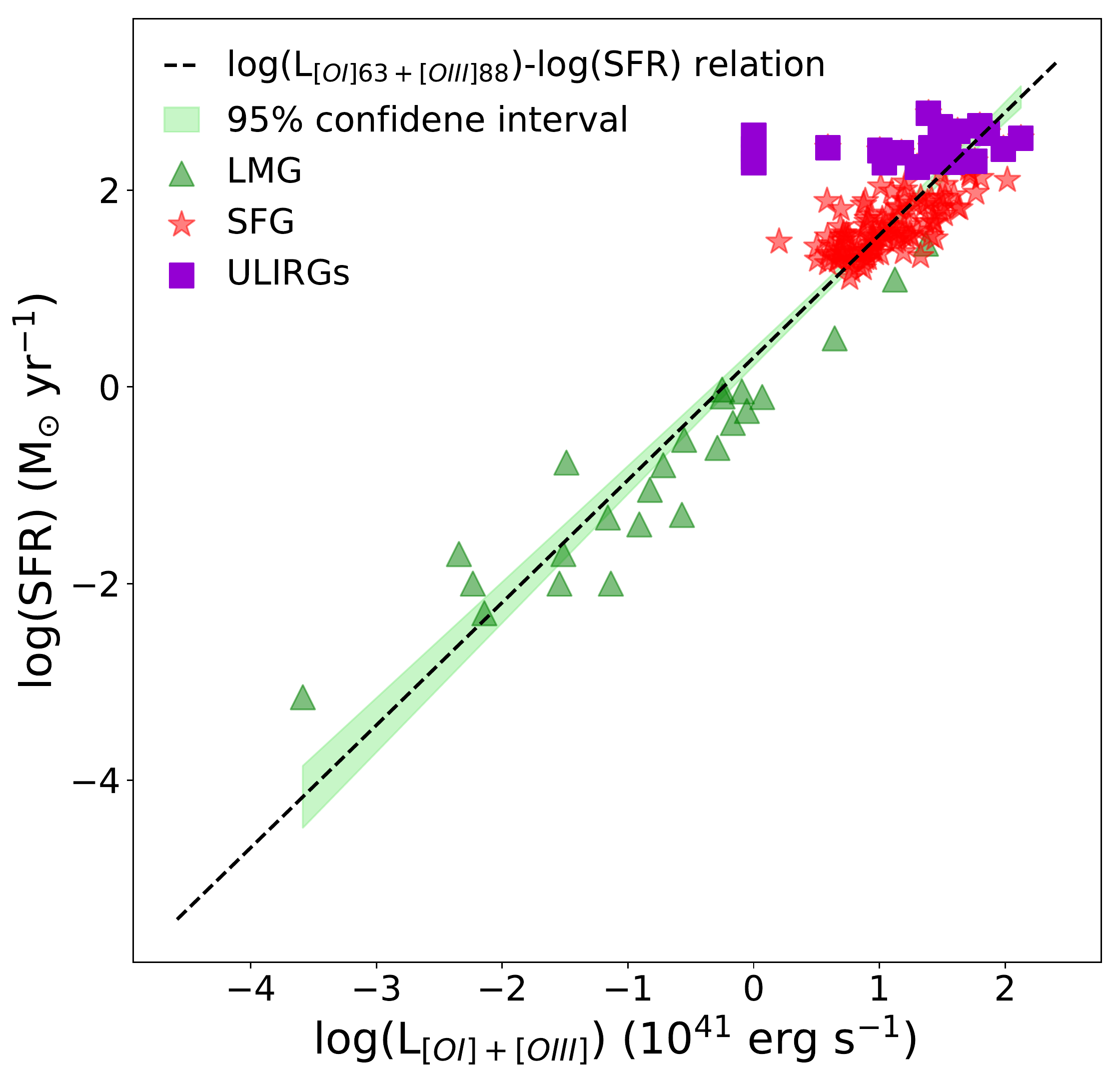}
    \caption{Correlation between the [OI]63$\mu$m and [OIII]88$\mu$m summed emission line luminosities, in units of $10^{41}\, \rm{erg\,s^{-1}}$, and the SFR derived from the total IR luminosity (black dashed line) for a composite sample of SFG (red stars) and from the H$_{\alpha}$ luminosity (corrected for the IR luminosity) for LMG (green triangles). Purple squares indicate the sample of ULIRGs included in the SFG sample.}
    \label{fig:o1_o3_sfr}
\end{figure}
Besides C$^{+}$, also O and O$^{2+}$ are two important coolants of the ISM. The [OI]63$\mu$m line traces the warm and/or dense PDRs, while the [OIII]88$\mu$m emission line originates from diffuse, highly ionised regions near young, hot stars.

The [OIII] line can be an important SFR tracer in LMG \citep{delooze2014}, where PDRs are weak or absent in the ISM and the ionisation field is stronger. In order to trace the SFR in SFG, however, a tracer that can account for PDRs needs to be included. For this reason, combining these two lines can provide an accurate estimate of the SFR, both in SFG and LMG, probing both neutral and ionised medium.

Fig.\,\ref{fig:o1_o3_sfr} shows the correlation obtained for a sample of 151 objects, 24 LMG and 124 SFG, of which 22 ULIRGs. The correlation, with a Pearson correlation coefficient of $r=0.94$, can be expressed by:
\begin{multline}
 \log\left(\frac{SFR}{\rm M_{\odot}\,yr^{-1}}\right)= (0.30 \pm 0.04)\\ +(1.25 \pm 0.04) \log\left( \frac{L_{\rm [OI]+[OIII]}}{\rm 10^{41}\,erg\,s^{-1}}\right)
\end{multline}

\subsubsection{Neon- and sulfur-based SFR tracers}
\label{sec:sfr_ne_s}
\begin{figure*}%[h]
    \centering
    \includegraphics[width=0.66\columnwidth]{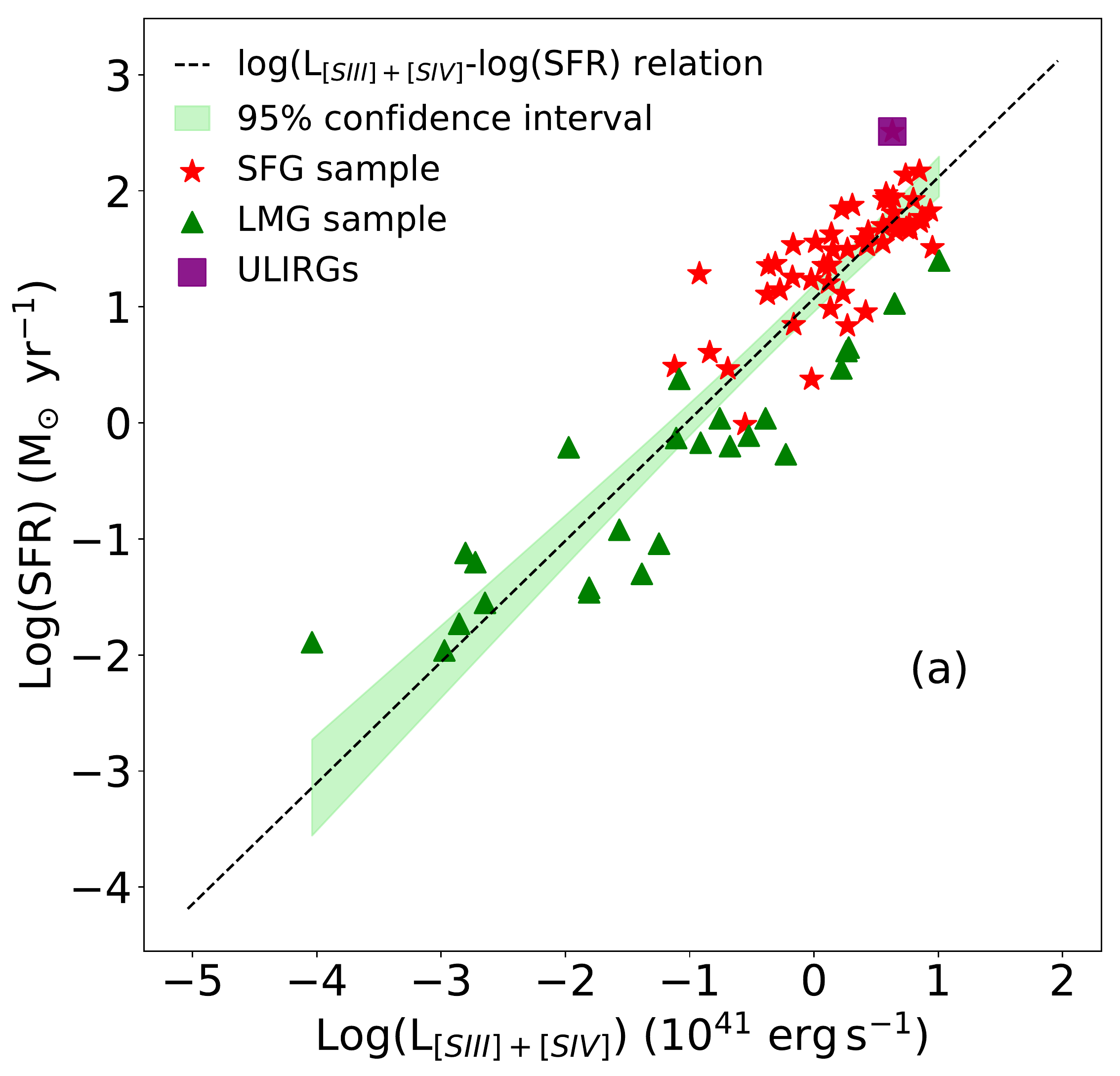}~
    \includegraphics[width=0.66\columnwidth]{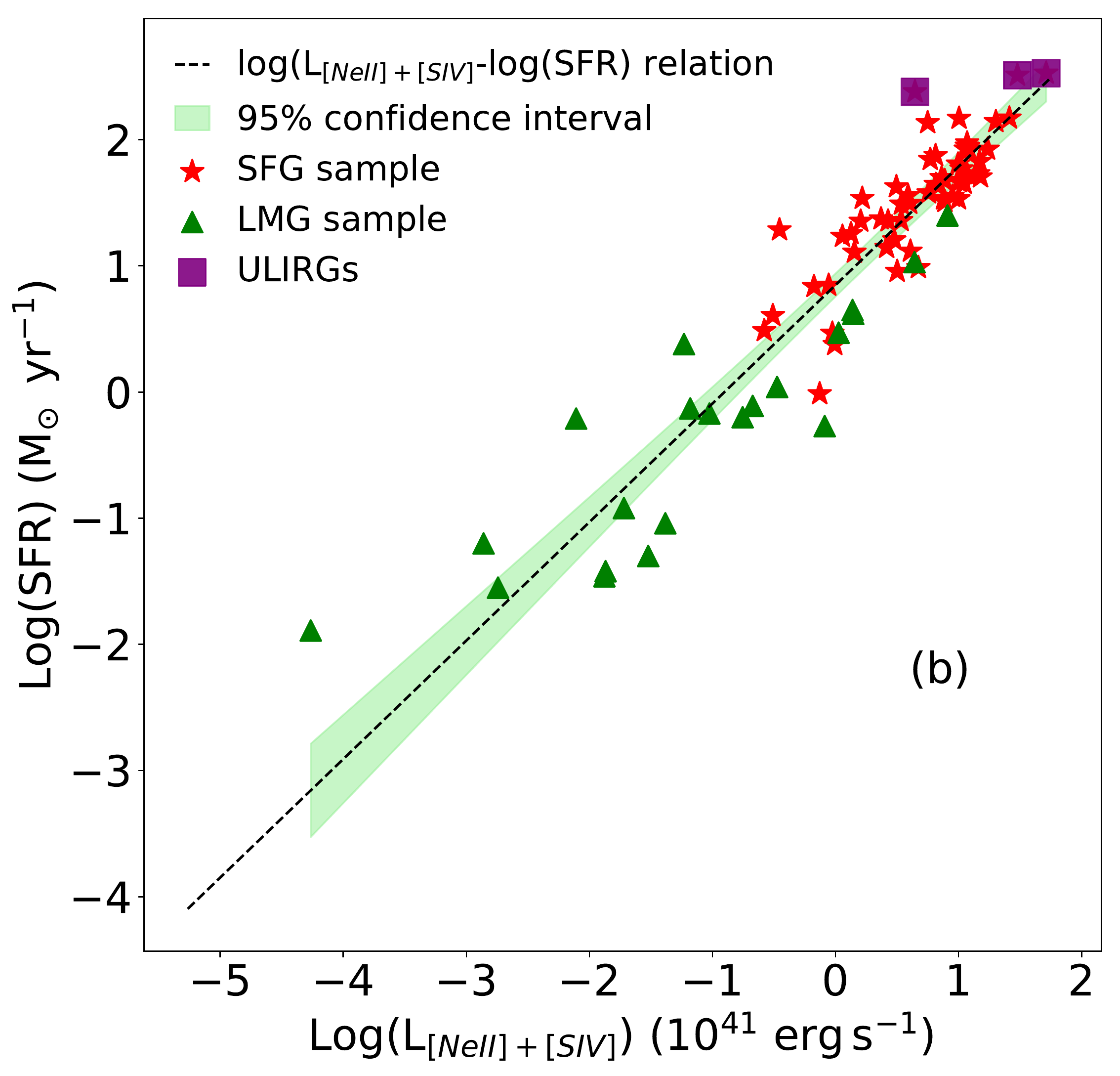}
    \includegraphics[width=0.66\columnwidth]{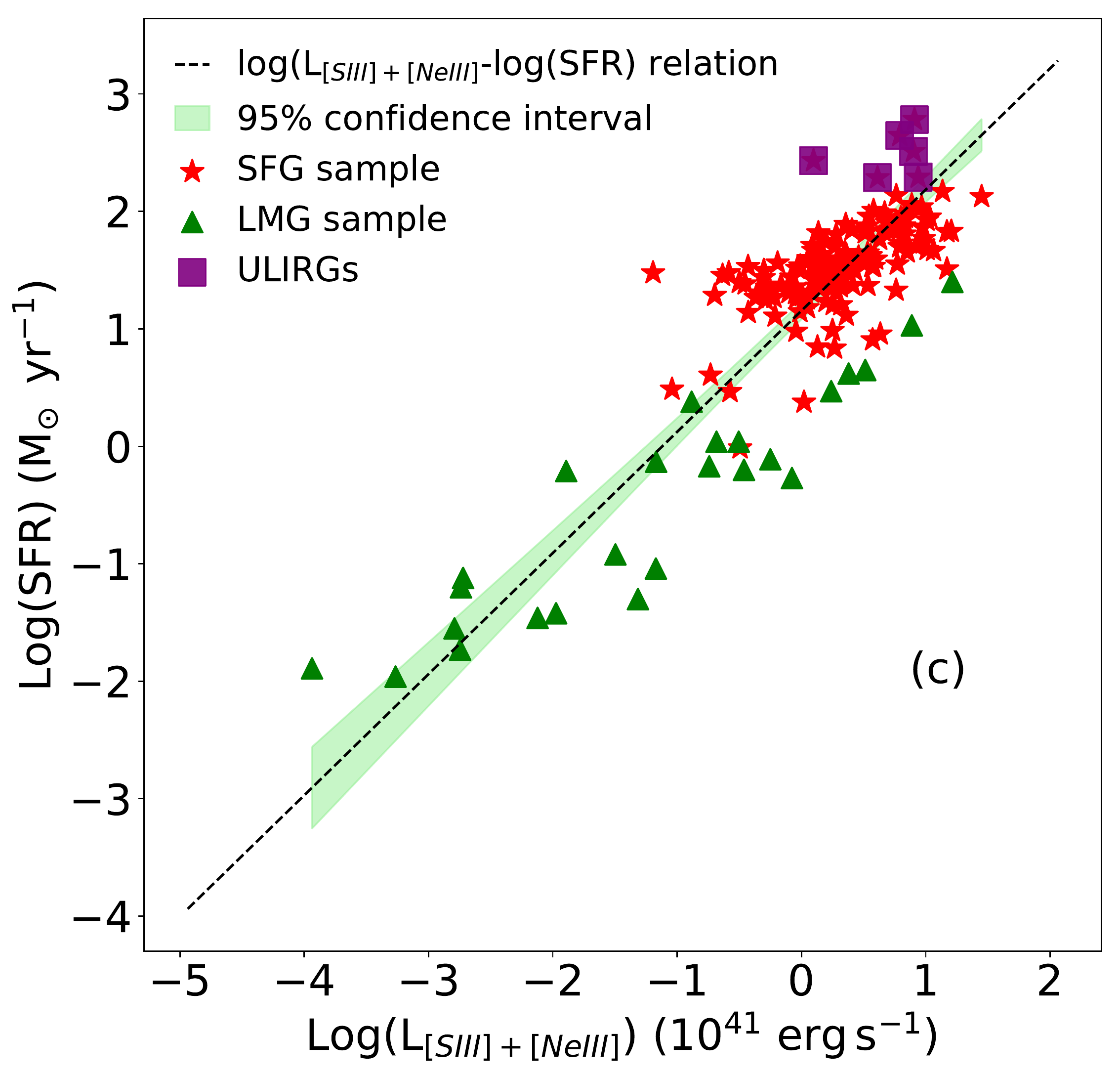}
    \caption{{\bf (a: left)} Correlation between the [SIV]10.5$\, \rm{\micron}$ and [SIII]18.7$\, \rm{\micron}$ summed emission lines luminosity, expressed in units of $10^{41}\, \rm{erg\,s^{-1}}$ and the SFR derived from the total IR luminosity (black dashed line) for a composite catalogue of SFG (red star) and from the H$_{\alpha}$ luminosity (corrected for the IR luminosity) for LMG (green triangles).
    The shaded green area in the three plots indicates the 95$\%$ confidence interval.
    {\bf (b: centre)} Correlation between the [NeII]12.8$\, \rm{\micron}$ and [SIV]10.5$\, \rm{\micron}$ summed luminosities and the SFR derived from the total IR luminosity (black dashed line) for SFG (red star) and from the H$_{\alpha}$ luminosity (corrected for the IR luminosity) for LMG (green triangles). 
    {\bf (c: right)} Correlation between the [NeIII]15.6$\, \rm{\micron}$ and [SIII]18.7$\, \rm{\micron}$ summed emission lines and the SFR derived from the total IR luminosity (black dashed line) for SFG (red star) and from the H$_{\alpha}$ luminosity (corrected for the IR luminosity) for LMG (green triangles).  In all three figures, the purple squares highlight the ULIRG population in the SFG sample.}
    \label{fig:SFR2}
\end{figure*}

A promising SFR tracer, proposed by \citet{ho2007,zhuang2019}, is the sum of the [NeII]12.8$\, \rm{\micron}$ and the [NeIII]15.6$\, \rm{\micron}$ emission line fluxes. By adding the two lines, this tracer is fairly independent from the effects related to the hardness of the radiation field, which are stronger at lower metallicities affecting single line diagnostics.
For instance, while the [NeII]12.8$\, \rm{\micron}$ line intensity scales consistently in LMG and SFG, the [NeIII]15.6$\, \rm{\micron}$ line becomes remarkably brighter in LMG (Fig.\,\ref{fig:corr_c2_ne2_ne3}). Although some dependency on density is still expected, the high critical density of the sulphur ($> 7 \times 10^3\, \rm{cm^{-3}}$; see Table\,\ref{tbl_lines}) and the neon lines ($> 5 \times 10^4\, \rm{cm^{-3}}$) used in this Section guarantees a minor effect on these tracers for the vast majority of the galaxy population.

In Fig.\,\ref{fig:SFR1}c we show the correlation found between the summed luminosity of the [NeII] and [NeIII] emission lines and the SFR. This relation, obtained from data of 203 local star forming galaxies, of which 182 SFG and 21 LMG, can be expressed as:
\begin{multline}
 \log\left(\frac{SFR}{\rm M_{\odot}\,yr^{-1}}\right)= (0.88 \pm 0.03)\\ +(0.96 \pm 0.03) \log\left( \frac{L_{\rm [NeII]+[NeIII]}}{\rm 10^{41}\,erg\,s^{-1}}\right)
\end{multline}
where $\log$(L$_{[NeII]+[NeIII]}$) is the luminosity corresponding to the sum of the fluxes of the [NeII]12.8$\, \rm{\micron}$ and the [NeIII]15.6$\, \rm{\micron}$ lines. This relation has a Pearson coefficient of $r$=0.91. 
We report in Fig.\,\ref{fig:SFR1}c the comparison between the theoretical relation obtained by \citet{zhuang2019}, and our empirical one. The relation by \citeauthor{zhuang2019} is derived following the assumption that the neon lines trace all the ionising photons in a star forming region. Our relation, derived from observational data, shows a lower slope, and thus a larger offset, thus suggesting a lower efficiency in reprocessing the ionising photons by the neon transitions. 

Analogously to the case of the two neon lines, we also take into consideration the sum of the two sulfur lines, i.e. [SIII]18.7$\, \rm{\micron}$ and [SIV]10.5$\, \rm{\micron}$, as a tracer of the SFR. We plot in Fig.\,\ref{fig:SFR2}a the correlation of the luminosity derived by summing  [SIII]18.7$\, \rm{\micron}$ and [SIV]10.5$\, \rm{\micron}$ with the SFR derived from the total IR luminosity:
\begin{multline}
 \log\left(\frac{SFR}{\rm M_{\odot}\,yr^{-1}}\right)= (1.07 \pm 0.06)\\ +(1.04 \pm 0.05) \log\left( \frac{L_{\rm [SIII]+[SIV]}}{\rm 10^{41}\,erg\,s^{-1}}\right)
\end{multline}

For this relation were used 52 SFG and 25 LMG, obtaining a relation with a Pearson $r$ coefficient of $r$=0.90.

As shown in Table \ref{tbl_lines}, Ne$^{+}$ has a ionisation potential (IP) of 21.56 eV, while Ne$^{2+}$ has a IP of $40.96\, \rm{eV}$ and a potential of the next stage at $63.45\, \rm{eV}$, thus covering the $\sim$ 20 -- 60 eV interval in energy. This is roughly the same ionisation interval covered by the %[NeII]+[NeIII] pair can be covered either by 
[NeII]+[SIV], with [SIV] having a IP of 34.79 eV and the next stage IP at 47.22 eV, or by [NeIII]+[SIII], with [SIII]18.7$\, \rm{\micron}$ that has a IP of 23.34 eV. Analyzing the former pair, we find the linear correlation between {L$_{\rm [NeII]+[SIV]}$} and SFR shown in Fig.\,\ref{fig:SFR2}b and expressed as:
\begin{multline}
\log\left(\frac{SFR}{\rm M_{\odot}\,yr^{-1}}\right)= (0.85 \pm 0.05)\\ +(0.94 \pm 0.04) \log\left( \frac{L_{\rm [NeII]+[SIV]}}{\rm 10^{41}\,erg\,s^{-1}}\right)
\end{multline}

This relation has been calculated using 77 galaxies, of which 56 SFG and 21 LMG, and has a Pearson correlation coefficient $r$ = 0.93. The small number of SFG used for determining this relation is due to the lack of [SIV] measurements available in literature, due to the intrinsic line faintness in star forming galaxies  (see Table \ref{tab:calibrations}) and the position of the line in the silicate absorption band in the 9-11$\, \rm{\micron}$ range.

The relation between SFR and the [NeIII]+[SIII] emission sum is shown in Fig.\,\ref{fig:SFR2}c and can be expressed as:
\begin{multline}
\log\left(\frac{SFR}{\rm M_{\odot}\,yr^{-1}}\right)= (1.16 \pm 0.04)\\ +(1.03 \pm 0.04) \log\left( \frac{L_{\rm [NeIII]+[SIII]}}{\rm 10^{41}\,erg\,s^{-1}}\right)
\end{multline}

This relation has been calculated using 165 galaxies, of which 140 SFG and 25 LMG, and has a Pearson correlation coefficient $r$ = 0.86.

\subsubsection{PAH SFR tracer}\label{sec:pah_sfr}

\begin{figure*}%[h!]
    \centering
    \includegraphics[width=0.685\columnwidth]{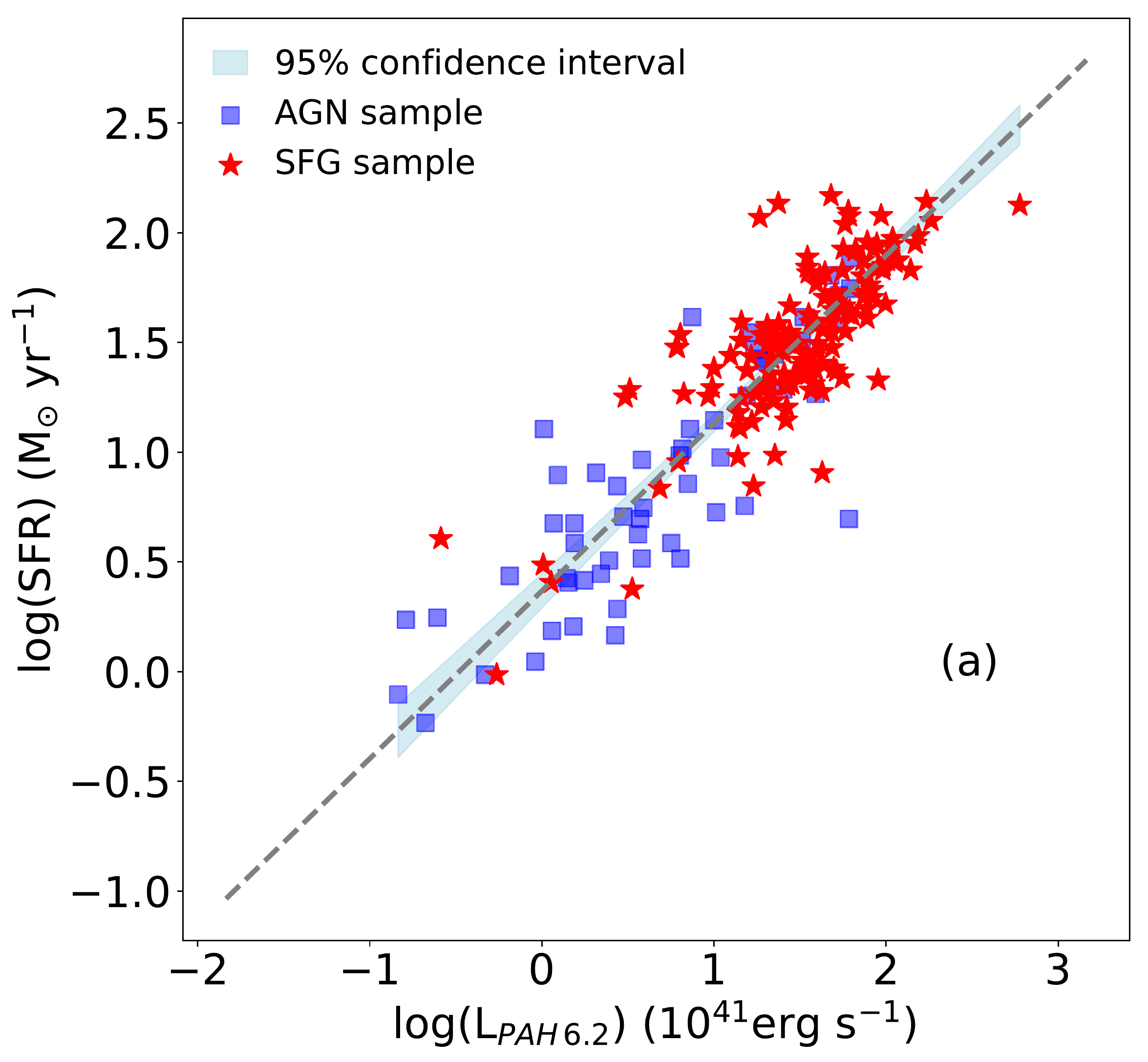}~
    \includegraphics[width=0.665\columnwidth]{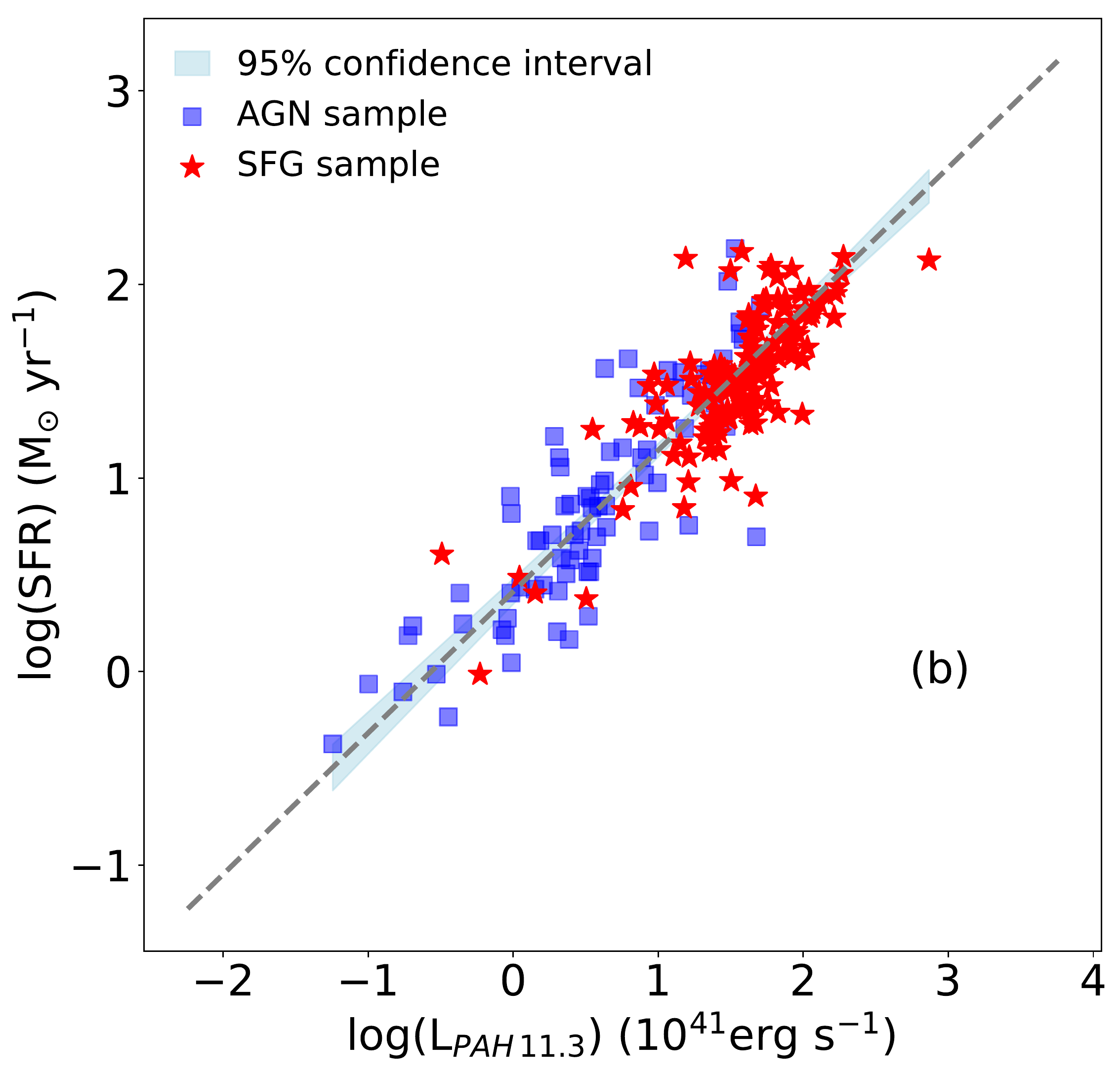}
    \includegraphics[width=0.645\columnwidth]{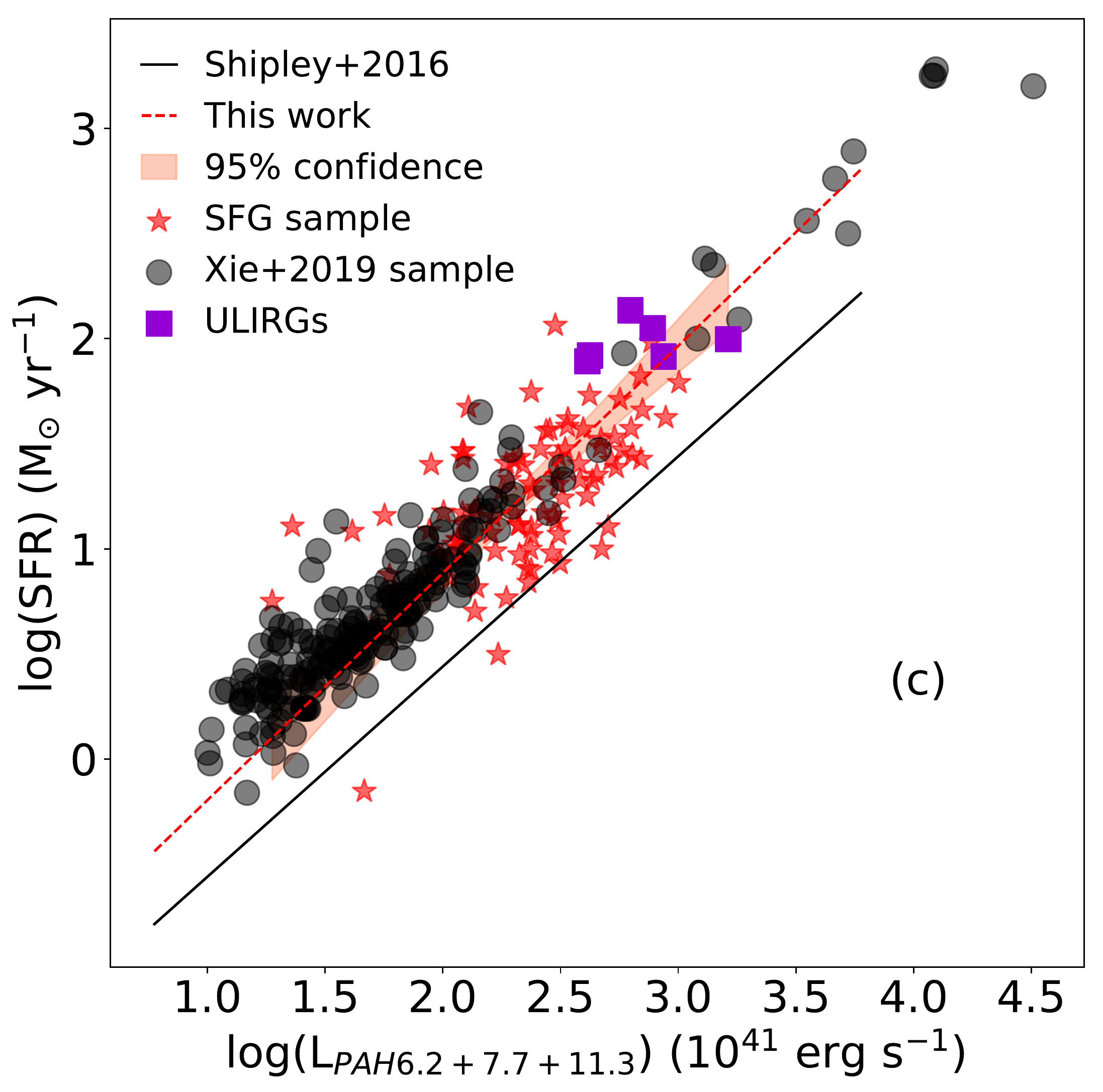}
     
    \caption{{\bf (a: left)} Correlation between the PAH emission feature at 6.2$\, \rm{\micron}$, expressed in units of $10^{41}\, \rm{erg\,s^{-1}}$ and the SFR derived from the total IR luminosity (black line) for a composite catalogue of SFG (red star) and AGN dominated galaxies (blue square).
    The shaded area in the three plots indicates the 95$\%$ confidence interval.
    {\bf (b: centre)} Correlation between the PAH emission feature at 11.3$\, \rm{\micron}$, expressed in units of $10^{41}\, \rm{erg\,s^{-1}}$ and the SFR derived from the total IR luminosity (black line) for a composite catalogue of SFG (red star) and AGN dominated galaxies(blue square). In panels (a) and (b) we exclude ULIRGs from both AGN and SFG samples due to the known PAH deficit in these sources, plus AGN with luminosities above $10^{45}\, \rm{erg\,s^{-1}}$ that could dominate the IR continuum used to estimate the SFR. {\bf (c: right)} Comparison between the relation between PAH total luminosity and SFR derived by \citet{shipley2016} (black solid line line) and using our sample (red dashed line). Red stars indicate SFG, purple squares indicate the ULIRG sub-sample included in the GOALS sample, and grey circles represent a sample of local galaxies used in \citet{xie2019}.}
    \label{fig:SFR_pah}
\end{figure*}

In the mid-IR range, the emission features due to the PAH molecules arise from PDR around \textsc{Hii} regions embedding young stars \citep[e.g., ][]{draine2007}, and can be used as SFR tracers. We analyse the PAH features at 6.2 and 11.3$\, \rm{\micron}$, considering the same sample of 155 SFG and the sample of 103 AGN galaxies described in section \ref{sec:data} for the PAH analysis. As discussed in Section \ref{sec:cor}, we have not computed the correlation between the PAH and the IR luminosity in LMG, therefore we exclude the LMG for the SFR determination with the PAH.

The use of the PAH as a measure of the SFR was originally proposed by \citet{wu2005}, who used the {\it Spitzer}-IRAC camera \citep{fazio2004} 8$\, \rm{\micron}$ band, whose flux density is dominated by the strongest PAH feature, i.e that at 7.7$\, \rm{\micron}$. They derive a calibration of the 8$\, \rm{\micron}$ SFR using the radio VLA emission at 1.4GHz and the H$\alpha$ luminosity from the SFR-radio luminosity relation given by \citet{yun2001} and the SFR-H$\alpha$ luminosity from  \citet{kennicutt1998}, respectively.

More recently, the relatively bright PAH features at 6.2, 7.7 and 11.3$\, \rm{\micron}$ have been used to derive a total PAH luminosity and has been correlated to the extinction corrected H$\alpha$ luminosity of a sample of 227 galaxies \citep{shipley2016}. For star-forming galaxies (105 galaxies), the total PAH luminosity correlates linearly with the extinction-corrected H$\alpha$ luminosity.

In Fig.\,\ref{fig:SFR_pah} we show the correlation of the single PAH features and the SFR: panel (a) shows the relation calculated for the 6.2$\, \rm{\micron}$ feature using 142 SFG and 56 AGN, while panel (b) shows the relation for the 11.3 $\, \rm{\micron}$ feature, for which were used 142 SFG and 77 AGN. When computing these correlations, we excluded all ULIRGs from the original sample of AGN and SFG, following the approach by \citet{pope2013} to avoid saturation effects in the PAH-to-continuum luminosities. Moreover, when considering the AGN sample, we discarded all sources with a total IR luminosity greater than $L_{IR}=10^{45}\, \rm{erg\,s^{-1}}$, where the total IR luminosity, and therefore the derived SFR, could be severely contaminated by the AGN. Nevertheless, these are a few sources that do not affect the derived fit parameter when they are included in the fit. The two relations are described respectively by the following equations, which also report the Pearson $r$ coefficient relative to each equation:
\begin{multline}
    \log\left(\frac{SFR}{\rm M_{\odot}\,yr^{-1}}\right)=(0.37 \pm 0.04)\\+(0.76 \pm 0.03)\log\left(\frac{L_{PAH6.2}}{\rm 10^{41}\,erg\,s^{-1}}\right), r=0.87
\end{multline}

\begin{multline}
    \log\left(\frac {SFR}{\rm M_{\odot}\,yr^{-1}}\right)=(0.41 \pm 0.03)\\+(0.73 \pm 0.02)\log \left( \frac{L_{PAH11.3}}{\rm 10^{41}\,erg\,s^{-1}}\right), r=0.89
\end{multline}

Both results show a shallow slope, significantly lower than unity. This is a consequence of including the AGN sample. Considering only the SFG sample, the slopes would increase, becoming 0.87 for the PAH feature at 6.2$\mu$m, and of 0.90 for the 11.3$\mu$m feature.
On the other hand, while we excluded all ULIRGs, including these objects does not result in a significant change in the slope, that remains comparable within the errors. In particular, including the entire AGN and SFG samples for the PAH feature at 6.2$\mu$m results in a slope of 0.81$\pm$0.03, while for the PAH feature at 11.3$\mu$m the slope would be of 0.78$\pm$0.03.

We compared our results with the one proposed by \citet{shipley2016}: in order to properly determine the SFR using the H$_{\alpha}$ and 24$\, \rm{\micron}$ luminosities, as done by these authors, we selected only the SFG, excluding the AGN, leaving our sample composed of 100 sources with the 6.2, 7.7, 11.3 PAH features, the 24$\, \rm{\micron}$ and the H$_{\alpha}$ data.% While the \citeauthor{shipley2016} sample is composed of galaxies with L$_{IR}$ in the 10$^{9}$-10$^{12}$L$_{\odot}$, our sample is composed only of LIRGs and ULIRGs, thus representing a more limited population. %nevertheless, while 
We find that  %there is a difference of 
our estimate of the SFR is higher by $\sim$0.25 dex than that derived by \citet{shipley2016} and almost linear in slope, (see Fig.\,\ref{fig:SFR_pah}c). We obtain:
\begin{multline}
    \log\left(\frac{SFR}{\rm M_{\odot}\,yr^{-1}}\right)=(-1.28 \pm 0.23)+\\(1.08 \pm 0.10)\log\left(\frac{L_{\Sigma PAH}}{\rm 10^{41}\,erg\,s^{-1}}\right)
\end{multline}
with a Pearson r coefficient of $r$=0.66.
While the slope of our result is comparable, within the error, to that obtained by \citeauthor{shipley2016}, we attribute the difference in the intercept to the significant difference in sample characteristics. In particular, the disagreement arises when we extrapolate the results obtained by \citeauthor{shipley2016} for local galaxies. The sample used by \citeauthor{shipley2016} is in fact composed of galaxies with redshift in the $z\sim 0.2-0.6$ range, while our sample has a mean redshift of $z\sim 0.027$. Moreover, while our sample is primarily composed of LIRGs and ULIRGs, the sample used by these authors is mainly constituted of galaxies in the $L_{IR}\sim 10^{9}-10^{12}L_{\odot}$ interval. We additionally compared our derived relation to a sample of local SFG described by \citet{xie2019}, for which was available in the literature the SFR derived from H$\alpha$, corrected by the flux at 24$\mu$m. This sample is composed of local galaxies and, while the bulk of these galaxies present a lower total PAH luminosity, it follows the same relation derived from the GOALS sample. This suggests that the difference observed between our result and the result by \citeauthor{shipley2016} is indeed due to an intrinsic difference of the sample used.

\subsubsection{\texorpdfstring{H$_{2}$}{H2} SFR tracers}

\begin{figure*}%[h!]
    \centering
    \includegraphics[width=0.66\columnwidth]{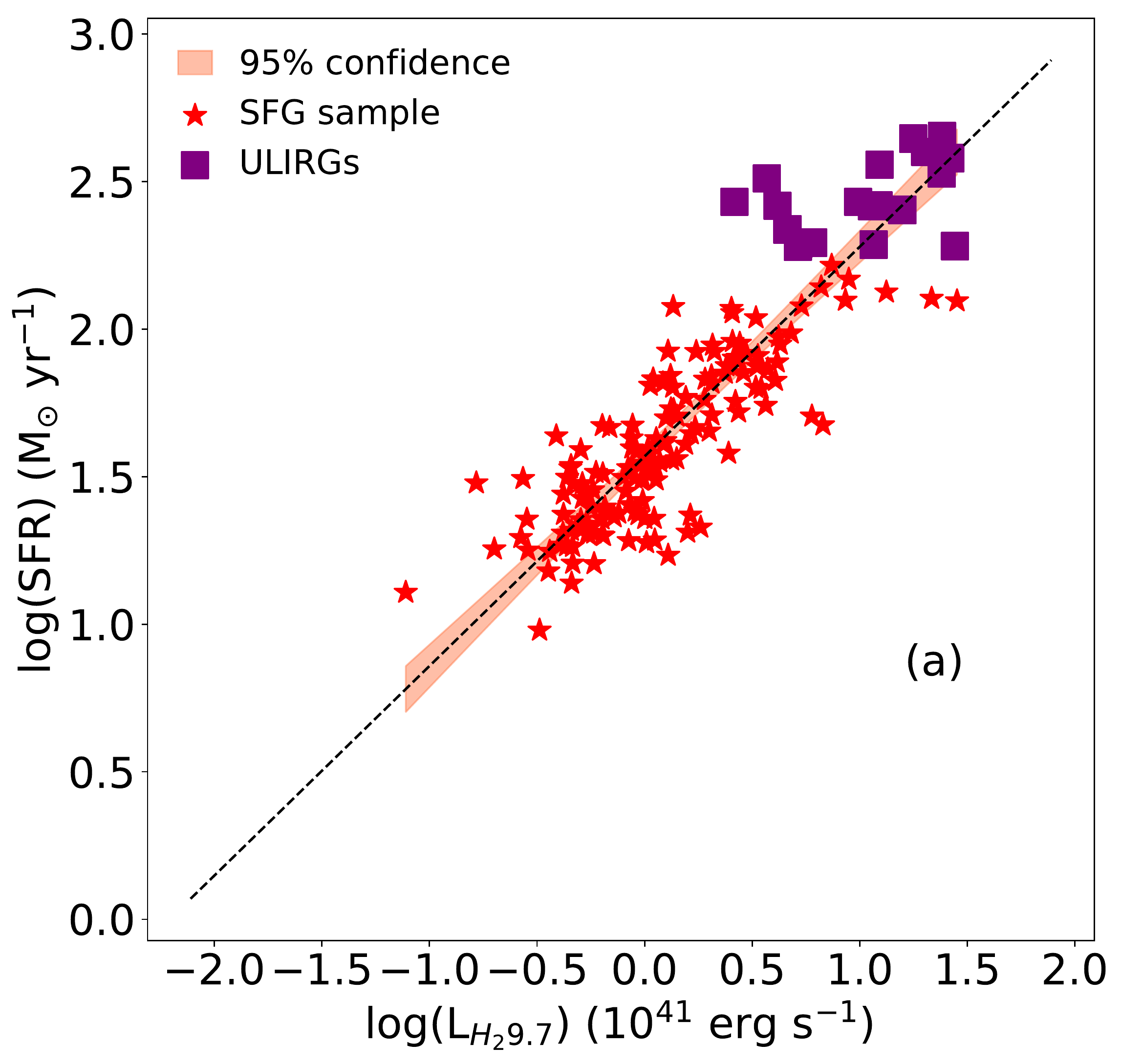}~
    \includegraphics[width=0.66\columnwidth]{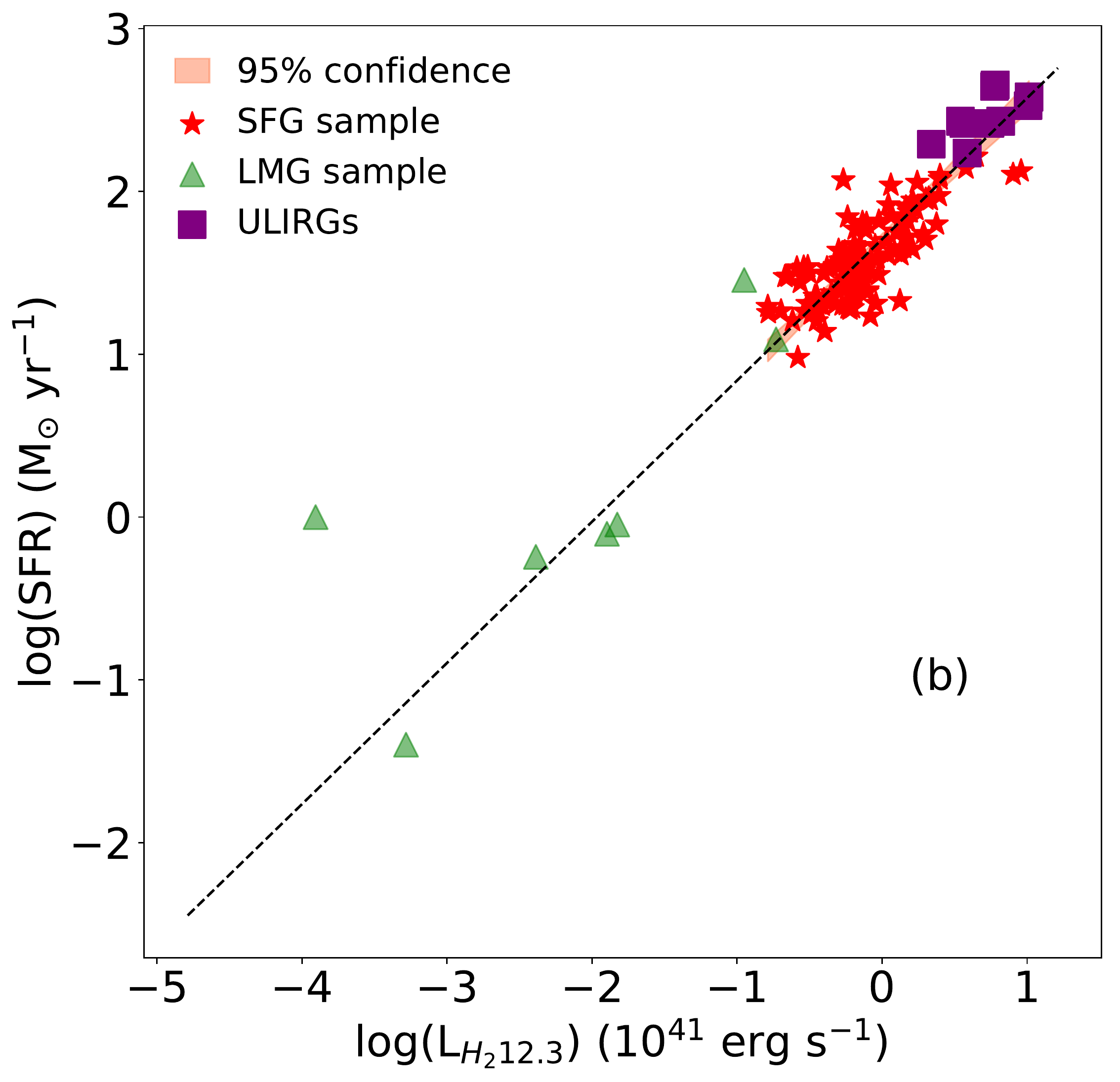}
    \includegraphics[width=0.66\columnwidth]{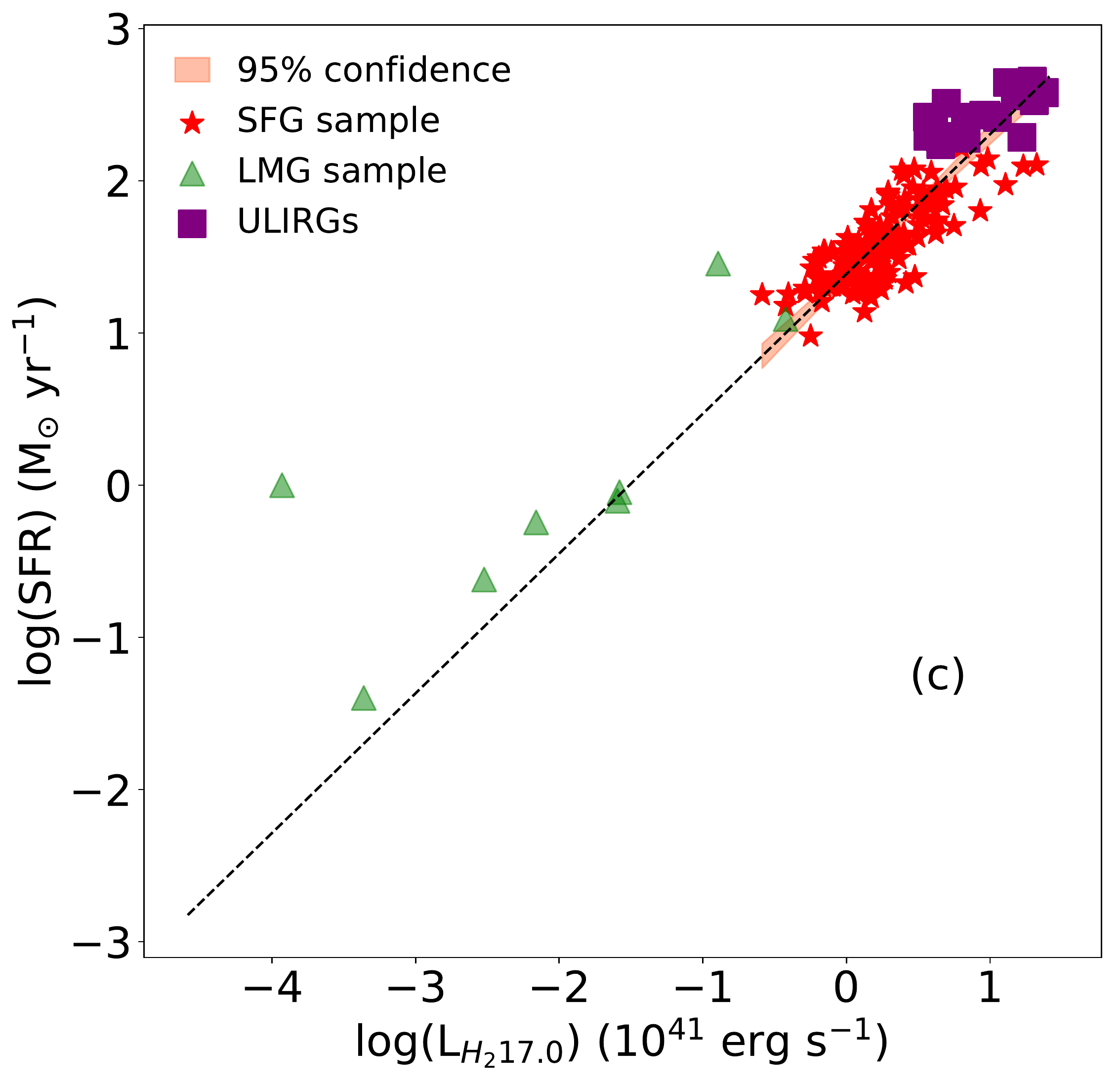}
     
    \caption{{\bf (a: left)} Correlation between the H$_{2}$ molecular line at 9.67$\, \rm{\micron}$, expressed in units of $10^{41}\, \rm{erg\,s^{-1}}$ and the SFR derived from the total IR luminosity (black line) for a catalogue of SFG (red star).
    {\bf (b: centre)} Correlation between the H$_{2}$ molecular line at 12.28$\, \rm{\micron}$, expressed in units of $10^{41}\, \rm{erg\,s^{-1}}$ and the SFR derived from the total IR luminosity (black line) for a catalogue of SFG (red star). The green triangle show LMG detection, not used to derive the correlation. {\bf (c: right)} Correlation between the H$_{2}$ molecular line at 17.03$\, \rm{\micron}$, expressed in units of $10^{41}\, \rm{erg\,s^{-1}}$ and the SFR derived from the total IR luminosity (black line) for a catalogue of SFG (red star).
    In all three figures, the shaded ares shows the 95$\%$ confidence interval of the relations, and the purple squares highlight the ULIRG population in the SFG sample.}
    \label{fig:SFR_h2}
\end{figure*}

Molecular hydrogen is the most abundant molecule in the Universe and can be found in various environments. It can be excited by UV fluorescence \citep{black1987}, shocks \citep{hollenbach1989} or X-ray illumination \citep{maloney1997}, thus probing various astrophysical environments. H$_{2}$ forms on the surface of dust grains, affecting the ISM chemistry, and acts as a coolant. It is particularly important in all processes that regulate star formation and galaxy evolution, where the principal mechanism of these lines is associated with the UV radiation from massive stars.

Comparing theoretical models to observations, \citet{rigopoulou2002} found evidence that an important fraction of H$_{2}$ emission in SFG can originate in PDRs. In this section we test the use of different H$_{2}$ molecular lines as SFR tracers. In Fig.\,\ref{fig:SFR_h2}a we show the correlation between the H$_{2}$ (S(3)) molecular line at 9.67$\, \rm{\micron}$ and the SFR determined from the total L$_{IR}$ for the SFG in the GOALS sample. We find a relation expressed by:
\begin{multline}
    \log\left(\frac{SFR}{\rm M_{\odot}\,yr^{-1}}\right)=(1.57 \pm 0.02)\\+(0.71 \pm 0.03)\log\left(\frac{L_{S(3)}}{\rm 10^{41}\,erg\,s^{-1}}\right)
\end{multline}
determined using 168 SFG, with a Pearson $r$ coefficient of r=0.87. Fig.\,\ref{fig:SFR_h2}b shows the correlation between the H$_{2}$ (S(2)) line at 12.28$\, \rm{\micron}$ and the SFR, determined with a sample of 126 SFG, with a Pearson coefficient of $r$=0.86 and expressed by:
\begin{multline}
    \log\left(\frac{SFR}{\rm M_{\odot}\,yr^{-1}}\right)=(1.71 \pm 0.02)\\+(0.87 \pm 0.04)\log\left(\frac{L_{S(2)}}{\rm 10^{41}\,erg\,s^{-1}}\right)
\end{multline}

Finally, Fig.\,\ref{fig:SFR_h2}c shows the correlation between the H$_{2}$ (S(1)) line at 17.03$\, \rm{\micron}$ and the SFR, derived from a sample of 154 SFG, with r=0.85 and expressed by:
\begin{multline}
    \log\left(\frac{SFR}{\rm M_{\odot}\,yr^{-1}}\right)=(1.39 \pm 0.02)\\+(0.92 \pm 0.04)\log\left(\frac{L_{S(1)}}{\rm 10^{41}\,erg\,s^{-1}}\right)
\end{multline}

In Fig.\,\ref{fig:SFR_h2}b and c  we also report, in green, seven detections in LMG of S(2) and S(1), respectively. We did not include the LMG sample in the correlation due to the small number of sources, limiting ourselves to a comparison of the results. While a good correlation is lacking for the S(1) line, for the S(2) line we have a good agreement between the LMG population and the SFG one. This can be of particular interest in those LMG in which the CO emission is considerably lower than what it would correspond to the estimated SFR, implying the presence of CO-dark molecular gas that is not traced by the CO emission in LMG \citep{togi2016}. 
We speculate that the sub-linear slopes obtained for the S(2) and S(3) lines might be linked to the different gas excitation temperature associated with each line. In particular, by moving to higher excitation lines, and thus to shorter wavelengths, the H$_2$ transitions are originated by an increasingly warmer gas in a thinner layer of the molecular gas clouds. Tracing the warmest material might cause the flattening of the slope, since colder star-forming clouds are not detected by the higher transitions. Additionally, S(2) and S(3) lines could have a larger contribution from other excitation mechanisms, as suggested by the different excitation temperatures measured for these transitions in the Boltzmann diagrams for nearby galaxies \citep[e.g.][]{tommasin2010}.

We have not included the AGN sample in this Section due to the different excitation mechanisms that can contribute to the H$_{2}$ rotational lines in these sources, such as shocks or X-ray illumination. For the same reason H$_2$ lines are neither used as BHAR tracers in Section\,\ref{sec:BHAR}, since they respond to the excitation temperature rather than the hardness of the ionising continuum. This makes quite difficult to establish a clear connexion between the line intensities and the SFR or the accreted mass onto the black hole when both contributions are present.

\subsection{BHAR tracers}\label{sec:BHAR}

\begin{figure*}
\centering
\includegraphics[width=0.66\columnwidth]{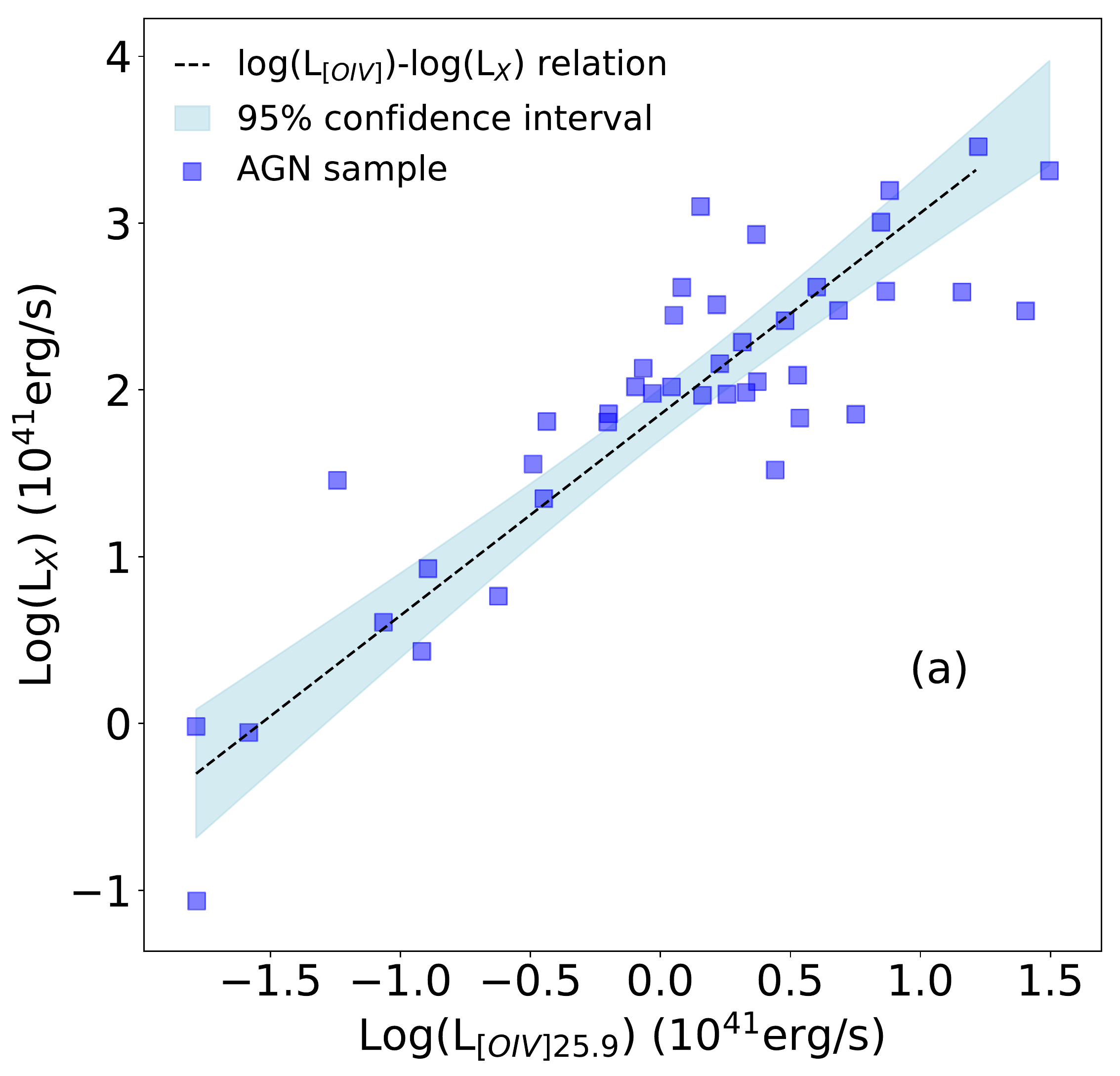}~
\includegraphics[width=0.66\columnwidth]{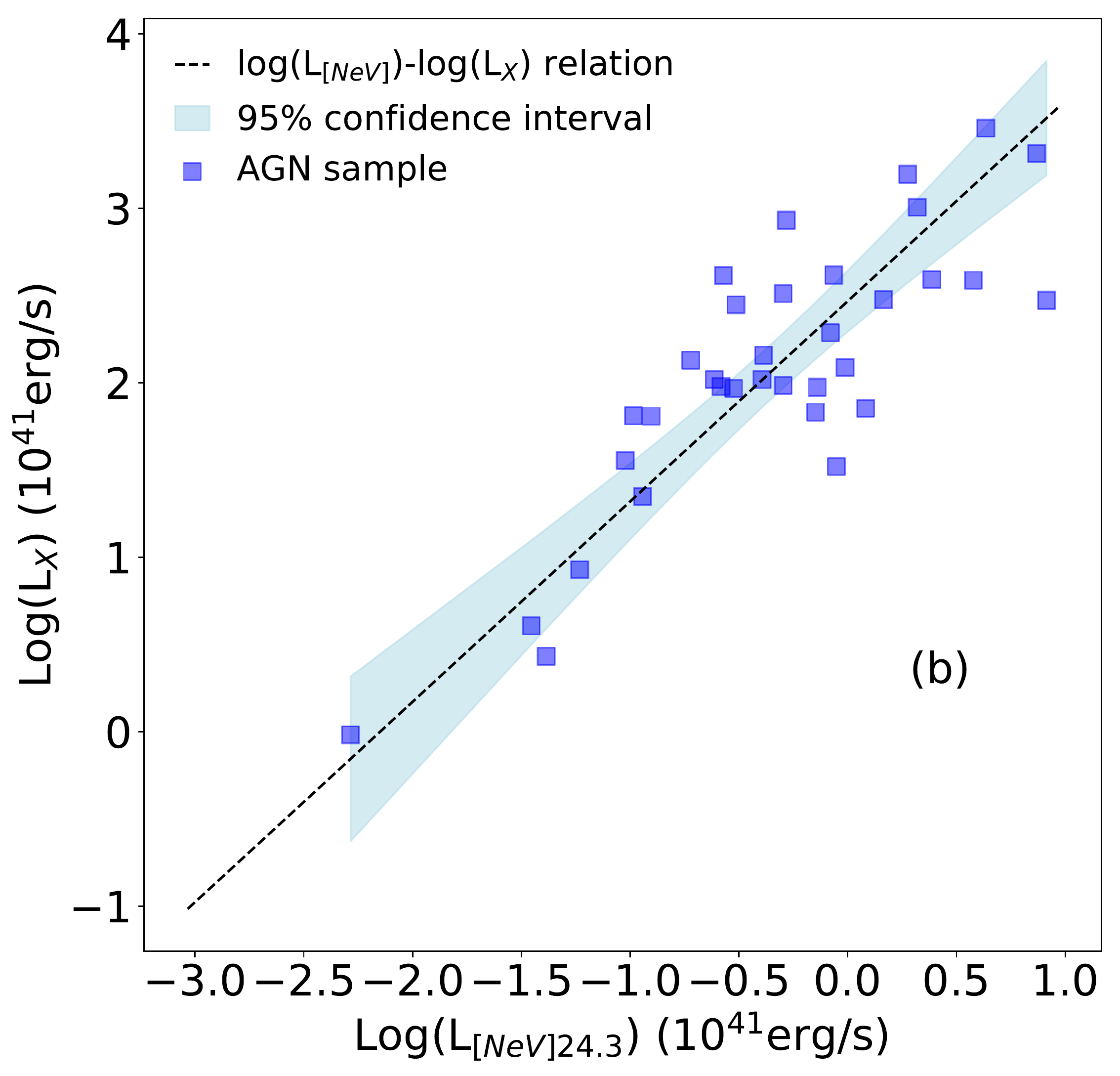}
\includegraphics[width=0.66\columnwidth]{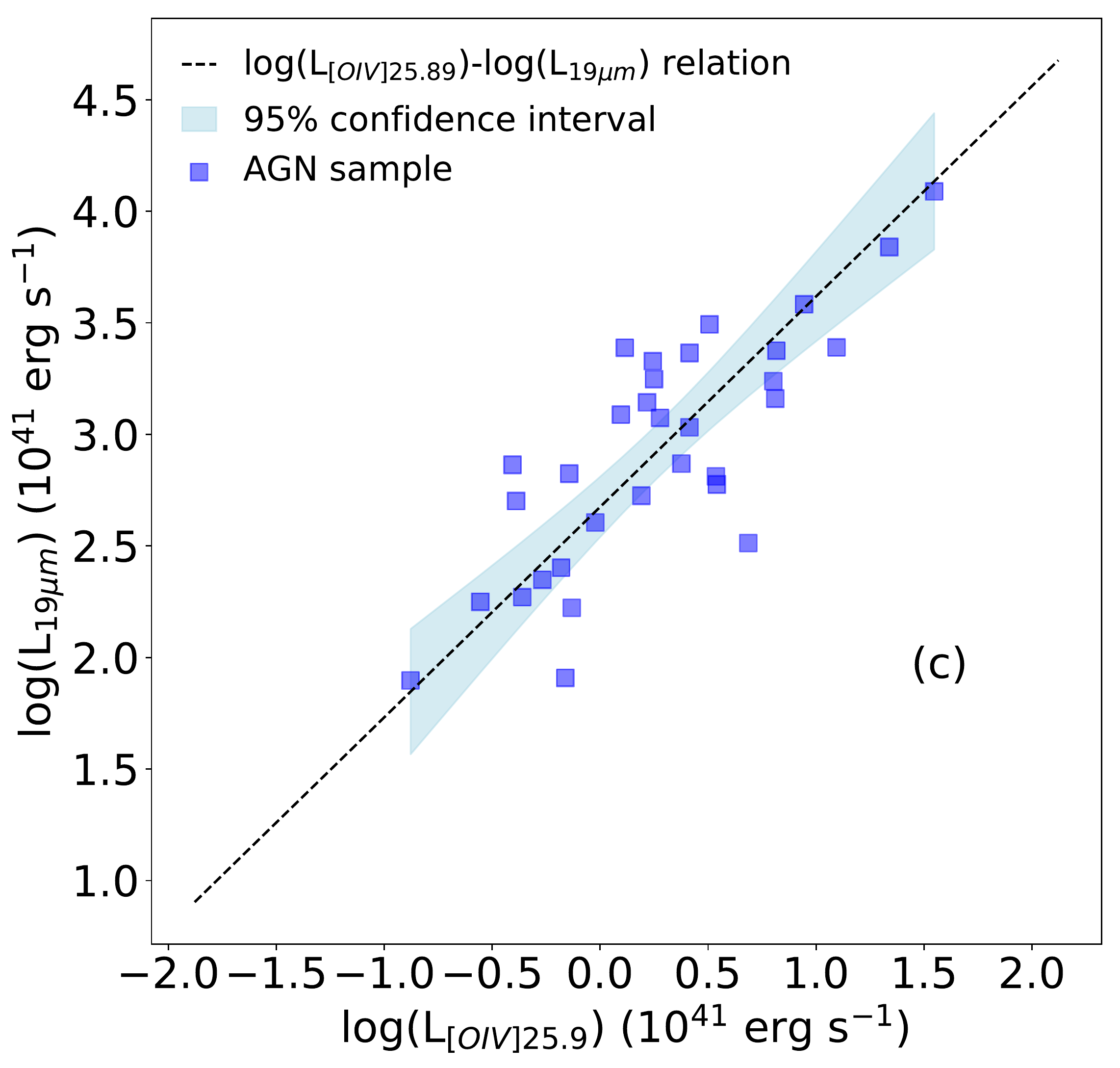}
\caption{{\bf (a: left)} Linear correlation between the [OIV]25.9$\, \rm{\micron}$ and the 2-10 keV X-ray luminosity. {\bf (b: centre)} Linear correlation between the [NeV]24.3$\, \rm{\micron}$ line luminosity and the 2-10 keV X-ray luminosity. {\bf (c: right)} Linear correlation between the [OIV]25.9$\, \rm{\micron}$ line luminosity and the 19$\, \rm{\micron}$ luminosity. All luminosities are expressed in units of $10^{41}\, \rm{erg\,s^{-1}}$.}
\label{fig:o4_x}
\end{figure*}

The [OIV]25.9$\, \rm{\micron}$ and the [NeV]24.3$\, \rm{\micron}$ lines can be used to trace AGN activity.
From our catalogue of AGN, derived from the 12$\, \rm{\micron}$ sample, we compiled from the literature the 2-10 keV X-rays fluxes corrected for absorption. We then selected all objects with hydrogen column density of N$_{H}$ $\leq$ 5$\times$10$^{23}\, \rm{cm^{-2}}$, because we want to exclude Compton-thick objects for which the 2-10 keV X-rays can be substantially absorbed,  thus obtaining a sub-catalogue of 42 objects. For these objects, we investigated the correlation between the [OIV]25.9$\, \rm{\micron}$ line luminosity and the 2-10 keV X-ray luminosity (see Fig.\,\ref{fig:o4_x}a).

We find a correlation expressed by the equation:
\begin{multline}\label{eq:X_o4}
    \log\left(\frac{L_{X}}{\rm 10^{41}\,erg\,s^{-1}}\right)=(1.85 \pm 0.07)\\+(1.21 \pm 0.10)\log\left(\frac{L_{[OIV]25.9}}{\rm 10^{41}\,erg\,s^{-1}}\right)
\end{multline}
with a Pearson $r$ coefficient of $r$=0.87. In order to obtain a measure of the BHAR, it is necessary to convert the luminosity in the 2-10 keV band to the bolometric luminosity of the object, and from there to the BHAR $(L_{AGN}=\eta \dot{M}_{BH}c^{2})$. Different studies have been carried out to determine the best bolometric correction to apply when considering the 2-10 keV luminosity, with those by \citet{marconi2004} and \citet{lusso2012} being the most used in the literature. The resulting bolometric luminosities obtained applying these corrections are different, giving us a range of possible values. We applied both correction to our data, and then from the resulting bolometric luminosities we calculated two linear relations linking the [OIV]25.9$\, \rm{\micron}$ line luminosity to the BHAR. Assuming a radiative efficiency of $\eta$=0.1, we obtained in equation \ref{eq:lusso_o4} the linear correlation applying the bolometric correction from \citet{lusso2012}, and in equation \ref{eq:marconi_o4} the one applying the correction from \citet{marconi2004}. The equations are reported each followed by its Pearson $r$ coefficient:
\begin{multline}\label{eq:lusso_o4}
    \log\left(\frac{\dot{M}_{BH}}{\rm M_{\odot}\,yr^{-1}}\right)=(-1.65 \pm 0.07)\\ +(1.04 \pm 0.09) \log\left(\frac{L_{[OIV]25.9}}{\rm 10^{41}\,erg\,s^{-1}}\right),\,\,r=0.86
\end{multline}

\begin{multline}\label{eq:marconi_o4}
   \log\left(\frac{\dot{M}_{BH}}{\rm M_{\odot}\,yr^{-1}}\right)=(-1.66 \pm 0.09)\\+(1.49 \pm 0.12)\log\left(\frac{L_{[OIV]25.9}}{\rm 10^{41}\,erg\,s^{-1}}\right),\,\,r=0.87
\end{multline}

In a similar way, we first calculated the linear correlation between the [NeV]24.3$\, \rm{\micron}$ line luminosity and the 2-10 keV luminosity (see right panel in Fig.\,\ref{fig:o4_x}). In this case, we have a total of 34 objects, due to a smaller number of [NeV] available data, and we obtained a linear relation described by the equation:
\begin{multline}\label{eq:X_ne5}
    \log\left(\frac{L_{X}}{\rm 10^{41}\,erg\,s^{-1}}\right)=(2.40 \pm 0.08)\\+(0.95 \pm 0.11)\log\left(\frac{L_{\rm [NeV]24.3}}{\rm 10^{41}\,erg\,s^{-1}}\right)
\end{multline}
with a Pearson coefficient of $r$=0.84. From this, we then applied the same bolometric corrections, obtaining linear correlations between the line luminosity and the BHAR, expressed by equations \ref{eq:lusso_ne5} and \ref{eq:marconi_ne5} for the \citet{lusso2012} and \citet{marconi2004} correction respectively:
\begin{multline}\label{eq:lusso_ne5}
    \log\left(\frac{\dot{M}_{BH}}{\rm M_{\odot}\,yr^{-1}}\right)=(-1.11 \pm 0.09)\\+(1.06 \pm 0.12)\log\left(\frac {L_{\rm [NeV]24.3}}{\rm 10^{41}\,erg\,s^{-1}}\right),\,\,r=0.83
\end{multline}

\begin{multline}\label{eq:marconi_ne5}
   \log\left(\frac{\dot{M}_{BH}}{\rm M_{\odot}\,yr^{-1}}\right)=(-0.89 \pm 0.11)\\+(1.47 \pm 0.15)\log\left(\frac{L_{\rm [NeV]24.3}}{\rm 10^{41}\,erg\,s^{-1}}\right),\,\,r=0.84
\end{multline}

As a general trend, we find that both the [NeV]24.3$\, \rm{\micron}$ and [OIV]25.9$\, \rm{\micron}$ lines correlate linearly with the 2-10 keV X-ray luminosity, thus providing a good proxy to measure the AGN activity, as shown in Fig.\,\ref{fig:o4_x} a and b. 

Following the work by \citet{tommasin2010}, we also analysed the correlation between the [NeV]24.3 and [OIV]25.9$\, \rm{\micron}$ lines with the luminosity at 19$\, \rm{\micron}$. The luminosity at 19$\mu$m (L$_{19{\mu}m}$) was used for two main reasons. On one hand, accurate {\it Spitzer}-IRS observations are available for the 19$\mu$m flux density, from two different apertures,
for the considered sample of galaxies. On the other hand, the 19$\mu$m photometry data are the only available accurate photometric data closest to the emission at 12$\mu$m. The 12$\mu$m emission is, in turn, the best proxy for the bolometric flux of an active galaxy \citep{spinoglio1995}. In particular, \citet{tommasin2010} found that the 19$\mu$m luminosity correlates with the [NeV]14.3$\mu$m line luminosity.

\begin{figure*}
\centering
\includegraphics[width=0.66\columnwidth]{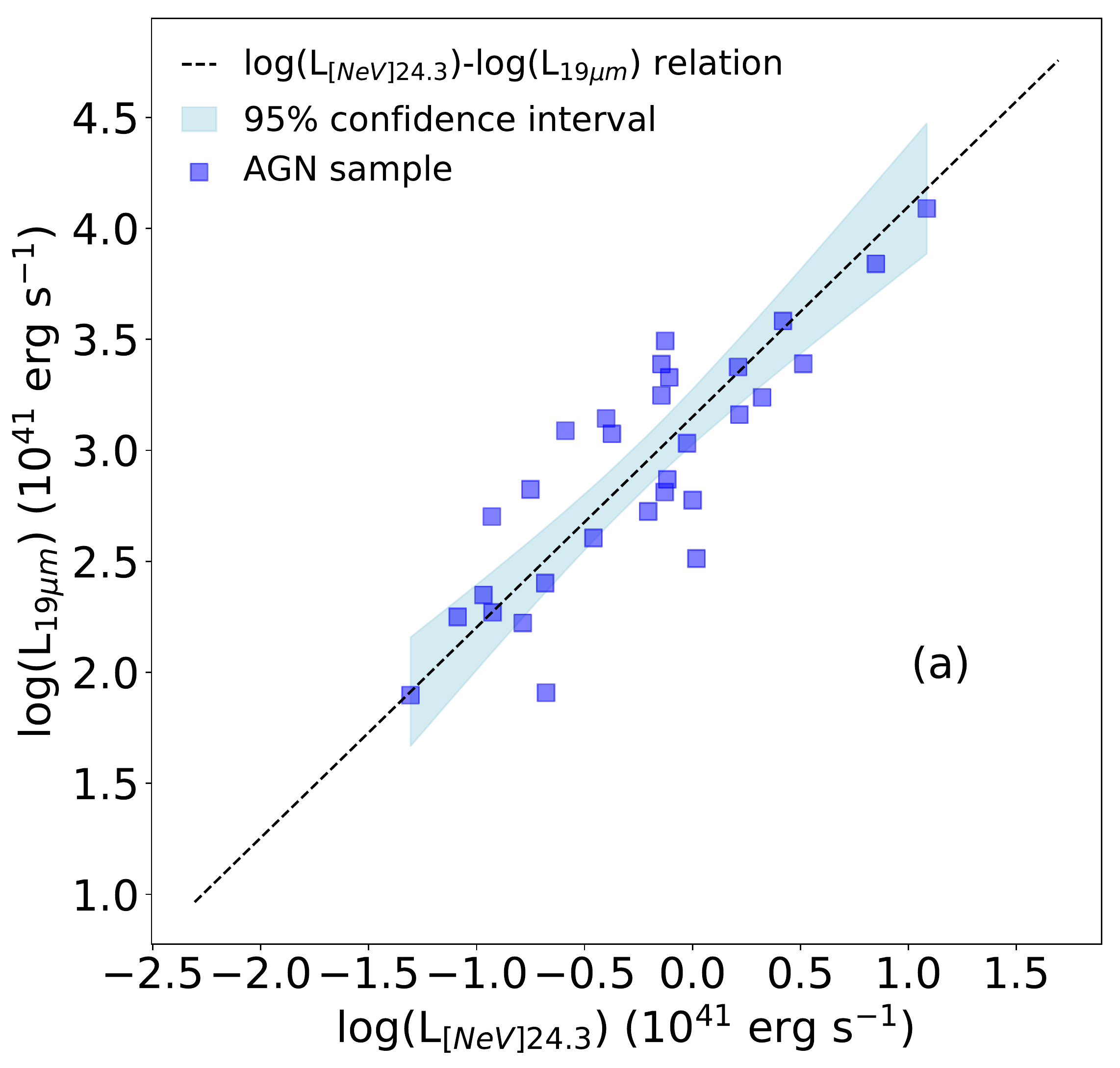}~
\includegraphics[width=0.66\columnwidth]{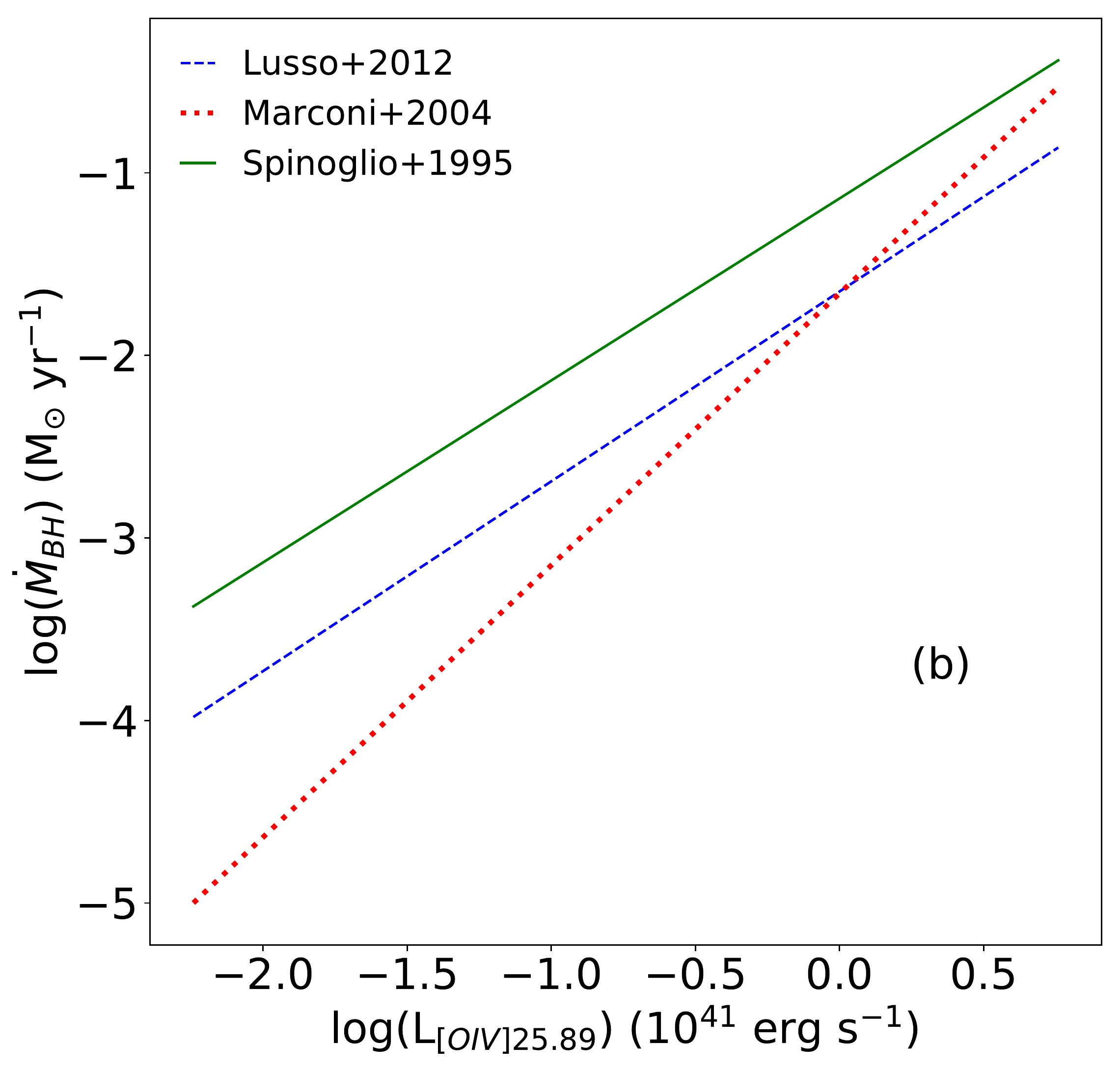}
\includegraphics[width=0.66\columnwidth]{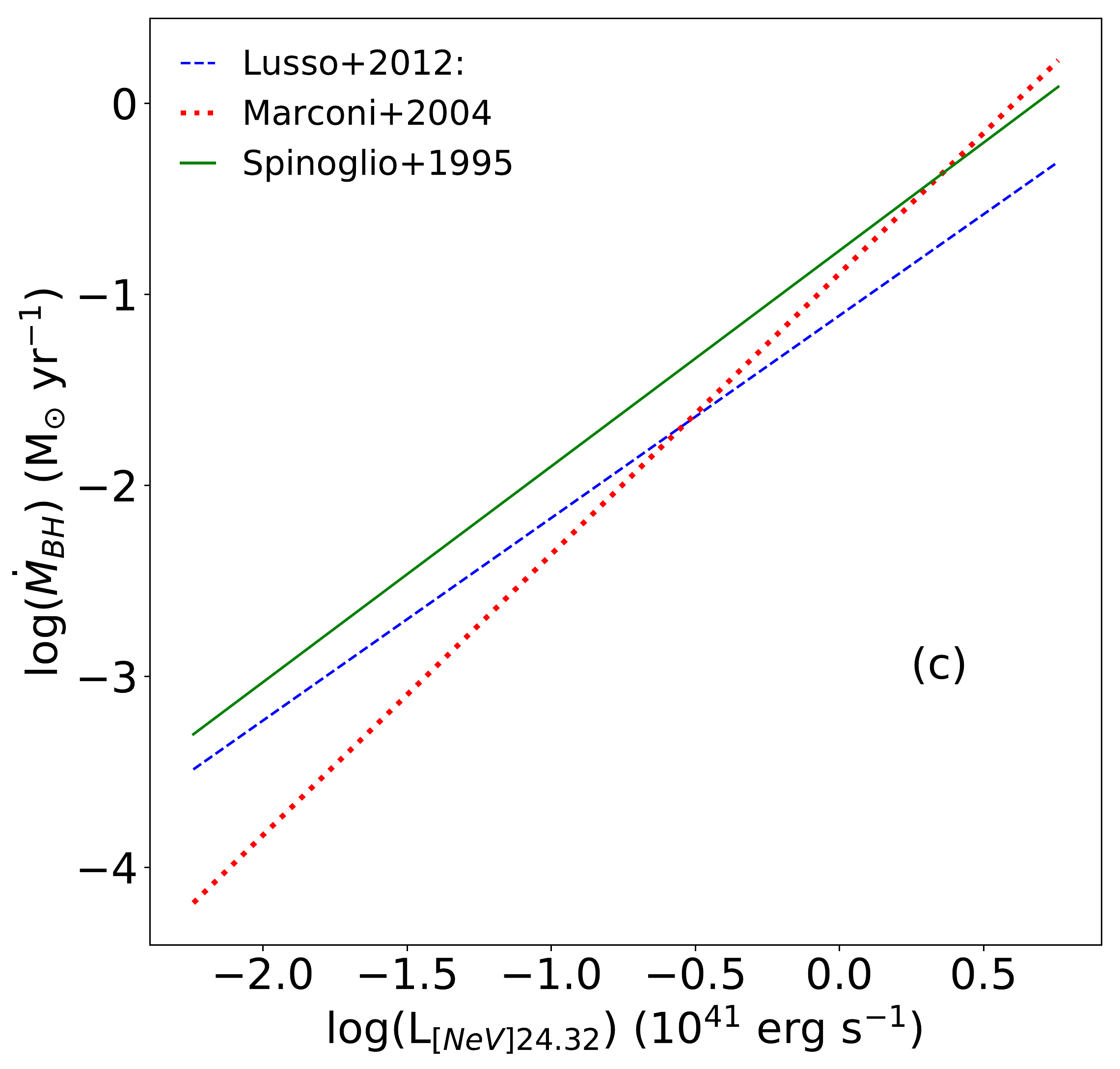}

\caption{{\bf (a: left)} Linear correlation between the [NeV]24.3$\, \rm{\micron}$ line luminosity and the 19$\, \rm{\micron}$ luminosity, expressed in units of $10^{41}\, \rm{erg\,s^{-1}}$. {\bf (b: centre)} Comparison of the three different relations between [OIV]25.9$\, \rm{\micron}$ line luminosity and the BHAR: the blue dashed line reports the relation obtained from the \citeauthor{lusso2012} bolometric correction, the red dotted line the relation from the \citeauthor{marconi2004} bolometric correction, and the green solid line the results from the \citeauthor{spinoglio1995} correction. {\bf (c: right)} Same as panel (a), for the [NeV]24.3$\, \rm{\micron}$ line.}
\label{fig:L_19_correlations}
\end{figure*}

The L$_{19{\mu}m}$ has been used by \citeauthor{tommasin2010} to compute the percentage of AGN and starburst components in 51 sources of their sample. Following these results, we selected all sources with an AGN component at 19$\, \rm{\micron}$ equal or above 85$\%$, and for this sub-sample of 35 objects we determined the correlation between the [OIV]25.9$\, \rm{\micron}$ and the [NeV]24.3$\, \rm{\micron}$ line luminosities with the 19$\mu$m luminosity. These correlations are shown in Fig. \ref{fig:o4_x}c and  Fig.\,\ref{fig:L_19_correlations}a, respectively, and expressed by Eq. \ref{eq:o4_L_19} and \ref{eq:ne5_L_19} respectively, where $r$ indicated the Pearson correlation coefficient, and $n$ the number of objects used to derive the correlation.
\begin{multline}\label{eq:o4_L_19}
   \log\left(\frac{L_{19{\mu}m}}{\rm 10^{41}\,erg\,s^{-1}}\right)=(2.72 \pm 0.06)\\+(0.77 \pm 0.10)\log\left(\frac{L_{\rm [OIV]25.9}}{\rm 10^{41}\,erg\,s^{-1}}\right),\,\,r=0.81,\,\,n=32
\end{multline}

\begin{multline}\label{eq:ne5_L_19}
  \log\left(\frac{L_{19{\mu}m}}{\rm 10^{41}\,erg\,s^{-1}}\right)=(3.12 \pm 0.06)\\+(0.81 \pm 0.10)\log\left(\frac{L_{\rm [NeV]24.3}}{\rm 10^{41}\,erg\,s^{-1}}\right),\,\,r=0.85,\,\,n=30
\end{multline}

We then used these relations, (equations \ref{eq:o4_L_19} and \ref{eq:ne5_L_19},) to determine the bolometric luminosity of our sources, and then the accretion rate. In order to calculate the bolometric luminosity, we used the relation, calculated by \citet{spinoglio1995}, that links the bolometric luminosity to the 12$\, \rm{\micron}$ luminosity. The 12$\, \rm{\micron}$ luminosity for our sample was determined using data from \citet{deo2009}: the authors report continuum measurements at 5.5$\, \rm{\micron}$, 14.7$\, \rm{\micron}$ and 20$\, \rm{\micron}$ taken from {\it Spitzer} low-resolution spectroscopic observations. Where possible, we interpolated the continuum slope and the 12$\, \rm{\micron}$ continuum flux using the 5.5$\, \rm{\micron}$ and 14.7$\, \rm{\micron}$ measurements, otherwise (but only in three cases) using the extrapolation from the 14.7$\, \rm{\micron}$ and 20$\, \rm{\micron}$ fluxes. We then matched the \citeauthor{deo2009} sample with the \citeauthor{tommasin2010} sample, calculated the bolometric luminosity starting from the monochromatic 12$\, \rm{\micron}$ luminosity for those sources with a 85$\%$ AGN component. Similarly to what has been done for the 2-10 keV luminosity, we determined the BHAR starting from the bolometric luminosity derived from the 12$\, \rm{\micron}$ luminosity, and its relation to the [NeV]24.3$\, \rm{\micron}$ and [OIV]25.9$\, \rm{\micron}$ line luminosities, obtaining the following relations:
\begin{multline}\label{eq:o4_AR_ir}
   \log\left(\frac{\dot{M}_{BH}}{\rm M_{\odot}\,yr^{-1}}\right)=(-1.14 \pm 0.07)\\+(0.67 \pm 0.15)\log\left(\frac{L_{\rm [OIV]25.9}}{\rm 10^{41}\,erg\,s^{-1}}\right),\,\,r=0.68,\,\,n=26
\end{multline}

\begin{multline}\label{eq:ne5_AR_it}
  \log\left(\frac{\dot{M}_{BH}}{\rm M_{\odot}\,yr^{-1}}\right)=(-0.77 \pm 0.10)\\+(0.77 \pm 0.16)\log\left(\frac{L_{\rm [NeV]24.3}}{\rm 10^{41}\,erg\,s^{-1}}\right),\,\,r=0.71,\,\,n=24
\end{multline}

The smaller number of objects used to derive these relations is due to the lack of data for the determination of the continuum at 12$\, \rm{\micron}$. 

Measuring the BHAR requires important approximations in terms of bolometric correction, which can yield significantly different results. Between the three proposed bolometric corrections, we note that the correlations obtained when applying the Marconi correction are steeper than those obtained from the Lusso and Spinoglio correction. In particular, we compare the results in Fig.\,\ref{fig:L_19_correlations}: for both the [OIV]25.9 $\mu$m  (panel b, at the centre) and [NeV]24.3$\, \rm{\micron}$ (panel c, on the right) lines, we note that the results derived from the corrections of \citet{spinoglio1995} and \citet{lusso2012} show a similar, flatter slope, indicating, within the errors, the expected linear relation between the line tracers and the BHAR.
It is important to note that the \citeauthor{marconi2004} and \citeauthor{lusso2012} bolometric corrections are based on a third-degree polynomial transformation of the X-ray luminosity. This necessarily leads to deviations from the slopes obtained in Eq.\,\ref{eq:X_o4} and \ref{eq:X_ne5}.

We refer to the Appendix \ref{app:bhar} for a discussion on the use of the  mid-ionisation lines of [NeIII]15.5 $\mu$m and [SIV]10.5 $\mu$m as alternative BHAR tracers.

\section{Discussion}
\label{sec:discussion}
\subsection{Applicability of our results to composite objects}

\begin{figure}
    \centering
    \includegraphics[width=\columnwidth]{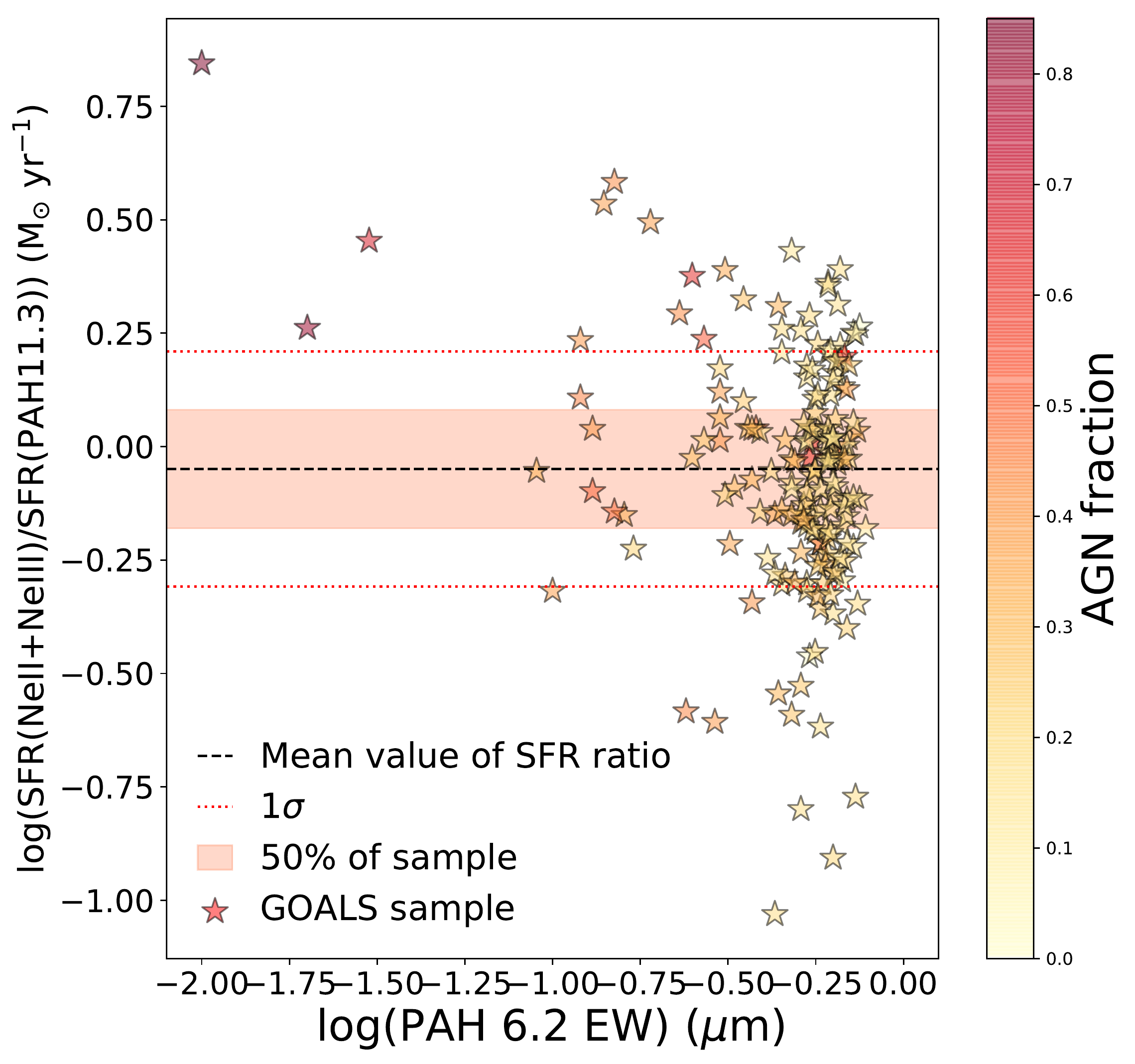}
    \caption{Comparison of the ratio between the SFR obtained using the [NeII]+[NeIII] tracer and the SFR obtained using the PAH feature at 11.3$\mu$m, against the EW of the PAH feature at 6.2$\mu$m for the GOALS sample. The colour gradient indicates the percentage of the AGN component in each object as determined by mid-IR tracers \citep{diazsantos2017}. The black dashed line shows the mean value of the SFR ratio for the population, with the shaded area indicating where the 50$\%$ of the entire sample is located. The red dotted lines represent the 1$\sigma$ interval around the mean value.}
    \label{fig:mixed_obj}
\end{figure}

The coexistence of AGN and star formation in galaxies is well known \citep[see, e.g.][for a review]{perez-torres2021} and therefore we expect that many galaxies are characterised by both components at work together. 
In general, when considering composite objects, with both a SF and an AGN component detectable, our results in Section\,\ref{sec:SFR} and Section\,\ref{sec:BHAR} can still be applied.

We use the GOALS sample to determine whether the results on the SFR tracers can be applied to composite objects (Section\,\ref{sec:SFR}). The GOALS sample is composed of SFG, but for part of the sample an AGN component is also present, if not always detected. For our calibrations, we have excluded all sources for which a [NeV] emission line is detected. Here we instead include all objects of the sample. In Fig.\,\ref{fig:mixed_obj} we plot the ratio of the SFR obtained using the [NeII]+[NeIII] tracer to the SFR obtained using the 11.3$\mu$m PAH feature, as a function of the Equivalent Width (EW) of the PAH feature at 6.2$\mu$m, as determined by \citet{stierwalt2014}. The mean ratio for the entire population is -0.049, with the median value equal to -0.023 and 50$\%$ of the sample included in the [-0.18, 0.06] interval. While there is some dispersion around zero, for the majority of the sample we obtain a similar SFR whether we use the [NeII]+[NeIII] tracer or the PAH tracer. This suggests that for mixed objects these two tracers are equivalent, and independent of the presence of an AGN. Significant differences are observable for the extreme objects in the sample: the sources with the larger AGN content are located in the top left corner, for which the SFR obtained using the [NeII]+[NeIII] tracer is significantly higher than the SFR obtained using the PAH. This is plausibly due to an increase of the [NeIII] emission related to the presence of the AGN.

\subsection{Comparison among the various tracers}\label{sec:best}

Summarising the results presented in Section\,\ref{sec:SFR}, among the various SFR tracers presented, for galaxies dominated by SF processes we suggest that the best tracers are the [CII]158$\mu$m line (slope $ \alpha = 0.89 \pm 0.02$ and correlation coefficient $r$ = 0.92) or the combination of the [OI] and [OIII] lines (slope $ \alpha = 1.25 \pm 0.04$ and $r$ = 0.94) for high-redshift galaxies observed from the ground with submillimeter telescopes.
For galaxies observed from space or airborne facilities (but also from the new generation of the very large optical ground based telescopes in the 8-13$\mu$m atmospheric window, for local galaxies), the best SFR tracers are the combination of the [NeII] and [NeIII] lines (slope $\alpha = 0.96 \pm 0.03$ and $r$ = 0.91) or the combination of the other mid-IR fine-structure lines. For galaxies containing an AGN, the PAH features can reliably be used and the PAH 11.3$\mu$m feature is probably the best SFR tracer (slope $\alpha = 0.75 \pm 0.04$ and $r$ = 0.79). As an alternative, the [NeII] and [NeIII] lines can still be used also in AGN, with the correction to the total neon flux that can be computed using one of the [NeV] lines, as suggested by \citet{zhuang2019}.

When measuring the BHAR, following the results presented in Section\,\ref{sec:BHAR}, the [NeV] lines at 14.3 and 24.3$\mu$m are exclusive probes of AGN activity, because their emission is a direct signature of the hard ionising spectrum due to the accretion process. However, these lines are fainter than the [OIV]25.9$\mu$m line, which can be detected more easily in faint objects. When using the [OIV] line as a BHAR tracer, it is important to keep into account the possible contamination due to SF processes, because its emission can be attributed also, to some extent, to starburst activity \citep{lutz1998}.

\subsection{Comparison with previous line calibrations}\label{sec:comp}

We briefly compare here the results of the line calibrations obtained in this work with those of  \citet{spinoglio2012,spinoglio2014} and \citet{gruppioni2016}, while leaving all the details of the comparisons to Appendix\,\ref{app:comparison}. We note that in this work, to derive the correlations, we used the orthogonal distance regression method, instead of the least-squares minimisation method, which was used by the other authors, because we consider the total IR luminosity and the line luminosity two independent variables.
%which keeps into account that measurement errors are present in both the dependent and independent variables, while the considered works for comparison use ordinary least square regression, which considers the error only in the dependent variable.

The AGN sample used in this work and in \citeauthor{spinoglio2012} is the same for the lines in the 10-35 $\, \rm{\micron}$ interval, and the differences in the correlation are only due to the different methods of analysis. We note that in this work, in order to obtain a correlation between the total L$_{IR}$ and the [OIV]25.9$\, \rm{\micron}$ line luminosity that better represents the AGN population, we did not use the entire sample of AGNs by \citet{tommasin2008,tommasin2010}, but a sub-sample of objects with an AGN component in the 19$\, \rm{\micron}$ luminosity of at least 85$\%$ (see Section\,\ref{sec:cor}).

For the SFG sample, while in \citeauthor{spinoglio2012} the sample of galaxies described by \citet{bernardsalas2009} was used, in this work we have expanded the same sample by including the LIRGs and ULIRGs sample of GOALS, as described in Section\,\ref{sec:data}. Nonetheless, we obtain comparable results, except in the cases of the PAH feature at 11.3$\, \rm{\micron}$ and of the H$_{2}$ line at 17.03$\, \rm{\micron}$, for which our results show line luminosities an order of magnitude higher. This is due to the presence, in our sample, of LIRGs and ULIRGs, which shift the relation toward a steeper slope.

For the lines at wavelengths in the 50-158 $\mu$m range, we refer to the Appendix \,\ref{app:comparison}, where we discuss the differences and plot the results of the different calibrations of \citeauthor{spinoglio2012} with respect to our results.

Because of the different analytical method used to determine the correlations between total and line luminosities, 
for the comparison with \citeauthor{spinoglio2012}, we have recomputed our calibrations using the least square fit and give the results in the Appendix \,\ref{app:comparison}. For the comparison with \citeauthor{gruppioni2016}, instead, we recomputed the correlations of the \citeauthor{gruppioni2016} sample, applying the orthogonal distance regression and then compare these with our results. %Given the available data, we only applied the method to the lines in the 10-35$\mu$m interval and the PAH features at 6.2 and 11.3$\mu$m. 
As a general trend, we find good agreement between our results and those by \citeauthor{gruppioni2016}, with comparable slopes within 2$\sigma$ of each other. For all the details and the plots of the results of the different calibrations, we refer to the Appendix \,\ref{app:comparison}.

We do not compare our results with those obtained by \citet{bonato2019}, since the methods of analysis are widely different. In particular, while in this work we calibrate the line luminosities leaving the slope of the correlations as a free parameter, in \citet{bonato2019} the slope of the relation was fixed to unity, thus giving raise to substantial differences in the results.

\subsection{Metallicity and SFR tracers}
\label{sec:metallicity_discussion}
In Section\,\ref{sec:SFR-CII} we revise the [CII]-SFR relation for a wide galaxy sample, from LMG to extreme ULIRGs. In Section\,\ref{sec:sfr_ne_s} we derive a measure of the SFR through the neon and sulfur mid-IR lines and propose new SFR tracers using different combinations of these lines. In this section we discuss the possible effects that metallicity, and the associated changes in the ISM of these galaxies, may have on these tracers.

First of all, for the sample of dwarf galaxies, we adopted SFR values derived from the observed H$_\alpha$ luminosity and corrected from the total IR luminosity \citep{remyruyer2015}. This is motivated by the underestimation of the SFR by the IR luminosity at very low metallicities ($12 + \log(O/H) \lesssim 8.5$; \citealt{lee2013}), due to the lower metal abundance in these galaxies compared to SFG. In principle, the lower dust to gas ratio of LMG should also have an impact on the observed intensities of the fine-structure lines. This is, however, balanced by the higher cooling rates in these transitions, as discussed by \citet{delooze2014}. 

Fig.\,\ref{fig:SFR1}a shows that the [CII]$158\, \rm{\micron}$ emission in LMG follows the trend found in SFG with no need to perform any additional correction for metallicity in these galaxies. Similarly, the different combinations of [NeII]$12.8\, \rm{\micron}$, [NeIII]$15.6\, \rm{\micron}$, [SIII]$18.7\, \rm{\micron}$, and [SIV]$10.5\, \rm{\micron}$, shown in Figs.\,\ref{fig:SFR1} and \ref{fig:SFR2}, follow the correlation of solar-like metallicity galaxies. The higher cooling rates in LMG are particularly evident when the neon and sulfur transitions are considered. While the [NeII]$12.8\, \rm{\micron}$ emission in LMG scales with $L_{\rm IR}$ similarly as for SFG (Fig.\,\ref{fig:corr_c2_ne2_ne3}b), the [NeIII]$15.6\, \rm{\micron}$ line becomes comparatively much brighter for a given IR luminosity, more than one order of magnitude above the correlation found for SFG, as can also be seen by the value of the constant $b$ in the best fit equation for the [NeIII]15.6$\, \rm{\micron}$ line in Table \ref{tbl_lines}. When considering the sulfur lines, this effect is even more pronounced (see Fig.\,\ref{fig:corr_app_4}a and Table\,\ref{tbl_lines}). This means that mid- to high-ionisation species such as Ne$^{2+}$ or S$^{3+}$ trace a contribution to the star-formation that is not revealed by either the Ne$^+$ and $S^{2+}$ low-ionisation gas or the IR emission. Thus, the combination of low and high-ionisation lines allows us to trace the total star formation in both low- and solar-metallicity galaxies \citep{ho2007,zhuang2019}.

In the case of the [CII]158$\, \rm{\micron}$ line, the Fig.\,\ref{fig:SFR1}a suggests that this transition still remains a dominant coolant of the ISM at low metallicities. Given its low ionisation potential (11.3 eV, see Table\,\ref{tbl_lines}), this line can originate from both neutral and ionised gas, and one could expect a decreasing contribution from the neutral component as the ionisation field becomes harder at low metallicities. However, \citet{cormier2019} demonstrated that the PDR contribution to the global [CII]$158\, \rm{\micron}$ emission is still dominant for the same LMG sample used in this work. This is also in line with the results of \citet{croxall2017}, suggesting that the [CII]$158\, \rm{\micron}$ emission linked to ionised gas is of the order of $\sim 10\%$ in LMG, and up to maximum $<40\%$ where a high-U is required. Moreover, analytical models developed in these studies show a decrease of [CII] emission from ionised gas with decreasing metallicity with $\sim 55-75\%$ of the [CII] emission arising from PDRs in LMG, reaching almost 100$\%$ when the metallicity decreases below 1/4 Z$_{\odot}$. Additionally, the thickness of the [CII] layer increases for molecular clouds exposed to the harder radiation fields typical of LMG. This is shown by the detection of higher [CII]/CO(1-0) ratios in local LMG when compared to SFG with solar or super-solar metallicity \citep[e.g.][]{madden1997,madden2000,hunter2001}, and it is supported by PDR models \citep{bolatto1999,rollig2006}.

While a detailed study of the ionised gas and PDR structure is out of the scope of the present work, the results discussed above suggest that both [CII]$158\, \rm{\micron}$ and the different combinations of neon and sulphur lines are robust star formation tracers, virtually independent of dust extinction, that can be applied to a wide diversity of environments with different physical conditions and metallicities. Specifically, the variations expected from the changes in the chemical abundances are mostly balanced by the increase in the cooling rates of these transitions.

\subsection{Comparison with high-z data}\label{sec:high-z_data}

\begin{figure*}
\centering
\includegraphics[width=0.67\columnwidth]{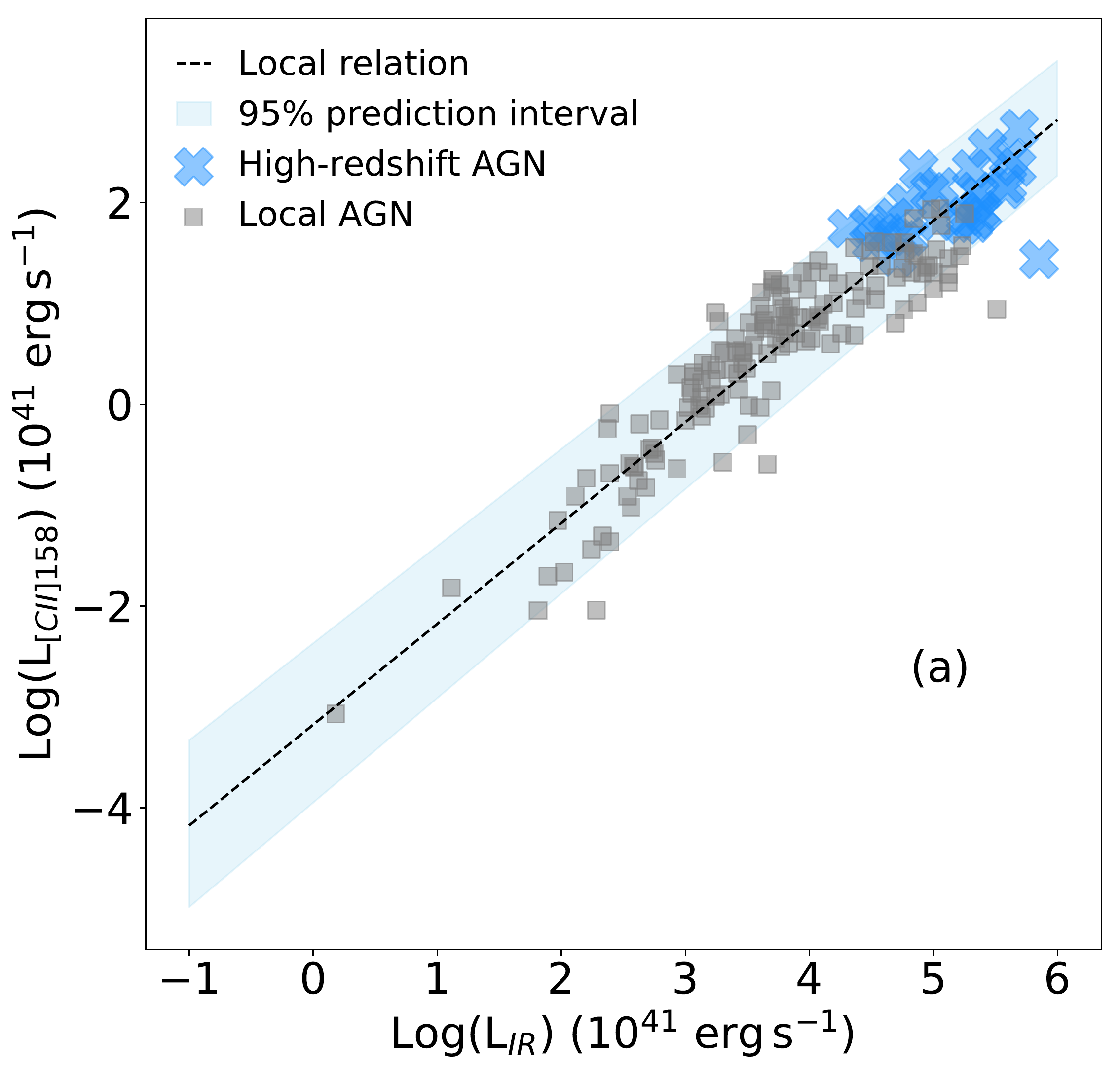}~
\includegraphics[width=0.67\columnwidth]{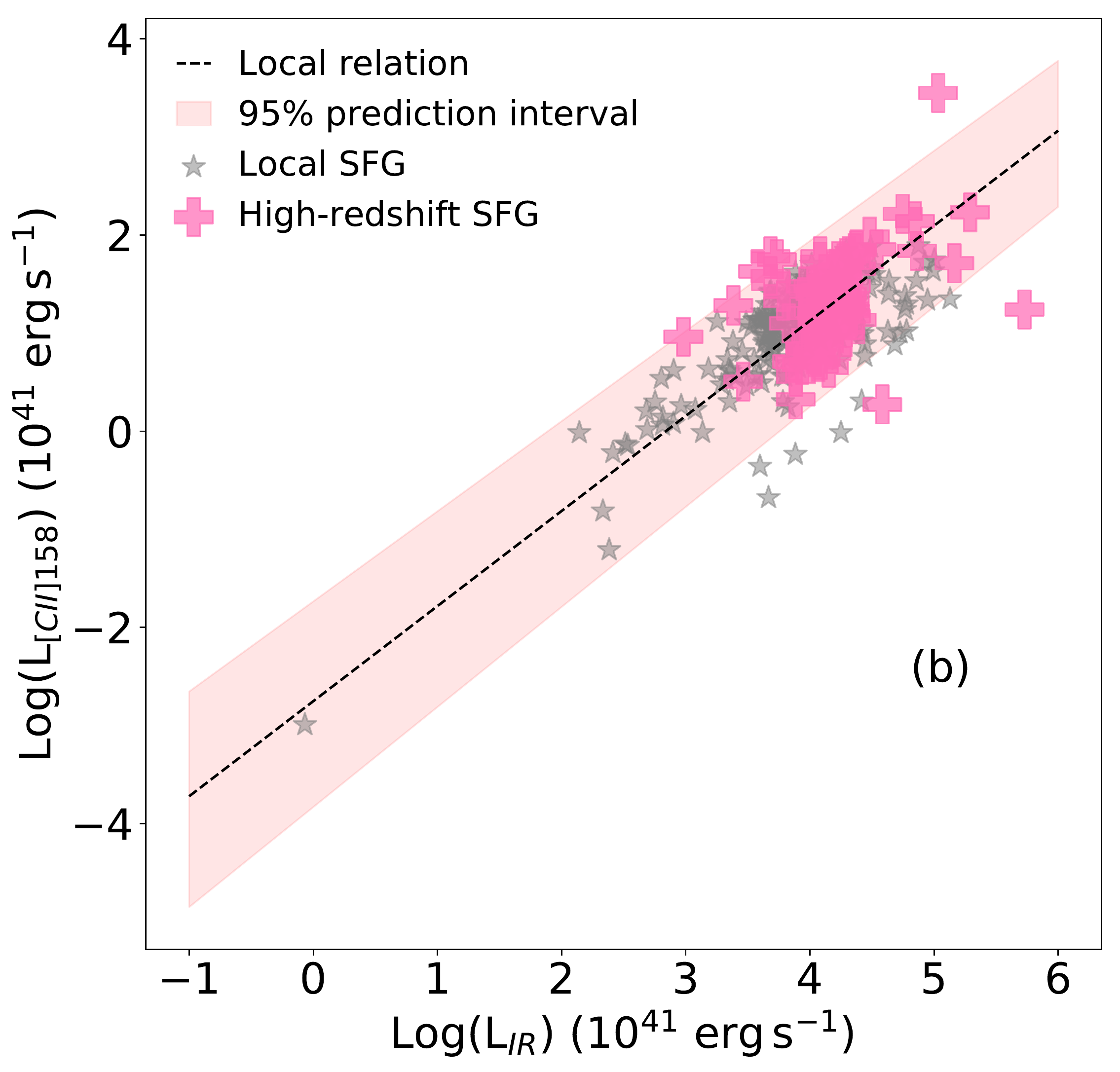}
\includegraphics[width=0.67\columnwidth]{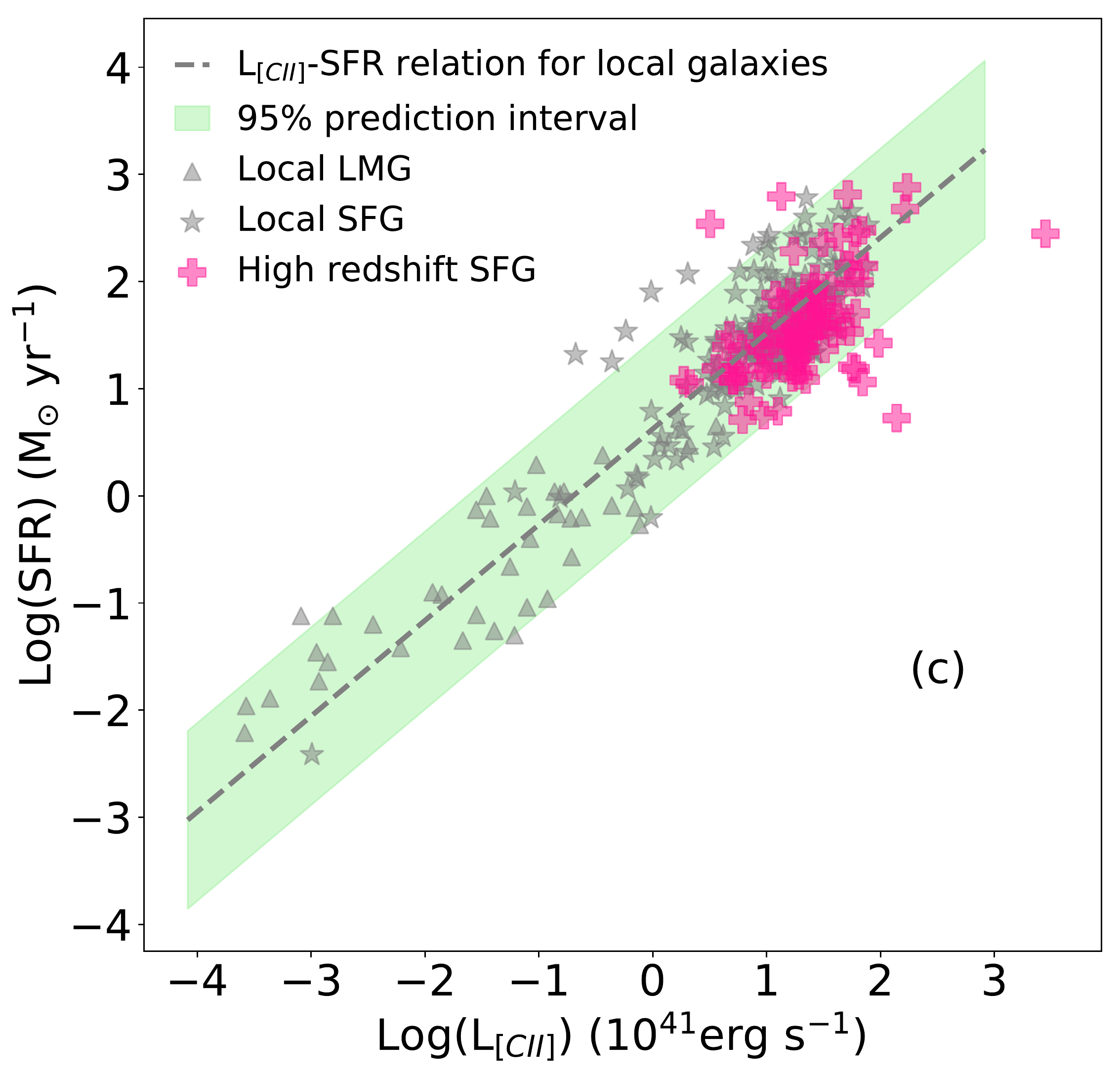}
\caption{{\bf (a: left)} Comparison between the L$_{IR}$-L$_{[CII]158}$ relation (black dashes line) for local AGN dominated galaxies (gray squares) and high redshift detections for QSOs (blue symbols). The shaded area shows the 95$\%$ prediction interval. {\bf (b: centre)} Same for high redshift starburst galaxies (pink symbols). {\bf (c: right)} Comparison of the local log(L$_{[CII]}$-log(SFR) relation (black dashes line) and L$_{[CII]}$-SFR values of high redshift sources (pink symbols). Grey dots represent local star forming galaxies. The shaded area shows the 95$\%$ prediction interval for the local relation.}
\label{fig:c2_158_hz}
\end{figure*}
\begin{figure*}
\centering
\includegraphics[width=0.67\columnwidth]{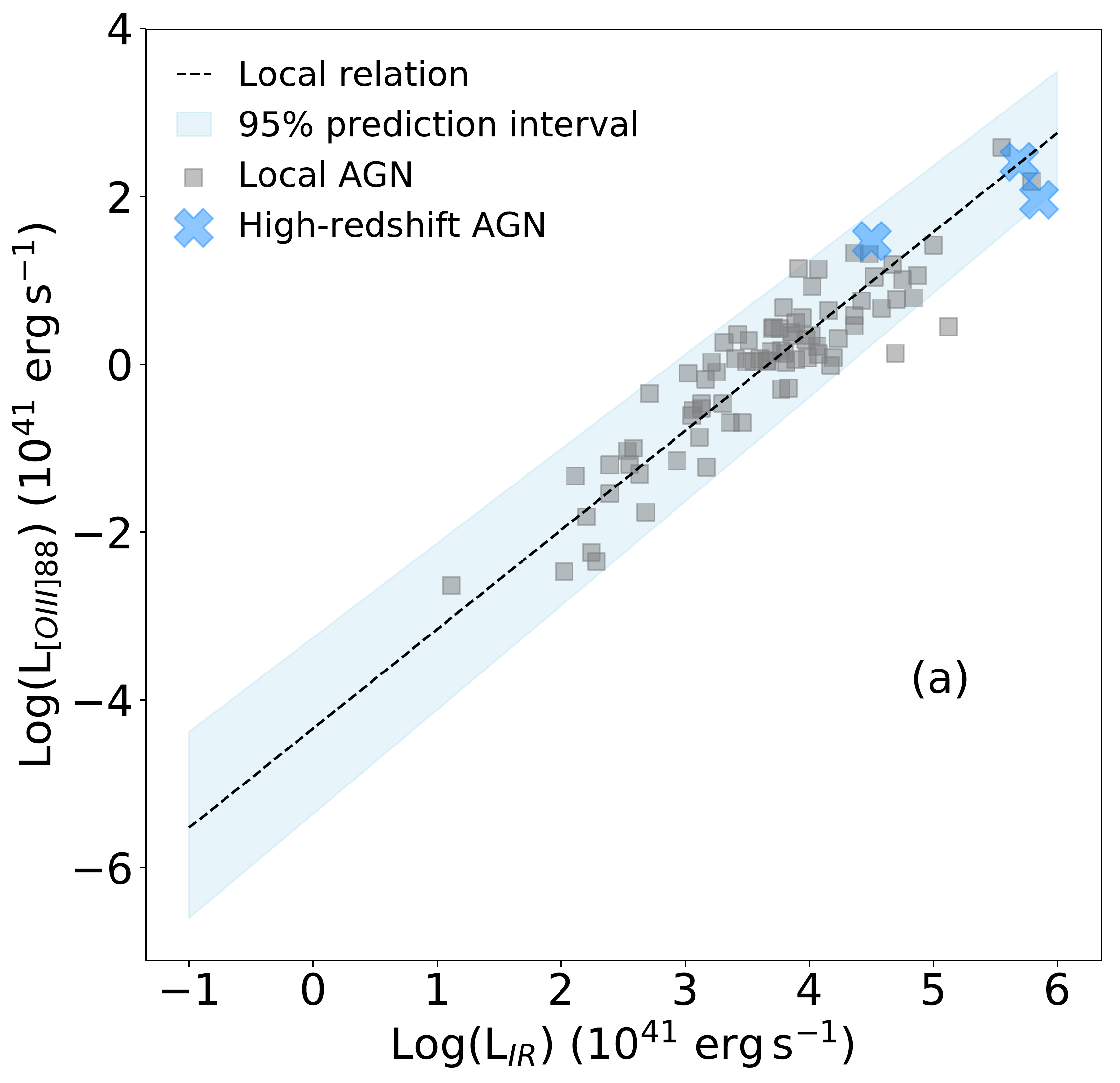}~
\includegraphics[width=0.67\columnwidth]{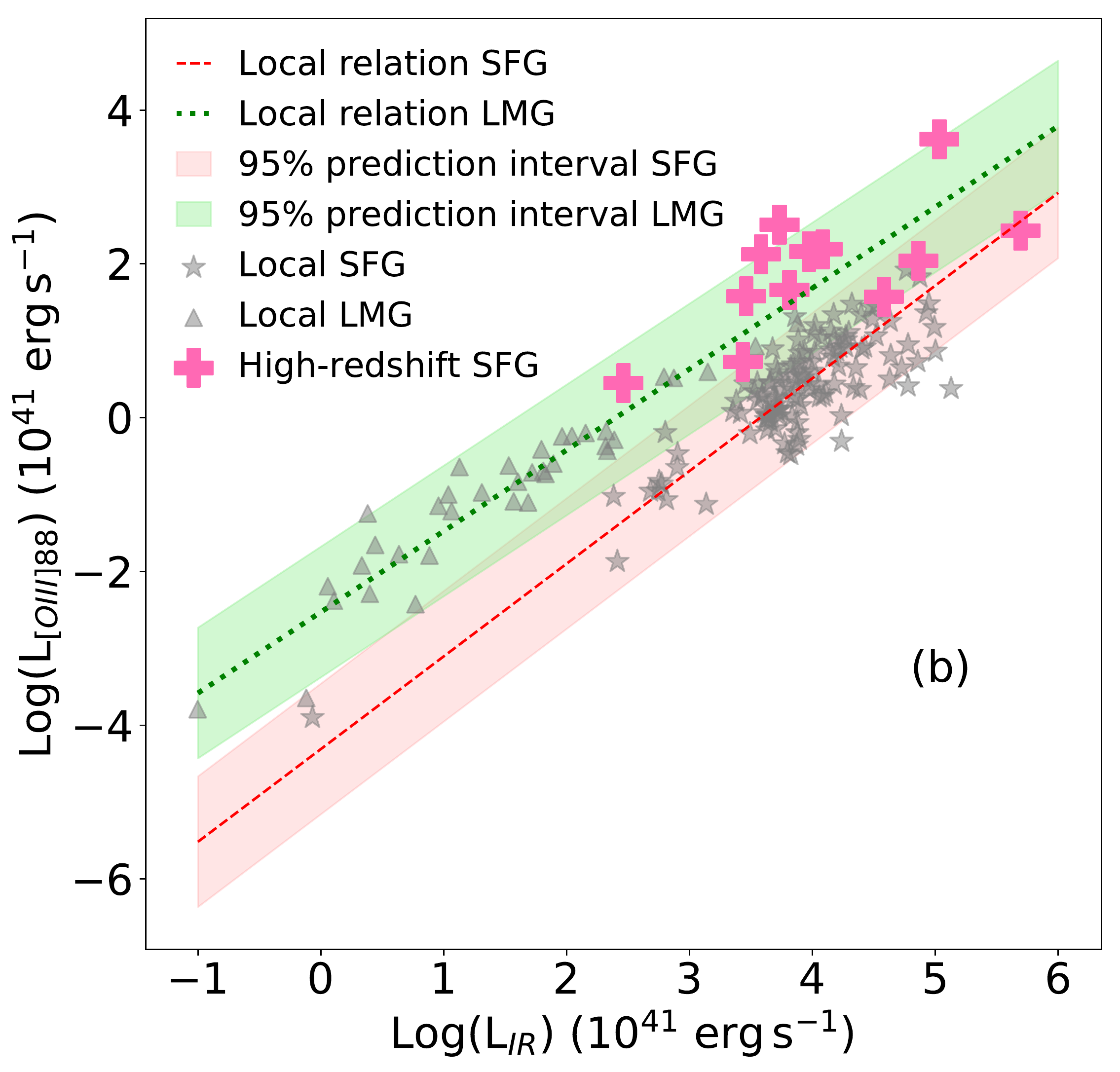}
\includegraphics[width=0.67\columnwidth]{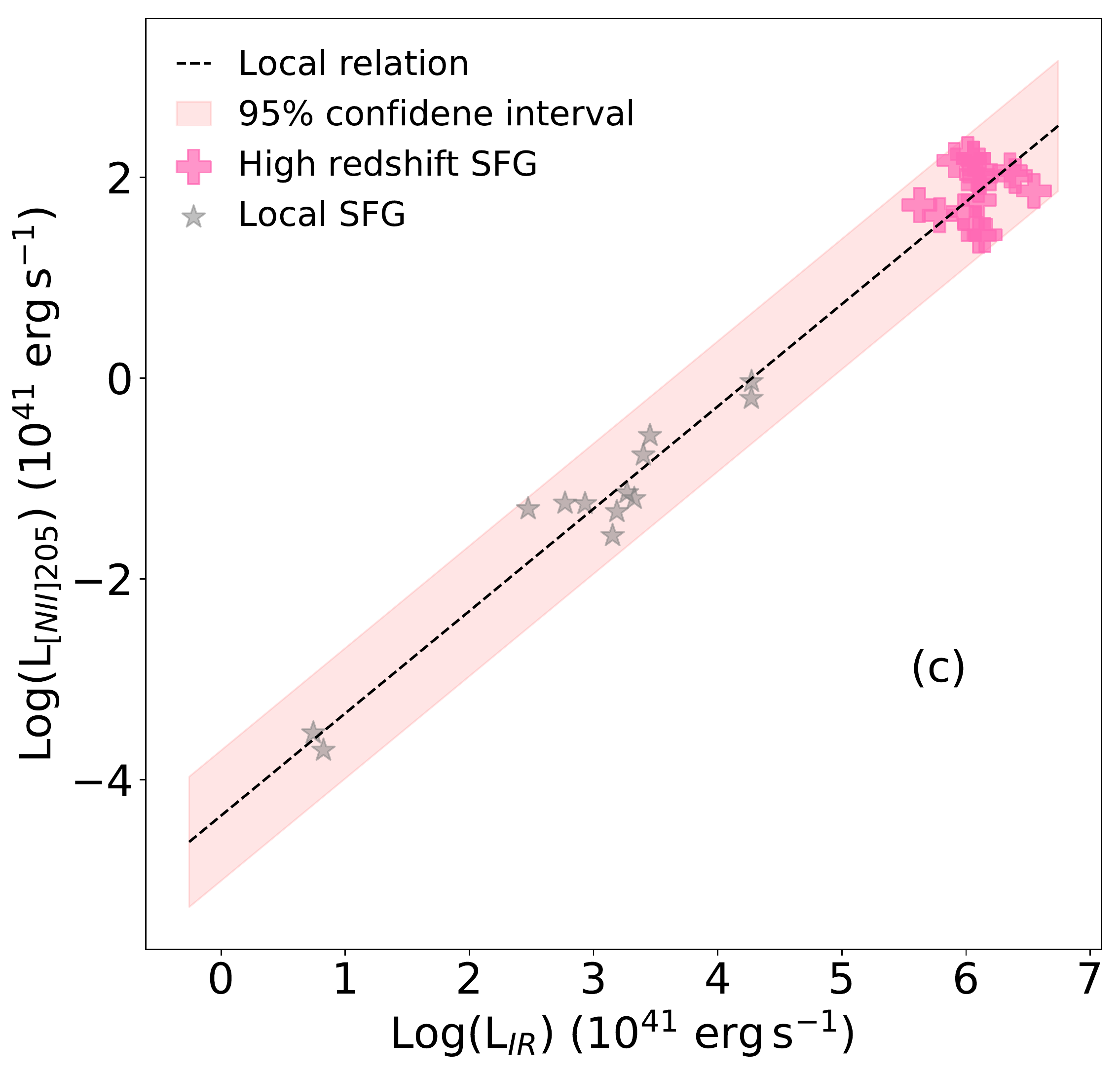}
\caption{{\bf (a: left)} Comparison between the L$_{IR}$-L$_{[OIII]88}$ relation (black dashes line) for local AGN dominated galaxies (gray squares) and high redshift detections for QSOs (blue symbols). {\bf (b: centre)} Comparison between the L$_{IR}$-L$_{[OIII]88}$ relation (red dashes line) for local SFG (gray stars) and high redshift detections for starburst galaxies (pink symbols). The red shaded area shows the 95$\%$ prediction interval of this relation. For comparison, the green dashed line shows the L$_{IR}$-L$_{[OIII]88}$ relation for local LMG, with gray squares indicating local LMG and the green shaded area giving the 95$\%$ prediction interval for this relation. {\bf (c: right)} Comparison of the local L$_{IR}$-L$_{[NII]205}$ relation (black dashes line) for high redshift detections of SFG (pink symbols). Grey stars represent local star forming galaxies. The shaded area in all the figures shows the 95$\%$ prediction interval for the local relation.}
\label{fig:o3_n2_hz}
\end{figure*}

We compare the calibrations described in Section\,\ref{sec:cor} obtained for local galaxies to high-redshift detections of sources obtained with ALMA. In particular, we consider detections of the [OIII]88$\, \rm{\micron}$, [NII]122$\, \rm{\micron}$, [CII]158$\, \rm{\micron}$ and [NII]205$\, \rm{\micron}$ lines at $z\geq3$. Sources identified as QSOs are compared to local AGN results, while sources for which a classification is not given in the literature, or are classified as starburst galaxies, are compared to local SFG.

Fig.\,\ref{fig:c2_158_hz}a shows the comparison between local and high redshift detections of the [CII]158$\, \rm{\micron}$ line for QSOs galaxies, and in particular 27 detections at z$\sim$6 reported in \citet{venemans2020}, plus the detections by \citet{walter2018} for one source at z$\sim$6.08 and by \citet{hashimoto2018} for two sources at z$\sim$7.1, for a total of 30 sources.
Fig.\,\ref{fig:c2_158_hz}b displays the comparison of local and high redshift SFG: 84 detections from the ALPINE catalogue \citep{faisst2020} plus other 9 detections \citep{inoue2016,vishwas2018,walter2018,debreuck2019,hashimoto2019,harikane2019,rybak2019} for a total of 93 objects with redshift in the 4.2$\lesssim$z$\lesssim$7.2 range. While some outliers are present, the bulk of the detections in both cases lies within the prediction interval, that we show in the figures at the 95\% level, suggesting that the relations derived for local galaxies hold for high redshift sources.

We compare in Fig.\,\ref{fig:c2_158_hz}c the L$_{[CII]}$-SFR relation extrapolated for local galaxies (see Section \ref{sec:SFR-CII}) with ALPINE detections of the [CII]158$\, \rm{\micron}$ emission line in starburst galaxies. We find that high redshift objects follow the same L$_{[CII]}$-SFR relation as local SFG and LMG, thus suggesting little or no evolution in the L$_{[CII]}$-SFR relation. In particular, no deficit of [CII] emission is seen for the highest SFR values of $\sim 100\, \rm{M_\odot \, yr^{-1}}$, suggesting that starbursts at high-z might not behave like local ULIRGs such as those shown in Fig.\,\ref{fig:SFR1} (purple squares). An analogues result was found by \citet{schaerer2020}, who analysed a large sample of galaxies at high redshift (z$\sim$ 4-6) observed by ALMA. \citet{leung2020} explored the possibility of tracing the SFR at the Epoch of Recombination with [CII]158$\, \rm{\micron}$ using simulated data. Although they find a good correlation, when compared to observed data, the simulated sample shows an average [CII] luminosity lower than the one obtained from ALMA pointed observations. A similar result was obtained for simulations of high redshift systems \citep{ferrara2019} showing the presence of under-luminous [CII] emission.

In Fig.\,\ref{fig:o3_n2_hz}a we show the comparison between the local L$_{IR}$-L$_{[OIII]88}$ relation for AGN and high redshift detections in QSOs \citep{hashimoto2018,walter2018}, while panel (b) shows the relation between local and high redshift SFG  \citep{inoue2016,vishwas2018,walter2018,debreuck2019,harikane2019,hashimoto2019,tamura2019}.
We note that, while the high-redshift detections of [OIII]88$\, \rm{\micron}$ in QSOs are comparable with the relation derived from local AGN dominated galaxies, in SFG the majority of high redshift detections appears to be one to two orders of magnitude brighter at comparable L$_{IR}$. These detections are better represented by the  L$_{IR}$-L$_{[OIII]88}$ relation derived for local LMG (see Fig.\,\ref{fig:o3_n2_hz}b). This suggests that the [OIII]88$\, \rm{\micron}$ line is produced in an environment with a  higher ionising spectrum, similar to the ISM of local LMG (see Section\,\ref{sec:cor}).

Finally, the high redshift detections in panel (c) show the local L$_{IR}$-L$_{[NII]205}$ relation for SFG and 17 high-redshift detections of [NII]205$\, \rm{\micron}$ \citep{cunningham2020} with the local correlation that well represents also the high redshift detections. We report in Fig.\,\ref{fig:corr_app_8} the comparison between the local L$_{IR}$-L$_{[NII]122}$ relation and high redshift SFG: in this case only one object \citep{debreuck2019} has been detected, while three \citep{harikane2019} objects have upper limits. Thus, only one object is located almost an order of magnitude higher than the local relation for SFG, and therefore there is no statistical evidence for a difference in the production mechanism of [NII] in high redshift objects, especially considering the results for the [NII]205$\, \rm{\micron}$ line.

\subsection{Application to present and future facilities} 
\label{sec:jwst_alma}
\begin{figure*}
    \centering
    \includegraphics[width=1\columnwidth]{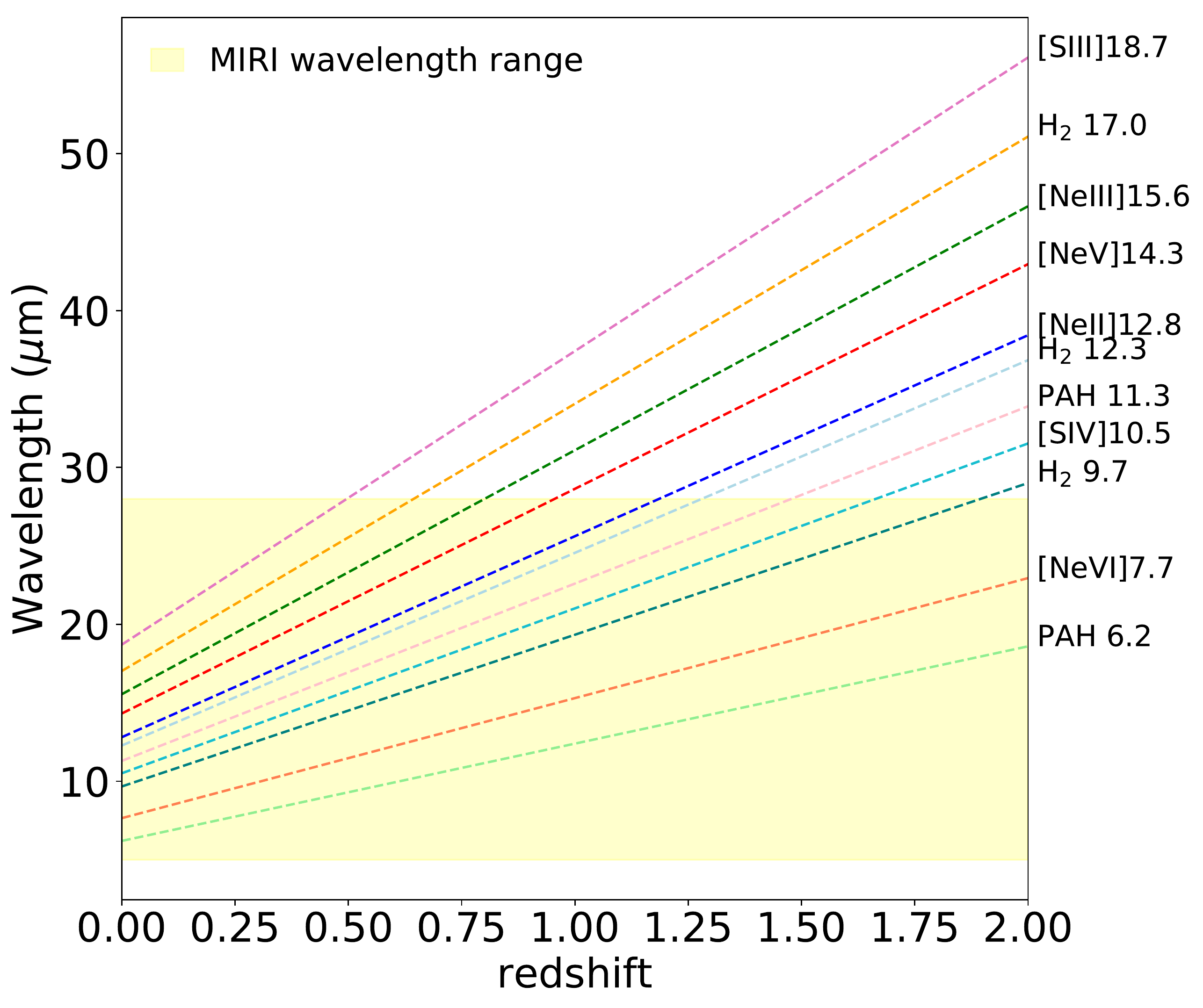}
    \includegraphics[width=1\columnwidth]{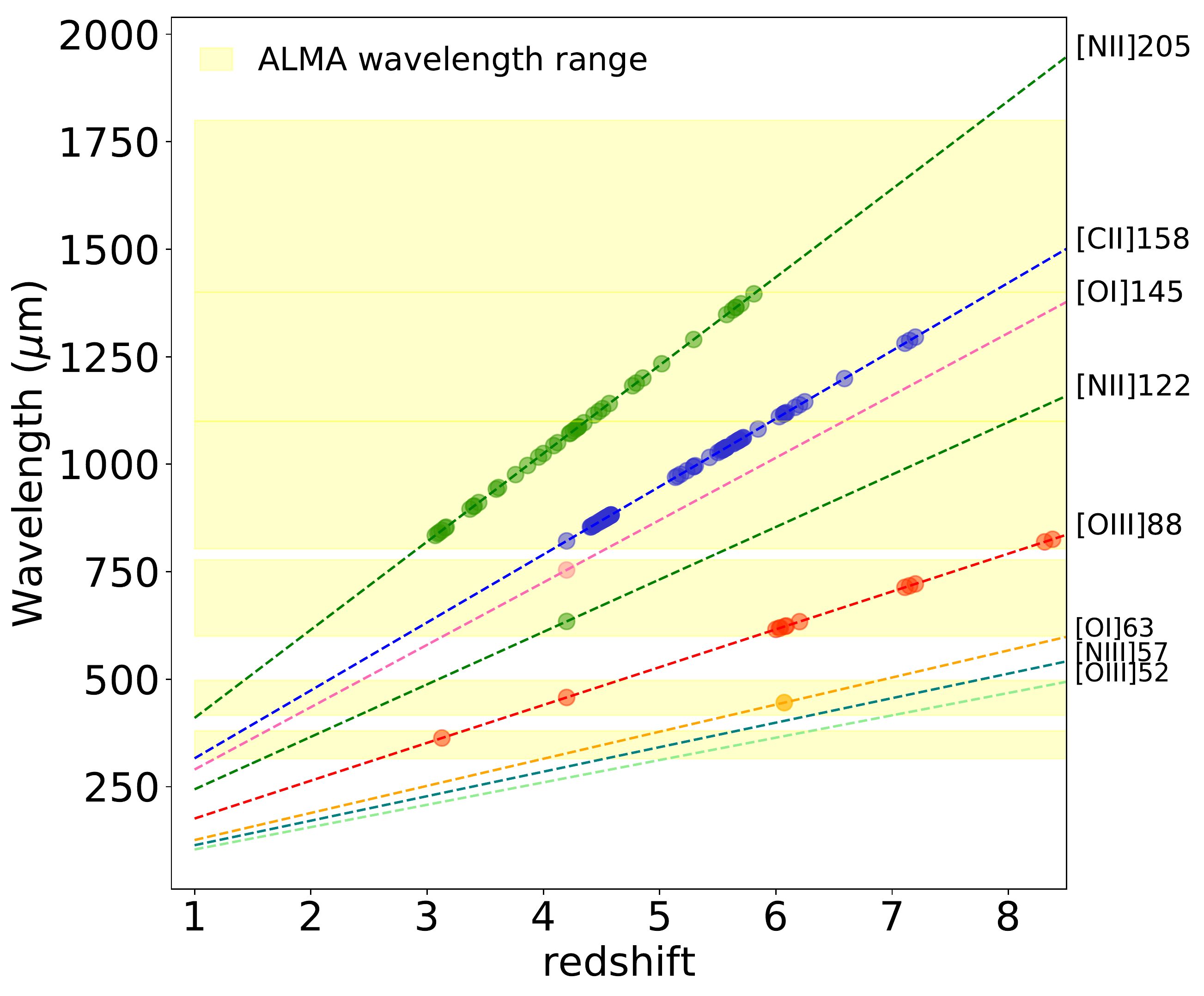}
    \caption{(Left) Observability of key mid-IR (dashed lines) lines compared to the \textit{JWST}-MIRI instrument (shaded area) in the 0-1.6 redshift interval. (Right) Observability of key far-IR (dashed lines) lines compared to the ALMA bands (shaded area) in the 1-8 redshift interval. Dots represent current detection for each line.}
    \label{fig:miri}
\end{figure*}

The lines and features presented in this work can be observed in galaxies by present and future IR/(sub)-mm facilities. The \textit{JWST} will be the next NASA observatory to explore the Universe in the near- and mid-IR spectral range. In particular, the \textit{JWST} Mid-InfraRed Instrument \citep[MIRI,][]{rieke2015,wright2015} will be able to obtain imaging and spectra with unprecedented sensitivity in the $4.9$--$28.9\, \rm{\micron}$ spectral range. In Fig.\,\ref{fig:miri}a we compare the MIRI wavelength interval to the observability of key mid-IR lines at various redshifts. In particular, from the analysis presented in Section\,\ref{sec:SFR}, it will be possible to study the SFR up to redshift z$\sim$3.5 with the PAH feature at 6.2$\, \rm{\micron}$, up to 
redshift z$\sim$1.5 with the PAH feature at 11.3$\, \rm{\micron}$, or up to z$\sim$0.8 using combinations of the neon and sulfur lines. 

On the other hand, while we do not report in Fig.\,\ref{fig:miri}a the [NeV]24.3 and [OIV]25.9$\, \rm{\micron}$ lines due to the limited wavelength range available for \textit{JWST}, we show the [NeV]14.3$\, \rm{\micron}$ line, for which the same analysis carried out in Section\,\ref{sec:BHAR} to derive the BHAR is applied and reported in Appendix \ref{app:bhar}. The [NeV]14.3$\, \rm{\micron}$ line will be observed by \textit{JWST}-MIRI up to redshift z$\lesssim$1, while the [NeV]24.3$\, \rm{\micron}$ line will be observed up to redshift z$\lesssim$0.15, and the [OIV]25.9 $\, \rm{\micron}$ line will be observed up to redshift z$\lesssim$0.1.

While \textit{JWST}-MIRI will be able to observe molecular and atomic lines up to redshift z$\sim$2, and the PAH up to higher redshift (z$\sim$3.5 for the 6.2$\mu$m feature), ALMA is able to trace the far-IR lines, like [CII]158$\, \rm{\micron}$, up to redshift z$\sim$8, as shown in Fig.\,\ref{fig:miri}b, providing important information on the evolution of galaxies with cosmic time. An analogues study can be found in Fig.\, 1 of  \citet{carilli2013}, where the CO transitions and other key tracers of the ISM are shown as a function of redshift versus frequency in terms of the observability by ALMA and JVLA \citep[Karl J. Jansky Very Large Array, ][]{perley2011}.
We also report in Fig.\,\ref{fig:miri}(b) the redshift of currently available data of ALMA detections: while lines like [CII]158$\, \rm{\micron}$ or [NII]205$\, \rm{\micron}$ are detected over a significant redshift interval, shorter wavelength lines, such as [OI]63$\, \rm{\micron}$, present very few results. This is due to the difficulty of observing in the bands at higher frequency, for which the atmospheric absorption allows good visibility only for 10$\%$ of the total observational time.

In a near future the extremely large ground-based telescopes under construction, with dedicated instruments for \textit{N}-band spectroscopy will be able to obtain observations of the mid-IR spectral range 8--13 $\mu$m and thus will test the [SIV]$_{\rm 10.5 \mu m}$ + [NeII]$_{\rm 12.8 \mu m}$ relation, for local galaxies and AGN. The ESO Extremely Large telescope (E-ELT; \citealt{gilmozzi2007}) is expected to obtain first light at the end of 2025, with the Mid-infrared ELT Imager and Spectrograph \citep[METIS, ][]{brandl2021} instrument covering the N band with imaging and low resolution spectroscopy, the Thirty Meter Telescope (TMT; \citealt{schock2009}) will have completed first light and will be ready for science at the end of 2027, and the Giant Magellan Telescope (GMT; \citealt{johns2012}) will be operational in 2029.

\section{Summary and conclusions}
\label{sec:conclusion}

In this work we systematically revise the calibration of lines and features in the 6-205$\mu$m spectral range. We report new line to L$_{IR}$ calibrations for three galaxy types: active galactic nuclei (AGN), low-metallicity galaxies (LMG) and star-forming galaxies (SFG), building well defined samples with available IR spectroscopy mainly obtained from {\it Herschel} and {\it Spitzer}. The main results of this work are:
\begin{itemize}
\item Statistically robust calibrations to the total IR luminosity have been obtained for the bright IR lines and features, including the PAH features, the H$_2$ pure rotational lines and the fine-structure lines, mostly for all the three galaxy populations, excluding the PAH features for the LMG and the high-ionisation fine structure lines for the SFG and LMG.
    \item The [CII]158$\, \rm{\micron}$ line can be used as SFR tracer for both LMG and SFG, covering 6 order of magnitudes in luminosity, independently of the source metallicity. Moreover, when compared to high redshift detections, the correlation obtained for local galaxies can still be applied.
    \item The sum of the two far-IR oxygen lines, the [OI]63$\mu$m line and the [OIII]88$\mu$m line, can be used as a tracer for the SFR.
    \item A combination of low- and intermediate-ionisation lines from neon and/or sulfur is also a robust proxy for the SFR, independent of the metallicity. In particular, the [SIV]$10.5\, \rm{\micron}$+[NeII]$12.8\, \rm{\micron}$ tracer will be accessible to ELTs facilities in the near future for galaxies in the local Universe.
    \item The brightest PAH features at 6.2 and 11.3 $\mu$m can be reliably used for SFR tracers in both SFG and AGN, while the lack of these features prevents their use in LMG.
    \item We present for the first time a correlation between the SFR and the H$_{2}$ molecular line fluxes at 9.7, 12.3 and 17.0$\mu$m.
    \item The [CII]158$\mu$m line, the combination of the two neon lines ([NeII]12.8$\mu$m and [NeIII]15.5$\mu$m) and, for {\bf solar-like metallicity galaxies} that may harbour an AGN, the PAH11.3$\mu$m feature are the best SFR tracers.
    \item The [NeV]14.3$\mu$m, [NeV]24.3$\mu$m and [OIV]25.9$\mu$m lines are good tracers of the BHAR, derived either from the 2-10 keV luminosity or from the 12~$\mu$m luminosity.
\end{itemize}

These results will assume particular relevance with the upcoming launch of the \textit{JWST}, which will observe with the MIRI instrument %neon and sulfur lines up to redshift $\sim$1, 
many of the mid-IR features in galaxies of the Local Universe and beyond, up to redshift of $\sim$1 for the brightest lines, 
thus allowing to measure with spectroscopy both the SFR and the BHAR. % as shown in Section\,\ref{sec:jwst_alma}.

%%%%%%%%%%%%%%%%%%%%%%%%%%%%%%%%%%%%%%%%%%%%%%%
\begin{acknowledgements}
We thank the anonymous referee for his/her comments which helped to improve the present article.
LS and JAFO acknowledge financial support by the Agenzia Spaziale Italiana (ASI) under the research contract 2018-31-HH.0.
\end{acknowledgements}
%%%%%%%%%%%%%%%%%%%%%%%%%%%%%%%%%%%%%%%%%%%%%%%

\bibliographystyle{pasa-mnras}
\bibliography{calibrations}

\onecolumn

\begin{appendix}

\begin{landscape}
\begin{table}
\section{Characteristics of IR fine-structure lines}
%\afterpage{%
%\clearpage

%\tabletypesize{\footnotesize}
%\setlength{\tabcolsep}{0.3em}
%\tablecolumns{11} 
%\tablewidth{0pt}
\caption{Fine-structure lines in the mid- to far-IR range. The columns correspond to the central wavelength, frequency, ionisation potential, excitation temperature, critical density, spectral and spatial resolution of the data presented in this work, and the number of AGN, starburst, and dwarfs galaxies with line detections above $3 \times$\,\textsc{rms}. Critical densities and excitation temperatures from: \citet{launay1977,tielens1985,greenhouse1993,sturm2002,cormier2012,goldsmith2012,farrah2013}.} \label{tbl_lines}
\centering
\begin{threeparttable}[b]
\begin{tabular}{lcccccccccc}
\bf Line & \bf $\lambda$ & \bf $\nu$ & \bf I.P. & \bf E & \bf $n_{\rm cr}$ & \bf Spec. Res. & \bf Ang. Res. \\
  & ($\, \rm{\micron}$)                   & (GHz)                    & (eV) & (K) & ($\rm{cm^{-3}}$) &($\rm{km\,s^{-1}}$)                        & (arcsec) \\
\hline \\[-0.3cm]
\nevi$^2$P$_{3/2}$--$^2$P$_{1/2}$    & 7.65   & 39188.56 & 126.21 & 1888 & $2.5 \times 10^5$           & $\sim 1500$  & $14 \times 20$     \\
H$_2$ (0,0) S(3)                     & 9.66   & 31034.41 & 4.48   & 2504 & $9 \times 10^{5}$\tnote{a}  & $\sim 500$   & $4.7 \times 11.3$ \\
\siv$^2$P$_{3/2}$--$^2$P$_{1/2}$   & 10.51 & 28524.50 & 34.79 & 1369 & $5.39 \times 10^4$ & $\sim 500$ & $4.7 \times 11.3$\\
H$_2$ (0,0) S(2)                     & 12.28  & 24413.07 & 4.48   & 1682 & $2 \times 10^{5}$\tnote{a}  & $\sim 500$   & $4.7 \times 11.3$  \\
\neii$^2$P$_{1/2}$--$^2$P$_{3/2}$  & 12.81 & 23403.00 & 21.56 & 1123 & $7.00 \times 10^5$ & $\sim 500$ & $4.7 \times 11.3$ \\
\nev$^3$P$_2$--$^3$P$_1$          & 14.32 & 20935.23 & 97.12 & 1892 & $3 \times 10^4$ & $\sim 500$ & $4.7 \times 11.3$ \\
\neiii$^3$P$_1$--$^3$P$_2$        & 15.56 & 19266.87 & 40.96 & 925 & $2.68 \times 10^5$ & $\sim 500$ & $4.7 \times 11.3$ \\
H$_2$ (0,0) S(1)                     & 17.03  & 17603.78 & 4.48   & 1015 & $2 \times 10^{4}$\tnote{a}  & $\sim 500$   & $4.7 \times 11.3$  \\
\siii$^3$P$_2$--$^3$P$_1$         & 18.71 & 16023.11 & 23.34 & 769 & $2.22 \times 10^4$ & $\sim 500$ & $4.7 \times 11.3$ \\
\nev$^3$P$_1$--$^3$P$_0$          & 24.32 & 12326.99 & 97.12 & 596 & $5.0 \times 10^5$ & $\sim 500$ & $11.1 \times 22.3$\\
\oiv$^2$P$_{3/2}$--$^2$P$_{1/2}$   & 25.89 & 11579.47 & 54.94 & 555 & $10^4$ & $\sim 500$ & $11.1 \times 22.3$  \\
\feii a$^6$D$_{7/2}$--a$^6$D$_{9/2}$ & 25.98  & 11539.35 & 7.9    & 553  & $2.2 \times 10^6$           & $\sim 500$   & $11.1 \times 22.3$ \\
H$_2$ (0,0) S(1)                     & 17.03  & 17603.78 & 4.48   & 1015 & $2 \times 10^{4}$\tnote{a}  & $\sim 500$   & $4.7 \times 11.3$  \\
\siii$^3$P$_1$--$^3$P$_0$          & 33.48 & 8954.37 & 23.34 & 430 & $7.04 \times 10^3$ & $\sim 500$ & $11.1 \times 22.3$ \\
\sitwo$^2$P$_{3/2}$--$^2$P$_{1/2}$   & 34.81 & 8612.25 & 8.15 & 413 & $3.4 \times 10^5$\tnote{a},$10^3$ & $\sim 500$ & $11.1 \times 22.3$ \\
\oiii$^3$P$_2$--$^3$P$_1$          & 51.81 & 5787.57 & 35.12 & 441 & $3.6 \times 10^3$ & $\sim 105$ & 9.4\tnote{c} \\
\niii$^2$P$_{3/2}$--$^2$P$_{1/2}$   & 57.32 & 5230.43 & 29.60 & 251 & $3.0 \times 10^3$ & $\sim 105$  & 9.4   \\ 
\oi$^3$P$_2$--$^3$P$_1$            & 63.18 & 4744.77 & -- & 228 & $4.7 \times 10^5$\tnote{a} & $\sim 86$ & 9.4 \\
\oiii$^3$P$_1$--$^3$P$_0$          & 88.36 & 3393.01 & 35.12 & 163 & $510$ & $\sim 124$ & 9.4 \\
\nii$^3$P$_2$--$^3$P$_1$           &   121.90  & 2459.38 & 14.53 & 118 & $310$ & $\sim 290$  & 9.4 \\
\oi$^3$P$_1$--$^3$P$_0$            &   145.52  & 2060.07 & -- & 98 & $9.5 \times 10^4$\tnote{a} & $\sim 256$ & 10.3 \\
\cii$^2$P$_{3/2}$--$^2$P$_{1/2}$    &   157.74  & 1900.54 & 11.26 & 91 & $20$,$[2.2\tnote{a},4.4\tnote{b}\ ]\times 10^3$ & $\sim 238$ &   11.2 \\
\nii$^3$P$_1$--$^3$P$_0$             & 205.3  & 1460.27  & 14.53  & 70   & $48$                         & $\sim 297$  & $\sim16.8$\\
%\ci$^3$P$_{2}$--$^3$P$_{1}$        &   370.37  & 809.44 & -- & 38.9 & $2.8 \times 10^{3}$\tnote{a} & $\sim 536$ & $\sim 35.0$ & 59 & 13 & 1 \\
%\ci$^3$P$_{1}$--$^3$P$_{0}$         &   609.7  & 491.70 & -- & 23.6 &$4.7 \times 10^{2}$\tnote{a} & $\sim 882$ & $\sim 35.0$ & 32 & 12 & 0 \\
\hline

\end{tabular}
\begin{tablenotes}
    \item[a] Critical density for collisions with hydrogen atoms.
    \item[b] Critical density for collisions with H$_2$ molecules.
    \item[c] The beam size for \textit{Herschel}/PACS is dominated by the spaxel size ($9\farcs4$) below $\sim 120\, \rm{\micron}$.
\end{tablenotes}
\end{threeparttable}
\end{table}
\end{landscape}

\clearpage

\begin{figure*}
\section{Line correlations}\label{sec:app}
\centering
\includegraphics[width=0.33\columnwidth]{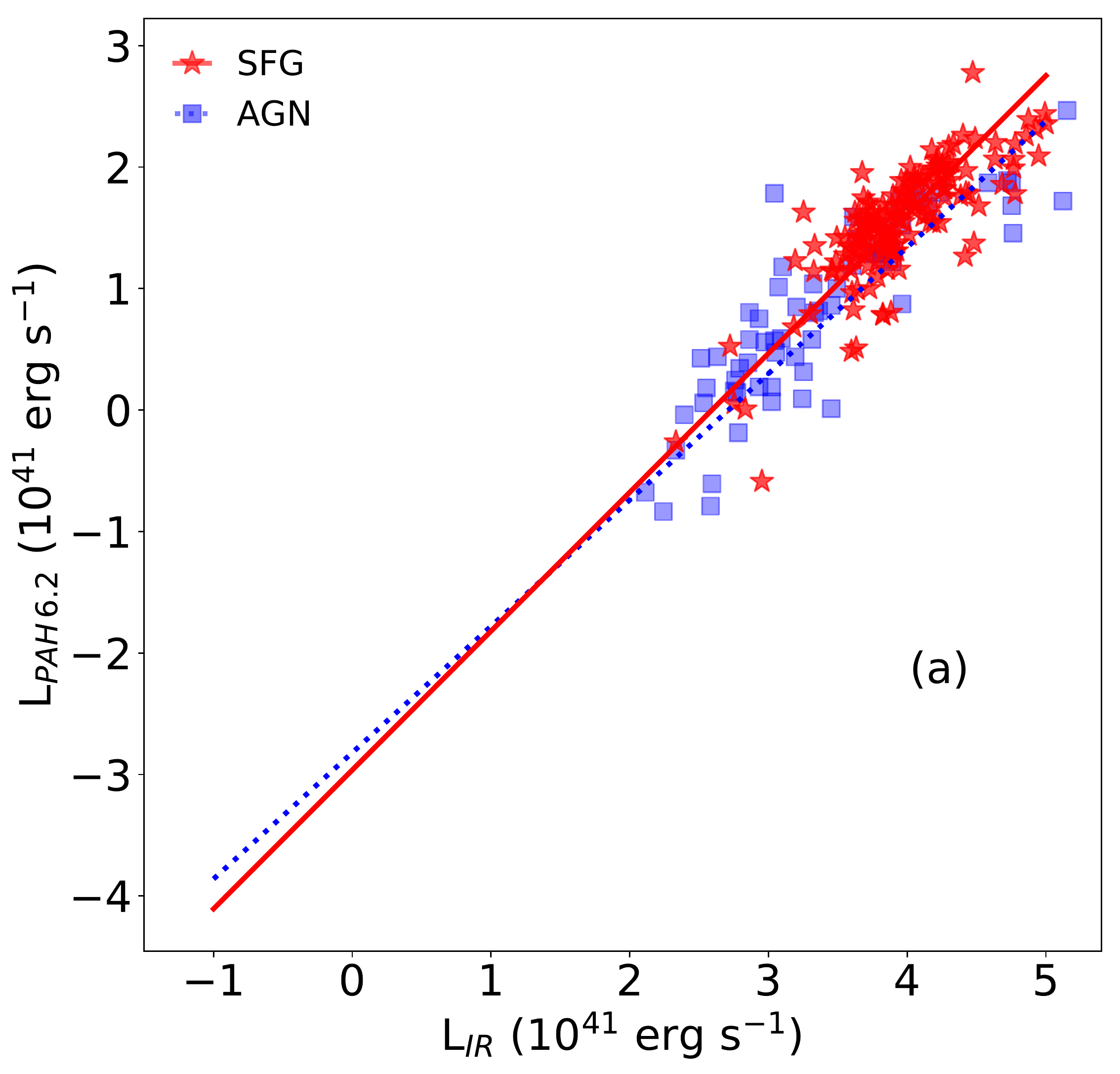}
\includegraphics[width=0.33\columnwidth]{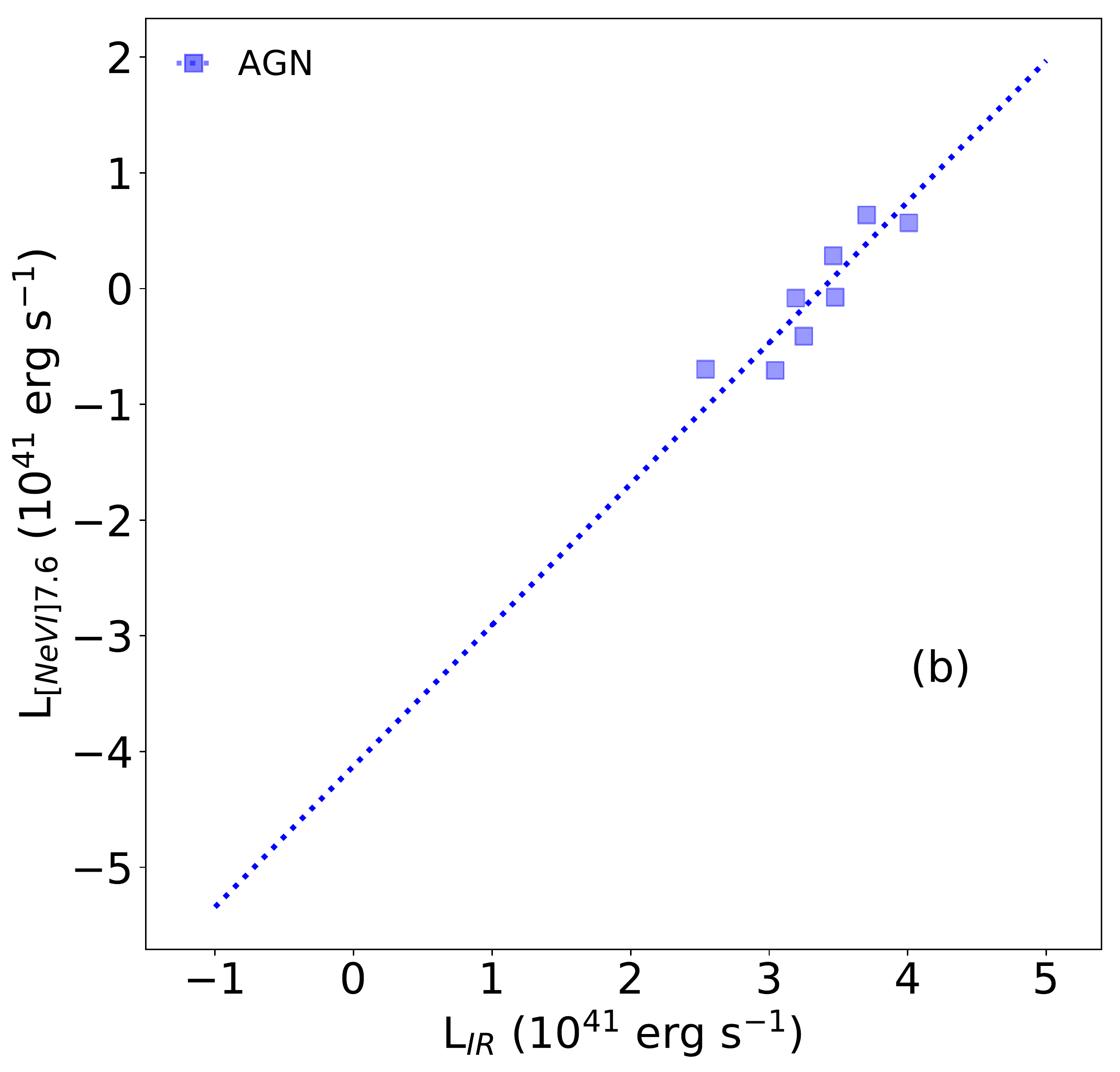}
\includegraphics[width=0.33\columnwidth]{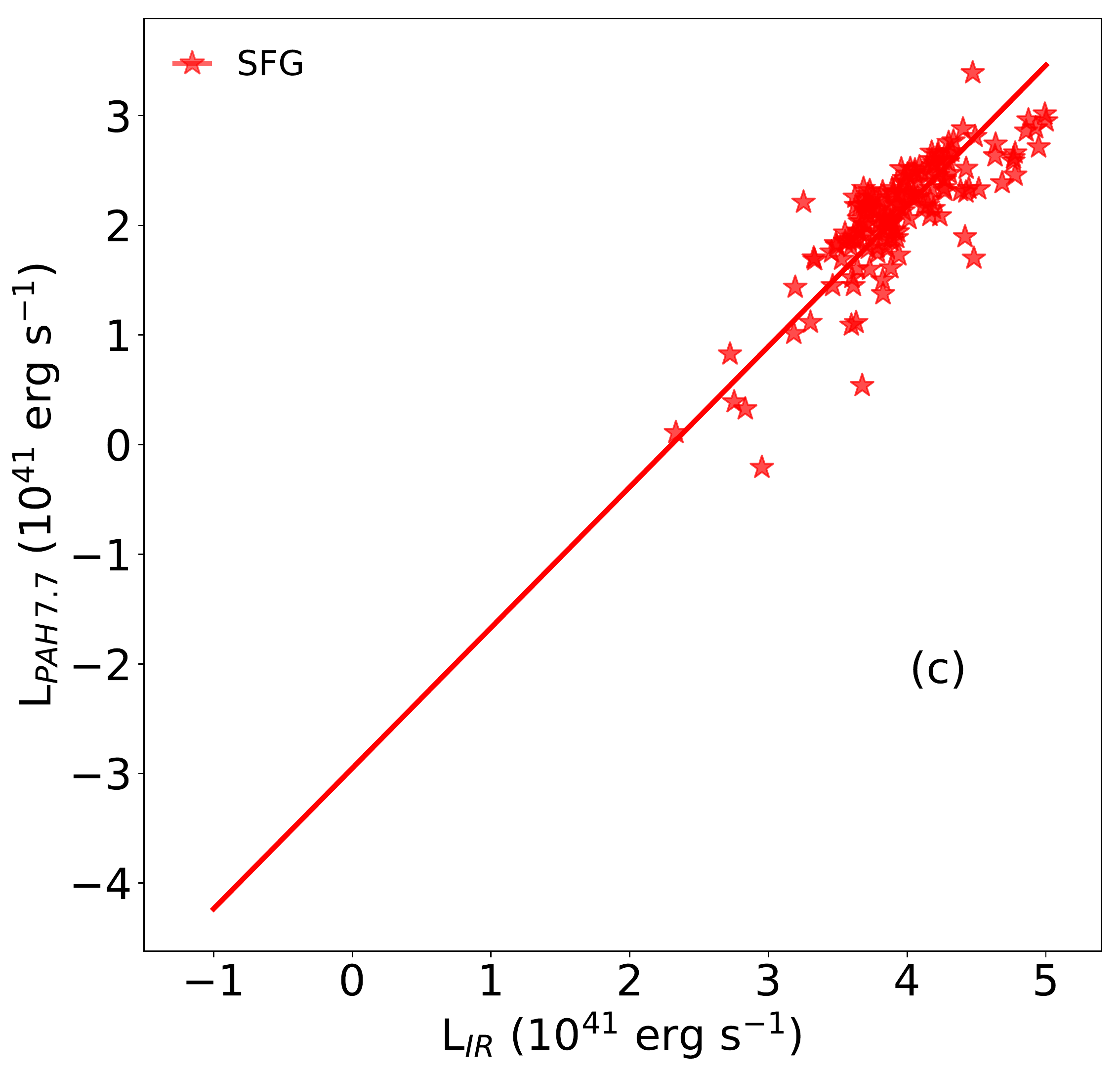}
\caption{{\bf (a: left)} The luminosity of the PAH feature at 6.2 $\, \rm{\micron}$ as a function of the total IR luminosity.  {\bf (b: centre)} The [NeVI]7.65$\, \rm{\micron}$ line luminosity as a function of the total IR luminosity. {\bf (c: right)} The luminosity of the PAH feature at 7.7 $\, \rm{\micron}$ as a function of the total IR luminosity. The same legend as in Fig.\,\ref{fig:corr_c2_ne2_ne3} was used.} 
\label{fig:corr_app_1}
\end{figure*}

\begin{figure*}
\centering
\includegraphics[width=0.33\columnwidth]{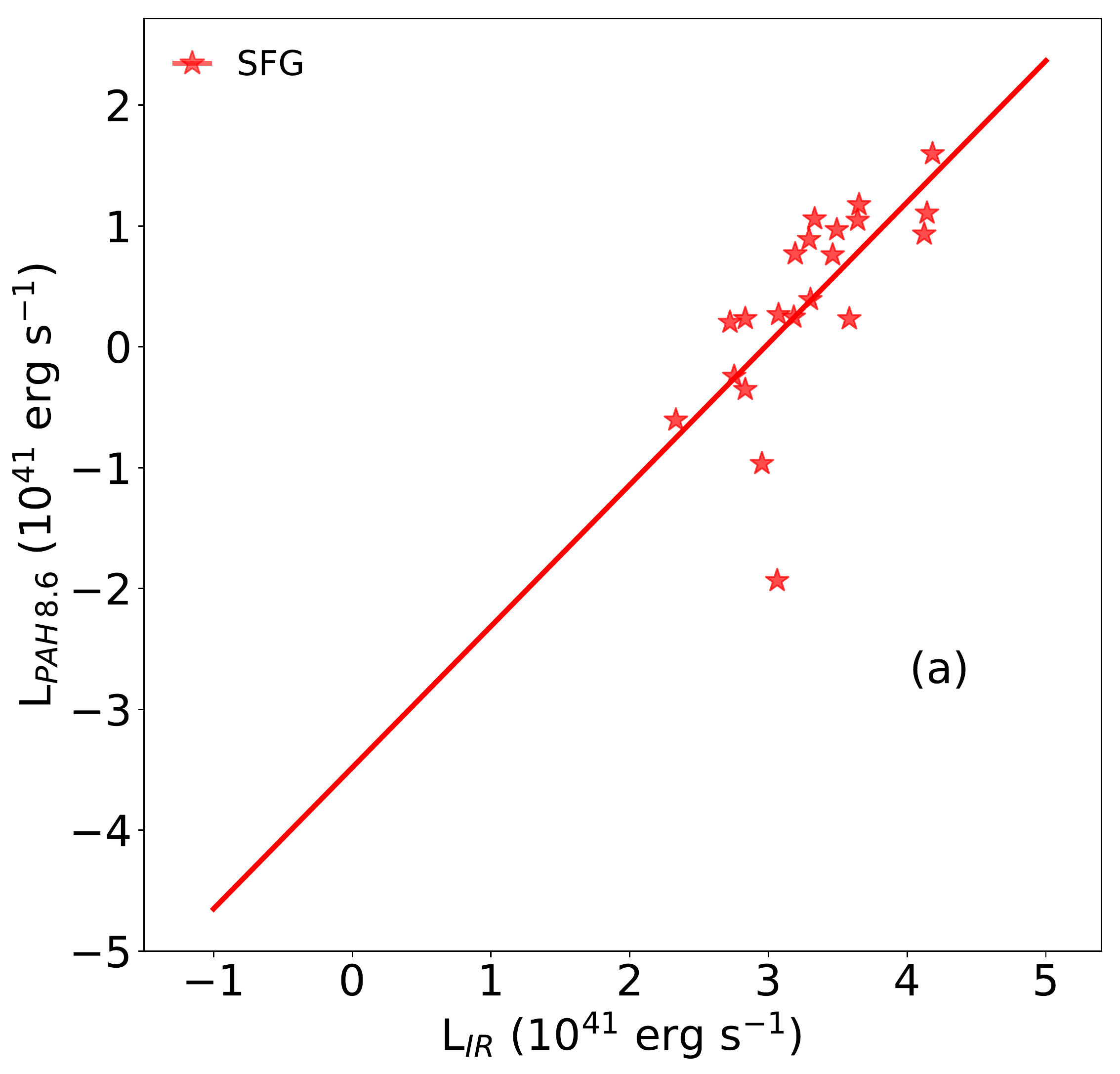}
\includegraphics[width=0.33\columnwidth]{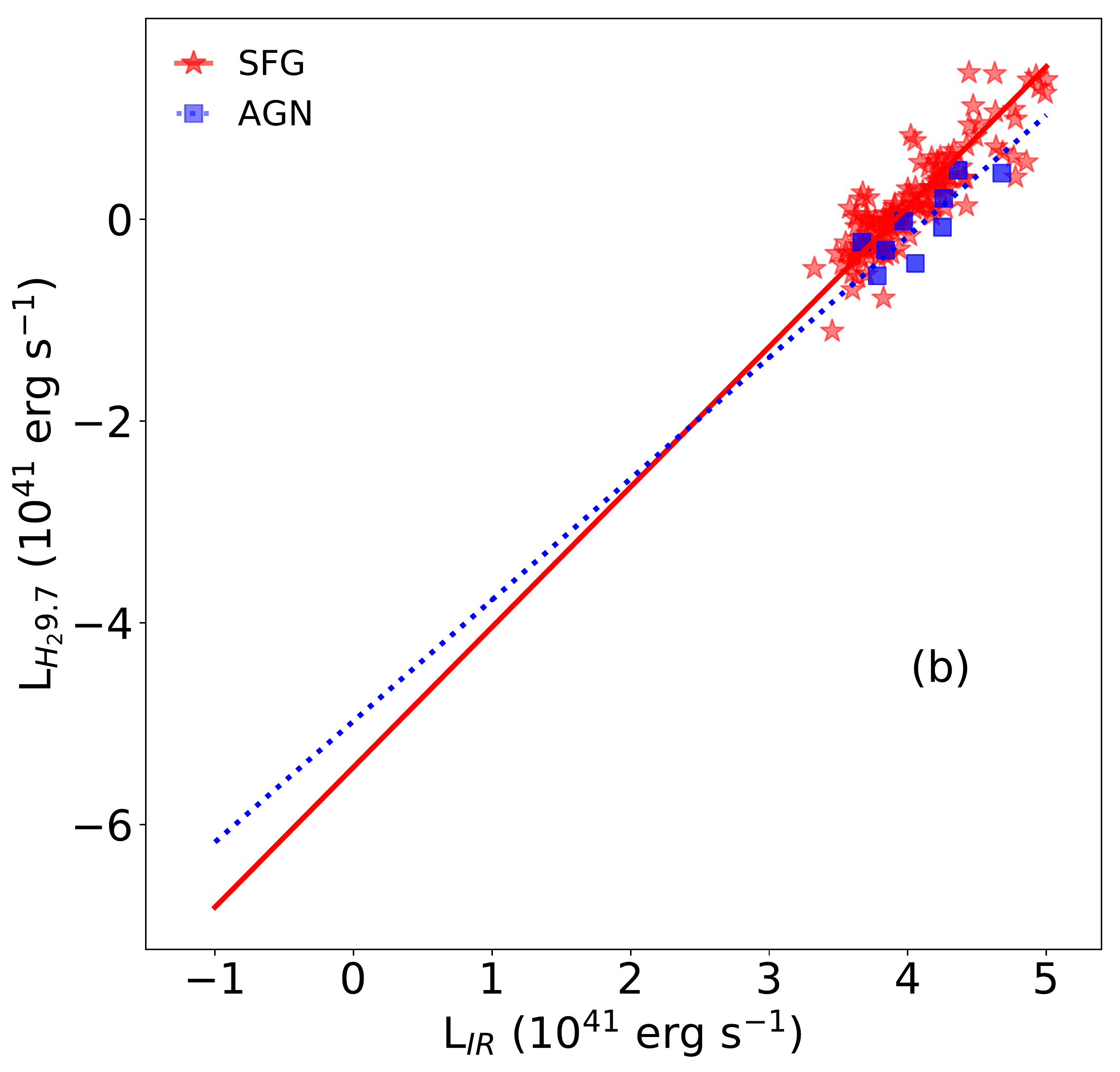}
\includegraphics[width=0.33\columnwidth]{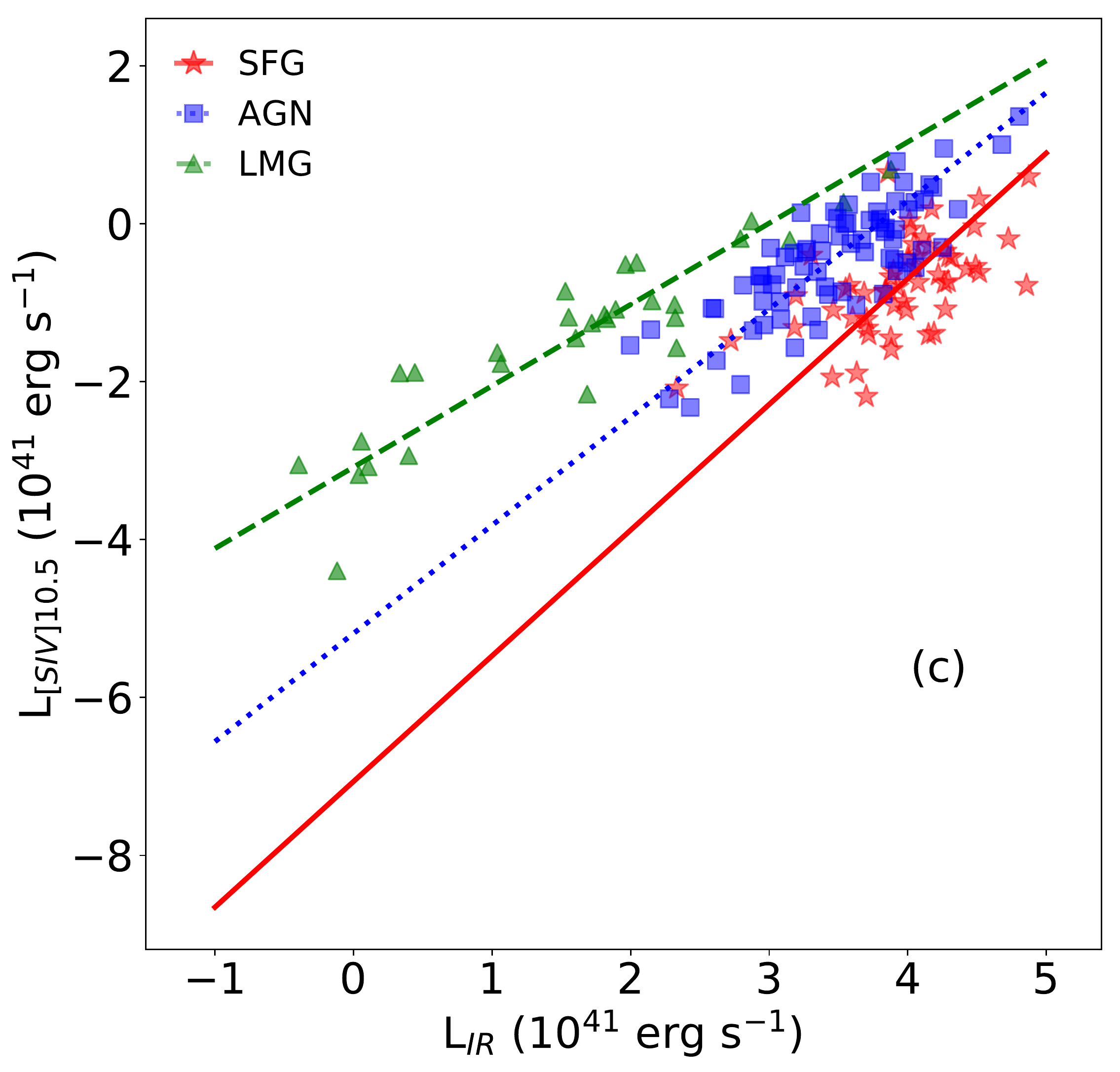}

\caption{{\bf (a: left)} The luminosity of the PAH feature at 8.6 $\, \rm{\micron}$ as a function of the total IR luminosity.  {\bf (b: centre)} The luminosity of the H$_{2}$ molecular line at 9.7 $\, \rm{\micron}$ as a function of the total IR luminosity.    {\bf (c: right)} The [SIV]10.5$\, \rm{\micron}$ line luminosity as a function of the total IR luminosity. The same legend as in Fig.\,\ref{fig:corr_c2_ne2_ne3} was used.} 
\label{fig:corr_app_2}
\end{figure*}

\begin{figure*}
\centering
\includegraphics[width=0.33\columnwidth]{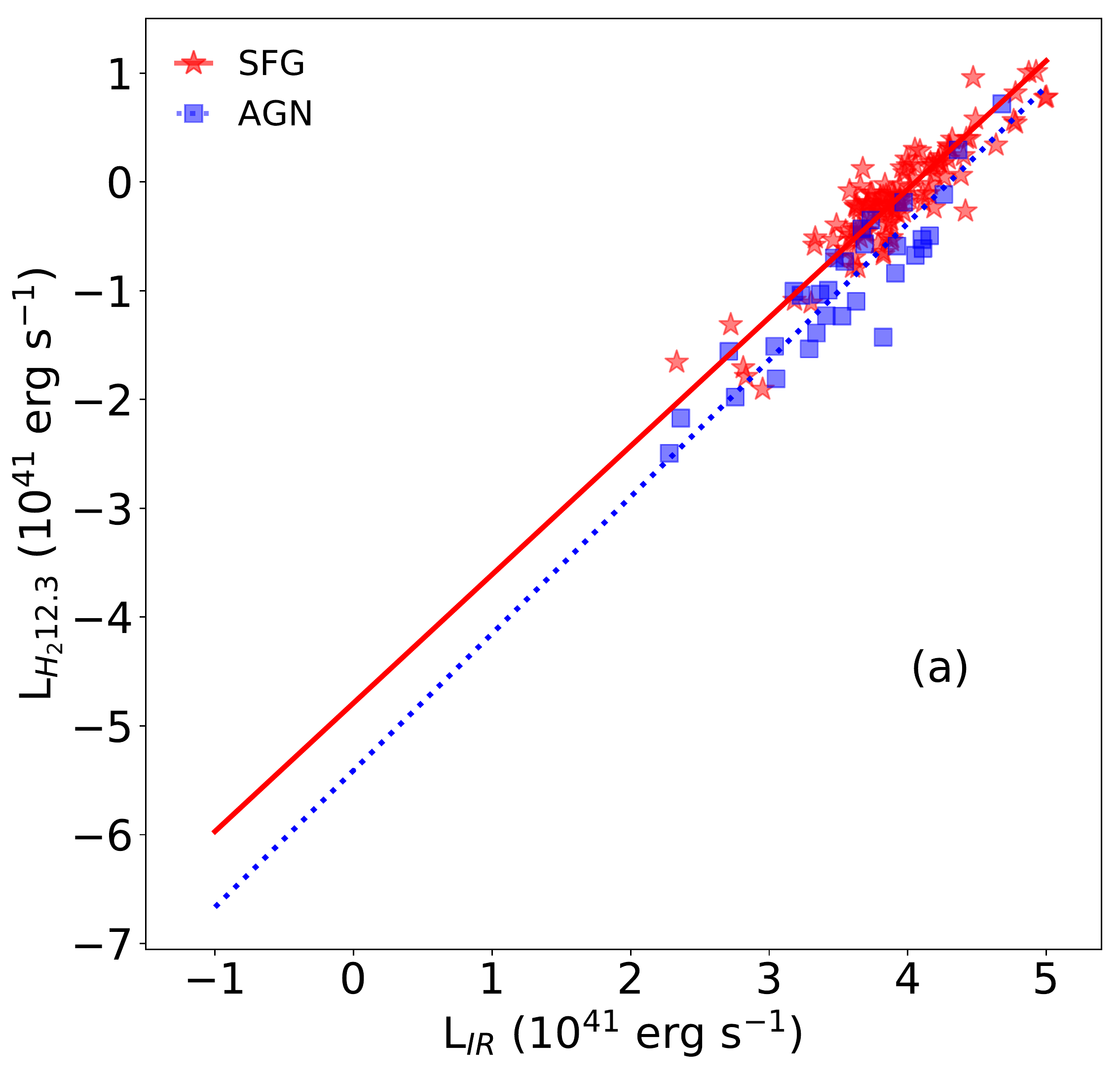}
\includegraphics[width=0.33\columnwidth]{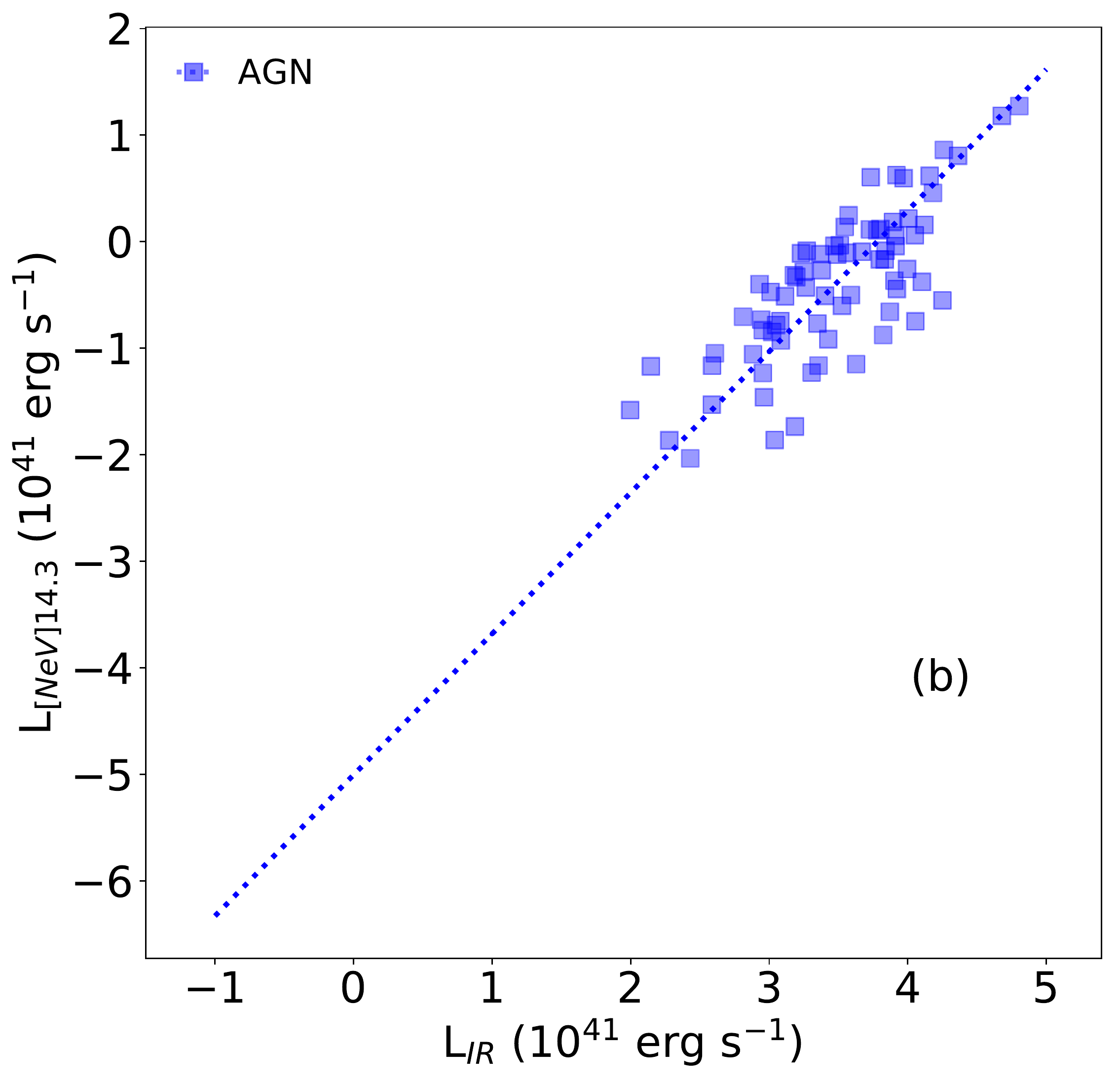}
\includegraphics[width=0.33\columnwidth]{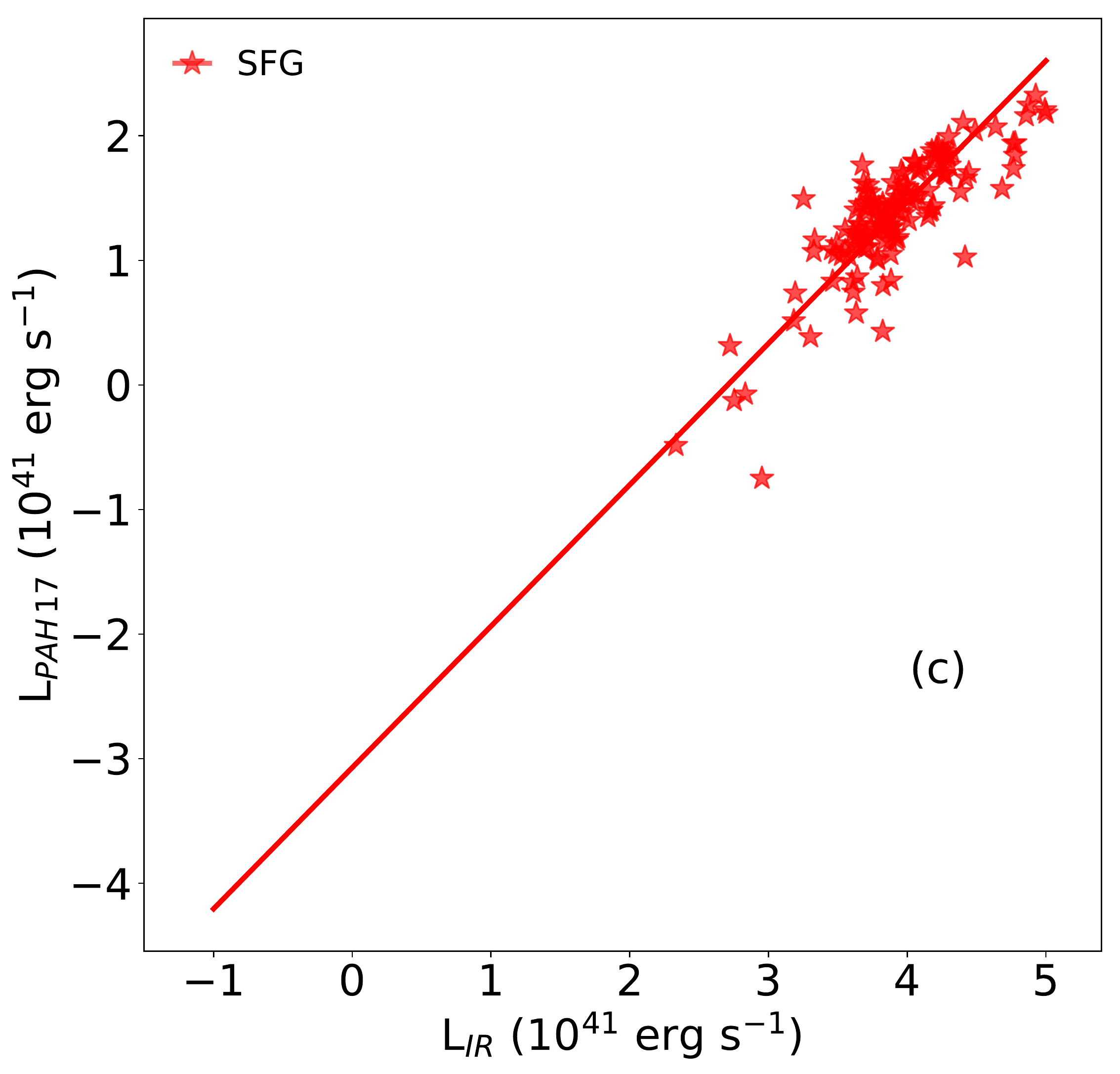}

\caption{{\bf (a: left)} The luminosity of the H$_{2}$ molecular line at 12.28 $\, \rm{\micron}$ as a function of the total IR luminosity.   {\bf (b: centre)} The [NeV]14.3$\, \rm{\micron}$ line luminosity as a function of the total IR luminosity. {\bf (c: right)} The luminosity of the PAH feature at 17 $\, \rm{\micron}$ as a function of the total IR luminosity. The same legend as in Fig.\,\ref{fig:corr_c2_ne2_ne3} was used.} 
\label{fig:corr_app_3}
\end{figure*}

\begin{figure*}
\centering
\includegraphics[width=0.33\columnwidth]{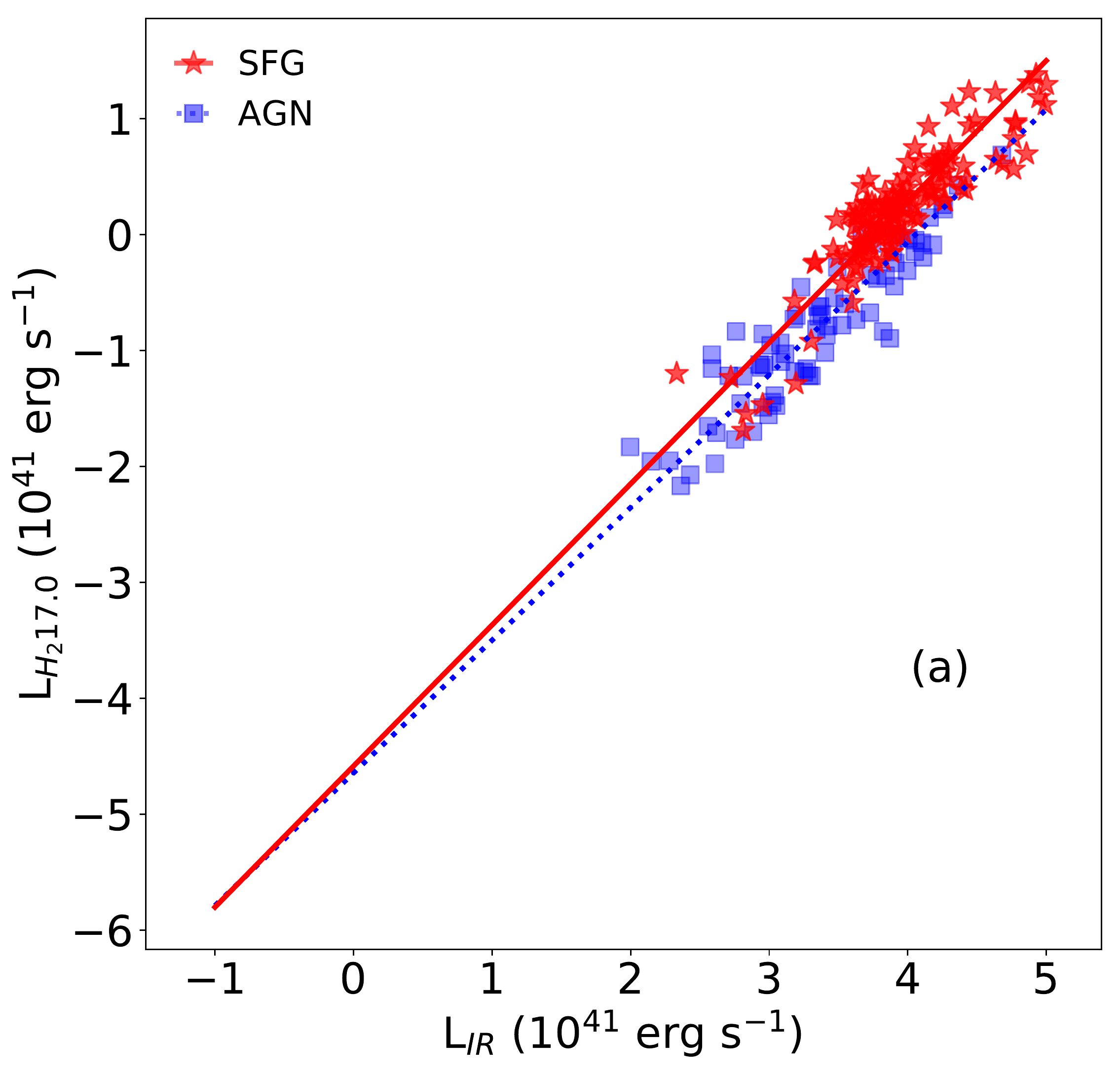}
\includegraphics[width=0.33\columnwidth]{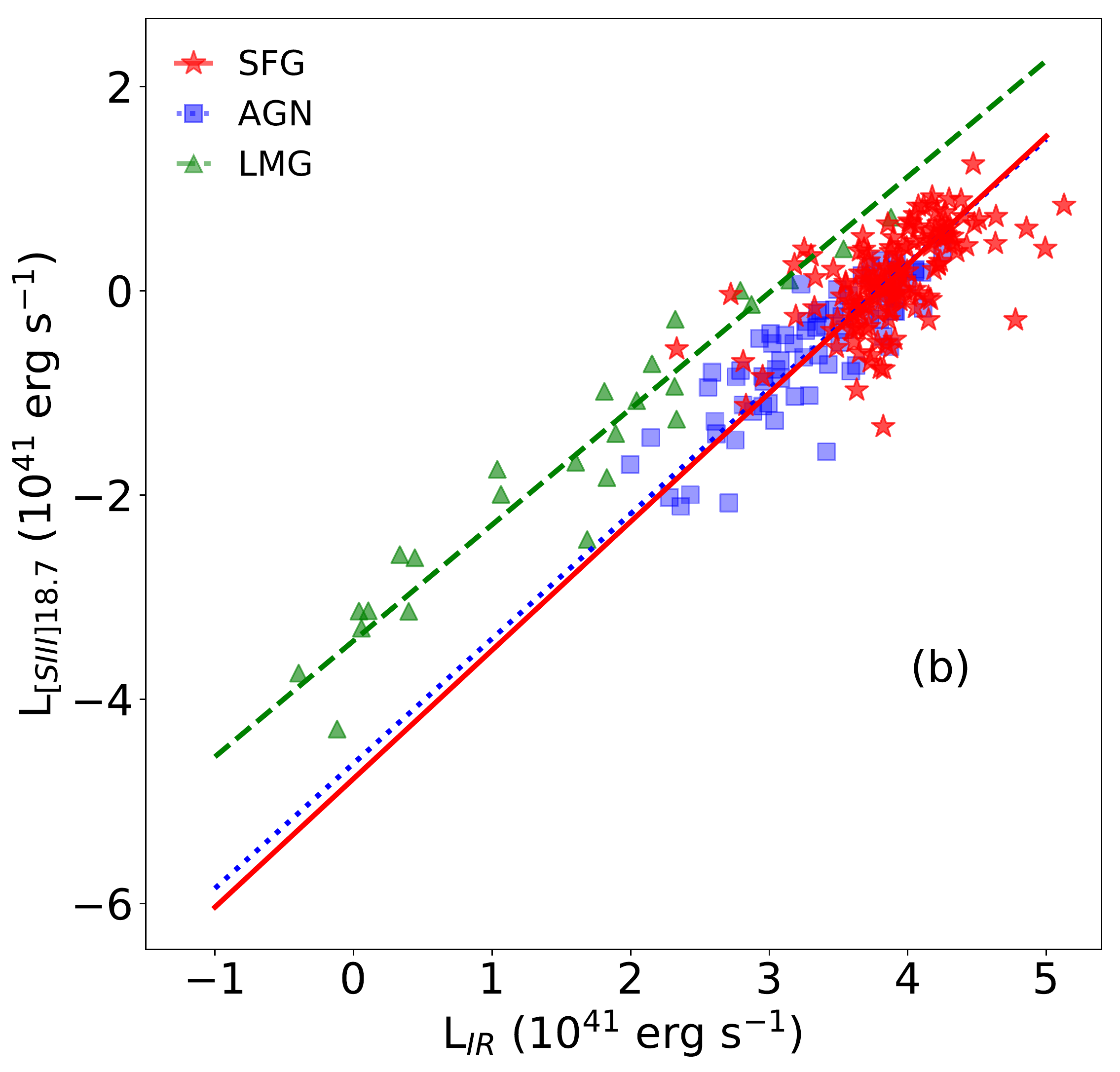}
\includegraphics[width=0.33\columnwidth]{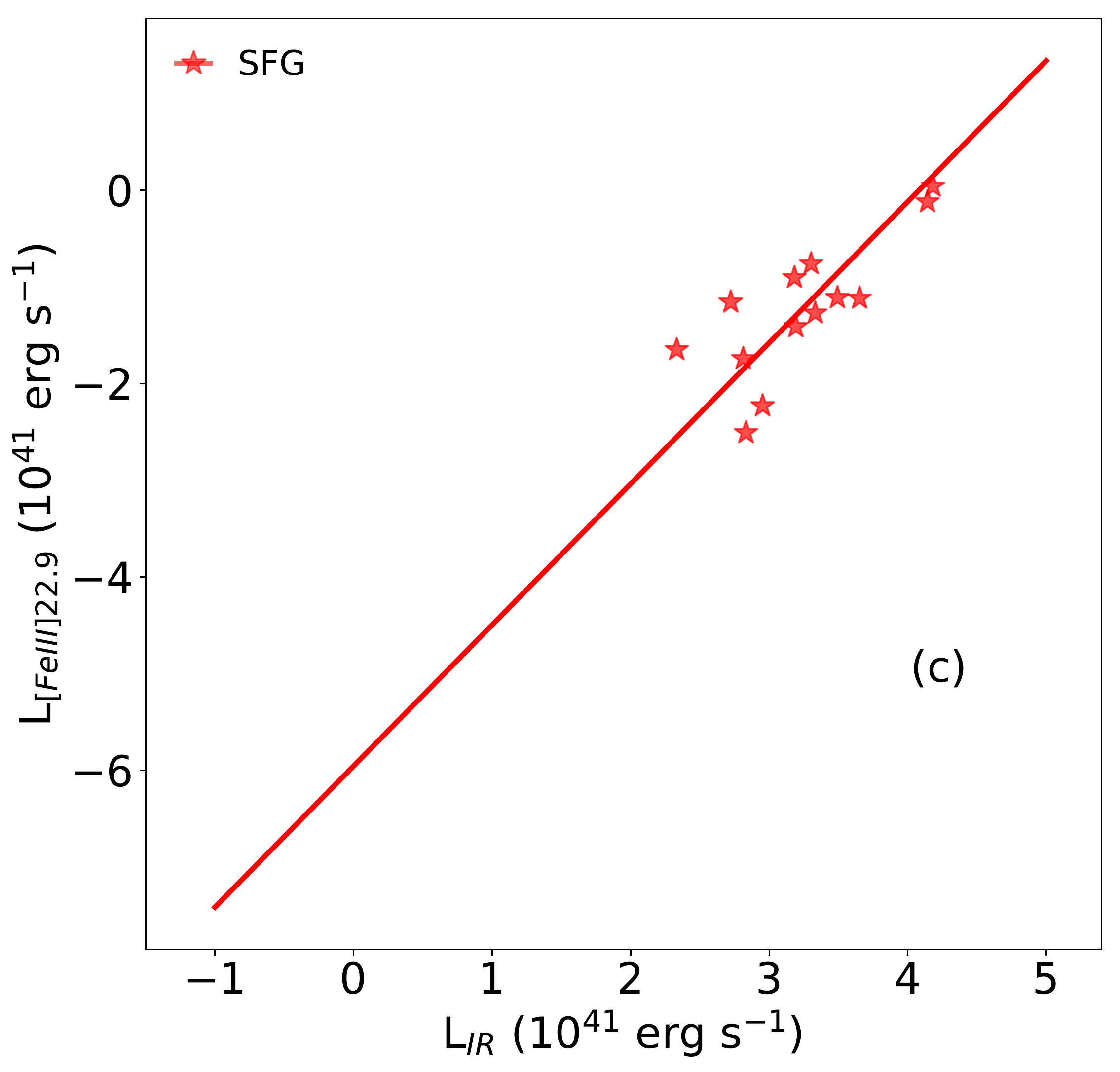}
\caption{{\bf (a: left)} The luminosity of the H$_{2}$ molecular line at 17.03 $\, \rm{\micron}$ as a function of the total IR luminosity.  {\bf (b: centre)} The [SIII]18.7$\, \rm{\micron}$ line luminosity as a function of the total IR luminosity. {\bf (c: right)} The [FeIII]22.93$\, \rm{\micron}$ line luminosity as a function of the total IR luminosity. The same legend as in Fig.\,\ref{fig:corr_c2_ne2_ne3} was used.} 
\label{fig:corr_app_4}
\end{figure*}

\begin{figure*}
\centering
\includegraphics[width=0.33\columnwidth]{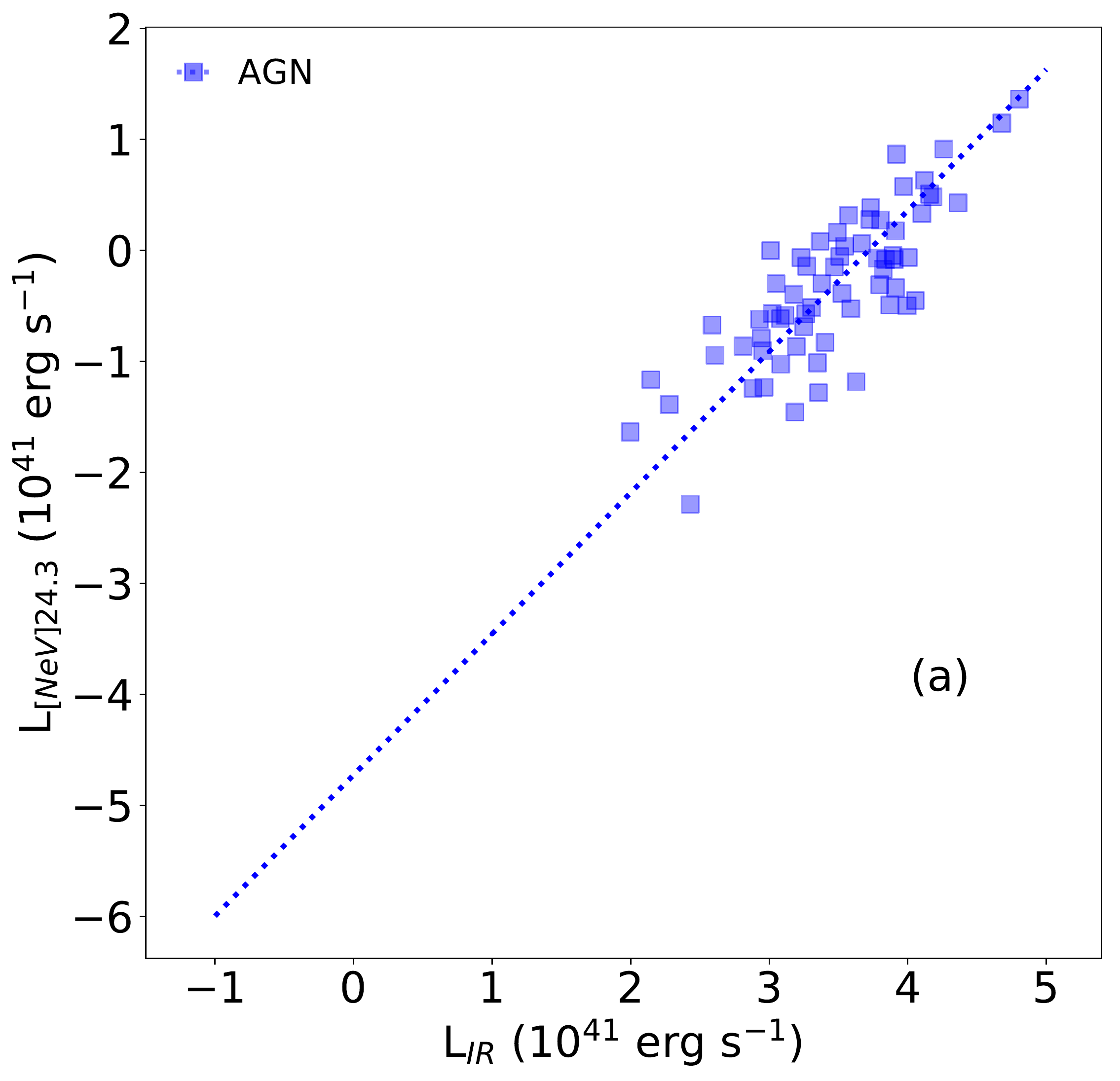}
\includegraphics[width=0.33\columnwidth]{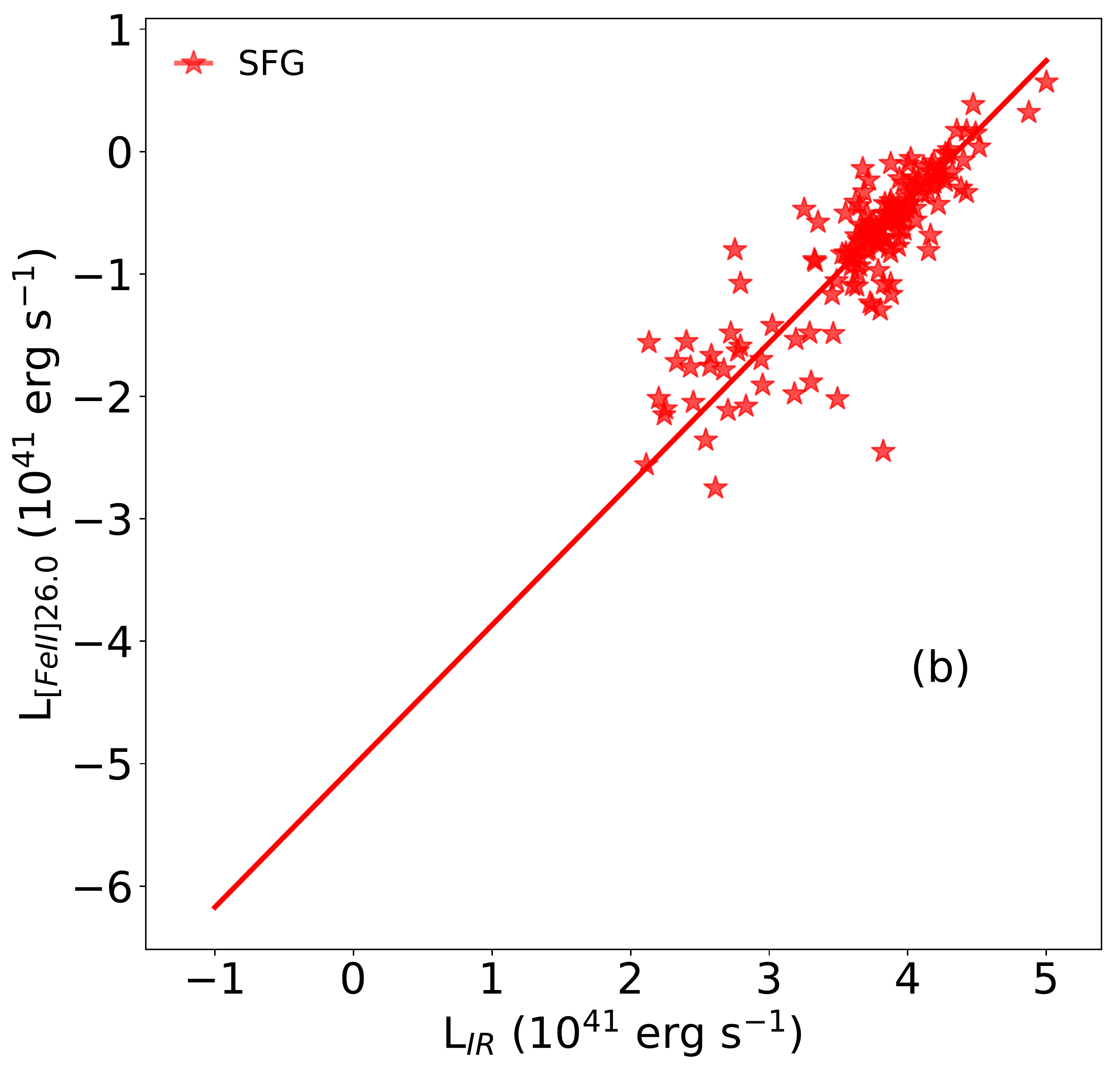}
\includegraphics[width=0.33\columnwidth]{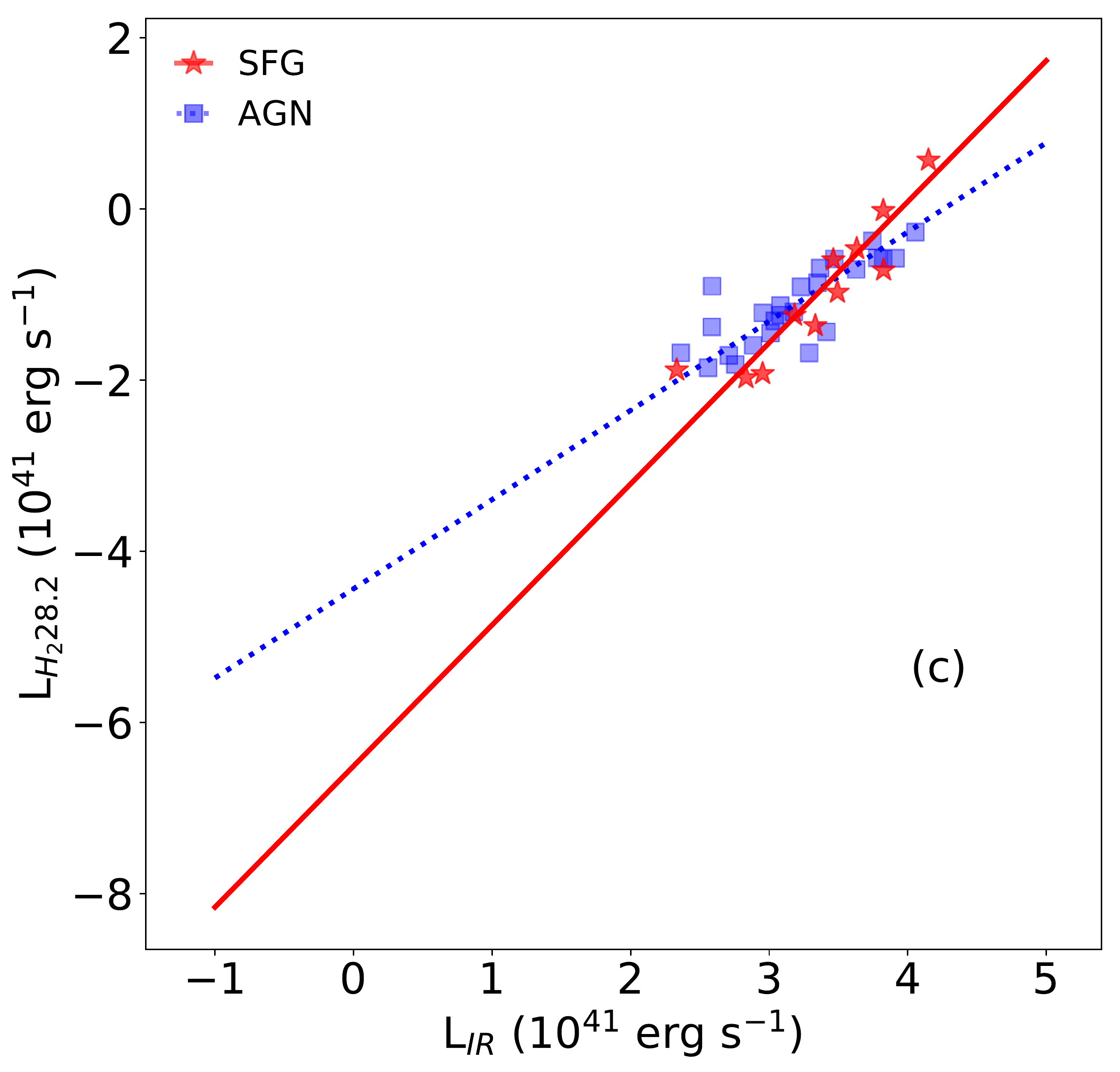}
\caption{{\bf (a: left)} The [NeV]24.3$\, \rm{\micron}$ line luminosity as a function of the total IR luminosity. {\bf (b: centre)} The [FeII]25.99$\, \rm{\micron}$ line luminosity as a function of the total IR luminosity. {\bf (c: right)} The luminosity of the H$_{2}$ molecular line at 28.22 $\, \rm{\micron}$ as a function of the total IR luminosity. The same legend as in Fig.\,\ref{fig:corr_c2_ne2_ne3} was used.} 
\label{fig:corr_app_5}
\end{figure*}

\begin{figure*}
\centering
\includegraphics[width=0.33\columnwidth]{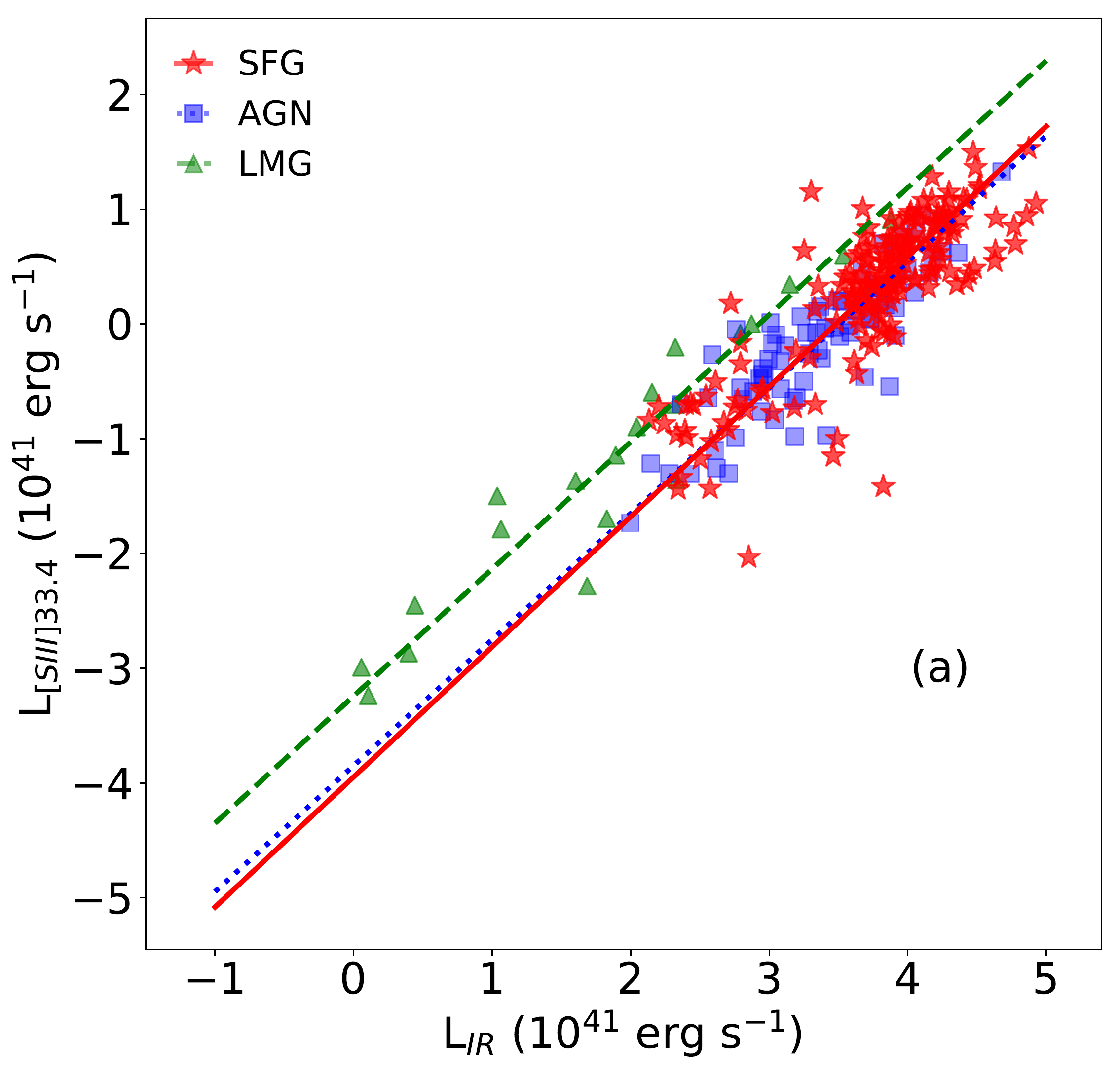}
\includegraphics[width=0.33\columnwidth]{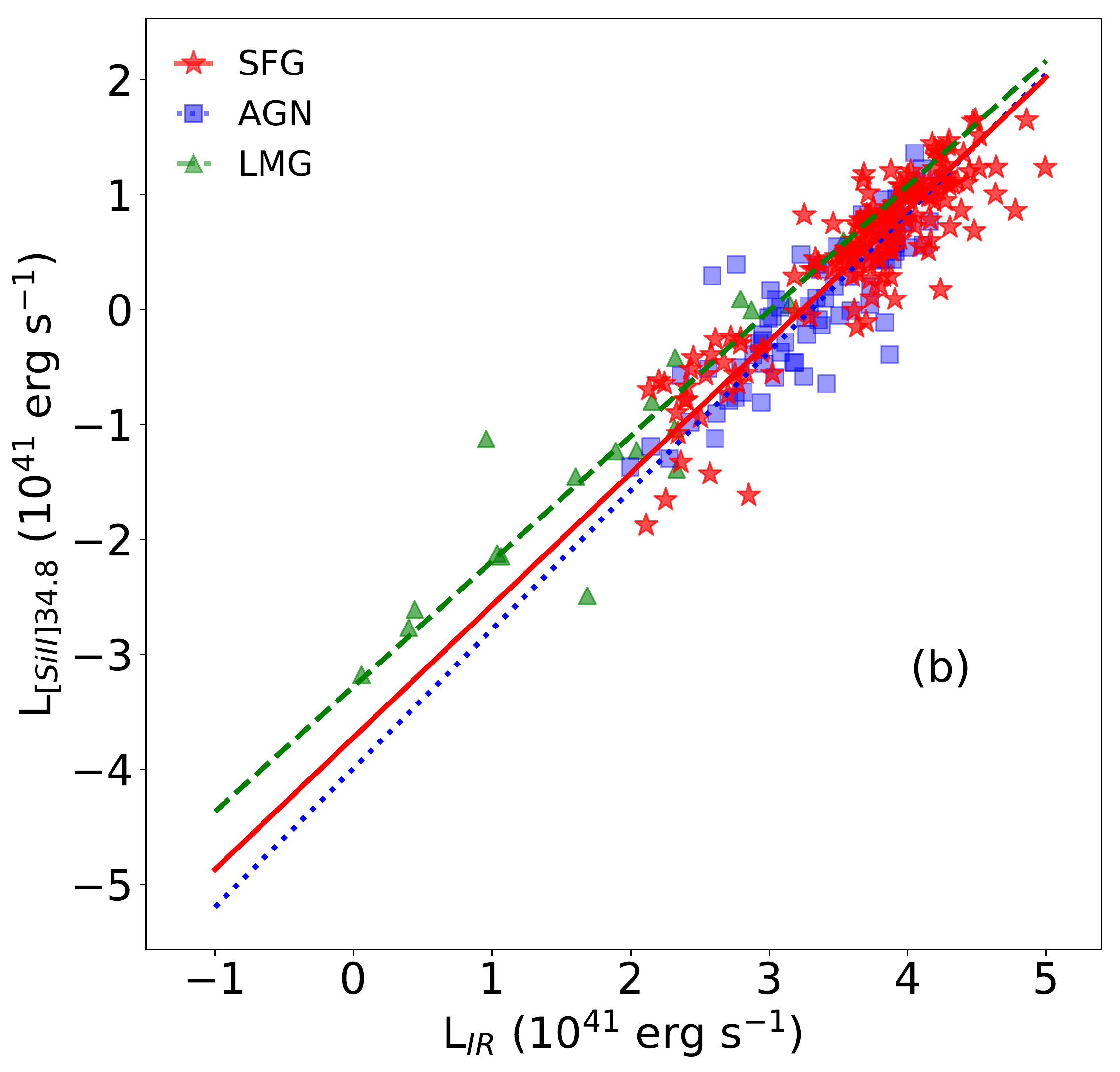}
\includegraphics[width=0.33\columnwidth]{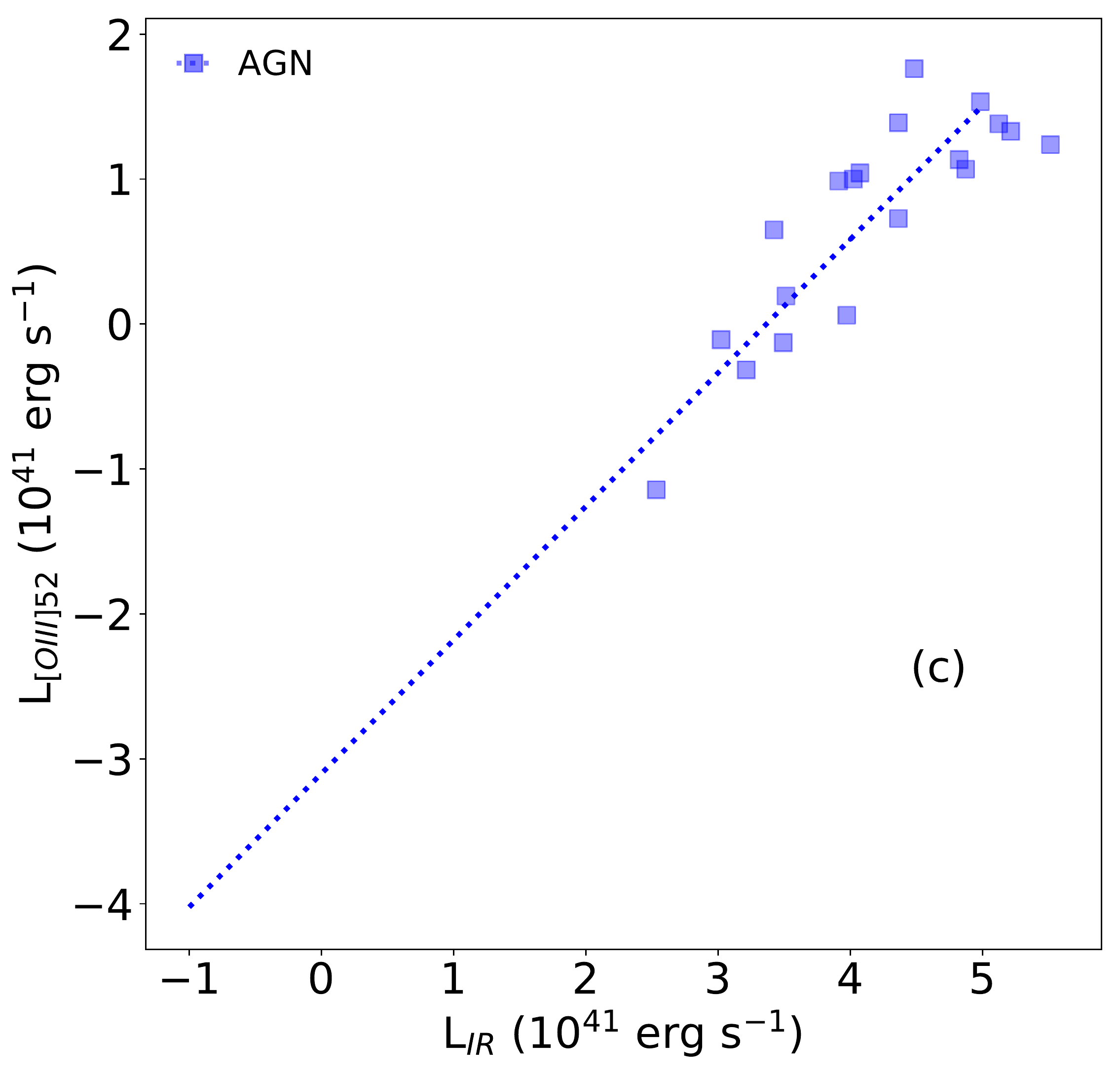}
\caption{{\bf (a: left)} The [SIII]33.5$\, \rm{\micron}$ line luminosity as a function of the total IR luminosity. {\bf (b: centre)} The [SiII]34.8$\, \rm{\micron}$ line luminosity as a function of the total IR luminosity.  {\bf (c: right)} The [OIII]52$\, \rm{\micron}$ line luminosity as a function of the total IR luminosity. The same legend as in Fig.\,\ref{fig:corr_c2_ne2_ne3} was used.} 
\label{fig:corr_app_6}
\end{figure*}
\newpage

\begin{figure*}
\centering
\includegraphics[width=0.33\columnwidth]{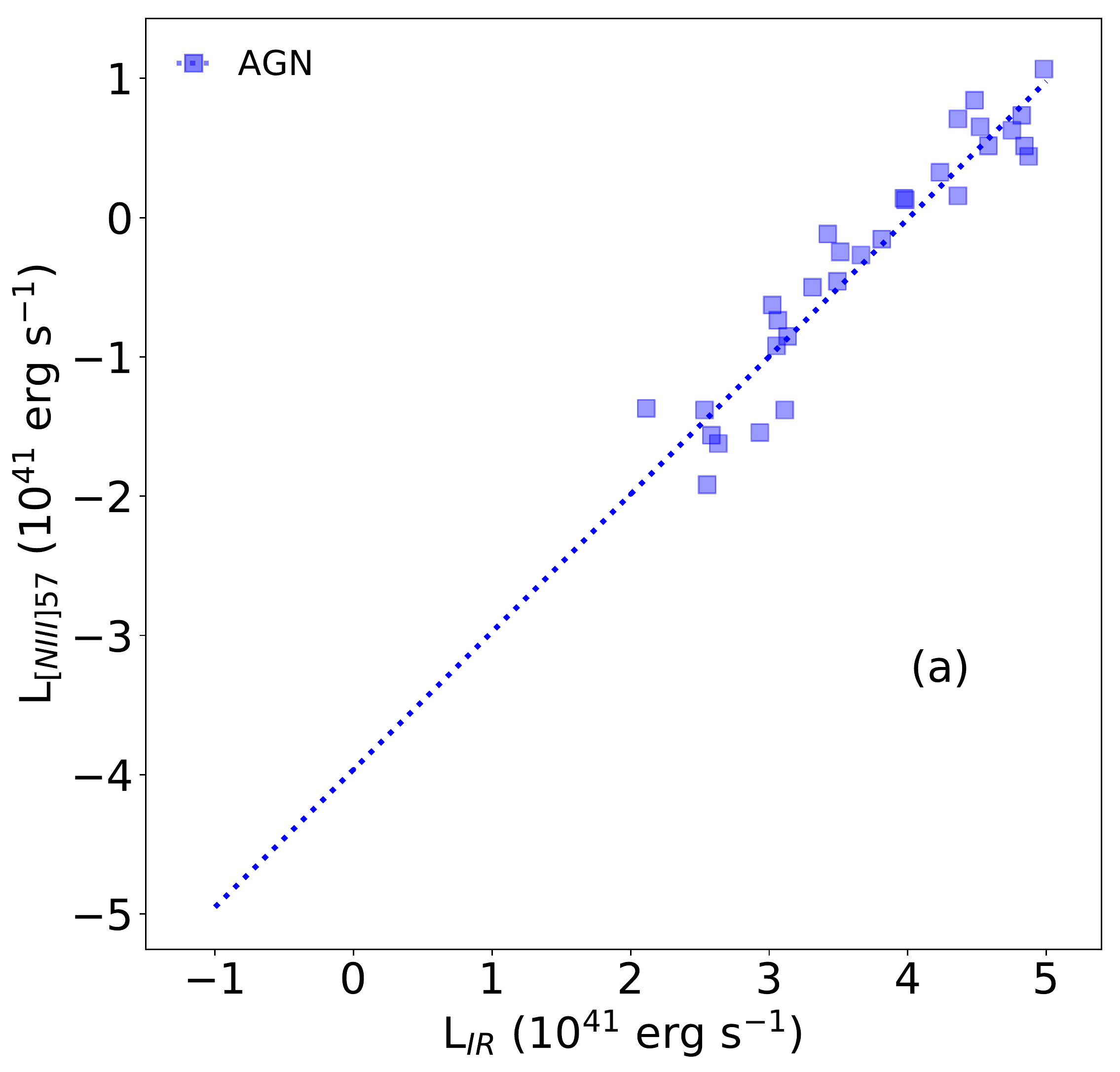}
\includegraphics[width=0.33\columnwidth]{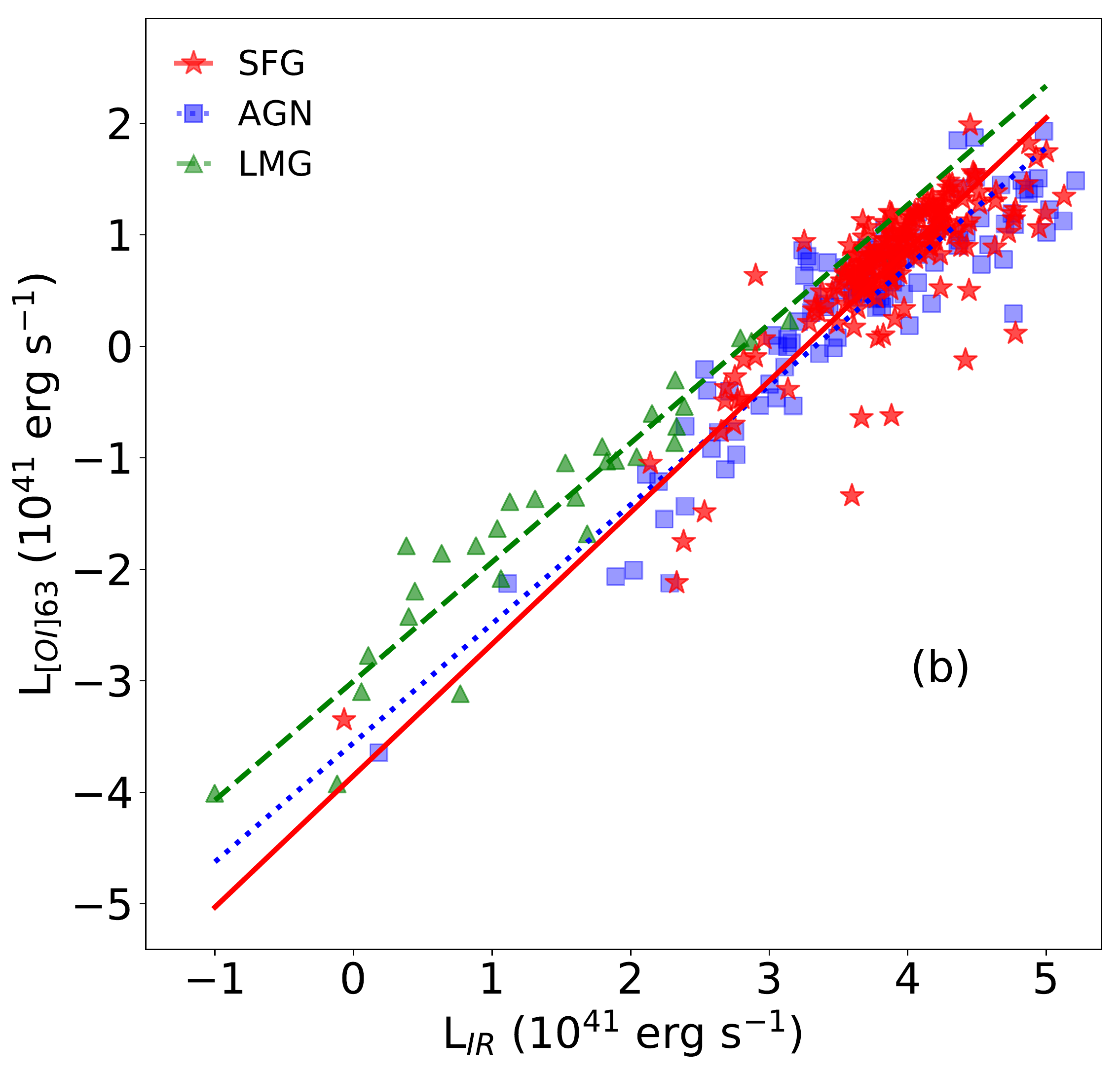}
\includegraphics[width=0.33\columnwidth]{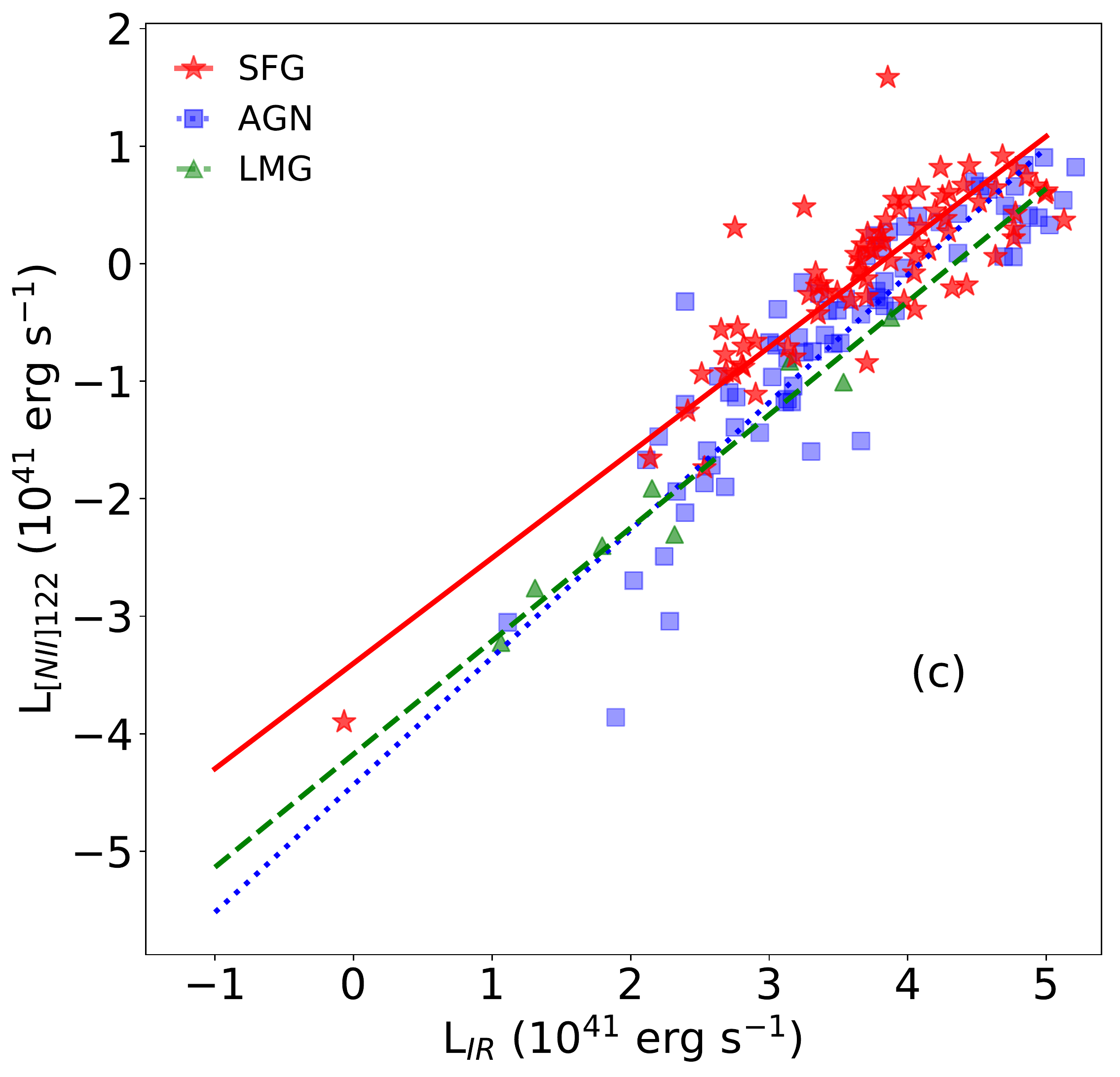}
\caption{{\bf (a: left)} The [NIII]57$\, \rm{\micron}$ line luminosity as a function of the total IR luminosity. {\bf (b: centre)} The [OI]63.18$\, \rm{\micron}$ line luminosity as a function of the total IR luminosity.  {\bf (c: right)} The [NII]122$\, \rm{\micron}$ line luminosity as a function of the total IR luminosity. The same legend as in Fig.\,\ref{fig:corr_c2_ne2_ne3} was used.} 
\label{fig:corr_app_7}
\end{figure*}

\begin{figure*}
\centering
\includegraphics[width=0.33\columnwidth]{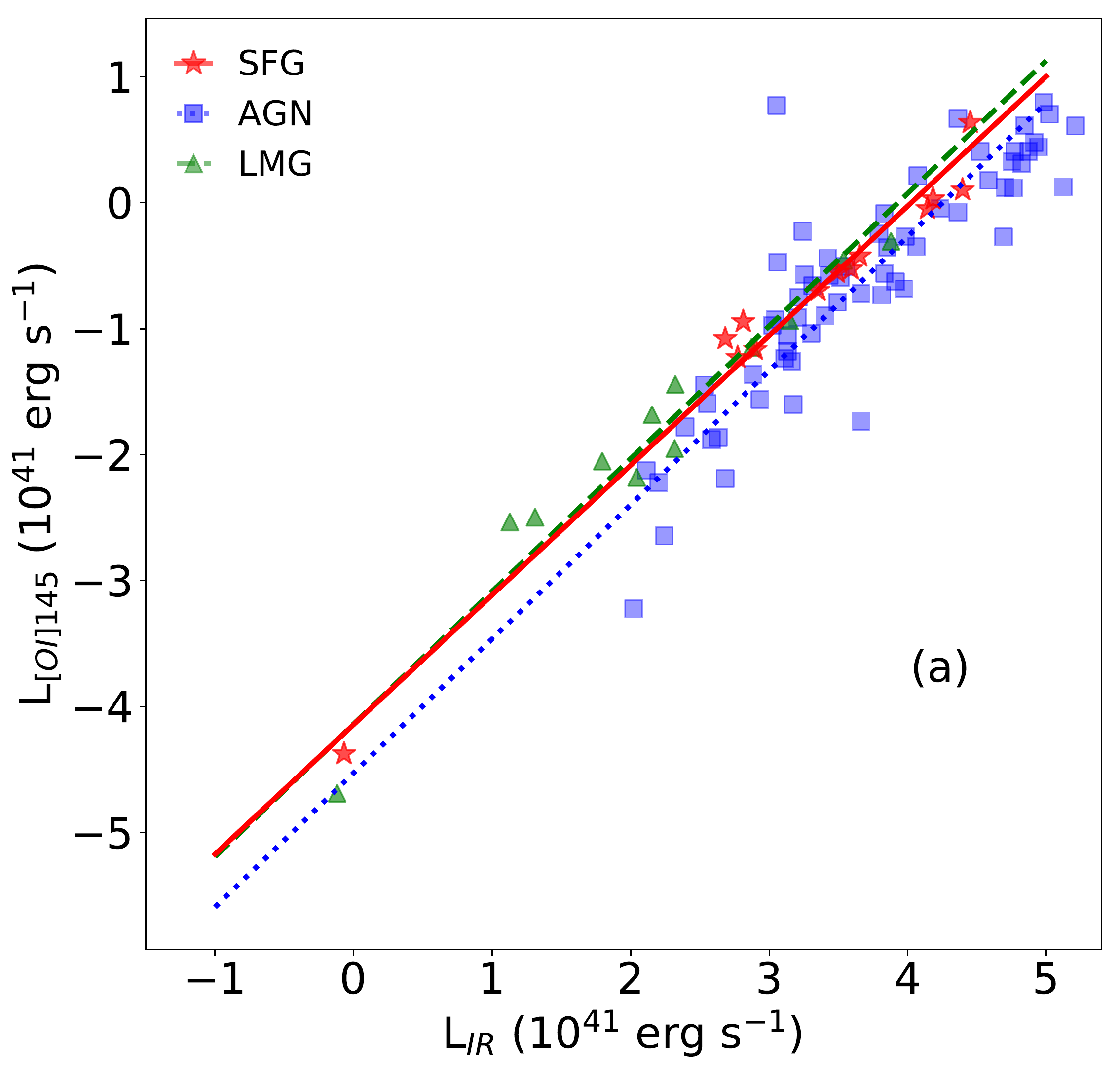}
\includegraphics[width=0.33\columnwidth]{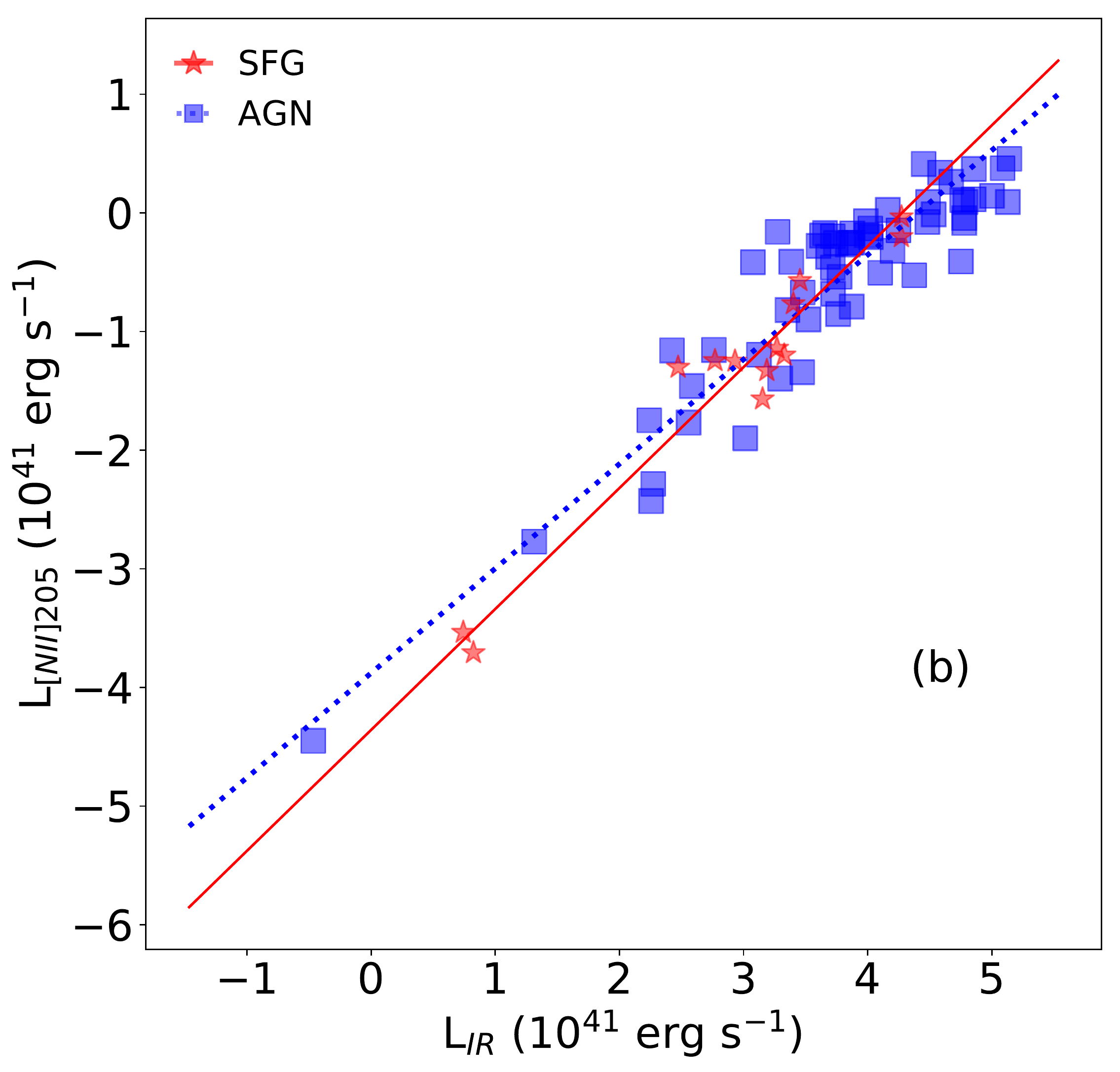}
\caption{{\bf (a: left)} The [OI]145$\, \rm{\micron}$ line luminosity as a function of the total IR luminosity. {\bf (b: right)} The [NII]122$\, \rm{\micron}$ line luminosity as a function of the total IR luminosity. The same legend as in Fig.\,\ref{fig:corr_c2_ne2_ne3} was used.}
\label{fig:corr_app_8}
\end{figure*}

\begin{figure*}
\centering
\includegraphics[width=0.33\columnwidth]{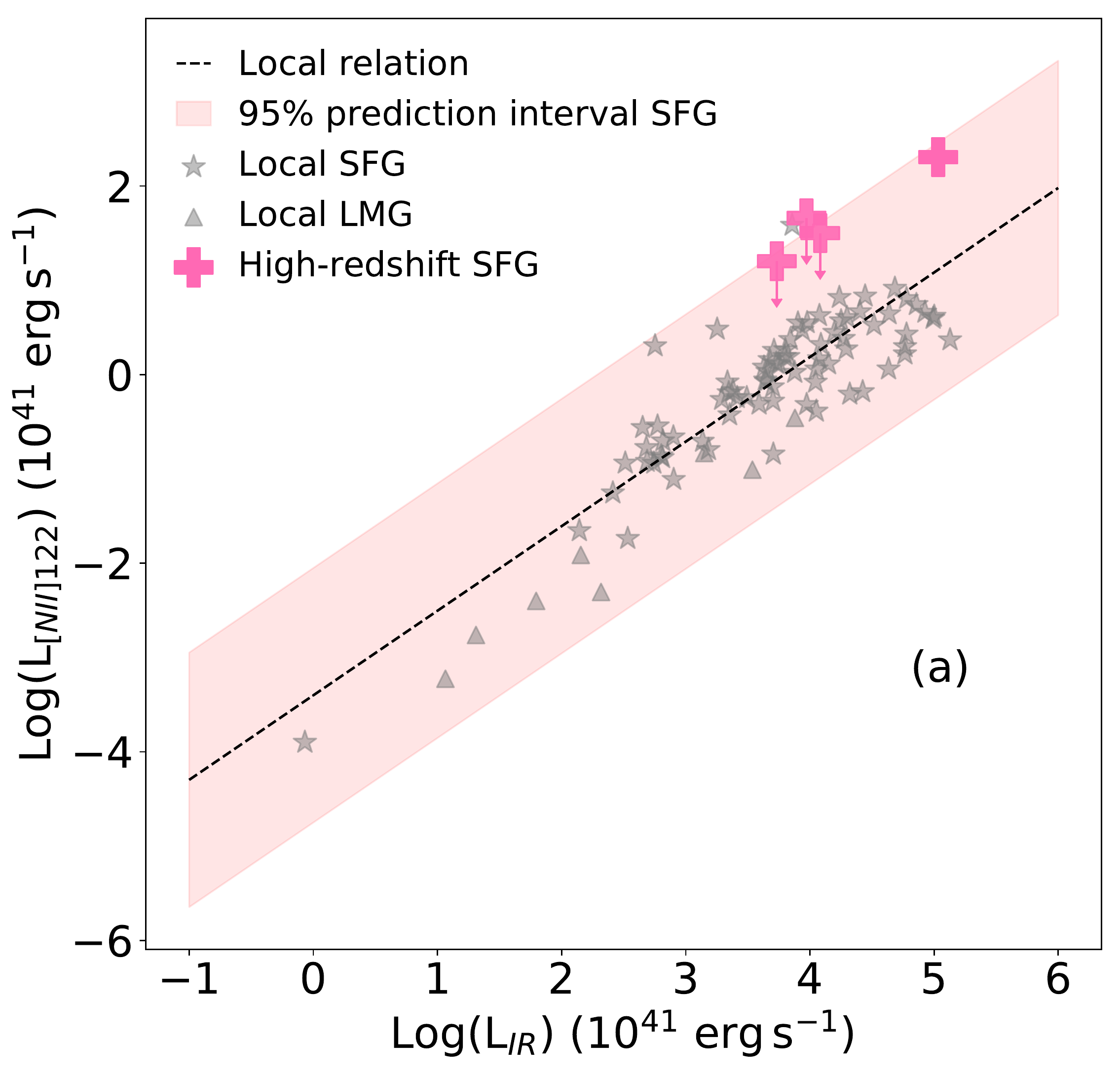}
\includegraphics[width=0.33\columnwidth]{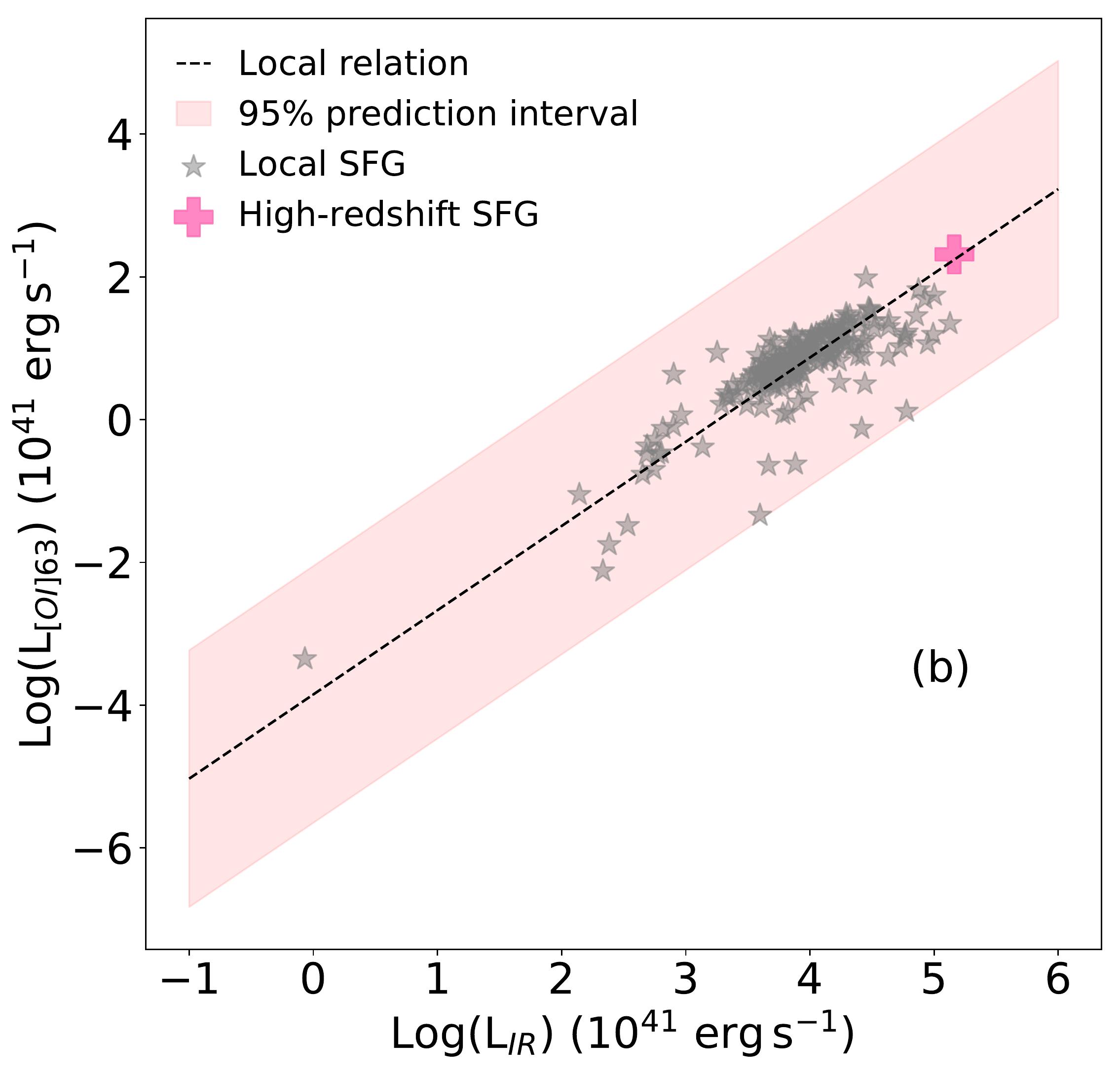}
\includegraphics[width=0.33\columnwidth]{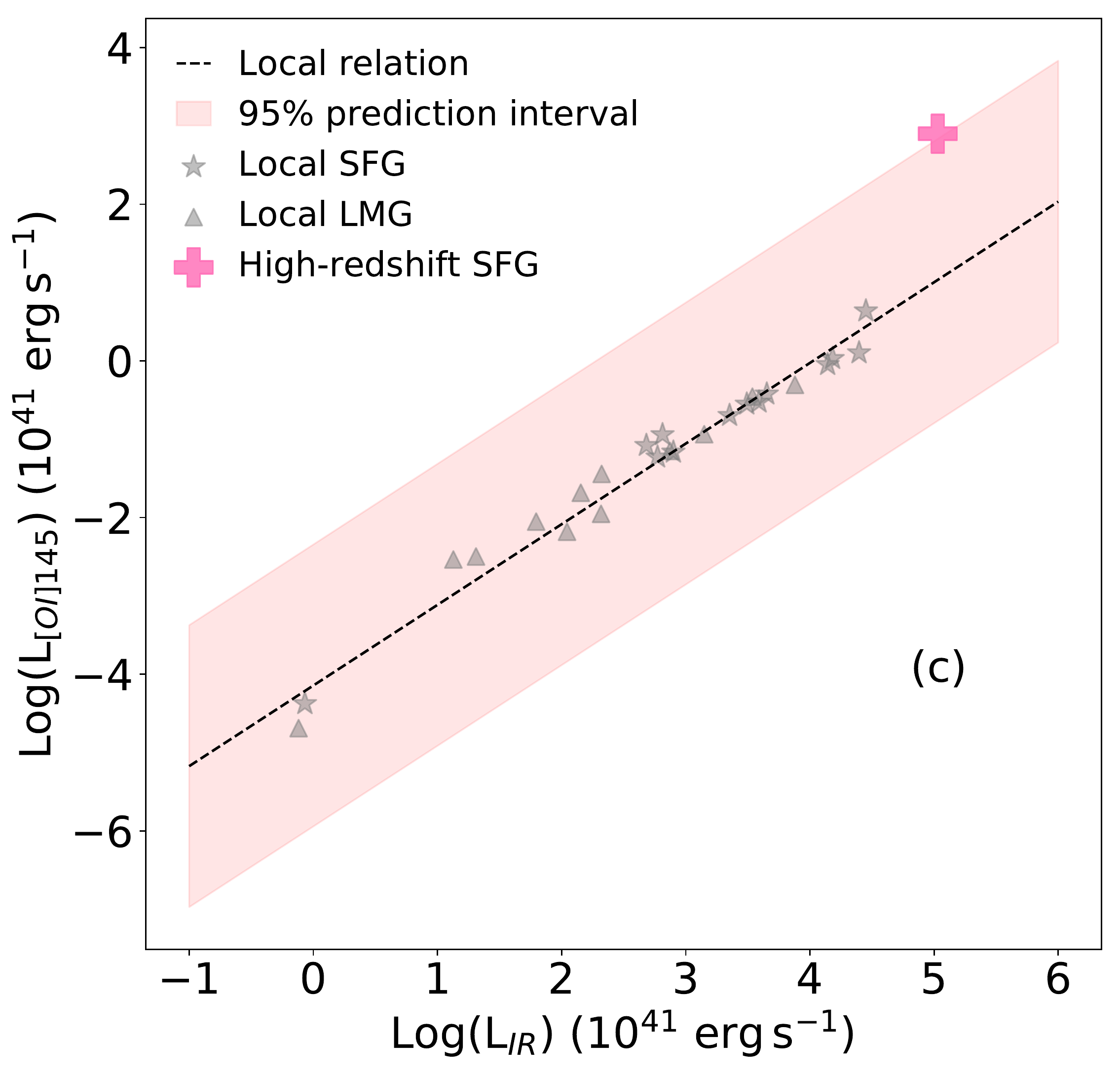}
\caption{{\bf (a: left)} Comparison between the local L$_{IR}$-L$_{[NII]122}$  relation for local SFG (black dashed line) and high redshift detections of [NII]122$\, \rm{\micron}$ line. The shaded area shows the 95$\%$ prediction interval for the local relation. The pink $+$ sign indicates a detection for the [NII]122$\, \rm{\micron}$ line \citep{debreuck2019}, while symbols with arrows indicate upper limits for high redshift sources \citep{harikane2019}. {\bf (b: centre)} Comparison between the local L$_{IR}$-L$_{[OI]63}$  relation for local SFG (black dashed line) and high redshift detections of [OI]63$\, \rm{\micron}$ line \citep[pink symbol][]{rybak2019}. {\bf (c: right)} Comparison between the local L$_{IR}$-L$_{[OI]145}$  relation for local SFG (black dashed line) and high redshift detections of [OI]145$\, \rm{\micron}$ line \citep[pink symbol][]{debreuck2019}. The shaded area shows the 95$\%$ prediction interval for the local relation. Grey stars show local SFG, while grey triangles show local LMG.}
\label{fig:corr_app_9}
\end{figure*}

\clearpage

\section{[NeV]14.3 \texorpdfstring{$\, \rm{\micron}$}{micron} as BHAR tracer and discussion on the use of the mid-ionisation lines}\label{app:bhar}

\begin{figure*}[b!!]
\centering
\includegraphics[width=0.33\columnwidth]{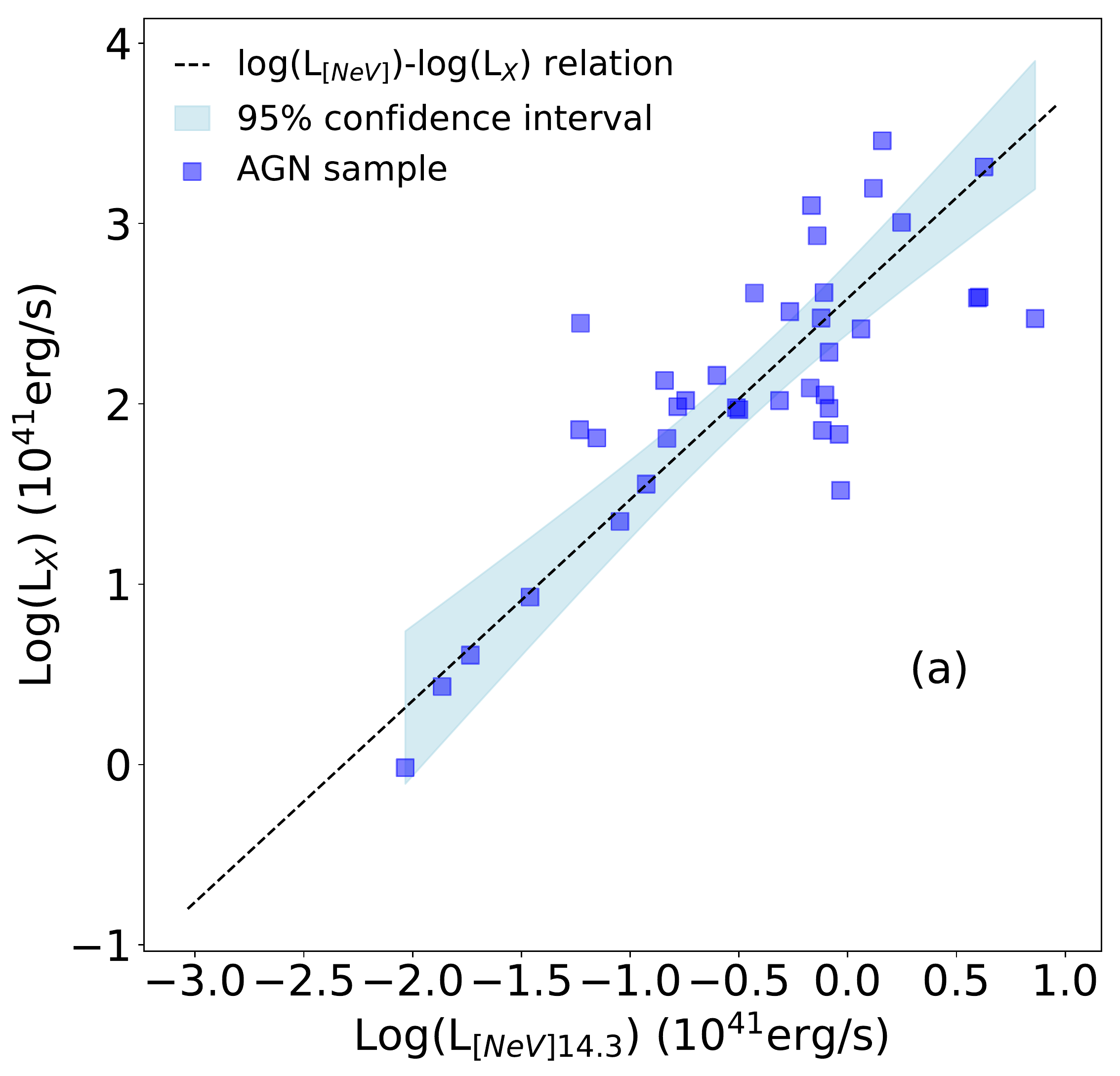}
\includegraphics[width=0.33\columnwidth]{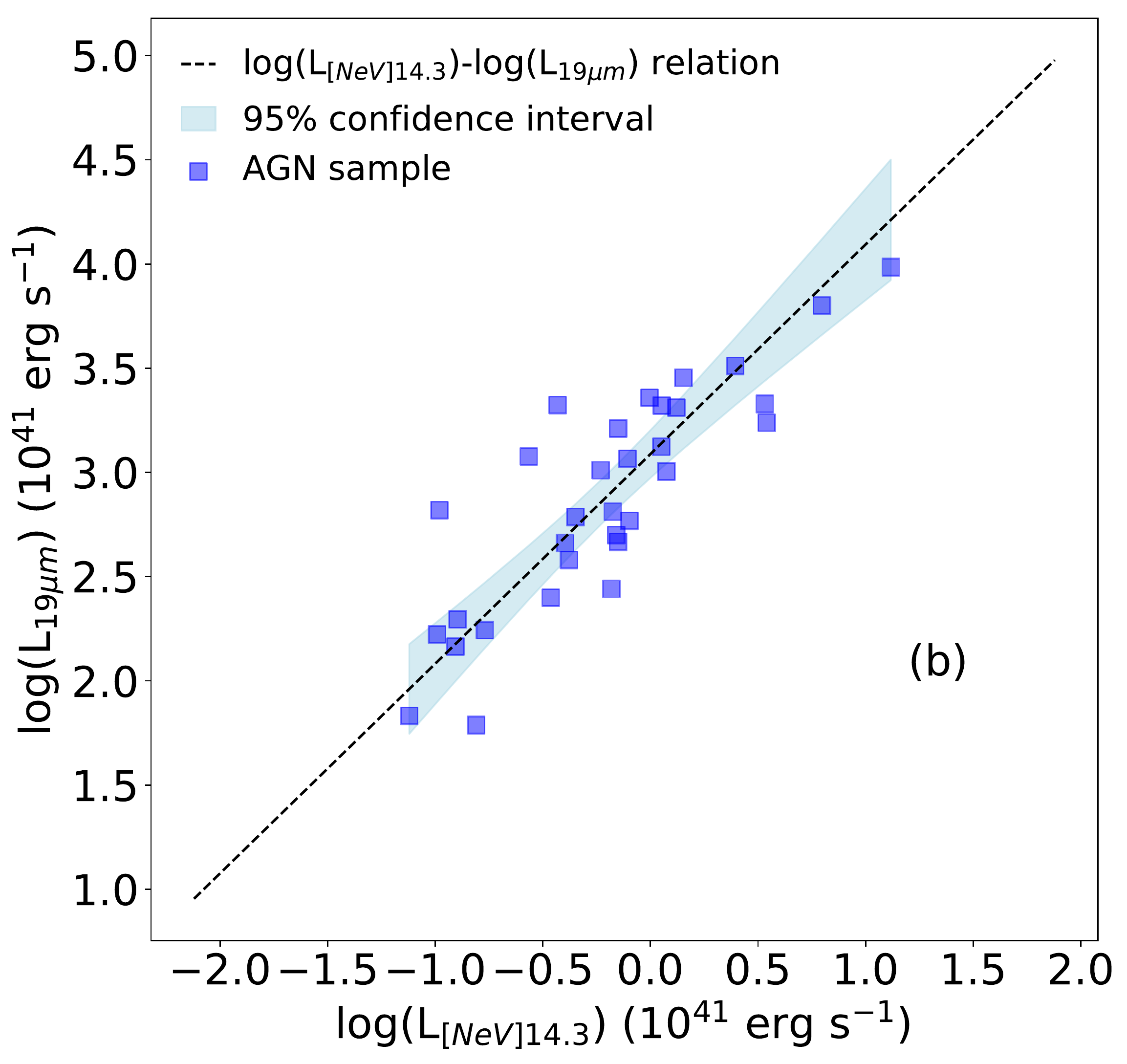}
\includegraphics[width=0.33\columnwidth]{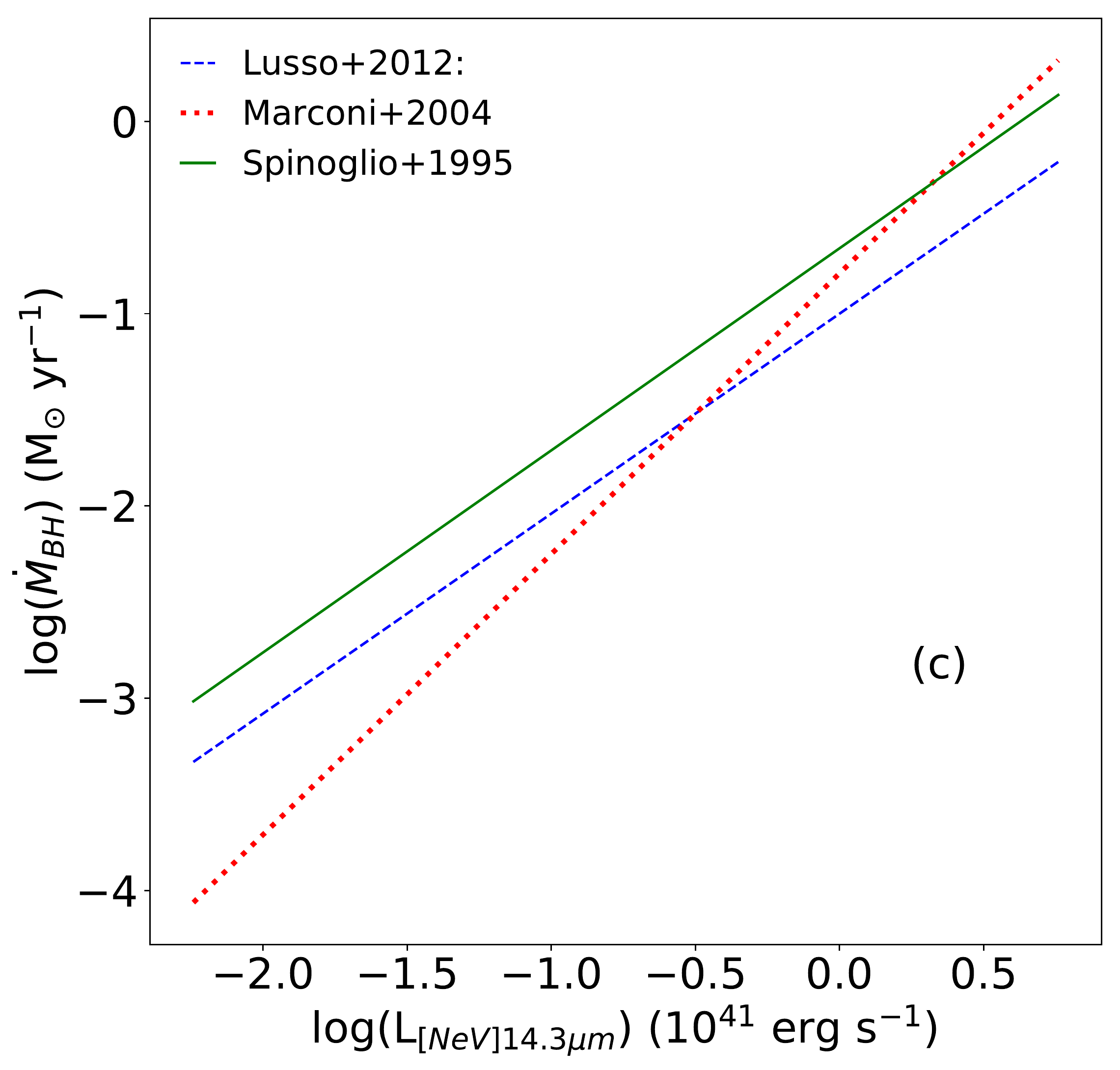}
\caption{{\bf (a: left)} Linear correlation between the [NeV]14.3$\, \rm{\micron}$ and the 2-10 keV X-ray luminosity. Blue squares show the AGN sample, and the shaded area shows the 95$\%$ confidence interval for the relation.{\bf (b: centre)} Linear correlation between the [NeV]14.3$\, \rm{\micron}$ line luminosity and the 19$\, \rm{\micron}$ luminosity.  {\bf (c: right)} Comparison between the [NeV]14.3$\, \rm{\micron}$ line luminosity and the BHAR derived from the bolometric correction by \citeauthor{spinoglio1995} (green solid line), \citeauthor{marconi2004} (red dotted line) and \citeauthor{lusso2012} (blue dashed line).}
\label{fig:corr_app_ne5}
\end{figure*}

In this section, following what has been done in Section\,\ref{sec:BHAR}, we report the results for the [NeV]14.3$\, \rm{\micron}$ line as BHAR tracer. In particular, starting from the total sample of \citet{tommasin2008,tommasin2010}, excluding Compton-thick objects we obtain a sample of 38 objects. The correlation between the [NeV]14.3$\, \rm{\micron}$ line luminosity and the 2-10 keV X-ray luminosity is shown in Fig\,\ref{fig:corr_app_ne5}a and is expressed by the equation:
\begin{equation}
    \log\left(\frac{L_{X}}{\rm 10^{41}\,erg\,s^{-1}}\right)=(2.58 \pm 0.10)+(1.11 \pm 0.13)\log\left(\frac{L_{\rm [NeV]14.3}}{\rm 10^{41}\,erg\,s^{-1}}\right)
\end{equation}
with a Pearson $r$ coefficient of $r=$0.79. From this result, we then apply the bolometric corrections by \citet{marconi2004} and \citet{lusso2012} to the 2-10 keV X-ray luminosity, in order to obtain the bolometric luminosity of the AGN. From the bolometric luminosities we then calculate the linear correlation linking the [NeV]14.3$\, \rm{\micron}$ line luminosity to the BHAR, with $L_{AGN}=\eta \dot{M}_{BH}c^{2}$. Assuming a radiative efficiency of $\eta =0.1$, we report in equation \ref{eq:lusso_ne5_14} the linear correlation applying the bolometric correction from \citet{lusso2012}, and in equation \ref{eq:marconi_ne5_14} the one applying the \citet{marconi2004} bolometric correction, and the relative Pearson r coefficient:
\begin{equation}\label{eq:lusso_ne5_14}
    \log\left(\frac{\dot{M}_{BH}}{\rm M_{\odot}\,yr^{-1}}\right)=(-1.0 \pm 0.10)+(1.04 \pm 0.13)\log\left(\frac {L_{\rm [NeV]14.3}}{\rm 10^{41}\,erg\,s^{-1}}\right),\,\,r=0.76
\end{equation}

\begin{equation}\label{eq:marconi_ne5_14}
   \log\left(\frac{\dot{M}_{BH}}{\rm M_{\odot}\,yr^{-1}}\right)=(-0.73 \pm 0.17)+(1.49 \pm 0.13)\log\left(\frac{L_{\rm [NeV]14.3}}{\rm 10^{41}\,erg\,s^{-1}}\right),\,\,r=0.78
\end{equation}

We have then analysed the correlation between the [NeV]$14.3\, \rm{\micron}$ line with the luminosity at 19$\, \rm{\micron}$ following the work by \citet{tommasin2010}. As in Section\,\ref{sec:BHAR}, we select only those sources with an AGN component at 19$\, \rm{\micron}$ equal or above 85$\%$, obtaining a subsample of 32 objects. The correlation between the [NeV]14.3$\, \rm{\micron}$ line luminosity and the 19$\, \rm{\micron}$ luminosity is shown in Fig.\,\ref{fig:corr_app_ne5}b, and expressed by:
\begin{equation}
    \log\left(\frac{L_{19}}{\rm 10^{41}\,erg\,s^{-1}}\right)=(3.09 \pm 0.06)+(1.01 \pm 0.10)\log\left(\frac{L_{\rm [NeV]14.3}}{\rm 10^{41}\,erg\,s^{-1}}\right)
\end{equation}
with a Pearson $r$ coefficient of r=0.85. We then calculate the bolometric luminosity from the monochromatic 12$\, \rm{\micron}$ luminosity extracted from \citet{deo2009} using the relation by \citet{spinoglio1995}. From the bolometric luminosity we determine the BHAR and its relation to the [NeV]14.3$\, \rm{\micron}$ line luminosity, obtaining:
\begin{equation}\label{eq:spinoglio_ne5_14}
   \log\left(\frac{\dot{M}_{BH}}{\rm M_{\odot}\,yr^{-1}}\right)=(-0.66 \pm 0.11)+(1.05 \pm 0.19)\log\left(\frac{L_{\rm [NeV]14.3}}{\rm 10^{41}\,erg\,s^{-1}}\right),\,\,r=0.68
\end{equation}
for a sample of 26 objects, where $r$ is the Pearson coefficient. The small number of objects used to derive this relation is due to the lack of data for the determination of the continuum at 12$\, \rm{\micron}$.

In Fig.\,\ref{fig:corr_app_ne5}c we compare the three different relation derived to determine the BHAR from the [NeV]14.3$\, \rm{\micron}$ line luminosity. as in Section\,\ref{sec:BHAR}, also for this line we find good agreement between the results obtained with the \citeauthor{lusso2012} and \citeauthor{spinoglio1995} bolometric corrections, while the \citeauthor{marconi2004} bolometric correction produces a significantly steeper relationship.

\begin{figure*}[hb!!]
    \centering
    \includegraphics[width=0.33\columnwidth]{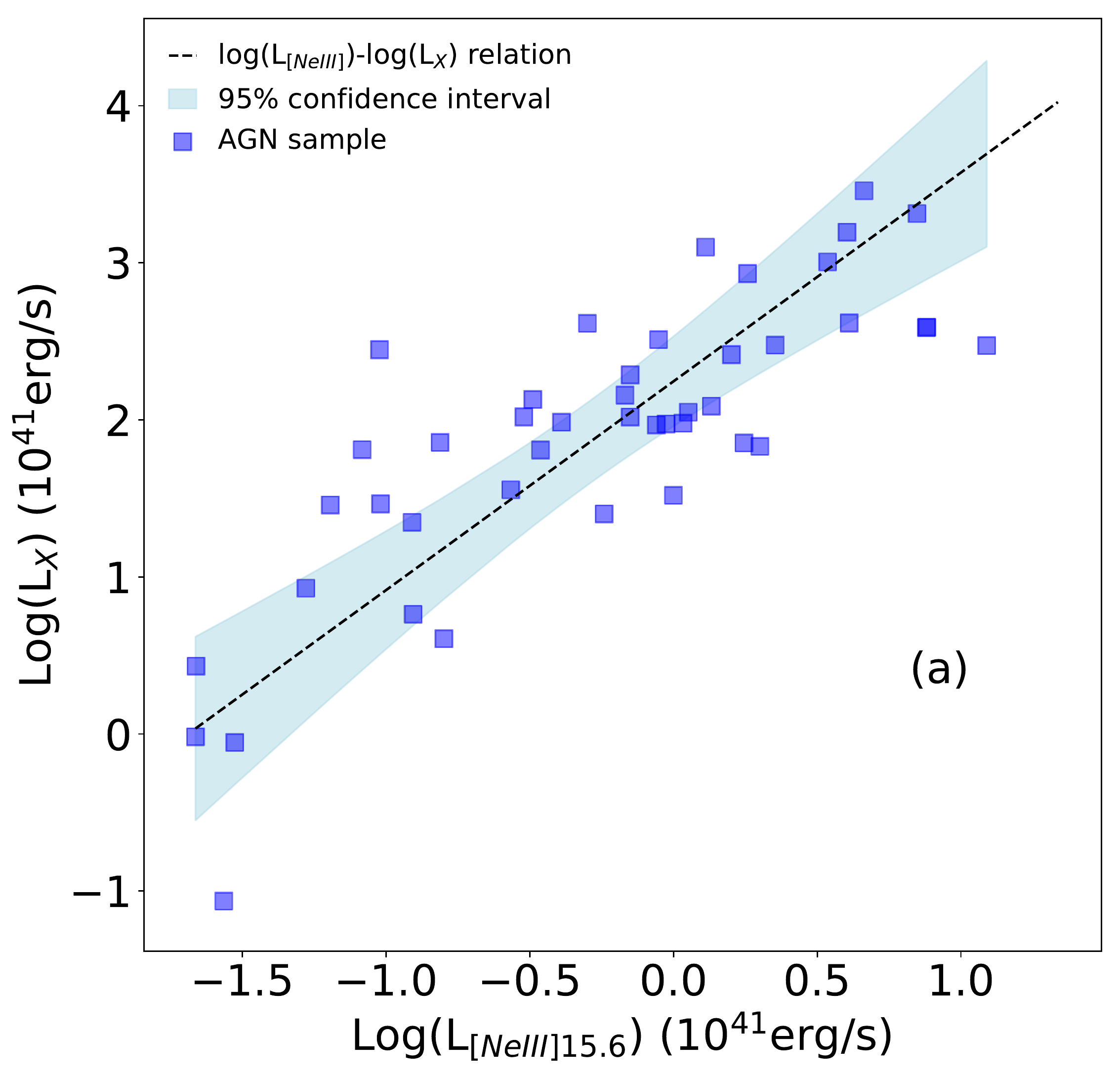}
    \includegraphics[width=0.33\columnwidth]{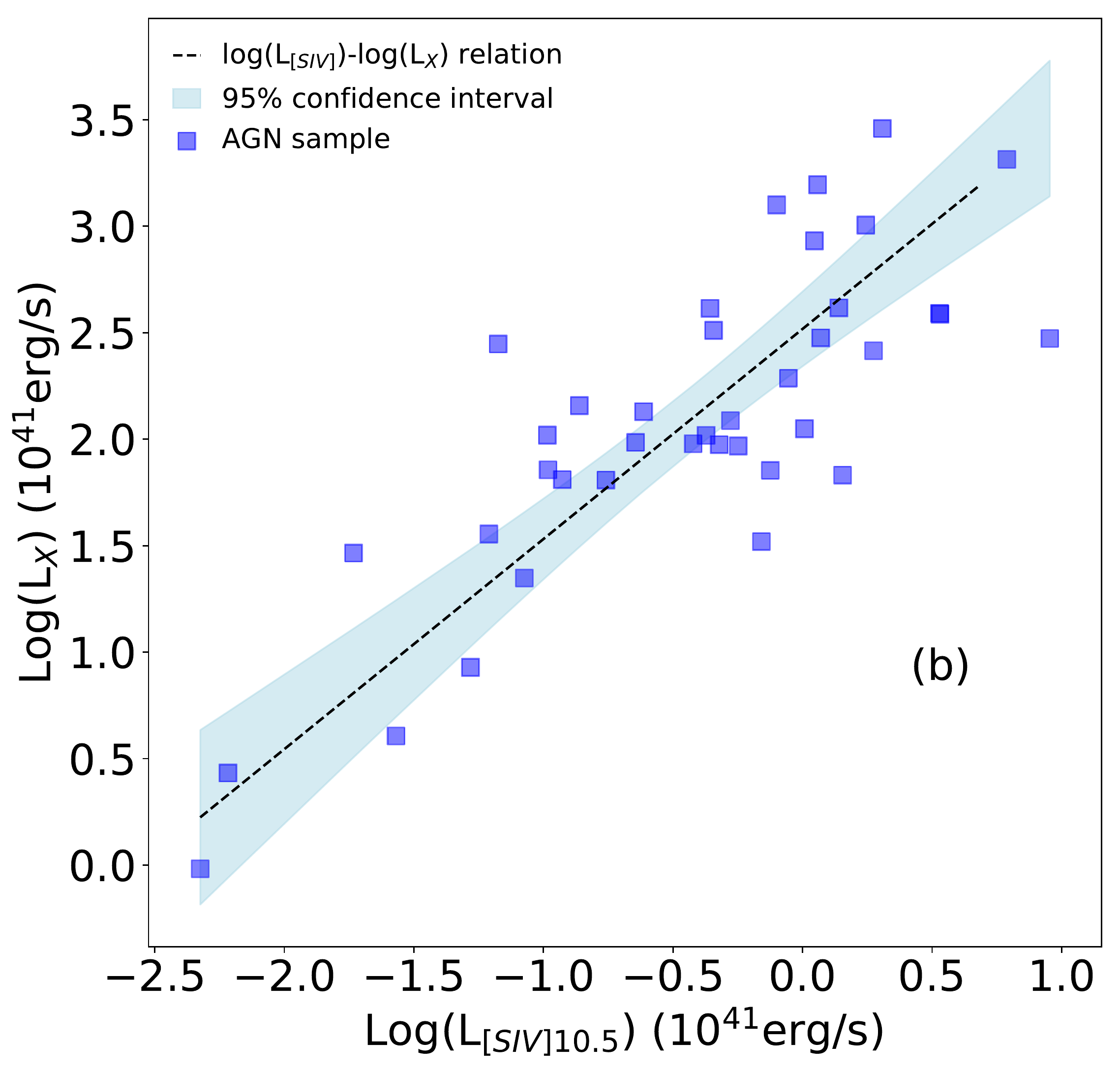}
    \caption{{\bf (a: left)} Linear correlation between the [NeIII]15.6$\, \rm{\micron}$ and the 2-10 keV X-ray luminosity. Blue squares show the AGN sample, and the shaded area shows the 95$\%$ confidence interval for the relation. {\bf (b: right)} Linear correlation between the [SIV]10.5$\, \rm{\micron}$ and the 2-10 keV X-ray luminosity}
    \label{fig:ne3_s4_x}
\end{figure*}

As presented in Section\,\ref{sec:BHAR} and here, high ionisation lines, with an ionisation potential above the threshold of doubly ionised helium (54.4 eV) can only be efficiently produced by AGN activity, and thus are optimal BHAR tracers. Mid-ionisation lines, however, are partially affected by AGN activity, and could in theory be used as tracers for the BHAR. For this reason, we have tested two mid-ionisation lines, namely the [SIV] 10.5$\mu$m line, and the [NeIII]15.6$\mu$m line, to evaluate their use as tracers for the X-ray luminosity in the 2-10 keV interval. Starting from the sample of \citet{tommasin2008,tommasin2010}, as in Sect.\,\ref{sec:BHAR}, we exclude all Compton-thick objects and obtain a sub-sample of 44 objects for the [NeIII] line, and of 39 objects for the [SIV] line. Fig.\,\ref{fig:ne3_s4_x}a shows the correlation between the [NeIII] line luminosity and the 2-10 keV X-ray luminosity, expressed as:
\begin{equation}
        \log\left(\frac{L_{X}}{\rm 10^{41}\,erg\,s^{-1}}\right)=(2.20 \pm 0.10)+(1.33 \pm 0.13)\log\left(\frac{L_{\rm [NeIII]15.6}}{\rm 10^{41}\,erg\,s^{-1}}\right)
\end{equation}
The correlation of the [SIV]10.5 $\mu$m line with the 2-10 keV X-ray luminosity, shown in Fig.\,\ref{fig:ne3_s4_x}b, can be expressed as:
 \begin{equation}
        \log\left(\frac{L_{X}}{\rm 10^{41}\,erg\,s^{-1}}\right)=(2.51 \pm 0.09)+(0.98 \pm 0.10)\log\left(\frac{L_{\rm [SIV]10.5}}{\rm 10^{41}\,erg\,s^{-1}}\right)
\end{equation}

For both correlations, we find a low Pearson correlation coefficient, equal to $r=0.63$, lower than the values found for [OIV] ($r=0.87$) and the [NeV] lines ($r=0.79$ and $r=0.84$, for the 14.3$\mu$m and 24.3$\mu$m lines, respectively). This suggests that, while mid-ionisation lines can correlate with the X-ray luminosity, they may suffer from contributions due to star formation activity, and thus are far less reliable as BHAR proxies. In particular, AGN sources in low-metallicity star-forming galaxies would have a significant contamination of the [SIV] and [NeIII] lines from the stellar population, as shown by the LMG in our sample (Figs.\,\ref{fig:corr_c2_ne2_ne3}c and \ref{fig:corr_app_2}c), in contrast with the [OIV] and [NeV] lines.

\clearpage

\begin{landscape}
\begin{table}
\section{Line and PAH calibrations}
\centering
\caption{New calibration obtained for fine-structure lines and PAH bands. For each class of objects, are reported the slope (a) and intercept (b) of the linear correlation with relative errors, the number of data from which the linear correlation was calculated (N) and the best-fit coefficients of determination (r).}\label{tab:calibrations}
\begin{tabular}{l|cccc|cccc|cccc}
\hline \\[0.03cm]%\hline
\bf Line/feature & \multicolumn{4}{c}{\bf AGN} & \multicolumn{4}{c}{\bf SFG} & \multicolumn{4}{c}{\bf LMG} \\[0.1cm]
%\hline\\[-0.3cm]
 & $a$ $\pm$ $\delta a$ & $b$ $\pm$ $\delta b$ & N & $r$ & $a$ $\pm$ $\delta a$ & $b$ $\pm$ $\delta b$ & N & $r$ & $a$ $\pm$ $\delta a$ & $b$ $\pm$ $\delta b$ & N & $r$  \\
\hline \\[0.05cm]
PAH 6.2 & 1.04$\pm$0.07 & -2.73$\pm$0.24 & 65 & 0.87 & 1.15$\pm$0.06 & -2.99$\pm$0.24 & 150 & 0.81 & --&-- &-- &-- \\
{[NeVI]}7.65 & 1.22$\pm$0.24 & -4.12$\pm$0.82 & 8 & 0.89 & -- & -- & -- & -- & -- & -- & --& -- \\
PAH 7.7 & --&-- &-- &-- &  1.29$\pm$0.07 & -2.98$\pm$0.27 & 150 & 0.81 & --&-- &-- &-- \\
PAH 8.6 & --&-- &-- &-- &  1.43$\pm$0.23 & -4.25$\pm$0.67 & 16 & 0.84 & --& --&-- &--\\
%ArII 6.99
H$_{2}$9.67 & 1.20$\pm$0.28 & -4.97$\pm$1.17 & 9 & 0.82 &1.39$\pm$0.06 & -5.43$\pm$0.26 & 137 & 0.87 & -- & -- & -- &  -- \\
{[SIV]}10.51 & 1.37$\pm$0.11 & -5.19$\pm$0.38 & 75 & 0.80 & 1.58$\pm$0.24& -7.00$\pm$0.94 & 51 & 0.59 & 1.03$\pm$0.08 & -3.08$\pm$0.16 & 29 & 0.91 \\
PAH 11.3 & 1.04$\pm$0.06 & -2.82$\pm$0.20 & 90 & 0.88 & 1.11$\pm$0.06 & -2.79$\pm$0.23 & 150 & 0.81 & -- & -- & -- & --\\
H$_{2}$12.28 & 1.25 $\pm$ 0.10 & -5.41 $\pm$ 0.37 & 31 & 0.91 & 1.18 $\pm$ 0.05 & -4.79$\pm$ 0.19 & 114 & 0.91 & -- & -- & --& -- \\
{[NeII]}12.81 & 1.19$\pm$0.07 & -4.23$\pm$0.26 & 86 & 0.85 &  1.04$\pm$0.43 & -3.48$\pm$0.17 & 186 & 0.84 & 1.37$\pm$0.10 & -4.36$\pm$0.20 & 21 & 0.95 \\
{[NeV]}14.32 & 1.32$\pm$0.11 & -5.01$\pm$0.39 & 74 & 0.78  & -- & -- & -- & -- & -- & -- & -- & -- \\
{[NeIII]}15.56 & 1.44 $\pm$0.10 & -5.16$\pm$0.36 & 88 & 0.80 & 1.13$\pm$0.06 & -4.60$\pm$0.21 & 182 & 0.80 &1.13$\pm$0.06 & -3.11$\pm$0.12 & 28 & 0.96 \\
PAH 17 & --&-- &-- &-- &  1.14$\pm$0.06 & -3.08$\pm$0.23 & 131& 0.84 & -- & -- & -- &  -- \\
H$_{2}$17.03 & 1.14$\pm$0.06 & -4.64$\pm$0.19 & 76 & 0.91 &1.22$\pm$0.05 & -4.58$\pm$0.20 & 135 & 0.89 & -- & -- & -- &  -- \\
{[SIII]}18.71 &1.22$\pm$0.09 & -4.62$\pm$0.30 & 70 & 0.83 &  1.26$\pm$0.11 & -4.79$\pm$0.42 & 140 & 0.60 & 1.14$\pm$0.06 & -3.42$\pm$0.12 & 25 & 0.96 \\
{[FeIII]}22.93&--&--&--&--  & 1.33$\pm$0.13 & -5.79$\pm$0.44 & 14 & 0.64 & -- & -- & -- & -- \\
{[NeV]}24.32 & 1.27$\pm$0.11 & -4.73$\pm$0.37 & 65 & 0.80 & -- & -- & -- & -- & -- & -- & -- & -- \\
{[OIV]}25.89 & 1.27$\pm$0.14 & -4.13$\pm$0.48 & 34 & 0.82& 1.15$\pm$0.05& -5.00$\pm$0.20 & 144 & 0.86 & 0.88$\pm$0.13 & -3.71$\pm$0.27 & 16 & 0.86 \\
{[FeII]}25.98&--&--&--&--  & 0.80$\pm$0.18 & -3.99$\pm$0.51 & 35 & 0.44& -- & -- & -- & -- \\
H$_{2}$28.22 & 1.04$\pm$0.14 & -4.44 $\pm$ 0.45 & 25 & 0.81 & 1.65$\pm$0.24 & -6.51$\pm$0.82 & 11 & 0.91 &--&--&--&--\\
{[SIII]}33.48 & 1.10$\pm$0.07 & -3.85$\pm$0.23& 75 & 0.87 & 1.13$\pm$0.05 & -3.94$\pm$0.20 & 170 & 0.83 & 1.11$\pm$0.08 & -3.24$\pm$0.17 & 20 & 0.96 \\
{[SiII]}34.81 & 1.21$\pm$0.08 & -3.99$\pm$0.28 & 73 & 0.84 & 1.15$\pm$0.04 & -3.72 $\pm$0.15 & 167 & 0.91& 1.05$\pm$0.08 & -3.14$\pm$0.18 & 17 & 0.95 \\
{[OIII]}51.81 & 0.92$\pm$0.13 & -3.10$\pm$0.54 & 19 & 0.85 & -- & -- & -- & -- & -- & -- & -- & -- \\
{[NIII]}57.32 & 0.98$\pm$0.06 & -3.96$\pm$0.24 & 31 & 0.94 & - & - & - & - & - & - & - & -\\
{[OI]}63.18 & 1.07$\pm$0.04 & -3.56$\pm$0.17 & 106 & 0.91 & 1.18$\pm$0.05 & -3.86$\pm$0.21 & 183 & 0.83 & 1.07$\pm$0.06 & -3.00$\pm$0.11 & 31 & 0.96\\
{[OIII]}88.36 & 1.18$\pm$0.06 & -4.34$\pm$0.22 & 81 & 0.91 & 1.20$\pm$0.06 & -4.27$\pm$0.25 & 117 & 0.85 & 1.05$\pm$0.06 & -2.53$\pm$0.10 & 37 & 0.95\\
{[NII]}121.9 & 1.08$\pm$0.06 & -4.43$\pm$0.22 & 77 & 0.89 & 0.90$\pm$0.06 & -3.39$\pm$0.22 & 75 & 0.86 & 0.96$\pm$0.09 & -4.17$\pm$0.23 & 8 & 0.97\\
{[OI]}145.5 & 1.06$\pm$0.07 & -4.53$\pm$ 0.26 & 64 & 0.88 & 1.03$\pm$0.05 & -4.14$\pm$0.16 & 13 & 0.99 & 1.05$\pm$0.07 & -4.14$\pm$ 0.18 & 12 & 0.97 \\
{[CII]}157.7 & 1.00$\pm$0.04 & -3.18$\pm$ 0.14 & 149 & 0.90 & 0.97$\pm$0.05 & -2.76$\pm$0.20 & 183 & 0.78 & 1.09$\pm$0.06 & -2.88$\pm$0.10 & 40 & 0.95\\
{[NII]}205 & 0.88$\pm$0.05 & -3.88$\pm$ 0.18 & 60 & 0.92 & 1.02$\pm$0.07 & -4.36$\pm$0.23 & 13 & 0.97 & -- & -- & -- & --\\

\hline
\end{tabular}
\end{table}
\end{landscape}

\clearpage

\section{Differences with previous results}
\label{app:comparison}

We analyse here the differences of the line calibrations obtained in this work with respect to those of  \citet{spinoglio2012,spinoglio2014} and \citet{gruppioni2016}.

\begin{figure*}[h!]
\centering
\includegraphics[width=0.33\columnwidth]{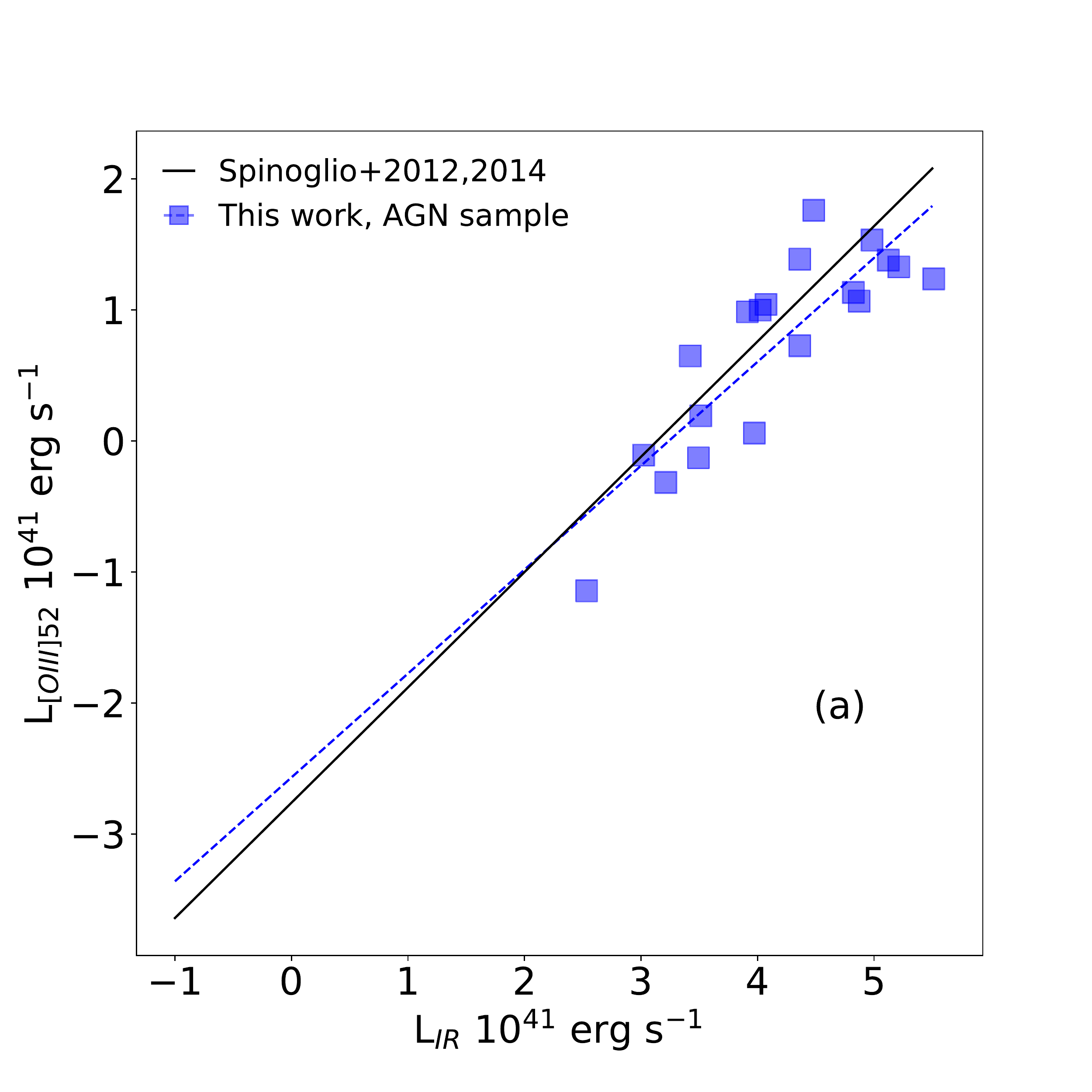}
\includegraphics[width=0.33\columnwidth]{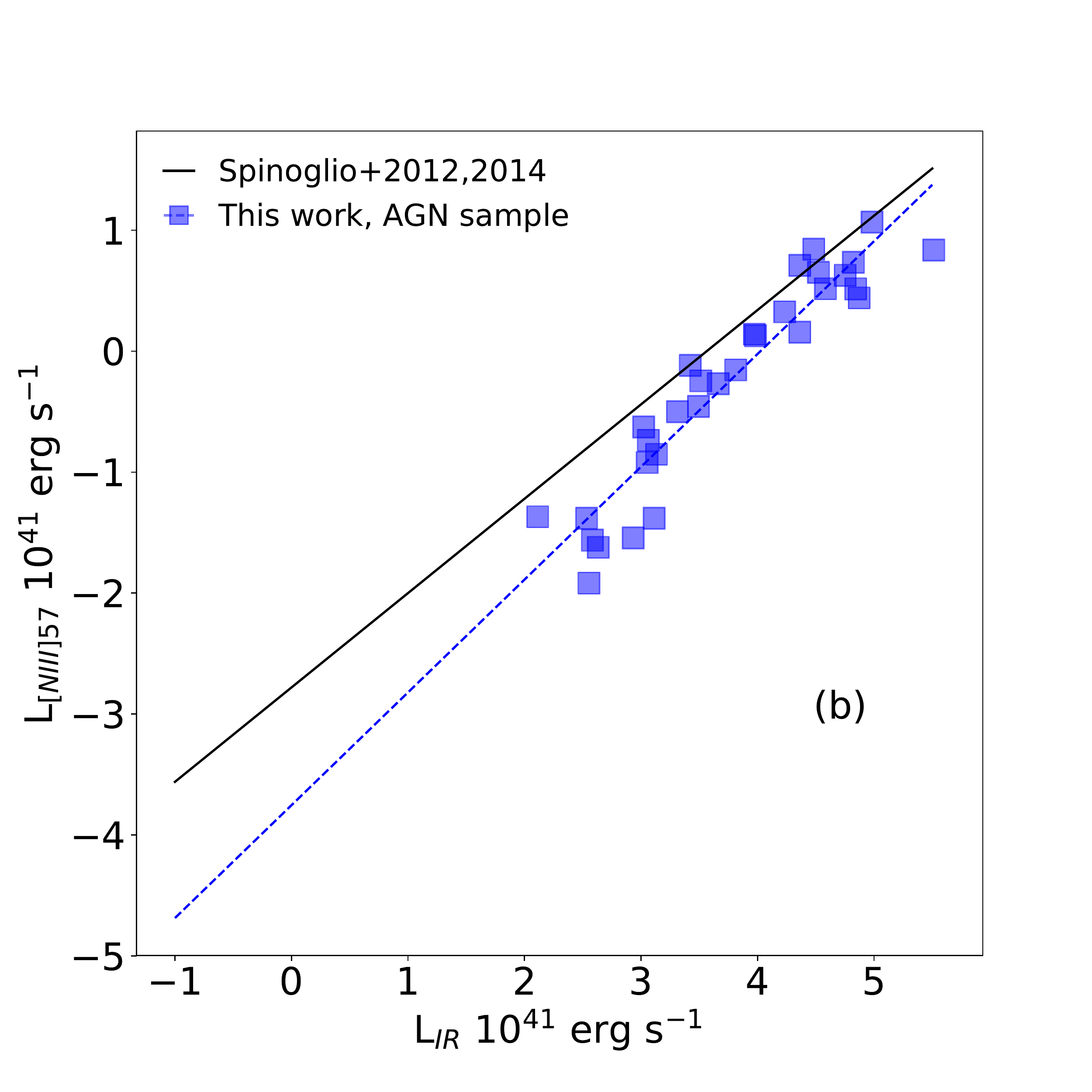}
\includegraphics[width=0.33\columnwidth]{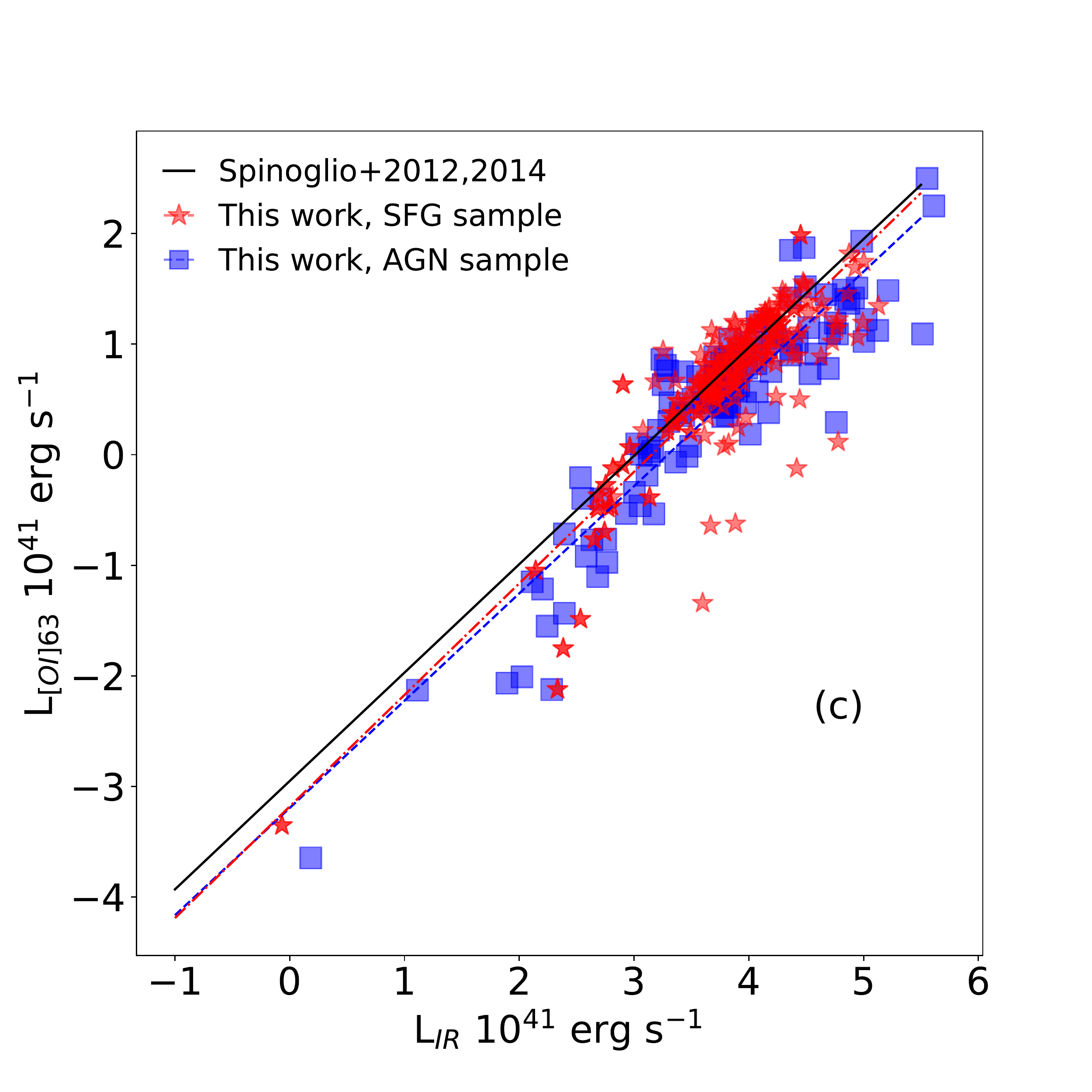}
\caption{{\bf (a: left)} The [OIII]52$\mu$m line luminosity versus the total IR luminosity. {\bf (b: centre)} The [NIII]57$\mu$m line luminosity versus the total IR luminosity.  {\bf (c: right)} The [OI]63$\mu$m line luminosity versus the total IR luminosity. In the figures, Blue squares represent AGN while red stars show the SFG sample. The dashed blue line represents the linear he dot-dashed red line the relation for SFG, and the black solid line the relation shows the relation obtained by \citet{spinoglio2012,spinoglio2014}.}
\label{fig:comp_sp2012_1}
\end{figure*}

\begin{figure*}[h!]
\centering
\includegraphics[width=0.33\columnwidth]{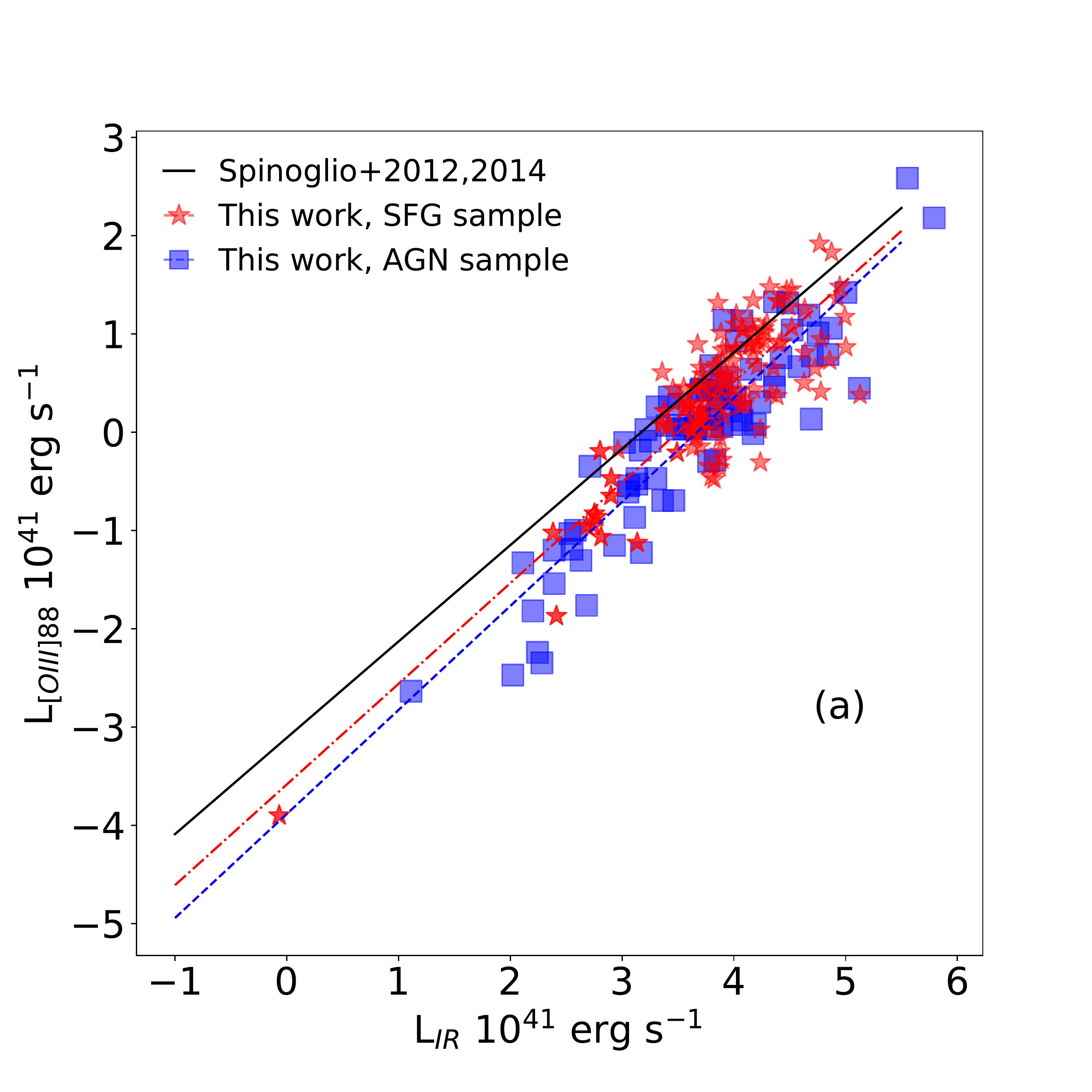}
\includegraphics[width=0.33\columnwidth]{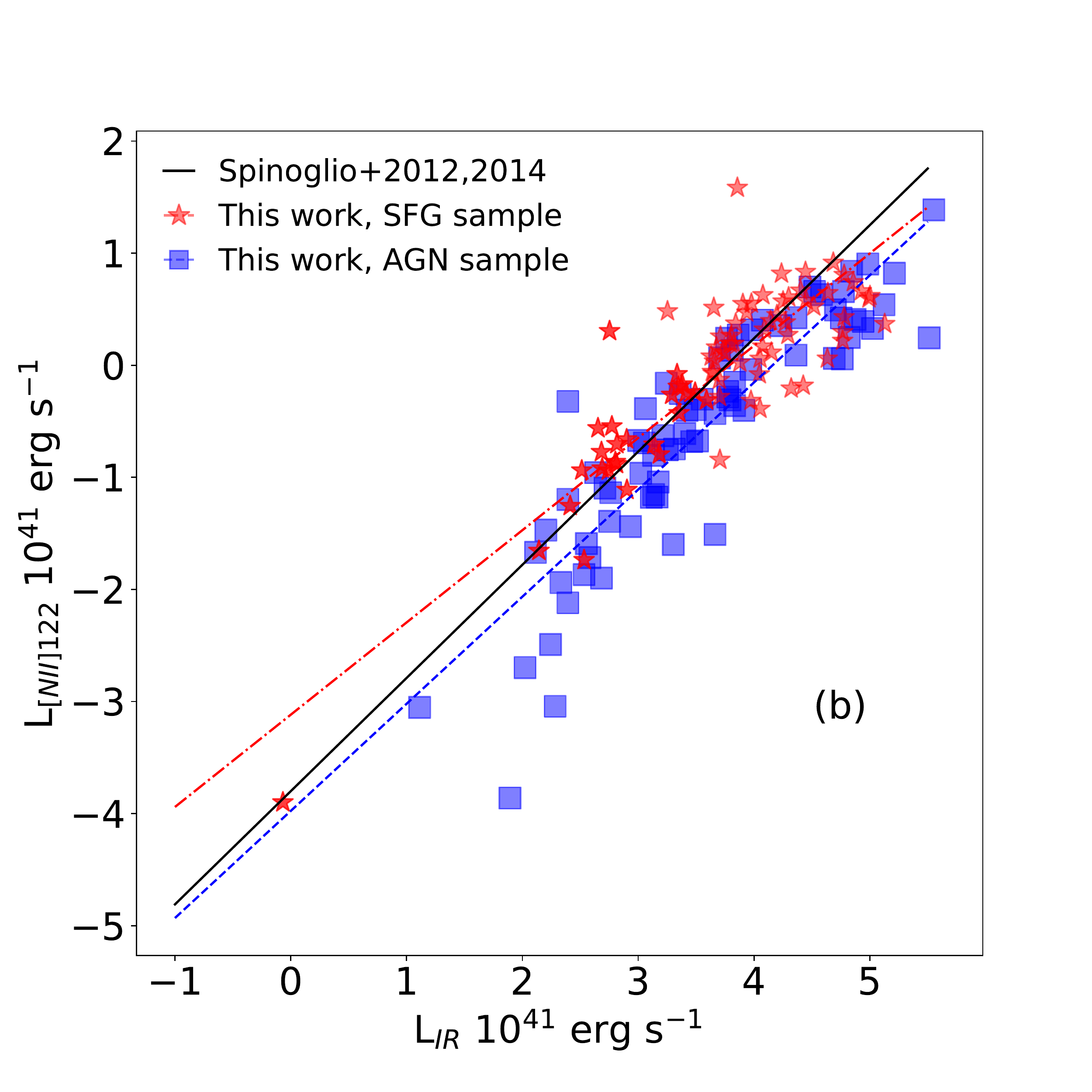}
\includegraphics[width=0.33\columnwidth]{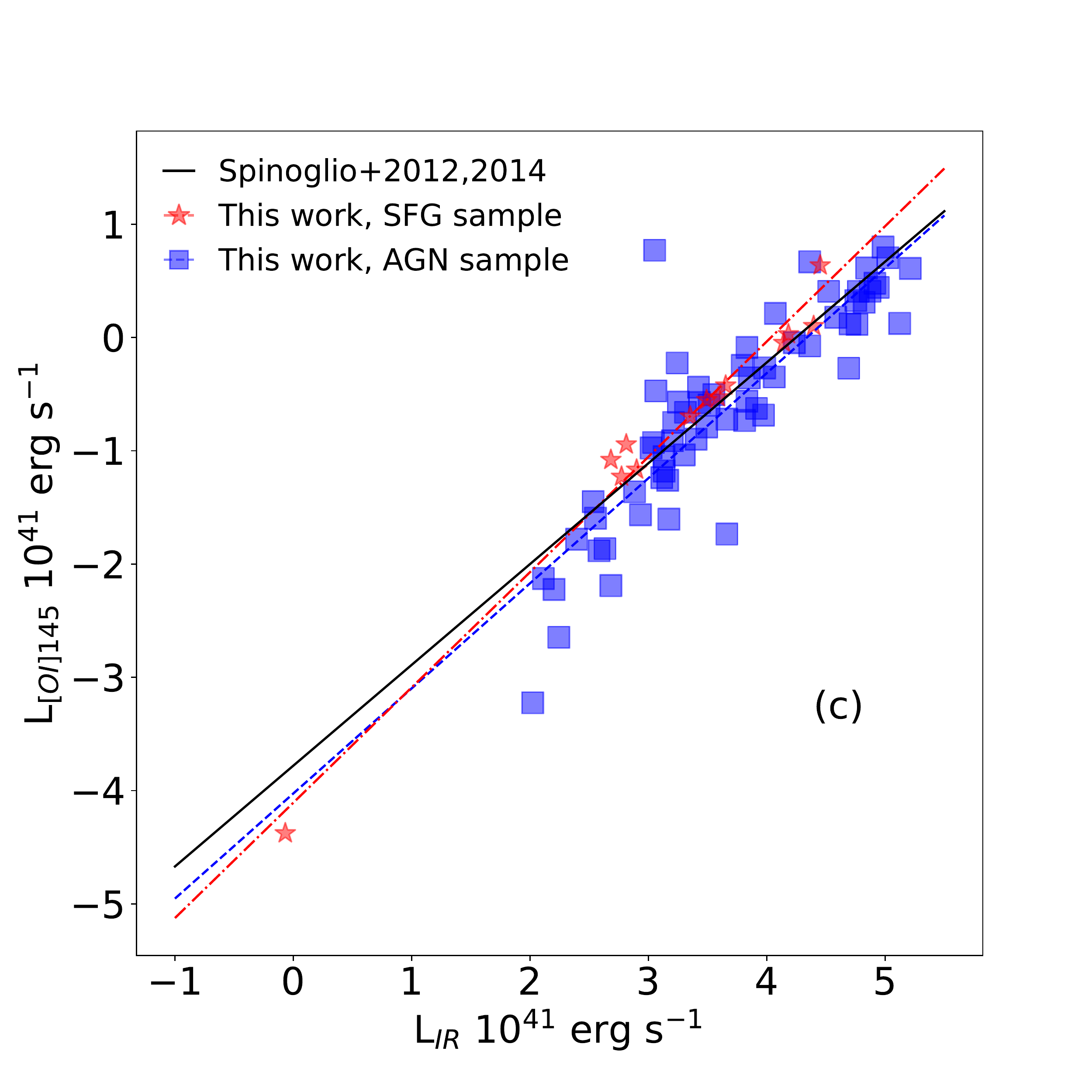}
\caption{{\bf (a: left)} The [OIII]88$\mu$m line luminosity versus the total IR luminosity. {\bf (b: centre)} The [NII]122$\mu$ m line luminosity versus the total IR luminosity.  {\bf (c: right)} The [OI]145$\mu$m line luminosity versus the total IR luminosity. The same legend as in Fig.\,\ref{fig:comp_sp2012_1} was used.}
\label{fig:comp_sp2012_2}
\end{figure*}

\begin{figure*}[!ht]
\centering
\includegraphics[width=0.33\columnwidth]{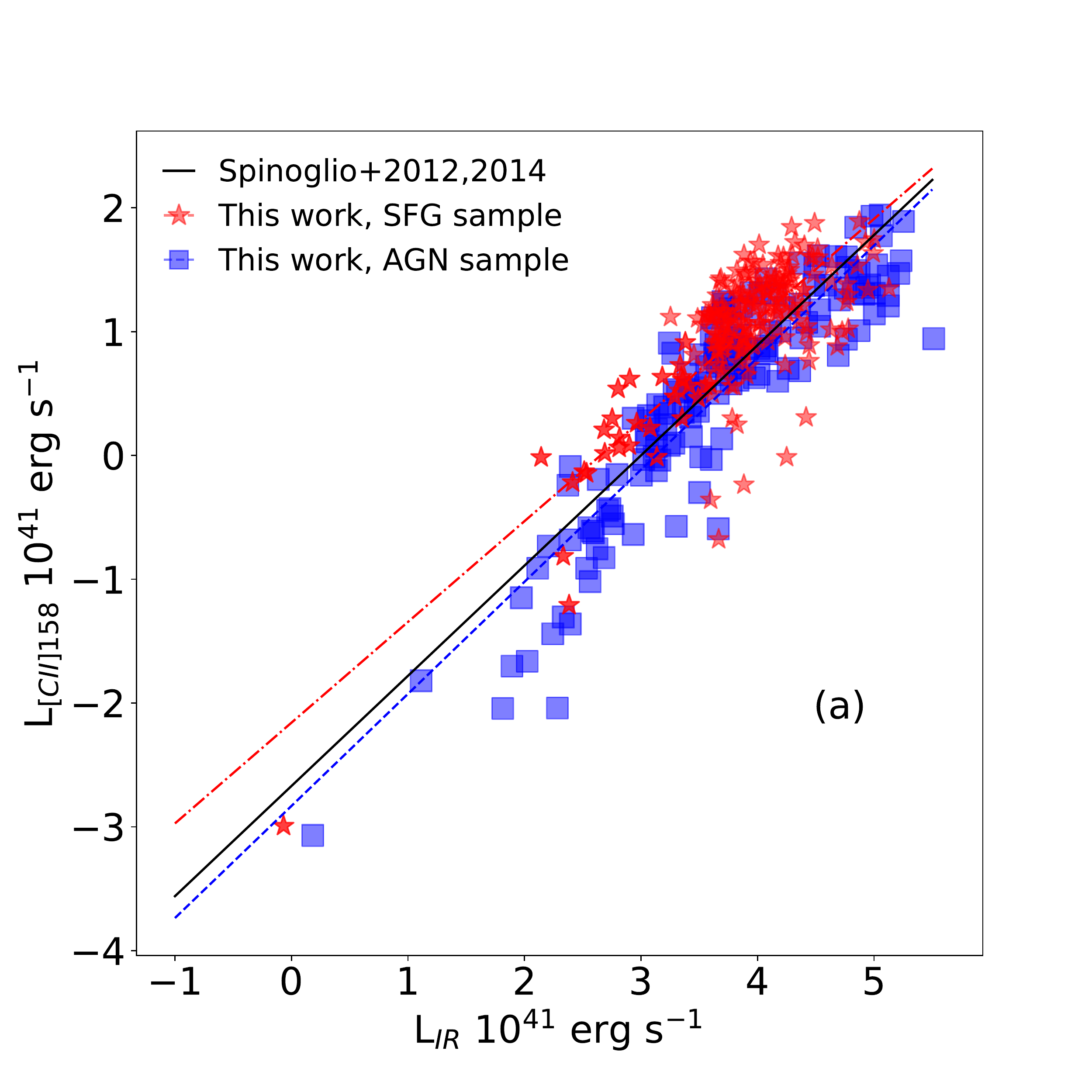}
\caption{{\bf (a:)} The [CII]158$\mu$m line luminosity versus the total IR luminosity. The same legend as in Fig.\,\ref{fig:comp_sp2012_1} was used}
\label{fig:comp_sp2012_3}
\end{figure*}

When considering the results by \citeauthor{spinoglio2012}, we only analyse the fine structure lines in the 50-160$\mu$m spectral interval, for which these authors consider an heterogeneous sample of both AGN and SFG, while in this work we consider three classes of galaxies. When comparing the results, we consider the calibrations obtained by \citeauthor{spinoglio2012} and compare them to our results for AGN and SFG. For this comparison, we apply to our samples the ordinary least square method, which was used in \citeauthor{spinoglio2012} to derive their correlations. Fig\,\ref{fig:comp_sp2012_1} shows the comparison for the [OIII]52$\mu$m, [NIII]57$\mu$m and [OI]63$\mu$m lines, Fig\,\ref{fig:comp_sp2012_2} shows the comparison for the [OIII]88$\mu$m, [NII]122$\mu$m and [OI]145$\mu$m lines, and Fig.\,\ref{fig:comp_sp2012_3} shows the comparison for the [CII]158$\mu$m line.

When compared to our AGN sample, the results are consistent within the errors, except for the [NII]57$\mu$m and [OIII]88$\mu$m lines, where the slopes are comparable within 2$\sigma$ of each other. If compared to the SFG results, we obtain results comparable within the errors except for [OIII]88$\mu$m and [NII]122$\mu$m lines, for which the slopes are comparable within 2$\sigma$ of each other.

When considering the results presented by \citet{gruppioni2016}, we take advantage of the published catalogue and apply the orthogonal distance regression to the sample presented by the authors. \citeauthor{gruppioni2016} present a sample of 76 AGN, divided in two sub-samples depending on the fraction of 5-40$\mu$m luminosity produced by the active nucleus [$f_{AGN}(5-40\mu m)$]. In particular, there are 30 objects with $f_{AGN}(5-40\mu m)\leq0.4$ and 43 objects with $f_{AGN}(5-40\mu m) >0.4$ .

\begin{figure*}[h!]
\centering
\includegraphics[width=0.33\columnwidth]{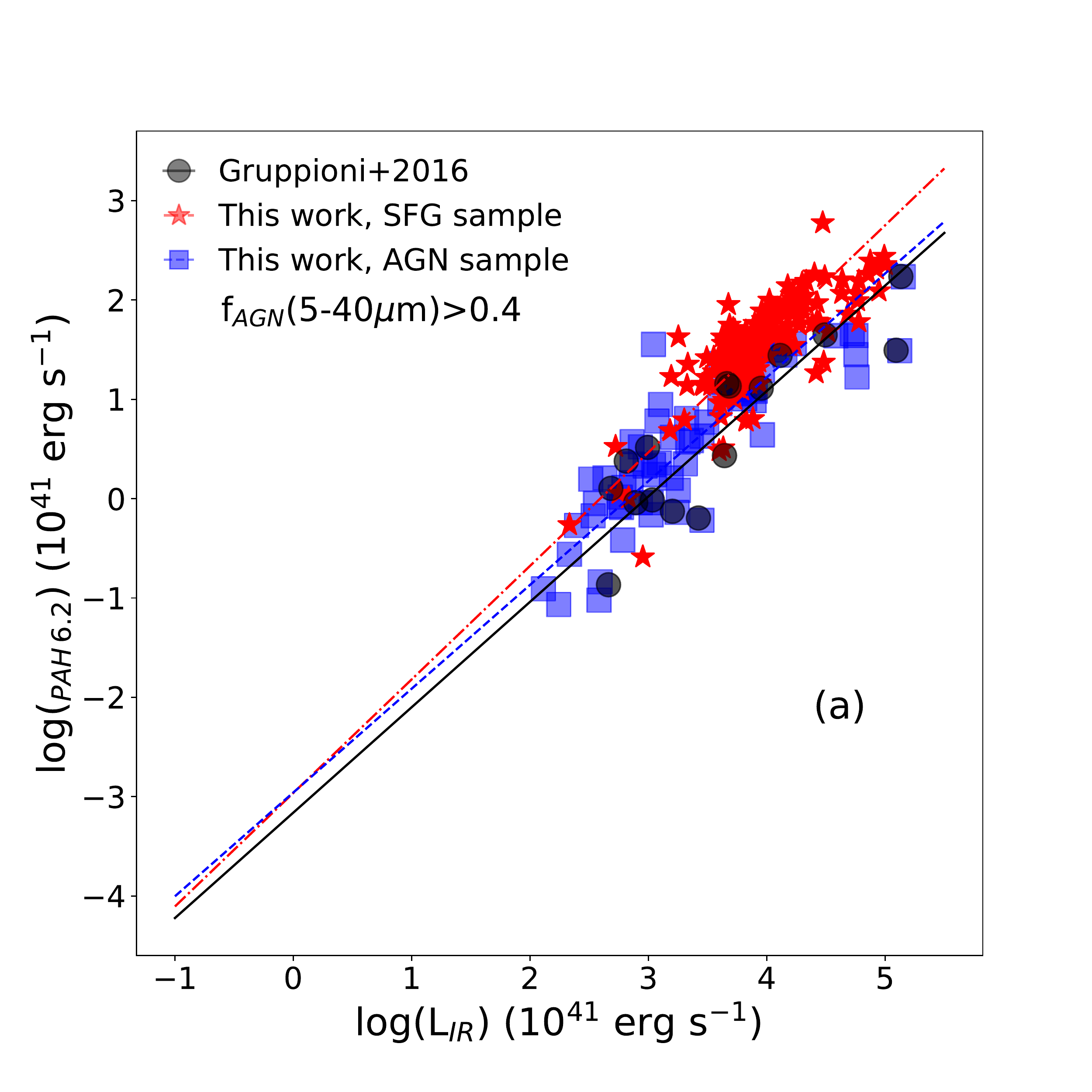}
\includegraphics[width=0.33\columnwidth]{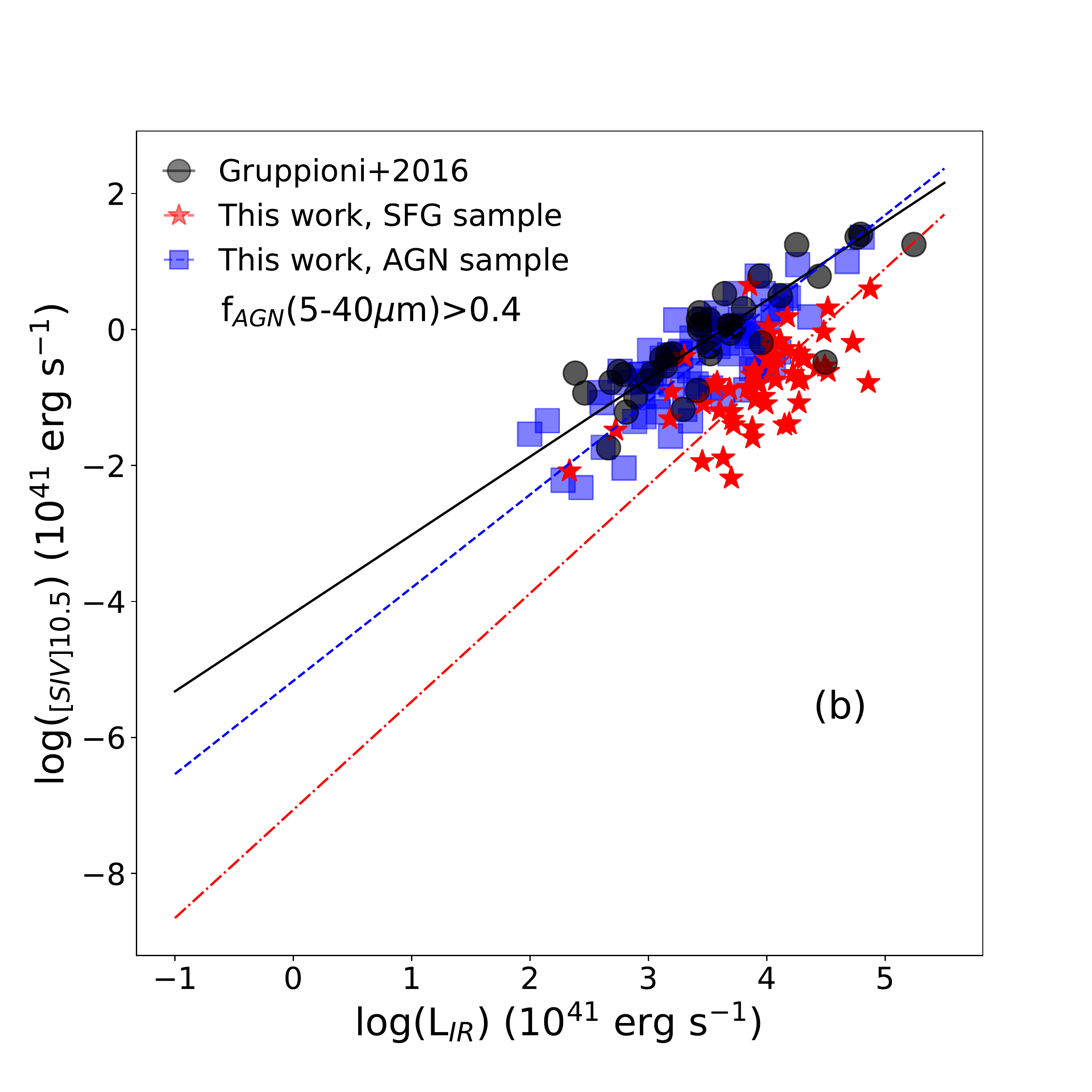}
\includegraphics[width=0.33\columnwidth]{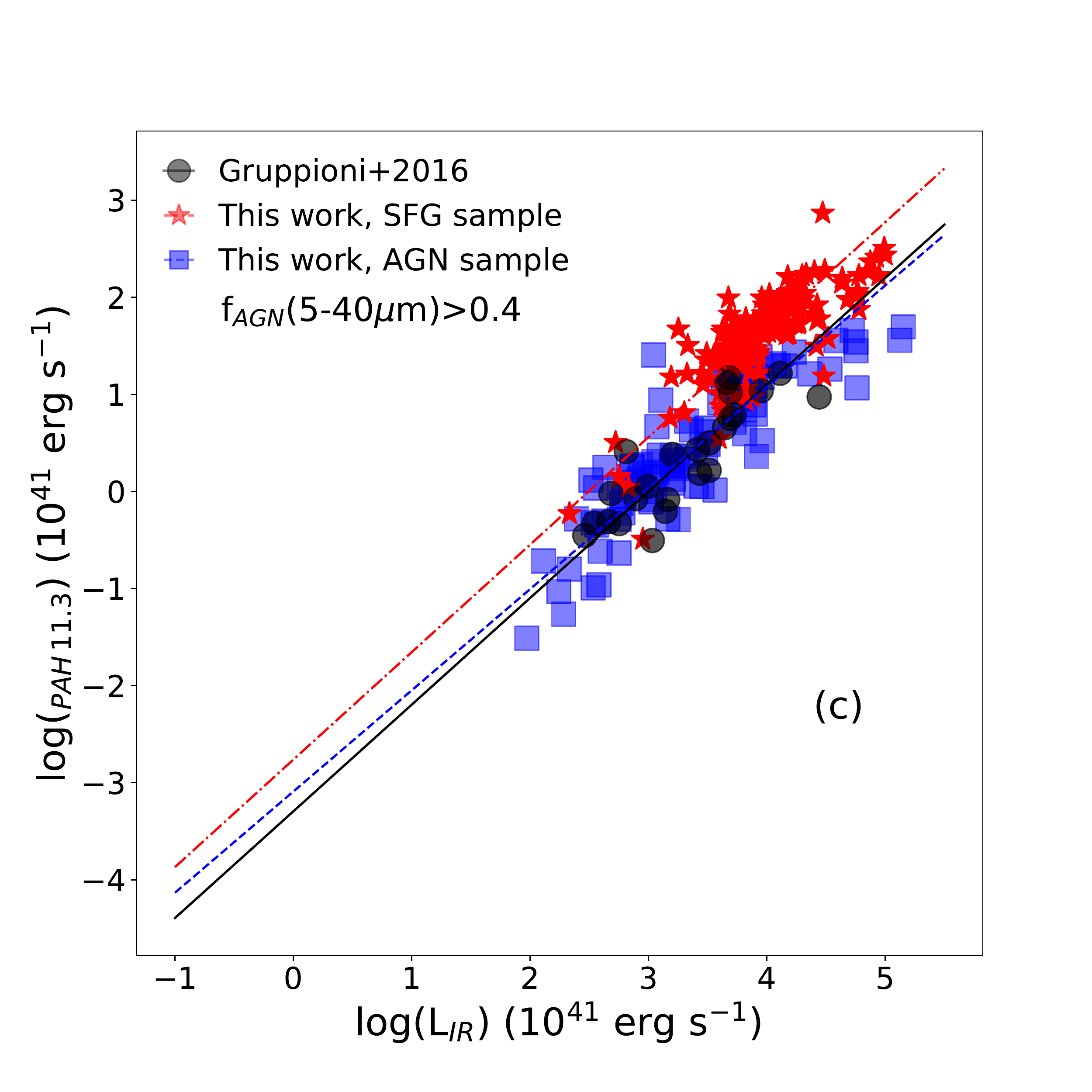}
\caption{{\bf (a: left)} The PAH 6.2$\mu$m luminosity versus the total IR luminosity. Blue squares represent AGN, red stars SFG and black circles are the sample by \citet{gruppioni2016}. The dashed blue line represents the linear relation for AGN, the dot-dashed red line the relation for SFG, and the black solid line the relation for the \citeauthor{gruppioni2016} sample. {\bf (b: centre)} The [SIV]10.5$\mu$m luminosity versus the total IR luminosity.  {\bf (c: right)} The PAH 11.3$\mu$m luminosity versus the total IR luminosity. In all three figures, the \citeauthor{gruppioni2016} sample was limited to AGN fractions $>$ 0.4 in the 5-40$\mu$m spectral interval (see Sect.\ref{sec:comp}).}
\label{fig:comp_1}
\end{figure*}

\begin{figure*}%[h!]
\centering
\includegraphics[width=0.33\columnwidth]{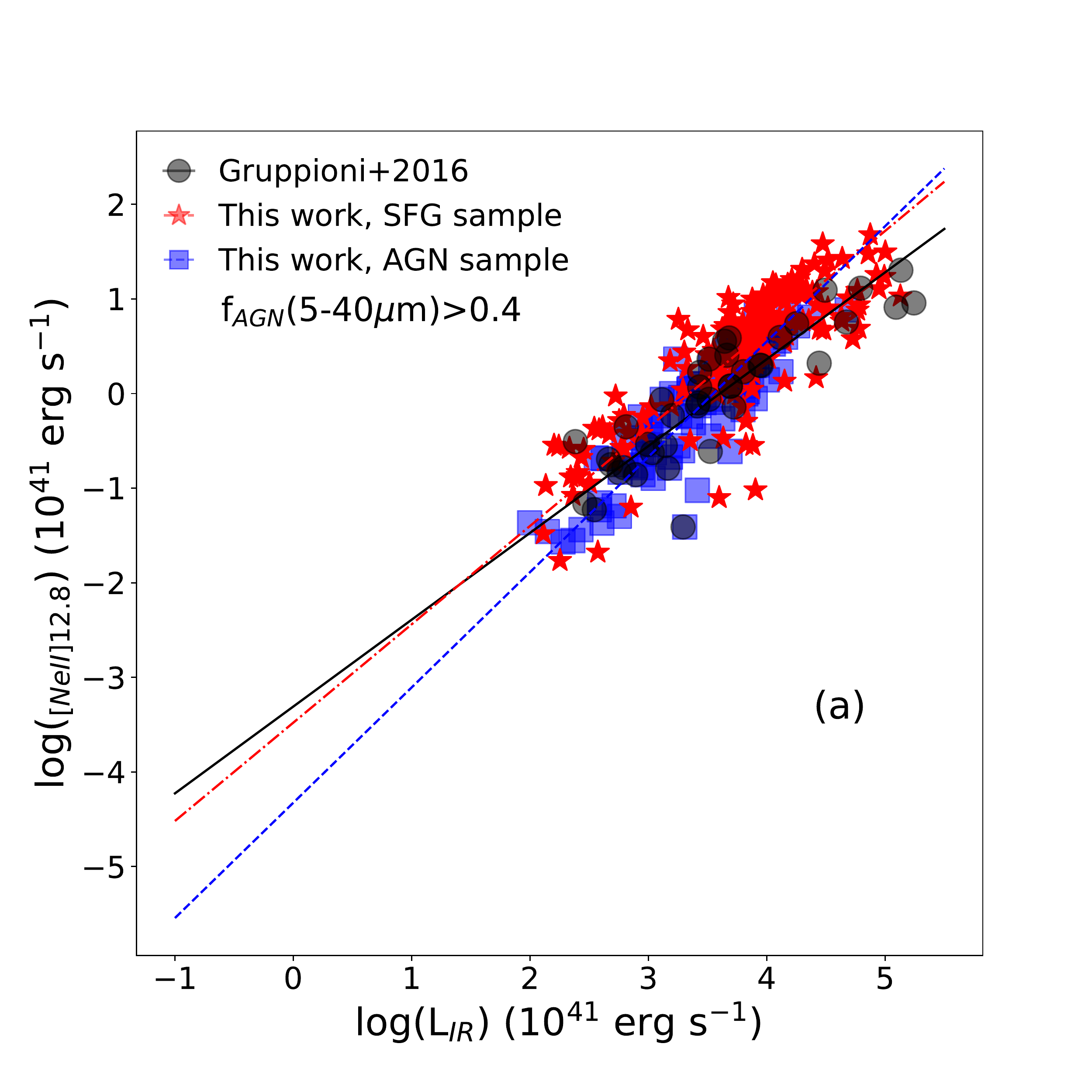}
\includegraphics[width=0.33\columnwidth]{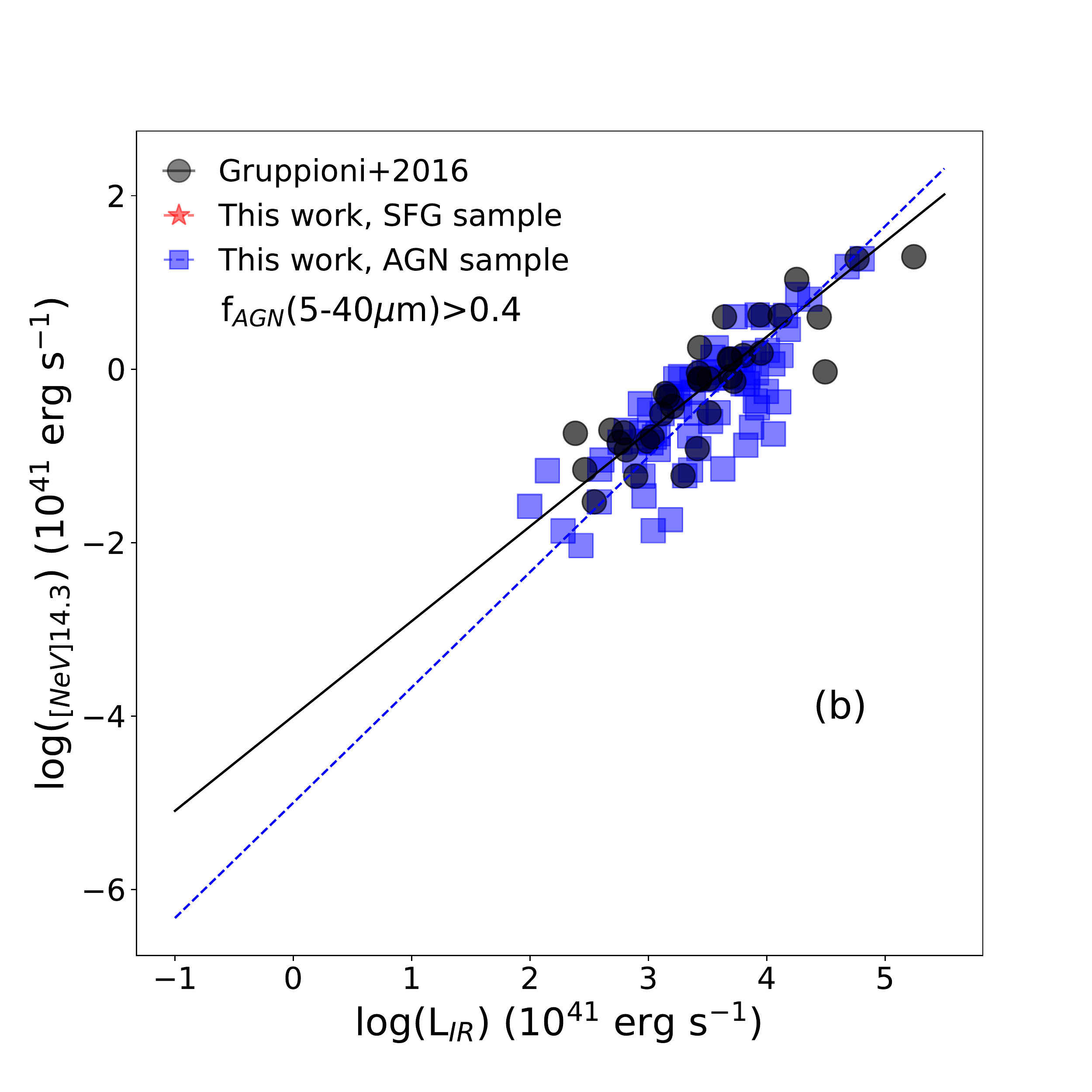}
\includegraphics[width=0.33\columnwidth]{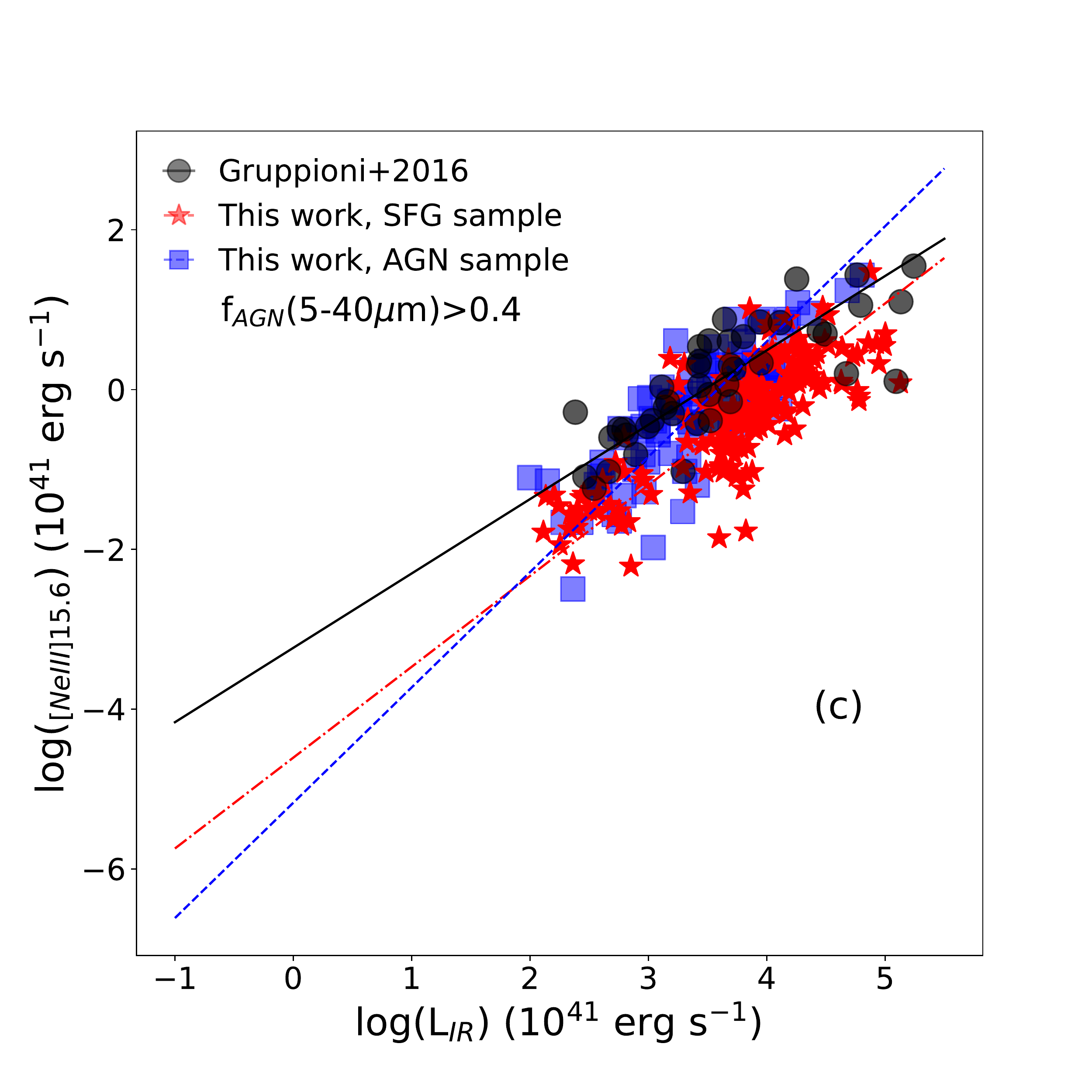}
\caption{{\bf (a: left)} The [NeII]12.8$\mu$m line luminosity versus the total IR luminosity. {\bf (b: centre)} The [NeV]14.3$\mu$m line versus the total IR luminosity.  {\bf (c: right)} The [NeIII]15.6$\mu$m line luminosity versus the total IR luminosity. In all three figures, the \citeauthor{gruppioni2016} sample was limited to AGN fractions $>$ 0.4 in the 5-40$\mu$m spectral interval (see Sect.\ref{sec:comp}). The same legend as in Fig.\,\ref{fig:comp_1} was used.}
\label{fig:comp_2}
\end{figure*}

\begin{figure*}%[h!]
\centering
\includegraphics[width=0.33\columnwidth]{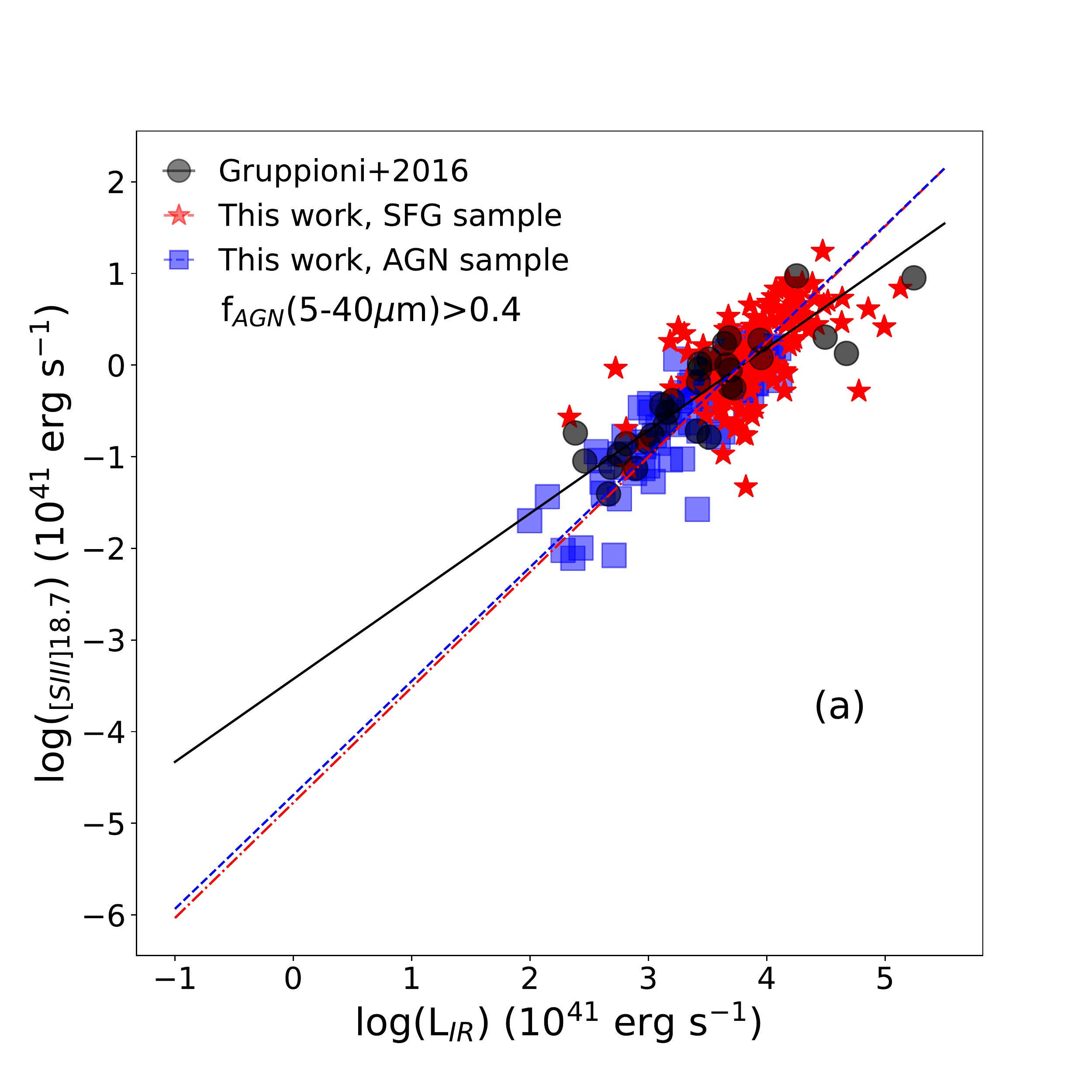}
\includegraphics[width=0.33\columnwidth]{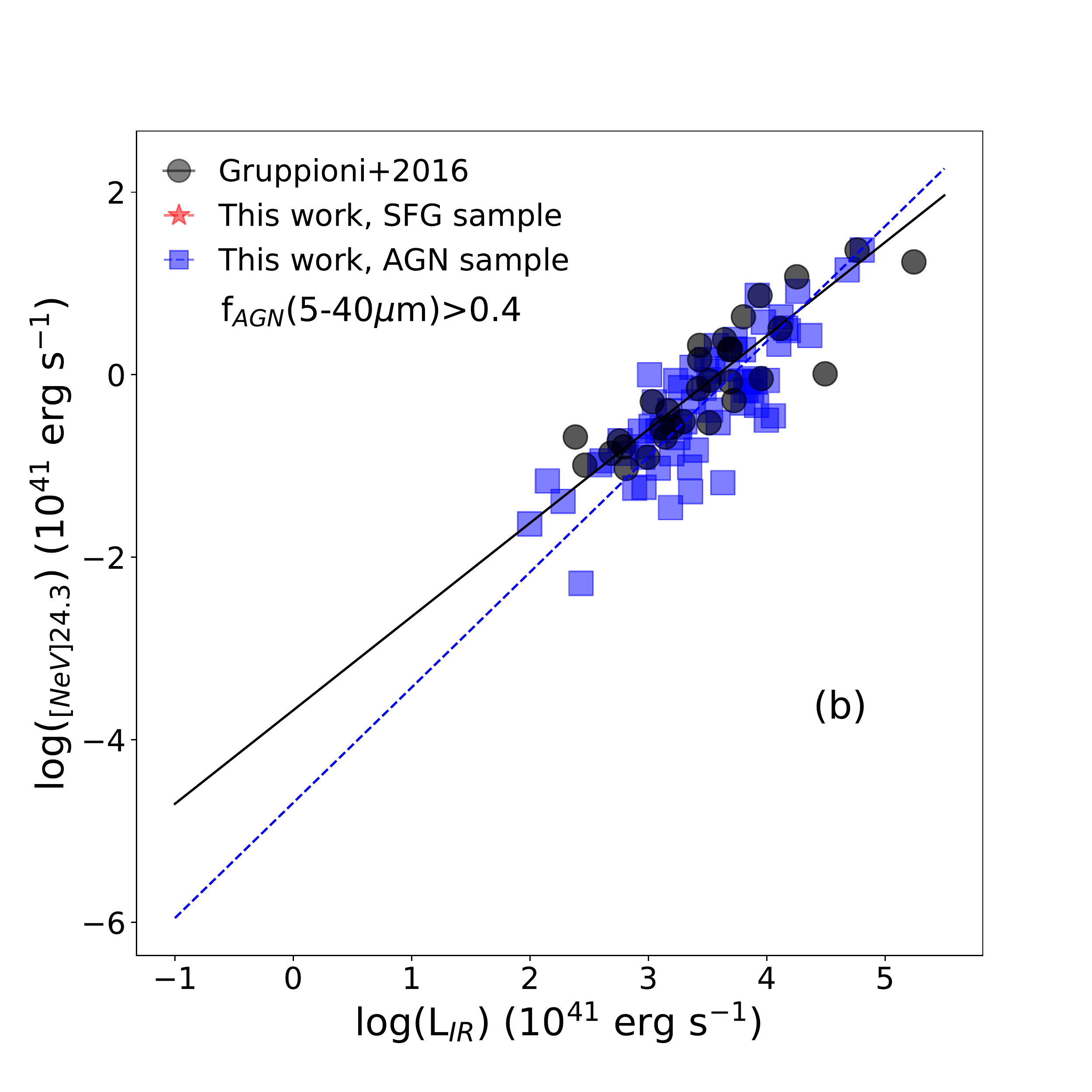}
\includegraphics[width=0.33\columnwidth]{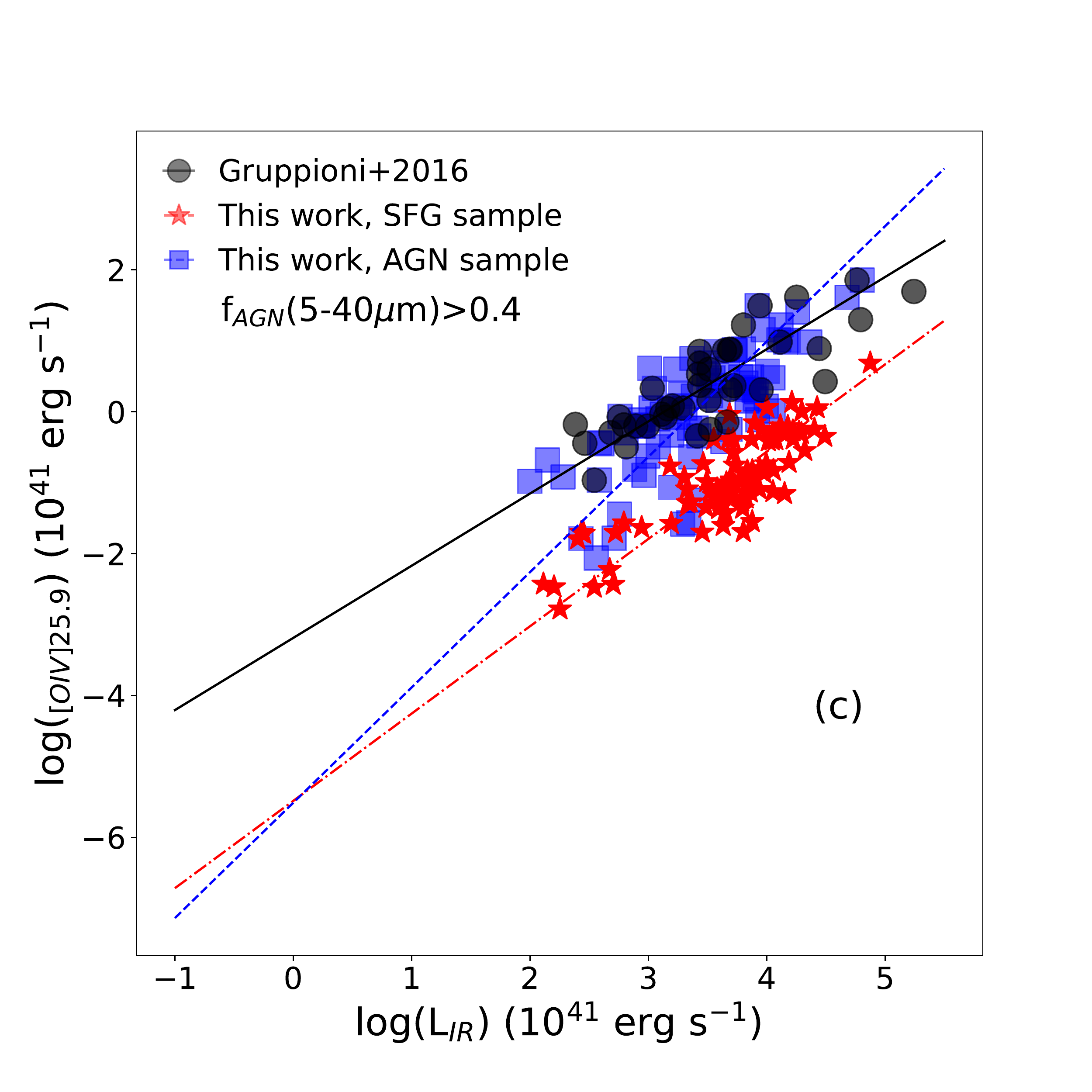}
\caption{{\bf (a: left)} The [SIII]18.7$\mu$m line luminosity versus the total IR luminosity. {\bf (b: centre)} The [NeV]24.3$\mu$m line luminosity versus the total IR luminosity.  {\bf (c: right)} The [OIV]55.9$\mu$m line luminosity versus the total IR luminosity. In all three figures, the \citeauthor{gruppioni2016} sample was limited to AGN fractions $>$ 0.4 in the 5-40$\mu$m spectral interval (see Sect.\ref{sec:comp}. The same legend as in Fig.\,\ref{fig:comp_1} was used.}
\label{fig:comp_3}
\end{figure*}

\begin{figure*}
\centering
\includegraphics[width=0.33\columnwidth]{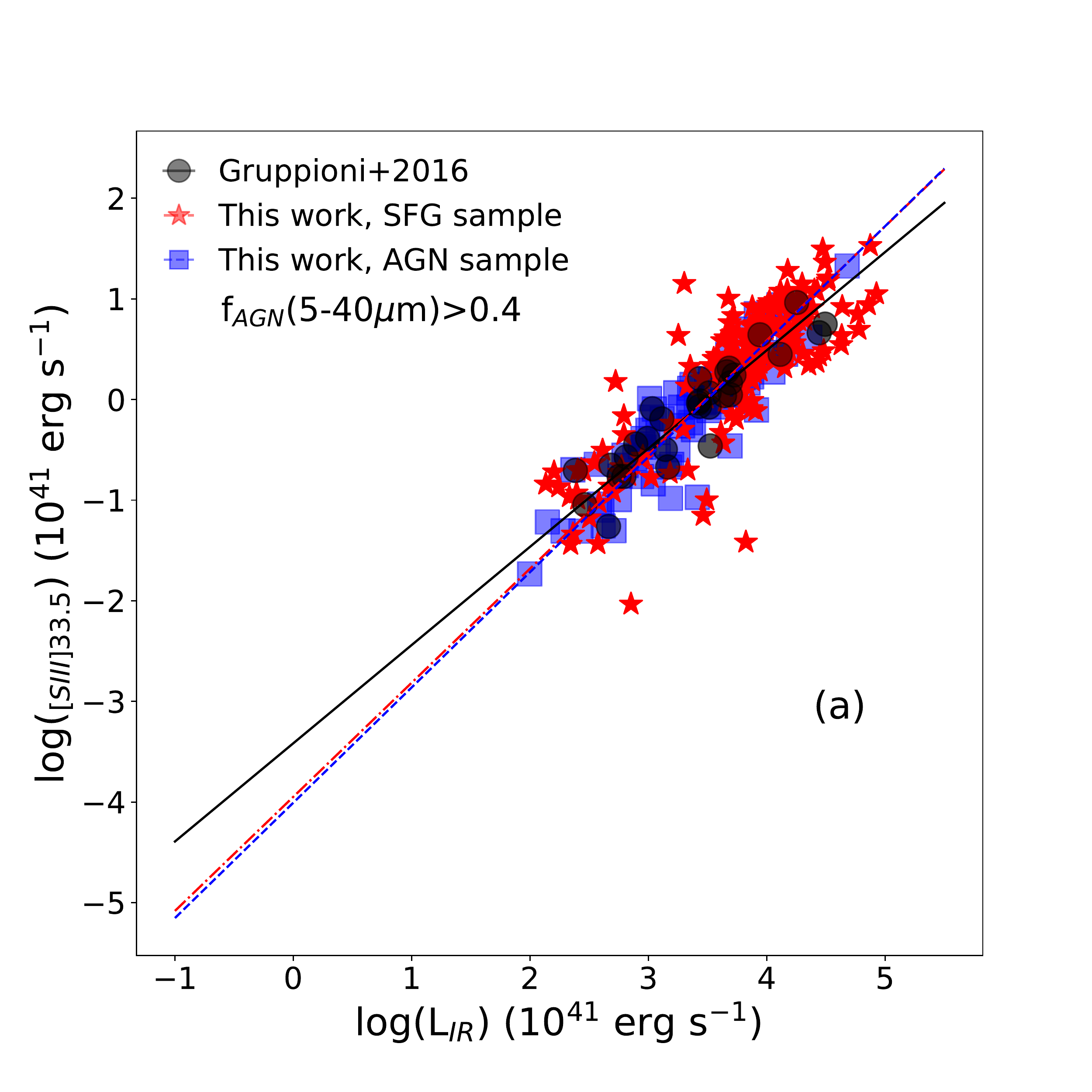}
\includegraphics[width=0.33\columnwidth]{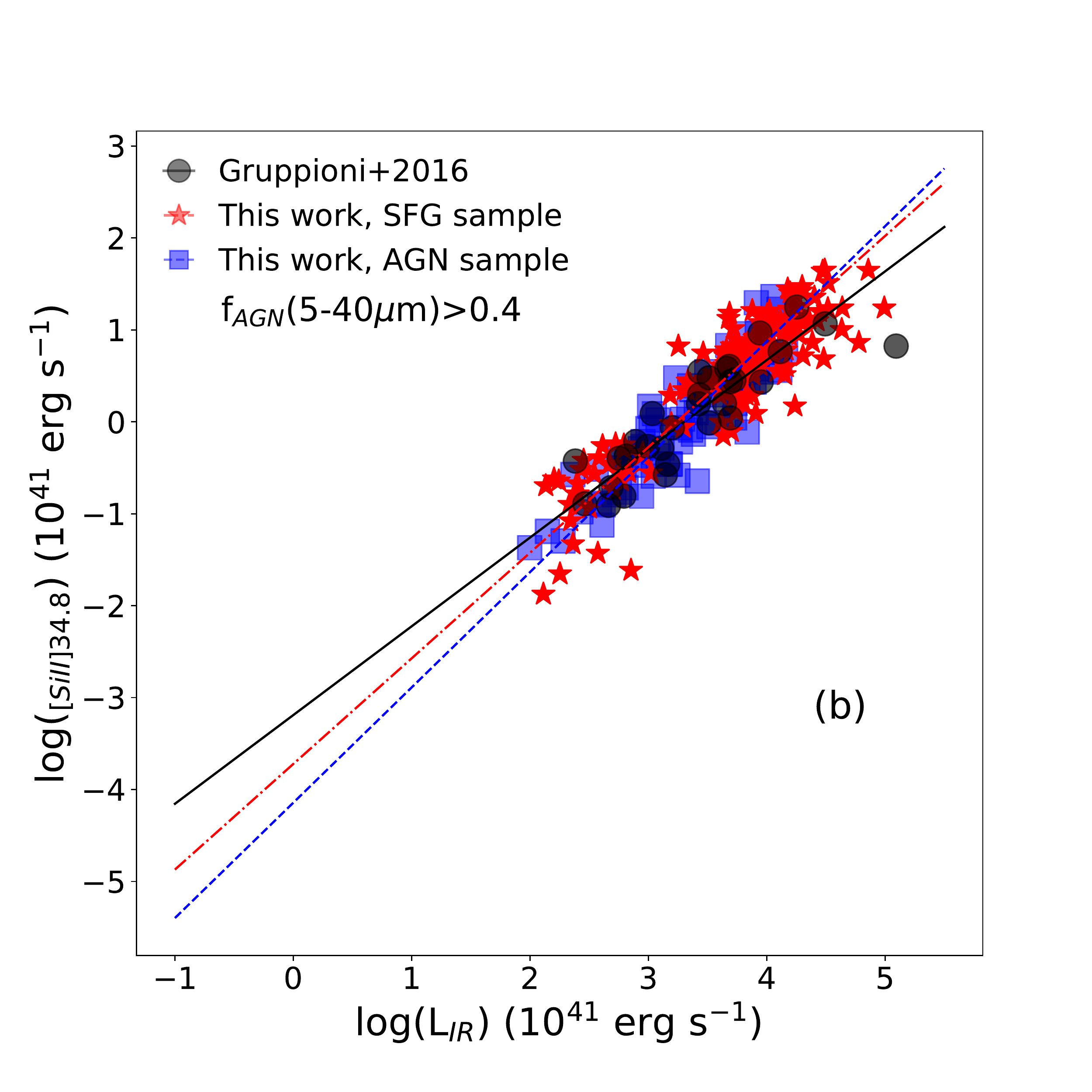}
\caption{{\bf (a: left)} The [SIII]33.5$\mu$m line luminosity versus the total IR luminosity. {\bf (b: right)} The [SiII]34.9$\mu$m line luminosity versus  the total IR luminosity.  In both figures, the \citeauthor{gruppioni2016} sample was limited to AGN fractions $>$ 0.4 in the 5-40$\mu$m spectral interval (see Sect.\ref{sec:comp}). The same legend as in Fig.\,\ref{fig:comp_1} was used.}
\label{fig:comp_4}
\end{figure*}

\begin{figure*}
\centering
\includegraphics[width=0.33\columnwidth]{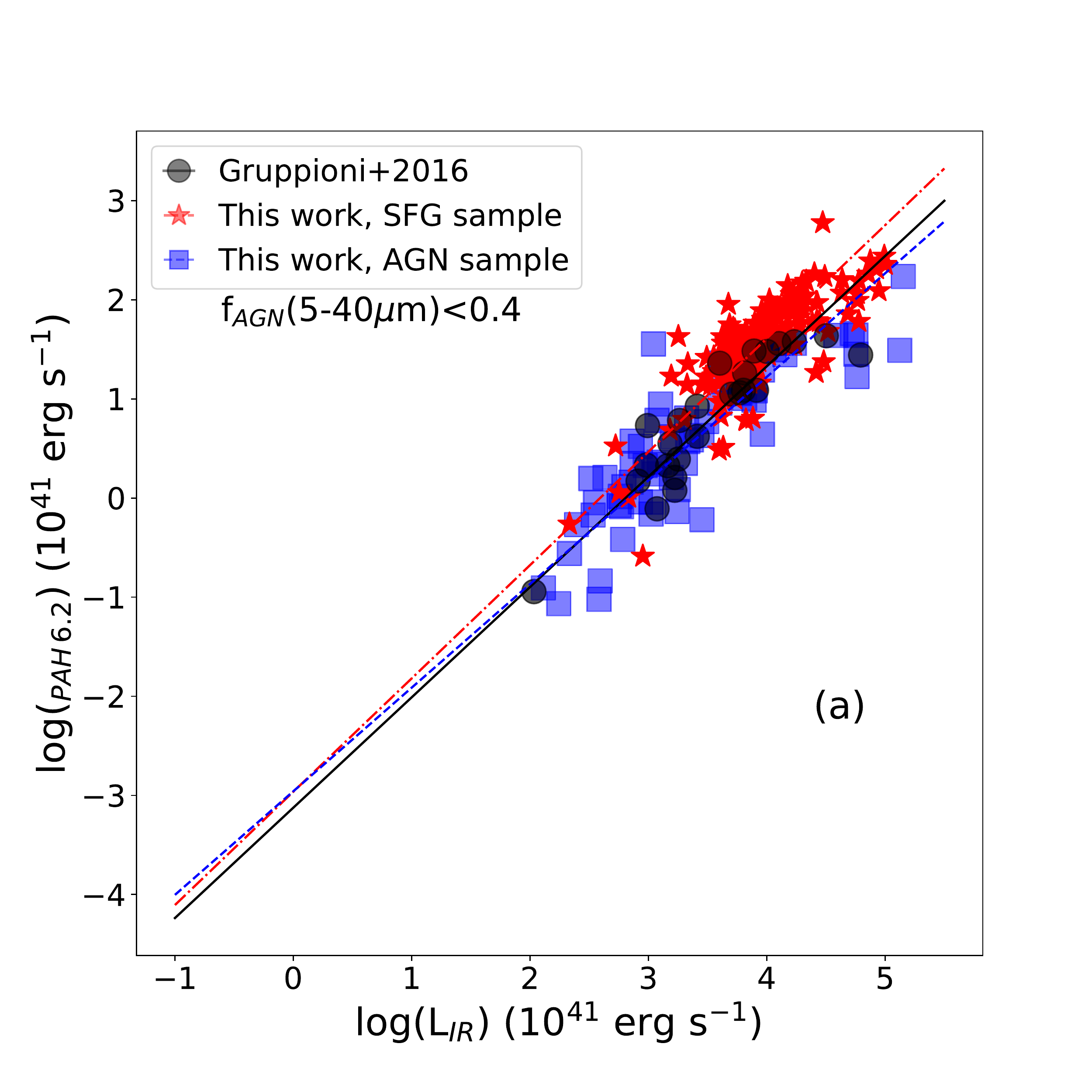}
\includegraphics[width=0.33\columnwidth]{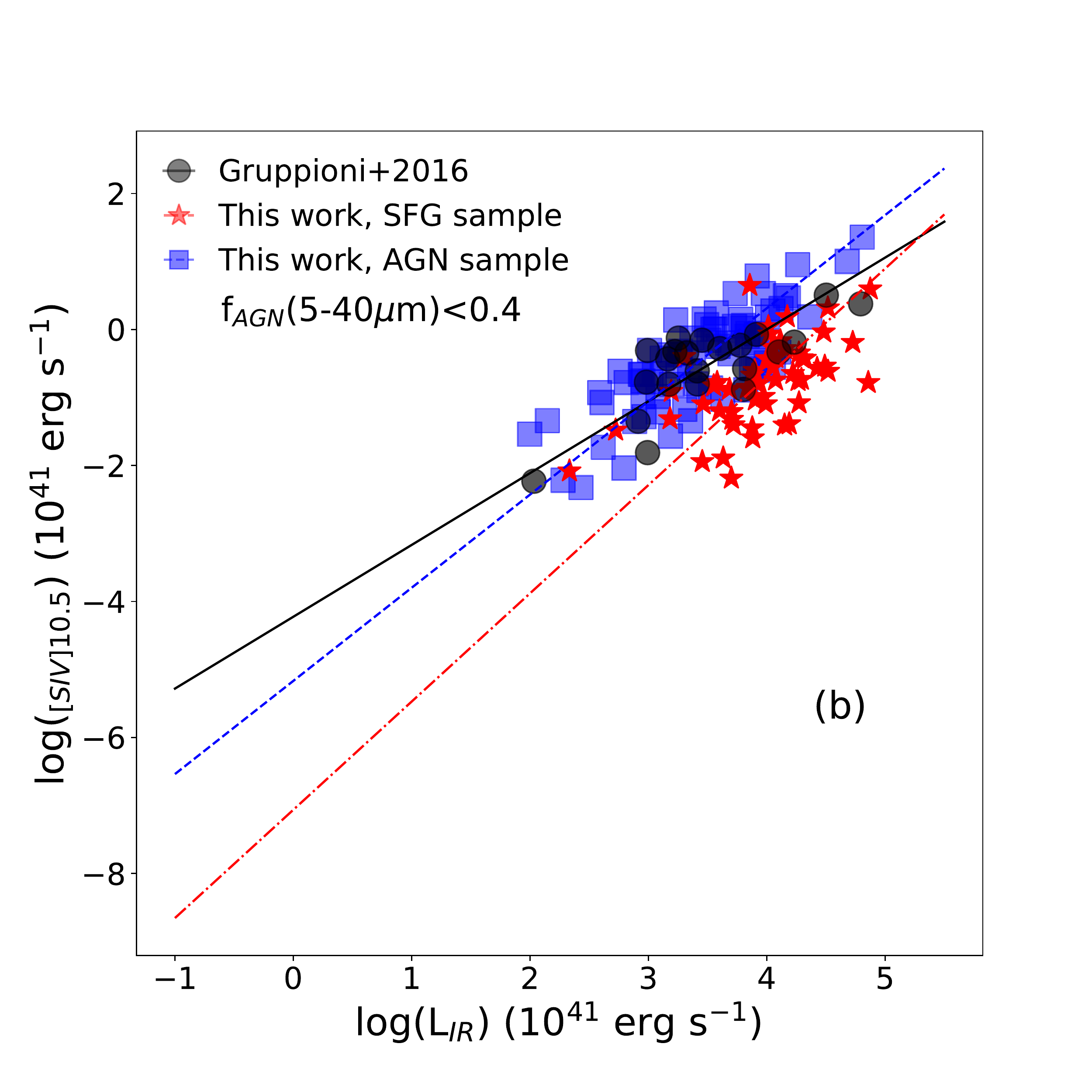}
\includegraphics[width=0.33\columnwidth]{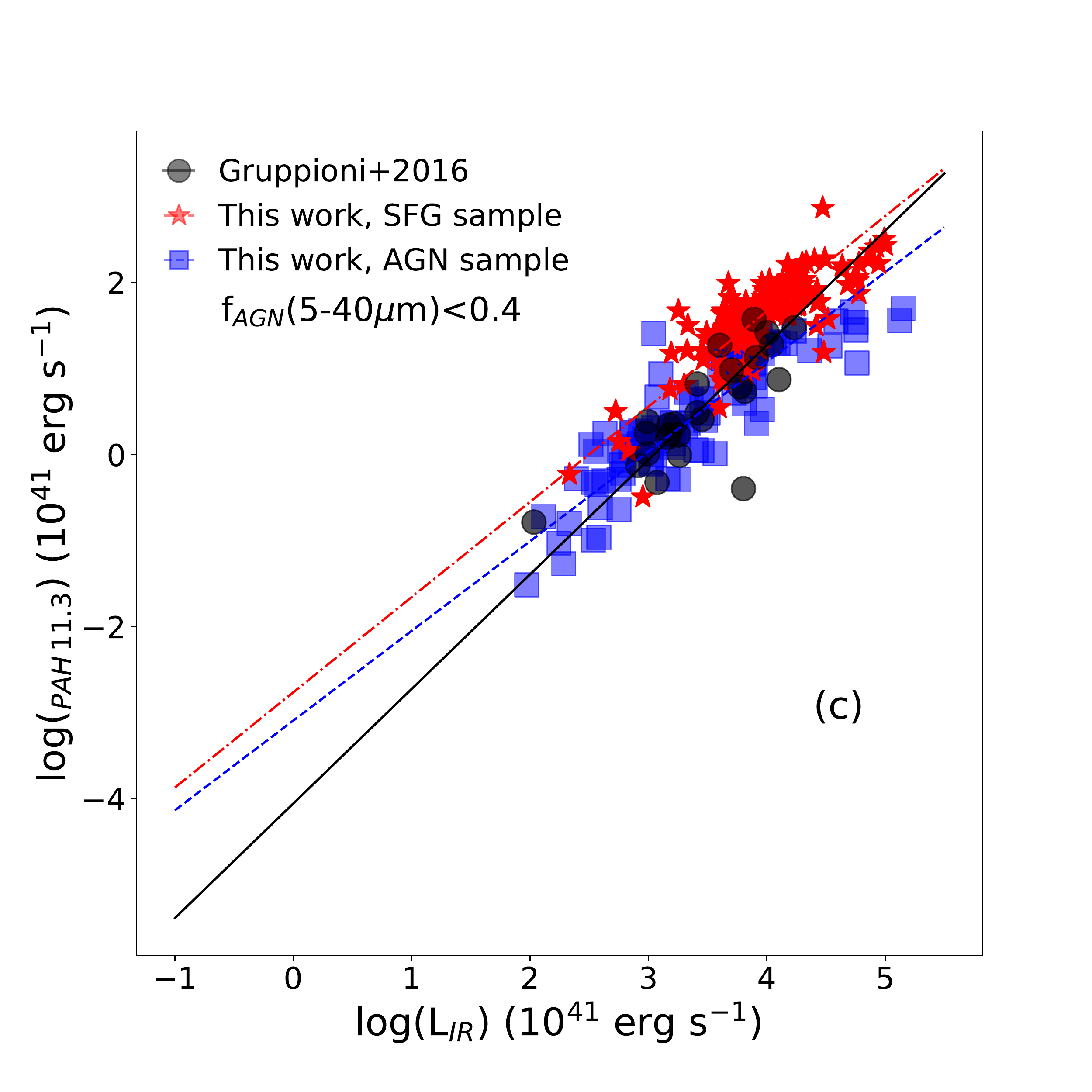}
\caption{{\bf (a: left)} The luminosity of the PAH feature at 6.2$\mu$m versus the total IR luminosity. %Blue squares represent detections in AGN, red stars SFG and black circles are the sample by \citet{gruppioni2016}. The dashed blue line represents the linear relation for AGN, the dot-dashed red line reation for SFG, and the black solid line the relation calculated for \citeauthor{gruppioni2016}. 
{\bf (b: centre)} The [SIV]10.5$\mu$m line luminosity as a function of the total IR luminosity.  {\bf (c: right)} The PAH 11.3$\mu$m luminosity versus the total IR luminosity. In all figures, the \citeauthor{gruppioni2016} sample was limited to AGN fractions $<$ 0.4 in the 5-40$\mu$m spectral interval (see Sect.\ref{sec:comp}). The same legend as in Fig.\,\ref{fig:comp_1} was used.}
\label{fig:comp_5}
\end{figure*}

\begin{figure*}
\centering
\includegraphics[width=0.33\columnwidth]{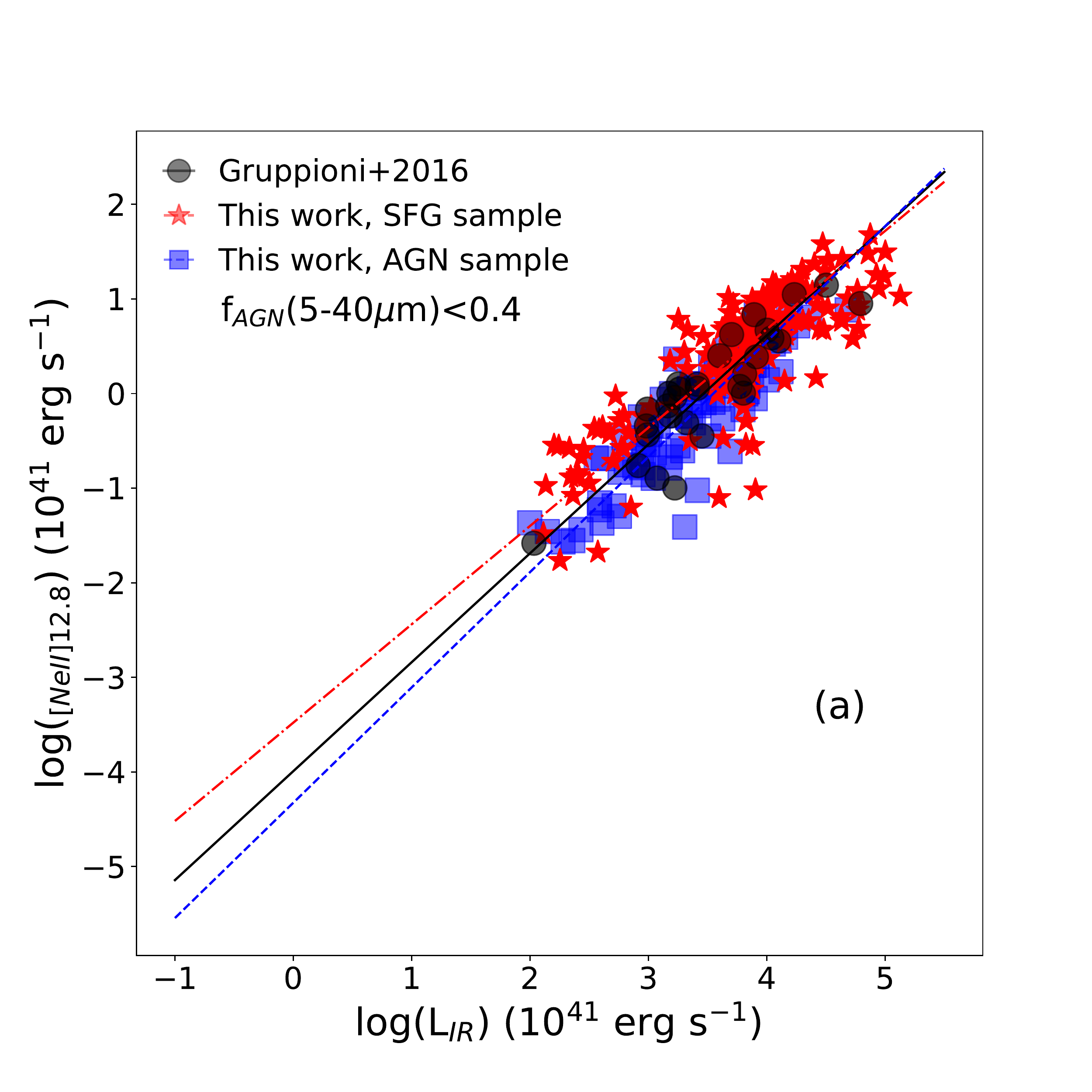}
\includegraphics[width=0.33\columnwidth]{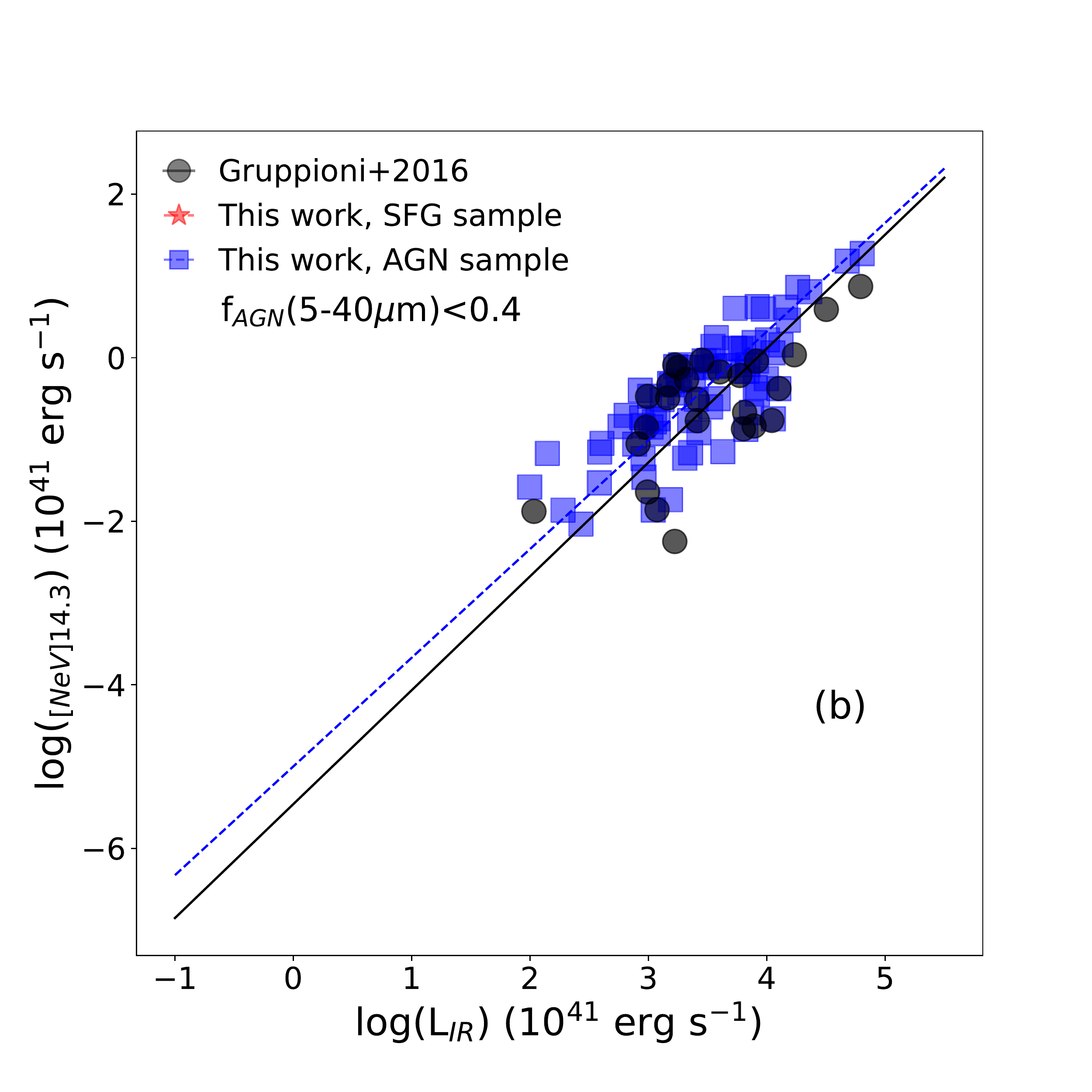}
\includegraphics[width=0.33\columnwidth]{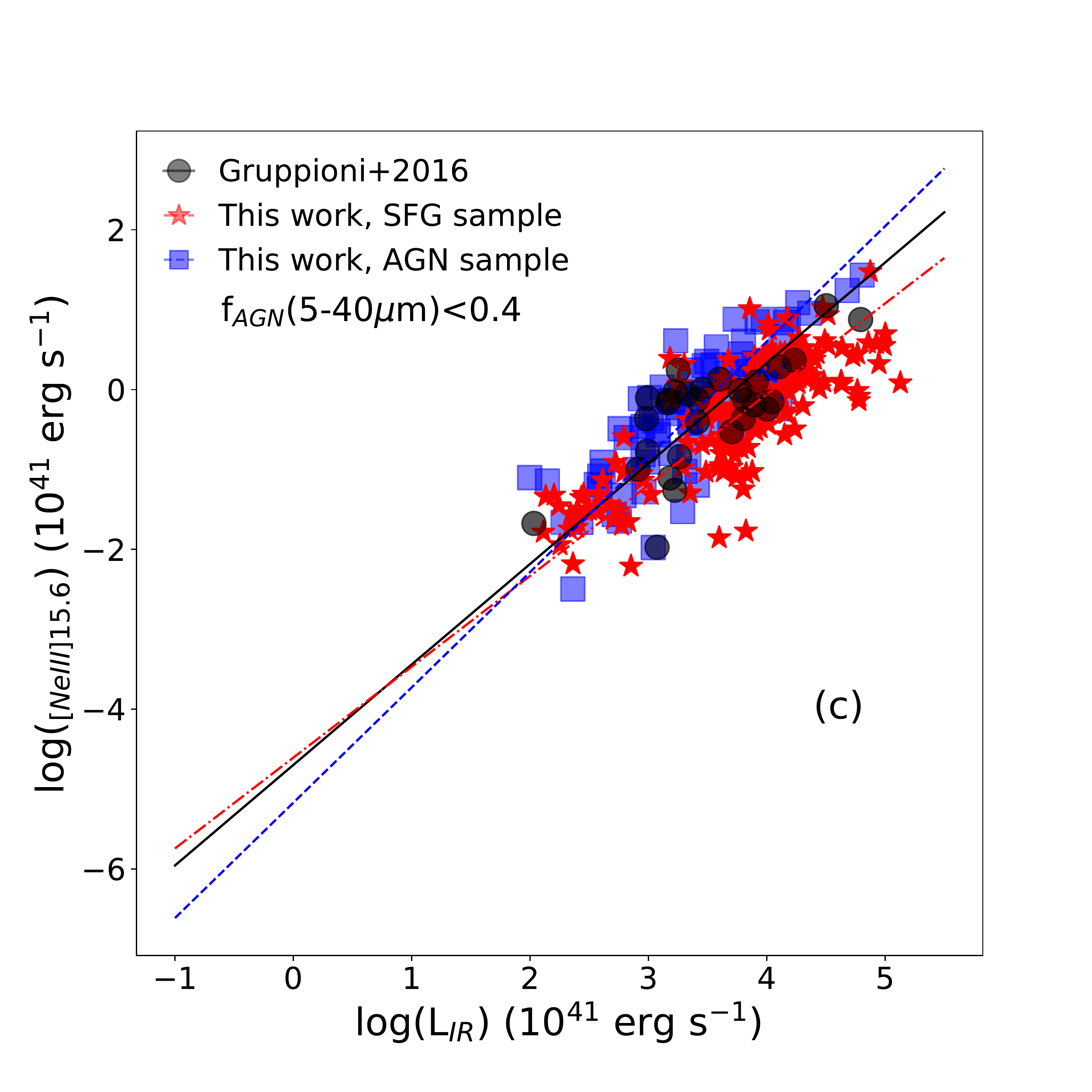}
\caption{{\bf (a: left)} The [NeII]12.8$\mu$m line luminosity versus the total IR luminosity. {\bf (b: centre)} The [NeV]14.3$\mu$m line luminosity versus  the total IR luminosity.  {\bf (c: right)} The [NeIII]15.6$\mu$m line luminosity versus the total IR luminosity. In all three figures, the \citeauthor{gruppioni2016} sample was limited to AGN fractions $<$ 0.4 in the 5-40$\mu$m spectral interval (see Sect.\ref{sec:comp}). The same legend as in Fig.\,\ref{fig:comp_1} was used.}
\label{fig:comp_6}
\end{figure*}

\begin{figure*}
\centering
\includegraphics[width=0.33\columnwidth]{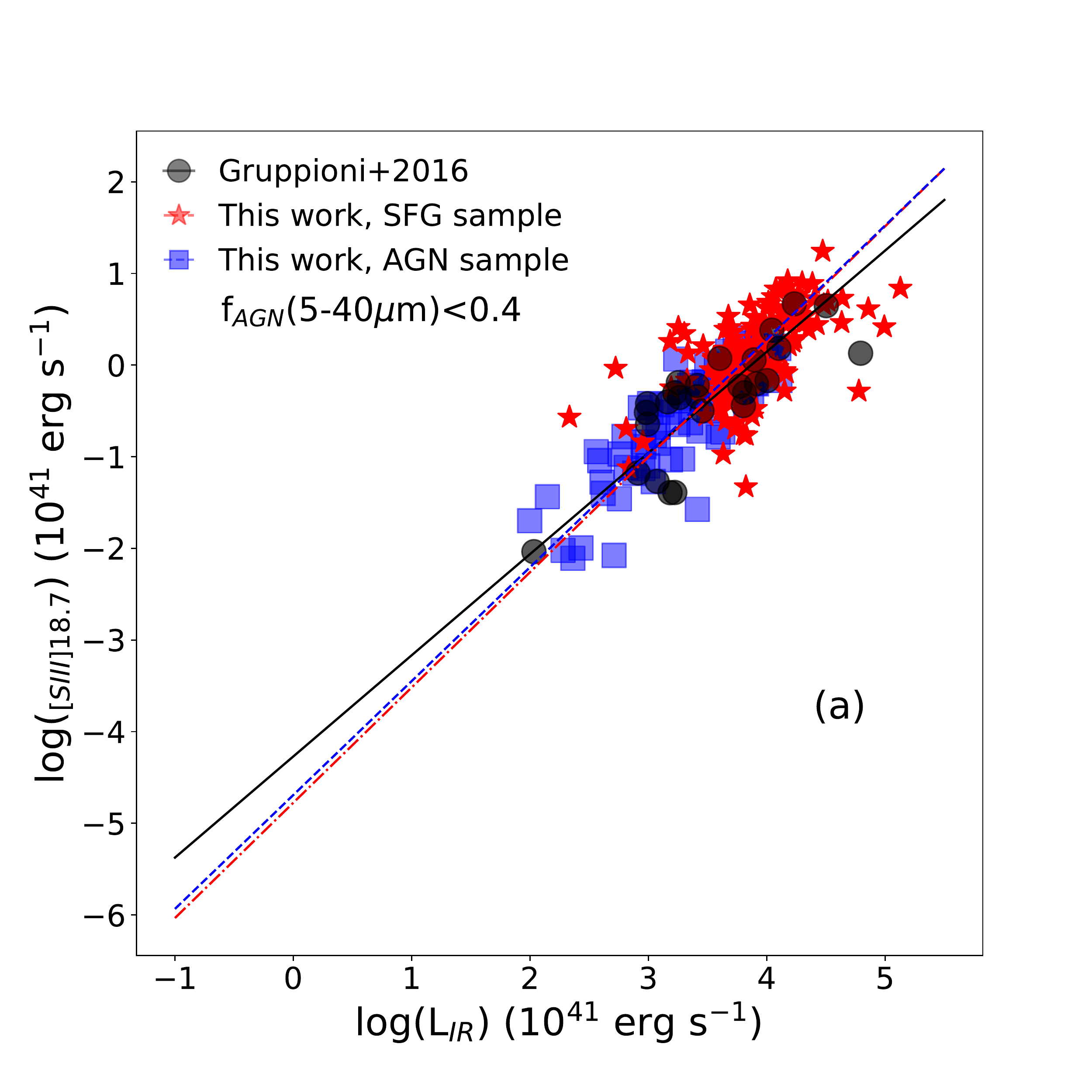}
\includegraphics[width=0.33\columnwidth]{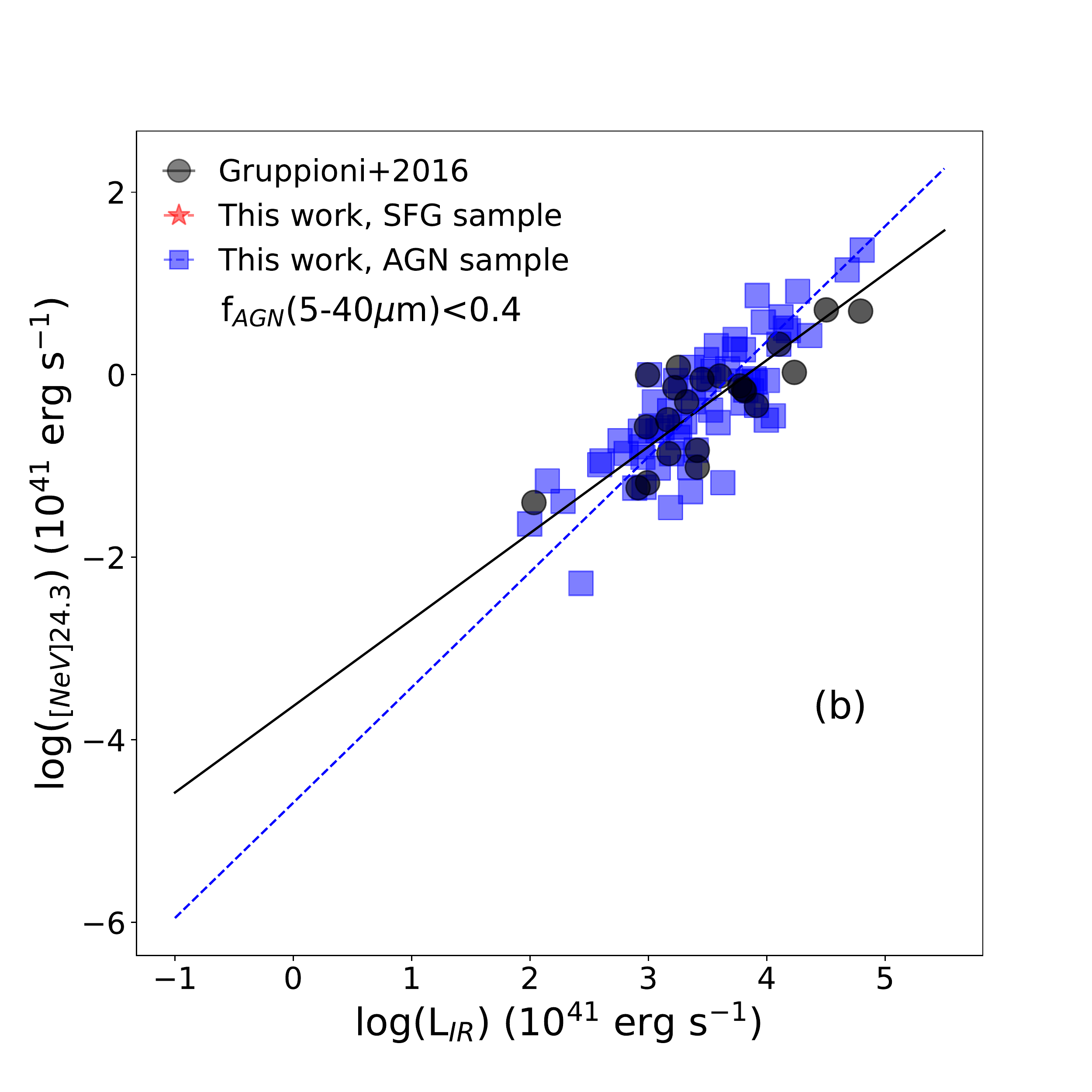}
\includegraphics[width=0.33\columnwidth]{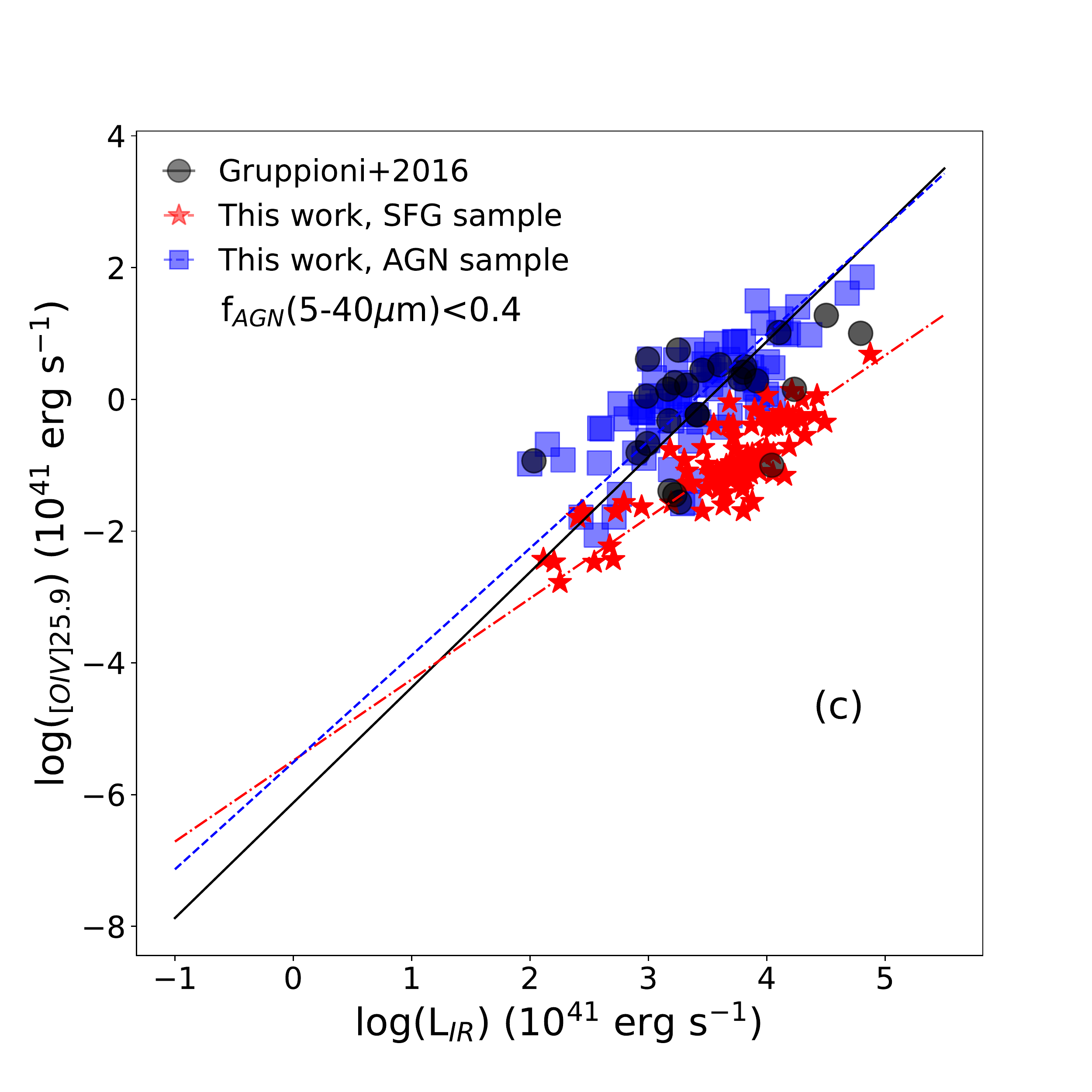}
\caption{{\bf (a: left)} The [SIII]18.7$\mu$m line luminosity versus the total IR luminosity. {\bf (b: centre)} The [NeV]24.3$\mu$m line luminosity versus the total IR luminosity.  {\bf (c: right)} The [OIV]55.9$\mu$m line luminosity versus the total IR luminosity. In all three figures, the \citeauthor{gruppioni2016} sample was limited to AGN fractions $<$ 0.4 in the 5-40$\mu$m spectral interval (see Sect.\ref{sec:comp}). The same legend as in Fig.\,\ref{fig:comp_1} was used.}
\label{fig:comp_7}
\end{figure*}

\begin{figure*}
\centering
\includegraphics[width=0.33\columnwidth]{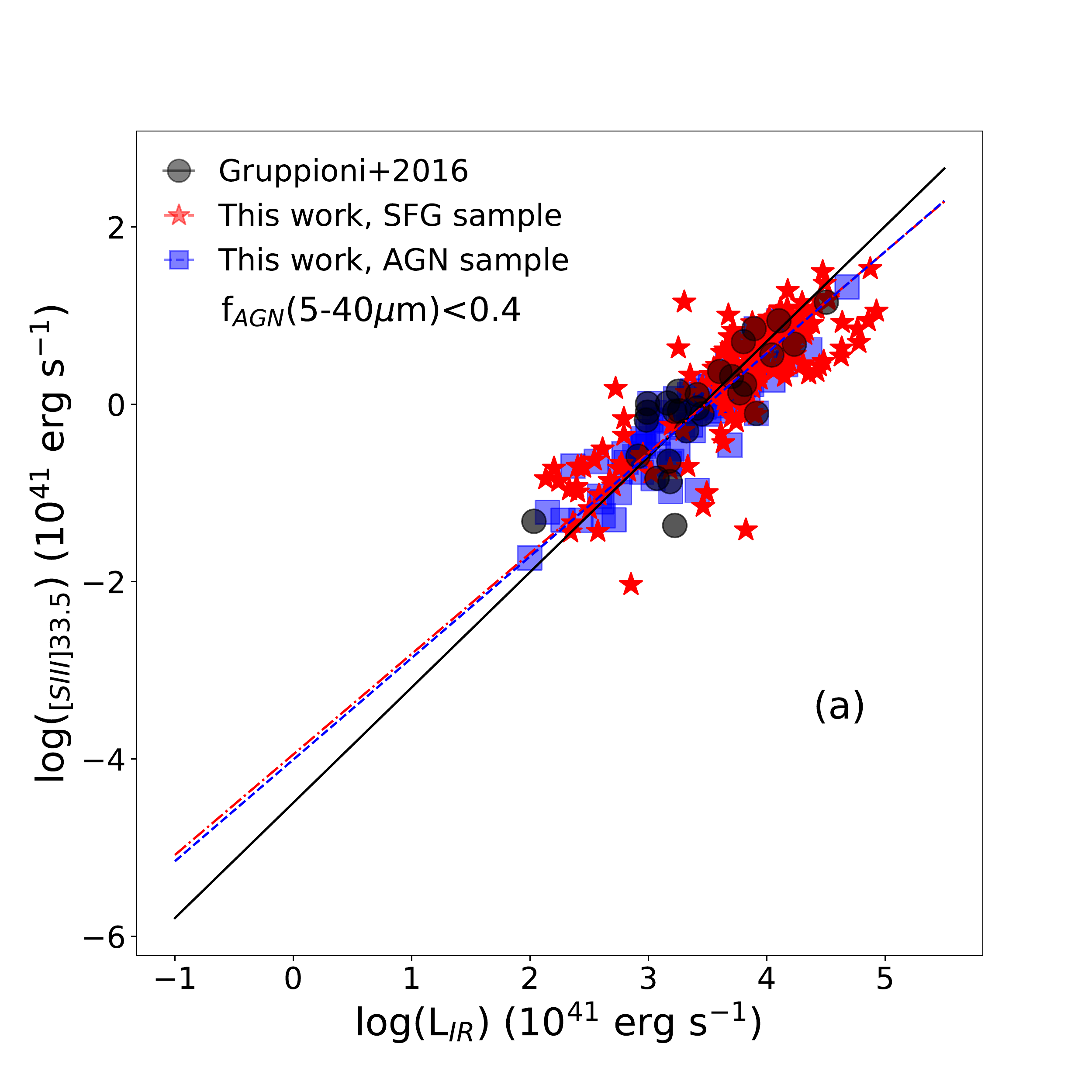}
\includegraphics[width=0.33\columnwidth]{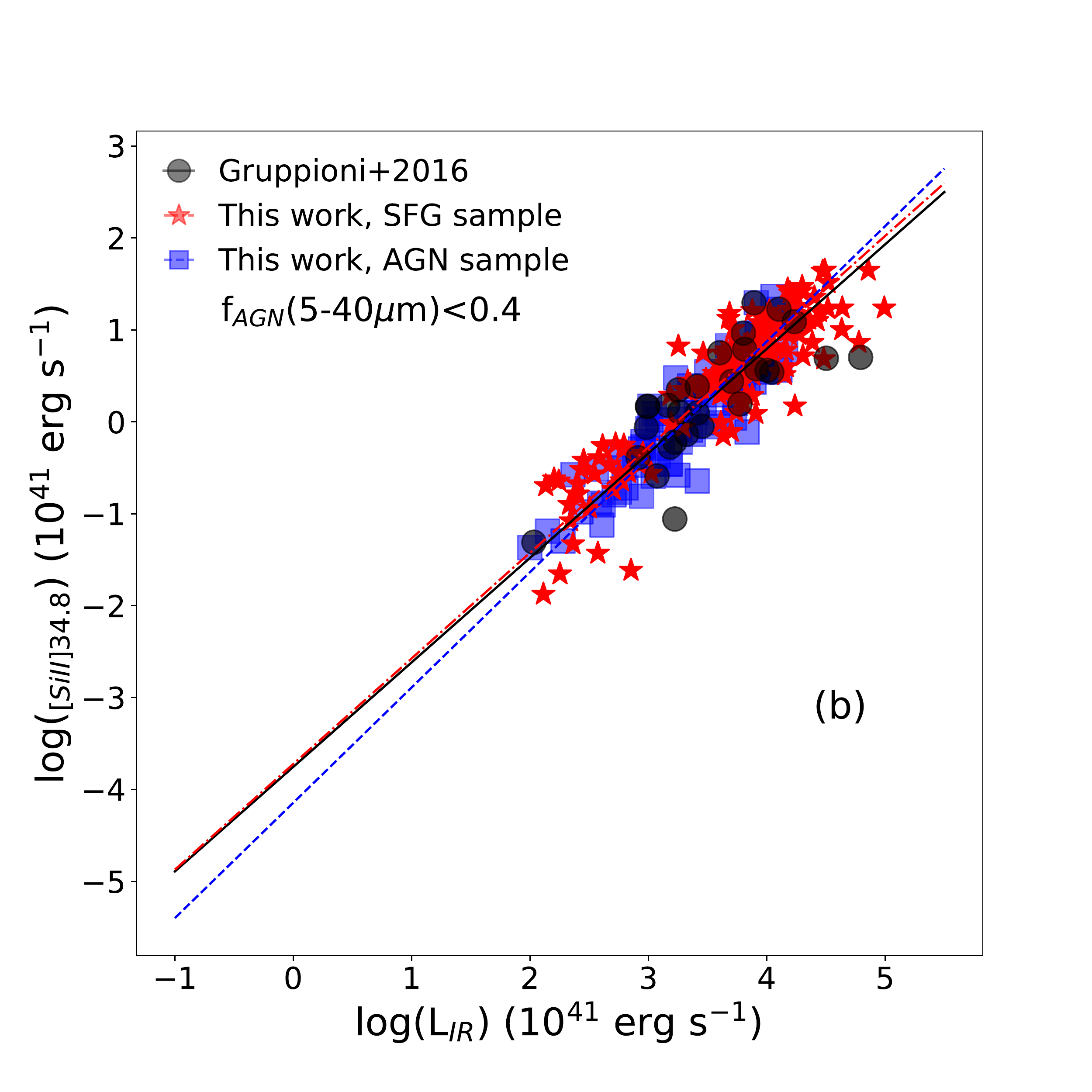}
\caption{{\bf (a: left)} The [SIII]33.5$\mu$m line luminosity versus the total IR luminosity. {\bf (b: right)} The [SiII]34.9$\mu$m line luminosity versus  the total IR luminosity.  In both figures, the \citeauthor{gruppioni2016} sample was limited to AGN fractions $<$ 0.4 in the 5-40$\mu$m spectral interval (see Sect.\ref{sec:comp}). The same legend as in Fig.\,\ref{fig:comp_1} was used.}
\label{fig:comp_8}
\end{figure*}

Figs. \ref{fig:comp_1}-\ref{fig:comp_4} show the comparison for the relations calculated for the $f_{AGN}$(5-40$\, \rm{\micron}$)$>0.4$ sub-sample, and Figs. \ref{fig:comp_5}-\ref{fig:comp_8} show the comparison for the relations calculated for the $f_{AGN}$(5-40$\, \rm{\micron}$)$\leq 0.4$ sub-sample. 

As a general trend, our sample of AGN  and the sample by \citeauthor{gruppioni2016} occupy the same region in the L$_{IR}$-L$_{line}$ space. We note here that, while the data to calculate the correlations were available in the literature, \citeauthor{gruppioni2016} only present the correlations for [NeV]14.3,  24.3$\mu$m, [NeIII]15.6$\mu$m and [OIV]25.9$\mu$m for the $f_{AGN}$(5-40$\, \rm{\micron}$)$>0.4$ sub-sample. For these relations, we find that our results show steeper slopes, but consistent within 3$\sigma$ of each other.

For the $f_{AGN}$(5-40$\, \rm{\micron}$)$\leq 0.4$ sub-sample, there is a better agreement between the relations, which are comparable within the errors. A significant difference in slope is present for the [NeV]24.3$\mu$m line, for which the relations are comparable within 3$\sigma$, and for the [SIV]10.5$\mu$m line, comparable within 2$\sigma$.

If we compare the results obtained using the \citeauthor{gruppioni2016} sample with the results we obtain using the SFG sample, we find comparable results only for the PAH features for the $f_{AGN}$(5-40$\, \rm{\micron}$)$>0.4$, and for all lines in the $f_{AGN}$(5-40$\, \rm{\micron}$)$\leq 0.4$ sub-sample, excluding the [SIV]10.5$\mu$m and [OIV]25.9$\mu$m lines.

\end{appendix}

\end{document}